\documentclass[12pt]{article}
\usepackage{graphicx,fancyhdr}
\newcommand{\smfrac}[2]{\mbox{$\frac{#1}{#2}$}}
\begin{document}
\def\K{\mathord{\cal K}}
\def\la{\langle}
\def\ra{\rangle}
\def\ltsim{\mathop{\,<\kern-1.05em\lower1.ex\hbox{$\sim$}\,}}
\def\gtsim{\mathop{\,>\kern-1.05em\lower1.ex\hbox{$\sim$}\,}}
\newpage
\begin{center}
\subsection*{MANY-BODY GREEN'S FUNCTION THEORY OF HEISENBERG FILMS \\}
\vspace{0.5cm}

P. Fr\"obrich$^1$,  P.J. Kuntz
\vspace{0.5cm}

Hahn-Meitner-Institut Berlin, Glienicker Stra{\ss}e 100, D-14109 Berlin,
Germany,\\
$^1$also: Institut f\"ur Theoretische Physik, Freie Universit\"at Berlin\\
Arnimallee 14, D-14195 Berlin, Germany\\
\vspace{2cm}
\end{center}
\vspace{2cm}
\subsubsection*{Abstract:}
The treatment of Heisenberg films with many-body Green's function theory (GFT)
is reviewed. The basic equations of GFT are derived in sufficient detail so
that the rest of the paper can be understood without having to consult
further literature.
The main part of the paper is concerned with applications of the formalism to
ferromagnetic, antiferromagnetic and coupled ferromagnetic-antiferromagnetic
Heisenberg films based on a
generalized Tyablikov
(RPA) decoupling of the exchange interaction and exchange anisotropy terms and
an Anderson-Callen decoupling for a weak single-ion anisotropy.
We not only give a consistent description of our own work but also refer
extensively to related investigations.
We discuss in particular the reorientation of the magnetization as a function
of
the temperature and film thickness. If the single-ion anisotropy is strong, it
can be treated
exactly by going to higher-order Green's functions. We also discuss the
extension of the theory beyond RPA. Finally the limitations of GFT are
pointed out.

 \newpage

\subsubsection*{Contents}

\noindent 1. Introduction and outline

\noindent 2. The Heisenberg exchange interaction

2.1. Direct exchange with orthogonal basis states (ferromagnetism)

2.2. Direct exchange with non-orthogonal basis states (antiferromagnetism)

\noindent 3. Basic equations of the
Green's function formalism

3.1. Definition of the double-time Green's function

3.2. The equations of motion

3.3. The eigenvector method and the standard spectral theorem

3.4. The proof of the standard spectral theorem

3.5. The singular value decomposition of ${\bf \Gamma}$ and its consequences

3.6. No advantage to using the anti-commutator instead of the commutator GF

3.7. The intrinsic energy, the specific heat and the free energy

 \noindent 4. The GF formalism for Heisenberg films

4.1. The ferromagnetic Heisenberg monolayer in a magnetic field

\hspace{1cm}4.1.1. The Tyablikov (RPA)-decoupling

\hspace{1cm}4.1.2. The Callen decoupling

\hspace{1cm}4.1.3. Mean-field theory (MFT)

\hspace{1cm}4.1.4. The Mermin-Wagner theorem

\hspace{1cm}4.1.5. Comparing with Quantum Monte Carlo calculations

\hspace{1cm}4.1.6. The effective (temperature-dependent) single-ion lattice
anisotropy

4.2. Ferromagnetic Heisenberg films with anisotropies for general spin S

\hspace{1cm}4.2.1. The Hamiltonian and the decoupling procedures

\hspace{1cm}4.2.2. Approximate treatment of the single-ion anisotropy

\hspace{1cm}4.2.3. Treating the exchange anisotropy

\hspace{1cm}4.2.4. Susceptibilities

\hspace{1cm}4.2.5. Exact treatment of the single-ion anisotropy

\hspace{1cm}4.2.6. The importance of spin waves in the Co-Cu-Ni trilayer

\hspace{1cm}4.2.7. The temperature dependence of the interlayer coupling

4.3. Antiferromagnetic and coupled ferromagnetic-antiferromagnetic Heisenberg

\hspace{1cm} films

\hspace{1cm}4.3.1. The antiferromagnetic spin $S=1/2$ Heisenberg monolayer

\hspace{1cm}4.3.2. A unified formulation for FM, AFM,
and AF-AFM multilayers

4.4. Working in the rotated frame

\hspace{1cm}4.4.1. The ferromagnetic film with an exact treatment of the
single-ion

\hspace{0,9cm}anisotropy

\hspace{1cm}4.4.2. Results of calculations in the rotated frame

\hspace{1cm}4.4.3. Discussion

\noindent 5. Beyond RPA

5.1. Field-induced reorientation of the magnetization of a Heisenberg
monolayer

\hspace{1cm}5.2. Limiting cases

\hspace{1cm}5.2.1. Ferromagnet in a magnetic field, no anisotropy

\hspace{1cm}5.2.2. Ferromagnet with no magnetic field and no exchange
anisotropy

\hspace{1cm}5.2.3. Ferromagnet with exchange anisotropy but no magnetic field.

5.3. The Tserkovnikov formulation of the GF theory

\hspace{1cm}5.3.1. The general formalism

\hspace{1cm}5.3.2. The Heisenberg monolayer in an external field.

\noindent 6. Conclusions

\noindent 7. Appendices

7.1. Appendix A: Calculating the intrinsic energy within GFT

7.2. Appendix B: The curve-following method

7.3. Appendix C: Reducing a 2-dimensional to a 1-dimensional integral for a

\hspace{3.2cm}square lattice

7.4. Appendix D: Treatment of the magnetic dipole-dipole interaction
\newpage
\subsection*{1. Introduction and outline}
Many-body GF theory is used in many fields in statistical mechanics ( e.g.
see the early reviews  \cite{Zub60} or  \cite{BoTy62}, or the
more
recent book \cite{GHE01}). Extensive applications of the formalism to the
theory of magnetism can be found in the books \cite{Tya67} and \cite{Nol86}.
In the sixties and seventies of the last century, emphasis was put on the
properties of bulk magnets. Since then, the advance in experimental
techniques has stimulated an increasing interest in magnetic systems
with reduced
dimension.

One main stream in current research is the attempt to describe 3D
magnetic
systems with strong electron-electron correlations in terms of electronic
structures with the help of ab initio
calculations: Density Functional
Theory (DFT),which is successful by itself for systems with weak
electron-electron correlations, must here be combined with
many-body techniques, as in Dynamical
Mean Field Theory (DMFT). For a recent review see e.g. Ref. \cite{Held05}.

The present paper is concerned with less ambitious models based on the
Heisenberg Hamiltonian with the inclusion of anisotropies. It provides
an overview
of many-body Green's function (GF) techniques applied to the magnetic
properties of layered two-dimensional
structures; i.e. it is concerned essentially with quasi-two-dimensional
Heisenberg films. The techniques developed in the present paper may also be
useful for treating cases in which a Heisenberg kind of Hamiltonian is derived
from a microscopic theory. Emphasis is put on the development of the formalism.
For a paper that discusses the relevant experimental situation in more detail
we refer to Ref. \cite{JB06}.

In Section 2, we derive the direct Heisenberg exchange interaction by
plausibility arguments: an orthogonal basis leads to
ferromagnetic
exchange and a non-orthogonal basis in the Heitler-London
framework allows antiferromagnetic exchange. In Section 3, we derive the
basic equations of the formalism for the double-time Green's function
in sufficient detail that it should not be necessary
to consult any further literature to understand the rest of the paper.
Section 4 deals with applications of the GF formalism to Heisenberg films.
In the pedagogical Section 4.1 a ferromagnetic spin $S=1/2$ Heisenberg
monolayer
in a magnetic field is treated. Section 4.2 deals with
ferromagnetic Heisenberg films with anisotropies for general spin and Section
4.3 considers
antiferromagnetic and coupled ferromagnetic-antiferromagnetic Heisenberg films.
Section 5 extends the formalism beyond the RPA approach. Section 6
presents our conclusions and points out some open problems and limitations
of Green's function theory (GFT).

 \subsection*{2. The Heisenberg exchange interaction}

The present article describes magnetic systems in terms of a
Heisenberg Hamiltonian and anisotropy terms. This is a phenomenological
approach, in which the strengths of the exchange interaction and
anisotropies are considered as
parameters which could be fitted to experiments. In this section, we
discuss the origin of the Heisenberg exchange Hamiltonian.

The exchange interaction is a manifestation of the Coulomb interaction and
quantum-mechanical indistinguishability (the Pauli principle). It is quite
complicated
to derive exchange Hamiltonians from first principles, but this  is often
possible by adopting adequate approximations. The form
of the Heisenberg exchange Hamiltonian is a gross simplification that can,
however, be made plausible for simple cases.
The {\em ferromagnetic} direct exchange can be derived from a two-electron
model by assuming
orthogonal basis states. The direct {\em antiferromagnetic} exchange can be
made plausible with  the Heitler-London scheme.
\subsubsection*{2.1. Direct exchange with orthogonal basis states
(ferromagnetism)}
Consider two electrons (e.g. in a $3d^2$ configuration) and a Hamiltonian
consisting of the sum of two single-electron Hamiltonians, $h_0({\bf r}_1)$ and
$h_0({\bf r}_2)$ and the Coulomb interaction:
\begin{equation}
{\cal H}=h_0({\bf r}_1)+h_0({\bf r}_2)+\frac{e^2}{|{\bf r}_1-{\bf r}_2|},
\label{2.1}
\end{equation}
where the single-electron problem is assumed to be solved:
\begin{eqnarray}
h_0({\bf r})\phi_a({\bf r}) &=&\epsilon_a \phi_a({\bf r}), \nonumber\\
\la \phi_a({\bf r})|\phi_b({\bf r})\ra&=&\delta_{ab}\ .
\label{2.2}
\end{eqnarray}
The electrons can couple to triplet ($S=1$) or singlet ($S=0$) states
with wave functions characterized by $|S,S^z\ra$:
\begin{eqnarray}
|1,1\ra=\psi_1\ , \ \ \ \ \ \
|1,0\ra&=&\frac{1}{\sqrt{2}}(\psi_2+\psi_3)\ ,  \ \ \ \ \
|1,-1\ra=\psi_4\ , \nonumber\\
|0,0\ra&=&\frac{1}{\sqrt{2}}(\psi_2 -\psi_3) \ ,
\label{2.3}
\end{eqnarray}
and
\begin{eqnarray}
\psi_1&=&\frac{1}{\sqrt{2}}\chi_\uparrow(s_1)\chi_\uparrow(s_2)
[\phi_a({\bf r}_1)\phi_b({\bf r}_2)-\phi_a({\bf r}_2)
\phi_b({\bf r}_1)]\ ,\nonumber\\
\psi_2&=&\frac{1}{\sqrt{2}}[\chi_\downarrow(s_1)\chi_\uparrow(s_2)
\phi_a({\bf r}_1)\phi_b({\bf r}_2)
-\chi_\uparrow(s_1)\chi_\downarrow(s_2) \phi_b({\bf r}_1)
\phi_a({\bf r}_2)]\ ,\nonumber\\
\psi_3&=&\frac{1}{\sqrt{2}}[\chi_\uparrow(s_1)\chi_\downarrow(s_2)
\phi_a({\bf r}_1)\phi_b({\bf r}_2)
-\chi_\downarrow(s_1)\chi_\uparrow(s_2) \phi_b({\bf r}_1)
\phi_a({\bf r}_2)]\ ,\nonumber\\
\psi_4&=&\frac{1}{\sqrt{2}}\chi_\downarrow(s_1)\chi_\downarrow(s_2)
[\phi_a({\bf r}_1)\phi_b({\bf r}_2)-\phi_a({\bf r}_2)
\phi_b({\bf r}_1)]\ .
\label{2.4}
\end{eqnarray}
Here $\chi_{\uparrow(\downarrow)}$ are the spin wave functions with spins up or
down.

Defining a Coulomb integal as
\begin{equation}
C_{ab}=e^2\int d{\bf r}_1\int d{\bf r}_2
\frac{|\phi_a({\bf r}_1)|^2|\phi_b({\bf r}_2)|^2}{|{\bf r}_1-{\bf r}_2|}\ ,
\label{2.6}
\end{equation}
and an exchange integral as
\begin{equation}
J_{ab}=e^2\int d{\bf r}_1\int d{\bf r}_2
\frac{\phi_a^*({\bf r}_1)\phi_b({\bf r}_1)
\phi_b^*({\bf r}_2)\phi_a({\bf r}_2)}{|{\bf r}_1-{\bf r}_2|}\ ,
\label{2.7}
\end{equation}

we may express the matrix elements of the Coulomb interaction as
\begin{eqnarray}
\la \psi_1|\frac{e^2}{|{\bf r}_1-{\bf r}_2|}|\psi_1\ra&=&C_{ab}-J_{ab}\
,\nonumber\\ \la \psi_2|\frac{e^2}{|{\bf r}_1-{\bf r}_2|}|\psi_2\ra&=&C_{ab}\
,\\ \la \psi_2|\frac{e^2}{|{\bf r}_1-{\bf r}_2|}|\psi_3\ra&=&-J_{ab}\ .
\nonumber
\label{2.5}
\end{eqnarray}
We then find for the Hamiltonian matrix
\begin{equation}
(\epsilon_a+\epsilon_b)\left( \begin{array}{llll}
1 & 0 & 0 & 0\\
0 & 1 & 0 & 0\\
0 & 0 & 1 & 0\\
0 & 0 & 0 & 1
\end{array}\right)
+
\left( \begin{array}{llll}
C_{ab}-J_{ab}&0&0&0\\
0&C_{ab}&-J_{ab}&0\\
0&-J_{ab}&C_{ab}&0\\
0&0&0&C_{ab}-J_{ab}
\end{array}\right)\ .
\label{2.8}
\end{equation}
There are three degenerate eigenvalues belonging to a
triplet state,
\begin{equation}
\epsilon_t=\epsilon_a+\epsilon_b+C_{ab}-J_{ab},
\label{2.9}
\end{equation}
and one eigenvalue belonging to a singlet state,
\begin{equation}
\epsilon_s=\epsilon_a+\epsilon_b+C_{ab}+J_{ab}\ .
\label{2.10}
\end{equation}
The exchange integral can be shown to be positive:
if we take $f({\bf r})=\phi_a({\bf r})\phi_b({\bf r})$ and perform
Fourier transforms, we have
\begin{eqnarray}
J_{ab}&=&\int d{\bf r}_1f^*({\bf r}_1)\int d{\bf r}_2 \frac{e^2}{|{\bf
r}_1-{\bf r}_2|}f({\bf r}_2)\nonumber\\
&=&\int d{\bf r}_1\frac{1}{(2\pi)^{3/2}}\int d{\bf k}'e^{-i{\bf k}'{\bf
r}_1}f^*({\bf k}') \nonumber\\
& &\int d{\bf r}_2\frac{4\pi e^2}{(2\pi)^3}
\int d{\bf k''}e^{i{\bf k}''({\bf r}_1-{\bf r}_2)}\frac{1}{k''^2}
\frac{1}{(2\pi)^{3/2}}\int d{\bf k}e^{i{\bf k}
{\bf r}_2}f({\bf k})\nonumber\\
&=&\int d{\bf k}|f({\bf k})|^2\frac{4\pi e^2}{k^2}\geq 0\ .
\label{2.11}
\end{eqnarray}
Because $J_{ab}$ is greater than zero, the triplet is lower in energy than the
singlet; i.e. in the lowest state the spins are parallel, which corresponds to
a ferromagnetic situation.

The action of the Hamiltonian can be expressed by spin operators.
We have
\begin{equation}
2{\bf S}_1{\bf S}_2+1/2=({\bf S}_1+{\bf S}_2)^2-1=
\left\{\begin{array}{ll}1 & \mbox{for the triplet}\\
-1 & \mbox{for the singlet}
\end{array}\right. \ .
\label{2.12}
\end{equation}
The action of the triplet and singlet can then be expressed by a single
Hamiltonian
\begin{equation}
{\cal H}=\frac{\epsilon_s+\epsilon_t}{2}-\frac{\epsilon_s-\epsilon_t}{2}
(2{\bf S}_1{\bf S}_2+\frac{1}{2})={\rm const}-2J_{ab}{\bf S}_1{\bf S}_2.
\label{2.13}
\end{equation}
Generalizing  the exchange interaction to a lattice, one may write
\begin{equation}
{\cal H}=-\frac{1}{2}\sum_{i\neq j}J_{ij}{\bf S}_i{\bf S}_j \ .
\label{2.14}
\end{equation}
This is the most familiar form of the Heisenberg exchange, where
$i$ and $j$ represent lattice site indices and the factor $\frac{1}{2}$
is introduced by convention.

\subsubsection*{2.2. Direct exchange with non-orthogonal states
(antiferromagnetism)}
Antiparallel spin alignment (antiferromagnetism) occurs in a two-center
system like  a hydrogen molecule in the Heitler-London approximation.
Consider two hydrogen atoms centred at ${\bf R}_a$ and ${\bf R}_b$
respectively, with a Hamiltonian in which each electron feels both protons and
the electron-electron and proton-proton interactions are included
\begin{eqnarray}
{\cal H}&=&H_{atom}({\bf r}_1-{\bf R}_a)+H_{atom}({\bf r}_2-{\bf
R}_b)\nonumber\\
& &-\frac{e^2}{|{\bf r}_1-{\bf R}_b|}
-\frac{e^2}{|{\bf r}_2-{\bf R}_a|}
+\frac{e^2}{|{\bf r}_1-{\bf r}_2|}
+\frac{e^2}{|{\bf R}_a-{\bf R}_b|}\ .
\label{2.15}
\end{eqnarray}
Each electron occupies a separate 1s-orbital centred on one of the atoms.
In the simplest approximation, the low-lying states are  assumed to be
described by four configurations
with spins $\uparrow\uparrow, \uparrow\downarrow,
\downarrow\uparrow,\downarrow\downarrow$.
We denote the orbital basis functions at atoms {\em a} and {\em b} as
$\phi_a({\bf r})$ and $\phi_b({\bf r})$. They are in general non-orthogonal:
\begin{equation}
\int \phi^*_a({\bf r})\phi_b({\bf r})d{\bf r}=l\neq 0 \ .
\label{2.16}
\end{equation}
Each orbital function is associated with one of two spin functions,
$\chi_{\uparrow}$ or $\chi_{\downarrow}$.
The Hamiltonian is diagonalized in the subspace of the following four
normalized spin-coupled functions
corresponding to triplet and singlet states $\psi(S,S^z)$:
\begin{eqnarray}
\psi(1,1)&=&\frac{1}{\sqrt{2(1-l^2)}}\chi_{\uparrow}(s_1)\chi_{\uparrow}(s_2)
[\phi_a({\bf r}_1)\phi_b({\bf r}_2)-\phi_a({\bf r}_2)\phi_b({\bf r}_1)]\ ,
\nonumber\\
\psi(1,0)&=&\frac{1}{\sqrt{2(1-l^2)}}[\chi_{\uparrow}(s_1)\chi_{\downarrow}(s_2
) + \chi_{\downarrow}(s_1)\chi_{\uparrow}(s_2)]
[\phi_a({\bf r}_1)\phi_b({\bf r}_2)-\phi_a({\bf r}_2)\phi_b({\bf r}_1)]\ ,
\nonumber\\
\psi(1,-1)&=&\frac{1}{\sqrt{2(1-l^2)}}\chi_{\downarrow}(s_1)\chi_{\downarrow}(s
_ 2 ) [\phi_a({\bf r}_1)\phi_b({\bf r}_2)-\phi_a({\bf r}_2)\phi_b({\bf r}_1)]\
, \nonumber\\
\psi(0,0)&=&\frac{1}{2\sqrt{(1+l^2)}}[\chi_{\uparrow}(s_1)\chi_{\downarrow}(s_2
) -\chi_{\downarrow}(s_1)\chi_{\uparrow}(s_2)]
[\phi_a({\bf r}_1)\phi_b({\bf r}_2)+\phi_a({\bf r}_2)\phi_b({\bf r}_1)]\ .
\nonumber\\
\label{2.17}
\end{eqnarray}
The triplet energy may again be written in terms of Coulomb and exchange
integrals:
\begin{equation}
\epsilon_t=\la \psi(1,1)|{\cal
H}|\psi(1,1)\ra=2\epsilon_{atom}+\frac{C_{ab}-I_{ab}}{1-l^2} \ .
\label{2.18}
\end{equation}
Here, $2\epsilon_{atom}$ comes from the one-electron part of the Hamiltonian.
The Coulomb integral (containing
terms where the electron
belonging to
one nucleus feels the attraction of the other nucleus) is
\begin{eqnarray}
C_{ab}&=&\int d{\bf r}_1\int d{\bf r}_2 |\phi_a({\bf
r}_1)|^2
\frac{e^2}{|{\bf r}_1-{\bf r}_2|}|\phi_b({\bf r}_2)|^2\nonumber\\
& &-\int d{\bf r}_1
\frac{e^2}{|{\bf r}_1-{\bf R}_b|}|\phi_a({\bf r}_1)|^2
-\int d{\bf r}_2
\frac{e^2}{|{\bf r}_2-{\bf R}_a|}|\phi_b({\bf r}_2)|^2 \ ,
\label{2.19}
\end{eqnarray}
and  the exchange integral is
\begin{eqnarray}
I_{ab}&=&\int d{\bf r}_1\int d{\bf r}_2 \phi_a^*({\bf r}_1)\phi_b({\bf r}_1)
\frac{e^2}{|{\bf r}_1-{\bf r}_2|}\phi_b^*({\bf r}_2)\phi_a({\bf r}_2)
\nonumber\\
 & &-l\int d{\bf r}_1
\frac{e^2}{|{\bf r}_1-{\bf R}_b|}\phi_a^*({\bf r}_1)\phi_b({\bf r}_1)
-l\int d{\bf r}_2
\frac{e^2}{|{\bf r}_2-{\bf R}_a|}\phi_b^*({\bf r}_2)\phi_a({\bf r}_2) \ .
\label{2.20}
\end{eqnarray}
The singlet eigenenergy is
\begin{equation}
\epsilon_s=\la\psi(0,0)|{\cal
H}|\psi(0,0)\ra=2\epsilon_{atom}+\frac{C_{ab}+I_{ab}}{1+l^2} \ .
\label{2.21}
\end{equation}
The singlet-triplet splitting is
\begin{equation}
\epsilon_t-\epsilon_s=2\frac{l^2C_{ab}-I_{ab}}{1-l^4}\ .
\label{2.22}
\end{equation}
As in the previous subsection, an
effective Hamiltonian may be defined as
\begin{equation}
{\cal H}=const+J_{12}{\bf S}_1{\bf S}_2\ ,
\label{2.23}
\end{equation}
with
\begin{equation}
J_{12}=2\frac{l^2C_{ab}-I_{ab}}{1-l^4}.
\label{2.24}
\end{equation}
Without the Coulomb term, one has ferromagnetic coupling; with a sufficiently
large overlap, the effective exchange coupling becomes
antiferromagnetic ( $J_{12}>0$); the
ground-state of the hydrogen molecule is a singlet state. Generalizing to
a many-electron system, one has the Heisenberg model for an antiferromagnetic
lattice.

Including ionic configurations where both electrons can sit on one or
the other atom leads to a hopping mechanism for
the electrons (kinetic exchange), which supports antiferromagnetic coupling,
see e.g.(\cite{Faz99}, p.60).

An antiferromagnetic Heisenberg model is also obtained from the Hubbard model
in the strong coupling limit, or from indirect exchange mechanisms like
the
RKKY scheme, leading to an effective Heisenberg model in
second-order
perturbation theory. Also, super-exchange or double exchange lead to Heisenberg
like terms or even to biquadratic terms, see e.g. (\cite{Nol86}).

In the present article, we do not try to give a better justification of the
Heisenberg model. Rather, we consider it  a
phenomenological
model that proves to be successful in describing many experimental data when
its parameters are fitted.
A Heisenberg model is adequate when the spins are localized (e.g. in the
rare earth elements). It should also be applicable to 3d-transition metal band
magnets because the magnetic moments are quasi-localized when integrating over
microscopically calculated spin densities. One also sees in experiments on bulk
transition-metal ferromagnets that the magnetization follows  the Bloch $T^{3/2}$
law at low temperatures. Above the Curie temperature, one
observes a Curie-Weiss behaviour of the magnetic susceptibility. Both features
follow from a Heisenberg type model.

\newpage
\subsection*{3. Basic equations of the Green's function
formalism}
In this section we place together the essential definitions and derivations of
the Green's function formalism which are necessary to understand the following
article without frequent recourse to the
literature. For further details of the basic features of the Green's function
formalism
as it is used in the present review, we recommend the article \cite{Zub60} and
the books \cite{GHE01} and \cite{Nol86}.

The double-time Green's functions (GF's), as they are exclusively used in the
present article, are defined in Section 3.1 and their equations of motion
 are given in Section 3.2. In Section 3.3 we discuss the eigenvector
method for determining the GF's.  Once the GF's are known,
the corresponding correlation functions (thermodynamic expectation values) are
determined by the standard spectral theorem, where, in general,
commutator and, in the case of zero eigenvalues of the equation-of-motion
matrix,  anti-commutator GF's have to be used. A proof of the standard
spectral theorem is given
in Section 3.4. In Section 3.5 we discuss the singular value decomposition of
the equation-of-motion matrix and show how a transformation can be found
to eliminate the null-space, obviating
the need for the anti-commutator GF.  This procedure is necessary whenever the
quantities associated with the null-space are momentum-dependent because the
standard
spectral theorem fails in this case. In Section 3.6 we show that there is no
advantage in starting the calculations with the anti-commutator GF instead of
the commutator GF.
In Section 3.7 we show how the intrinsic energy, the specific heat and the free
energy can be calculated with Green's function theory (GFT).

\subsubsection*{3.1. Definition of the double-time Green's function}
Because we deal later with multi-dimensional problems, we prefer to work with a
vector of Green's functions having components characterised by
the index $\alpha$:
\begin{equation}
G_{ij,\eta}^{\alpha}(t-t')=\la\la
A_i^{\alpha}(t);B_j(t')\ra\ra_\eta=-i\Theta(t-t')\la[A_i^{\alpha}(t),
B_j(t')]_{\eta}\ra\ . \label{3.1}
\end{equation}
Throughout the paper we deal exclusively with such double-time GF's.
In principle, either the commutator ($\eta=-1$) or anticommutator ($\eta=+1$)
of the Heisenberg operators $A_i(t)^{\alpha}$ and $B_j(t')$
can be used (but see Section 3.6); $i$ and $j$
are lattice site indices.

The operators obey the Heisenberg equation of motion, e.g.
\begin{equation}
A_i^{\alpha}(t)=e^{iH t}A_i^{\alpha}e^{-iHt}\ ,\ \ \
\dot{A}_i^{\alpha}=-i[A_i^{\alpha}, H]_{-1} \ .
\label{3.2}
\end{equation}
Here $H$ is the Hamiltonian under consideration. We set $\hbar=1$
throughout the paper.

For magnetic films, the $A_i^\alpha$ are spin operators
obeying the usual commutator rules. One has, for instance  (see Section 4.2.1),
\begin{eqnarray}
A_i^\alpha&=&(S_i^{+}, S_i^-,S_i^z), \\
B_j&=&(S_j^z)^m(S_j^-)^n\ \ {\rm with}\ m+n\leq 2S+1\ (m\geq 0,n\geq 1,
{\rm integer}).\nonumber
\label{3.3}
\end{eqnarray}
In eqn (\ref{3.1}), $i$ is the imaginary unit (when it is not an index) and
the step function $\Theta(t-t')$ is defined as
\begin{equation}
\Theta(t-t')= \left\{ \begin{array}{ll}
1 & \mbox{for $t>t'$}\\
0 & \mbox{for $t<t'$}\ .
\end{array}
\right.
\label{3.4}
\end{equation}
The double brackets $\la\la A_i(t)^{\alpha};B_j(t')\ra\ra_\eta$ are an
alternative notation for the Green's functions $G_{ij,\eta}^{\alpha}(t-t')$.
Single brackets denote  correlation functions, e.g.
$\la [A_i^{\alpha}(t),B_j(t')]_\eta\ra$, which are thermodynamic expectation
values %
\begin{equation}
\la ... \ra=\frac{1}{Z}\sum_n\la
n|e^{-\beta H}...|n\ra=\frac{1}{Z}Tr(e^{-\beta H}...)\ ,
\label{3.5}
\end{equation}
where
\begin{equation}
Z=\sum_n\la n|e^{-\beta H}|n\ra=Tr(e^{-\beta H})
\label{3.6}
\end{equation}
is the partition function with $\beta=1/(k_BT)$, $T$ the temperature and
$k_B$ the Boltzmann constant.

Usually, it is more convenient to work with the Fourier transforms of the
Green's functions in energy space,
\begin{eqnarray}
G_{ij}^{\alpha}(\omega)&=&\int_{-\infty}^\infty
d(t-t')G_{ij}^{\alpha}(t-t')e^{i\omega(t-t')},\nonumber\\
G_{ij}^{\alpha}(t-t')&=&\int_{-\infty}^\infty
\frac{d\omega}{2\pi}G_{ij}^{\alpha}(\omega)e^{-i\omega(t-t')},
\label{3.7}
\end{eqnarray}
and in momentum space,
\begin{eqnarray}
G_{\bf k}^{\alpha}(\omega)&=&\frac{1}{N}\sum_{ij}G_{ij}^{\alpha}(\omega)
e^{i{\bf k}({\bf R}_i-{\bf R}_j)},\nonumber\\
G_{ij}^{\alpha}(\omega)&=&\frac{1}{N}\sum_{\bf k}G_{\bf k}^{\alpha}(\omega)
e^{-i{\bf k}({\bf R}_i-{\bf R}_j)},\nonumber\\
\rm{with}\ \ \ \ \delta_{ij}&=&\frac{1}{N}\sum_{\bf k}e^{i{\bf k}({\bf
R}_i-{\bf R}_j)},\nonumber\\
\rm{and}\ \ \ \ \delta_{{\bf k}{\bf k}'}&=&\frac{1}{N}\sum_{i}e^{i({\bf k}-{\bf
k}'){\bf R}_i}\ .
\label{3.8}
\end{eqnarray}
Here the ${\bf R}_{i}$ are the lattice site positions and $N$ is the number
of lattice sites.

\subsubsection*{3.2. The equations of motion}
The Green's function vector has to be determined by its equation of motion.
This is  obtained by taking the time derivative of equation (\ref{3.1}),
\begin{equation}
i\frac{\partial}
{\partial{t}}G_{ij,\eta}^{\alpha}(t-t')=\delta(t-t')
\la[A_i^{\alpha}(t),B_j(t')]_{\eta}\ra +
\la\la[A_i^{\alpha},H]_{-1}(t);B_j(t')\ra\ra_\eta\ ,
\label{3.9}
\end{equation}
where the Heisenberg equation (\ref{3.2}) has been used together with
$\frac{\partial}{\partial{t}}\Theta(t-t')=\delta(t-t')$.
Eqn (\ref{3.9}) is a differential equation for determining the Green's
functions.
Because it is more convenient to work with algebraic equations, one usually
performs a Fourier transform to energy space (\ref{3.7}), characterized by the
index $\omega$:
\begin{equation}
\omega \la\la A_i^{\alpha};B_j\ra\ra_{\eta,\omega}=\la
[A_i^{\alpha},B_j]_\eta\ra+\la\la [A_i^{\alpha},
H]_-;B_j\ra\ra_{\eta,\omega}\ .
\label{3.10}
\end{equation}
Observe that on the right-hand side a higher-order Green's function
arises  which leads to
another equation of motion having
 even higher-order Green's functions
and so
on. In this way, an {\em exact} infinite hierarchy of equations of motion is
generated.
Only in rare cases does this hierarchy terminate automatically.
Usually, one has to terminate the hierarchy somewhere
in order to obtain a solvable closed system of equations:
the Green's function of some specified order must be factored
in such a way
as to contain only Green's functions
which already exist in the hierarchy up to the cut-off. This factorization is
called the decoupling procedure and
is the essential and most severe approximation in  GF theory.
Except for a few cases, it can be justified only by its success.

Very often one works with the lowest-order equation (\ref{3.10}) only.
The decoupling consists in this case in factoring the GF:
\begin{equation}
\la\la[A_i^{\alpha},H]_-;B_j\ra\ra_{\eta,\omega}\simeq
\sum_l\sum_\beta \Gamma_{il}^{\alpha\beta}\la\la
A_l^\beta;B_j\ra\ra_{\eta,\omega}\ .
\label{3.11}
\end{equation}
The right-hand side now has only GF's which are of the same
order as those already present. In this way one  arrives at a closed system
of equations of motion. The matrix
$\Gamma_{il}^{\alpha\beta}$ is in general {\em unsymmetric}.

Inclusion of the second-order equation of motion would require a decoupling
of the double-commutator GF $\la\la [[A_i^\alpha,H],H];B_j\ra\ra$, etc.

For periodic lattice structures, the equations of motion are simplified by
a Fourier transformation to momentum space (\ref{3.8}), which eliminates
the lattice site indices.
The equations of motion in compact
matrix notation are then
\begin{equation}
(\omega {\bf 1}-{\bf \Gamma}){\bf G}_{\eta}={\bf A}_\eta\ ,
\label{3.12}
\end{equation}
where ${\bf A}_\eta$  is the
inhomogeneity vector with components $A_\eta^\alpha=\la [A^\alpha,B]_\eta\ra$,
and ${\bf 1}$ is the unit matrix.

From Kramers rule one sees that the GF's have a pole structure
with
the eigenvalues of the matrix ${\bf \Gamma}$ as poles. In many applications,
use
is made only of these eigenvalues but we show in the next subsection that it
is of great advantage to use the eigenvectors of this matrix as well,
especially in treating multi-dimensional problems.

\subsubsection*{3.3. The eigenvector method and the standard spectral theorem}

In this section, we show how to take advantage of the eigenvectors of the
matrix ${\bf \Gamma}$ in transforming the GF's to a new set of GF's each
having but a single pole. This is particularly important in treating degenerate
eigenvalues of ${\bf \Gamma}$ because each eigenvalue can be associated with a
definite (transformed) GF. Also, the extra cost of finding the eigenvectors is
more than compensated by avoiding the effort of calculating determinants in a
Kramers-like treatment and by the clarity gained in the formulation.
We shall use the notation of
reference \cite{FJKE00}.

The first step is to diagonalize the  matrix
${\bf \Gamma}$
\begin{equation}
{\bf L\Gamma R}={\bf \Omega},
\label{3.13}
\end{equation}
where $\bf \Omega$ is the diagonal matrix of $N$ eigenvalues,
$\omega_\tau\  ({\tau=1,..., N})$, $N_0$ of which are zero and
$(N-N_0)$ are non-zero. The occurence of zero eigenvalues is not a rare case:
they arise as a consequence of the spin algebra of certain of the GF's and are
to be expected.
The matrix ${\bf R}$ contains the right eigenvectors as columns and its inverse
${\bf L}={\bf R}^{-1}$ contains
the left eigenvectors as rows. ${\bf L}$ is
constructed such that ${\bf LR}={\bf 1}$. We assume that the eigenvectors span
the whole space so that it is also true that ${\bf RL=1}$.

We now define new vectors by multiplying the original vectors with ${\bf L}$:
\begin{equation}
{\cal G}_\eta={\bf LG}_\eta\ \ \ {\rm and}\ \ \  {\cal A}_\eta={\bf LA}_\eta\ .
\label{3.14}
\end{equation}
Multiplying equation (\ref{3.12}) from the left by
${\bf L}$ and inserting ${\bf 1}={\bf RL}$
leads to
\begin{equation}
(\omega{\bf 1}-{\bf \Omega}){\cal G}_\eta={\cal A}_\eta.
\label{3.15}
\end{equation}
From this equation we see at once that each of the components
$\tau$ of this Green's function vector has but a single pole (!)
\begin{equation}
{(\cal G}_\eta)_\tau=\frac{({\cal A}_\eta)_\tau}{\omega-\omega_\tau}\ .
\label{3.16}
\end{equation}

This allows a direct application of the standard spectral theorem (see e.g.
\cite{Nol86,GHE01}; for its proof, see Section 3.4)
to each component of the Greens's function vector {\em separately}.
The spectral theorem relates
the correlation vector
\begin{equation}
{\cal C}={\bf LC_{k}}={\bf L}\la BA\ra_{\bf k}
\label{3.17}
\end{equation}
to the Green's function vector. The index ${\bf k}$
indicates that we work in momentum space.
Explicitly,
\begin{equation}
{\cal C}_\tau=\frac{i}{2\pi}\lim_{\delta\rightarrow
0}\int_{-\infty}^{\infty}d\omega \frac{\Big({\cal G}_\eta(\omega+i\delta)-{\cal
G}_\eta(\omega-i\delta)\Big)_\tau}{e^{\beta\omega}+\eta}
=\frac{({\cal A}_\eta)_\tau}{e^{\beta\omega_\tau}+\eta},
\label{3.18}
\end{equation}
where eqn (\ref{3.16}) and $\frac{1}{\omega\pm i\delta}=\frac{P}{\omega}\mp
i\pi\delta(\omega)$ have been used.

In general, we use the commutator GF's, in which case the
inhomogeneities
${\cal A}_{\eta=-1}$ are independent of the momentum ${\bf k}$, whereas the
${\cal A}_{\eta=+1}$ are not. Using the anti-commutator GF's ($\eta=+1$) leads
to problems connected with this ${\bf k}$-dependence (see Section 3.6).
The commutator GF's, on the other hand, lead to problems with zero
eigenvalues of the equation
of motion matrix ${\bf \Gamma}$ because there are then zeroes in the
denominator of eqn (\ref{3.18}). In this case, the
correlation vector must be split into two components ${\cal C}^1_\tau$ and
${\cal C}^0_{\tau_0}$ belonging to non-zero and zero eigenvalues respectively.
We then have
\begin{equation}
({\cal C}^1)_\tau=\frac{({\cal A}_{-1})_\tau}{e^{\beta\omega_\tau}-1},
\label{3.19}
\end{equation}
where $\omega_\tau\neq 0$.

For the correlation vector belonging to zero eigenvalues,
the anti-commutator GF is required (for a proof, see Section 3.4):
\begin{eqnarray}
({\cal C}^0)_{\tau_0}&=&\lim_{\omega\rightarrow 0}\frac{1}{2}\omega
({\cal G}_{\eta=+1})_{\tau_0} \nonumber \\
&=&\frac{1}{2}\lim_{\omega\rightarrow 0}
\frac{\omega({\cal A}_{+1})_{\tau_0}}{\omega-(\omega_{\tau_0}=0)} =
\frac{1}{2}({\cal A}_{+1})_{\tau_0}\nonumber \\
&=&\frac{1}{2}({\bf L}^0({\bf A}_{-1}+2{\bf C_k}))_{\tau_0}
=({\bf L}^0{\bf C_k})_{\tau_0}\ .
\label{3.20}
\end{eqnarray}
Here, the relation ${\bf A}_{+1}={\bf A}_{-1}+2{\bf C_k}$ has been used
together with the
fact that the commutator GF is regular at the origin (see eqn (\ref{3.43})).
By multiplying eqn (\ref{3.12}) with ${\bf L}^0$, using ${\bf L}^0{\bf
\Gamma}=0$ and taking the limit $\omega\rightarrow 0$ one obtains
\begin{equation}
\lim_{\omega\rightarrow 0}{\bf L}^0(\omega {\bf 1}-{\bf \Gamma}){\bf
G}_{-1}={\bf L}^0{\bf A}_{-1}=0\ .
\label{3.21}
\end{equation}
We call this the regularity condition.

We now partition all quantities with respect to the non-zero and zero
eigenvalue space
\begin{equation}
 {\bf R}=({\bf R}^1{\bf R}^0)\ , \ \ \ \ \   {\bf L} = \left( \begin{array}{c}
                         {\bf L}^1  \\ {\bf L}^0
                                           \end{array}      \right)\ , \ \ \ \
\ {\cal C} = \left( \begin{array}{c}
                         {\cal C}^1={\cal E}^1{\bf L}^1{\bf A}_{-1}
\\ {\cal C}^0={\bf L}^0{\bf C}_{\bf k}
                    \end{array}      \right)\ ,
\label{3.22}
\end{equation}
where ${\cal{E}}^1$ is a diagonal $(N-N_0)\times (N-N_0)$ matrix with elements
$1/(e^{\beta\omega_\tau}-1)$ on the diagonal ($\omega_\tau\neq 0$).

The original correlation vector in momentum space is then
\begin{equation}
{\bf C}_{\bf k}={\bf R}{\cal C}=({\bf R}^1{\bf R}^0)\left( \begin{array}{c}
                         {\cal C}^1  \\ {\cal C}^0
                                           \end{array}      \right)
={\bf R}^1{\cal E}^1{\bf
L}^1{\bf A_{-1}}+{\bf R}^0{\bf L}^0{\bf C}_{\bf k}\ .
\label{3.23}
\end{equation}
We are interested in the diagonal correlations ${\bf C}$ (without the
index ${\bf k}$) in configuration space
\begin{equation}
{\bf C}=\frac{1}{N}\sum_{\bf k}{\bf C_k}=\int d{\bf k}{\bf C_k},
\label{3.24}
\end{equation}
where the integration is over the first Brillouin zone.
This leads to a set of integral equations for the components $C_i$ (i=1,...,N)
which have to be solved self-consistently.
If the factor ${\bf R}^0{\bf L}^0$ is momentum
independent, one can take it outside the integration in the second term of eqn
(\ref{3.23}) to get the $C_i$ components  explicitly:
\begin{equation}
C_i=\int d{\bf k}\Big(\sum_{j=1}^{N-N_0}\sum_{l=1}^N R_{ij}^1{\cal
E}_{jj}^1L_{jl}^1(A_{-1})_l\Big)
+\sum_{j=1}^{N_0}\sum_{l=1}^N R_{ij}^0L_{jl}^0C_l.
\label{3.25}
\end{equation}
The components $C_i$ are obtained by iterating on the $C_i$ until Eqn
(\ref{3.25}) is satisfied.

If ${\bf R}^0{\bf L}^0$ is momentum-dependent, the standard
procedure fails because one cannot take
${\bf R^0L^0}$ outside the integration. Instead, one needs a more complicated
procedure that relies on the
singular value decomposition of the ${\bf \Gamma}$-matrix (see Section 3.5).
This leads to a formulation of the spectral theorem in which the null-space is
eliminated and only the commutator GF is needed, obviating the use of
the anti-commutator GF.

{\small
\noindent$\diamond\diamond\ \ \ $

\noindent Equation (\ref{3.23}) can also be derived without
the anticommutator GF in the following simple way:

\noindent Start with the spectral theorem for the commutator GF (\ref{3.18})
with $\eta=-1$
\begin{equation}
{\bf C_k=R{\cal E}LA}_{-1},
\label{3.26a}
\end{equation}
and make use of the decomposition (\ref{3.22}) and
${\cal E}_0=1/(e^{\beta 0}-1)=\infty$. Then
\begin{equation}
{\bf C_k}=({\bf R}^1{\bf R}^0)\left(\begin{array}{cc}
{\cal E}^1 & 0 \\ 0&{\cal E}^0 \end{array}\right)
\left(\begin{array}{c}
{\bf L}^1 \\ {\bf L}^0
\end{array}\right){\bf A}_{-1}={\bf R}^1{\cal E}^1{\bf L}^1{\bf A}_{-1}+
{\bf R}^0{\cal E}^0{\bf L}^0{\bf A}_{-1}.
\label{3.27a}
\end{equation}
The second term is undetermined because
${\cal E}^0{\bf L}^0{\bf A}_{-1}$ has the indeterminate form $\infty\times 0$,
see eqn(\ref{3.21}).
We get around this by multiplying the last
equation from the left by ${\bf R}^0{\bf L}^0$:
\begin{equation}
{\bf R}^0{\bf L}^0{\bf C_{k}} ={\bf R}^0{\bf L}^0 {\bf R}^1{\cal E}^1{\bf
L}^1{\bf A}_{-1}+ {\bf R}^0{\bf L}^0 {\bf R}^0{\cal E}^0{\bf L}^0{\bf A}_{-1}=
{\bf R}^0{\cal E}^0{\bf L}^0{\bf A}_{-1}.
\end{equation}
\label{3.28a}
The last term is obtained because ${\bf L}^0 {\bf R}^1=0$ and
${\bf R}^0{\bf L}^0{\bf R}^0{\bf L}^0={\bf R}^0{\bf L}^0$.
Thus the term ${\bf R}^0{\cal E}^0{\bf L}^0{\bf A}_{-1}$ in eqn (\ref{3.27a})
can be replaced by ${\bf R}^0{\bf L}^0{\bf C_k}$,
which completes the proof of eqn (\ref{3.23}).

\noindent$\diamond\diamond$}

\subsubsection*{3.4. The proof of the standard spectral
theorem}

The spectral theorem is the relation of greatest importance for the Green's
function formalism because it allows the calculation of the desired observables
(or more generally the correlation functions) from the corresponding Green's
functions. Although its proof can be found in text books
(e.g.
\cite{Nol86,GHE01}), we reproduce it here for the convenience of the reader.

Considering one component of the GF vector (\ref{3.1}) (for brevity we leave
out
the index $\alpha$), we introduce the spectral function $S_{ij,\eta}(t-t')$ by
\begin{equation}
G_{ij,\eta}(t-t')=-i\Theta(t-t')2\pi S_{ij,\eta}(t-t'),
\label{3.26}
\end{equation}
where, by comparing with eqn (\ref{3.1}),
\begin{equation}
S_{ij,\eta}(t-t')=\frac{1}{2\pi}\la [A_i(t),B_j(t')]_\eta\ra
=\frac{1}{2\pi}\la A_i(t)B_j(t')+\eta B_j(t')A_i(t)\ra\ .
\label{3.27}
\end{equation}
Inserting a complete set of eigenstates ($H|m\ra=\omega_m|m\ra$) yields
the following spectral representations for the correlations:
\begin{eqnarray}
\la A_i(t)B_j(t')\ra&=&\frac{1}{Z}\sum_{nm}\la n|B_j|m\ra\la m|A_i|n\ra
e^{-\beta\omega_n}e^{\beta(\omega_n-\omega_m)}e^{-i(\omega_n-\omega_m)(t-t')}\
, \label{3.28}\\
\la B_j(t')A_i(t)\ra&=&\frac{1}{Z}\sum_{nm}\la n|B_j|m\ra\la m|A_i|n\ra
e^{-\beta\omega_n}e^{-i(\omega_n-\omega_m)(t-t')}\ ,
\label{3.29}
\end{eqnarray}
and the spectral function,
\begin{equation}
S_{ij,\eta}(t-t')=\frac{1}{2\pi}\frac{1}{Z}\sum_{nm}\la n|B_j|m\ra\la
m|A_i|n\ra e^{-\beta\omega_n}(e^{\beta(\omega_n-\omega_m)}+\eta)
e^{-i(\omega_n-\omega_m)(t -t')},
\label{3.30}
\end{equation}
whose Fourier transform to energy space is
\begin{equation}
S_{ij,\eta}(\omega)=\frac{1}{Z}\sum_{nm}\la n|B_j|m\ra\la m|A_i|n\ra
e^{-\beta\omega_n}(e^{\beta\omega}+\eta)\delta(\omega-(\omega_n-\omega_m)).
\label{3.31}
\end{equation}
A relation between the energy representations of $S_{ij,\eta}(\omega)$ and
$G_{ij,\eta}(\omega)$ is derived by inserting in
\begin{equation}
G_{ij,\eta}(\omega)=-2\pi i\int_{-\infty}^\infty
d(t-t')e^{i\omega(t-t')}\Theta(t-t')S_{ij,\eta}(t-t')
\label{3.32}
\end{equation}
the following representation for the step function
\begin{equation}
\Theta(t-t')=\frac{i}{2\pi}\int_{-\infty}^\infty
dx\frac{e^{-ix(t-t')}}{x+i\eta},
\label{3.33}
\end{equation}
and the Fourier transform of $S_{ij,\eta}(t-t')$:
\begin{eqnarray}
G_{ij,\eta}(\omega)&=&\int_{-\infty}^\infty d\omega'\int_{-\infty}^\infty dx
\frac{1}{x+i\eta}\frac{1}{2\pi}\int_{-\infty}^\infty
d(t-t')e^{i(\omega-\omega'-x)(t-t')}S_{ij,\eta}(\omega')\nonumber\\
&=&\int_{-\infty}^\infty
d\omega'\frac{S_{ij,\eta}(\omega')}{\omega-\omega'+i\eta}\ .
\label{3.34}
\end{eqnarray}
With
\begin{equation}
G_{ij,\eta}(\omega+i\delta)-G_{ij,\eta}(\omega-i\delta)
=\int_{-\infty}^\infty
d\omega'S_{ij,\eta}(\omega')(\frac{1}{\omega-\omega'+i\delta}
-\frac{1}{\omega-\omega'-i\delta})
\label{3.35}
\end{equation}
and
\begin{equation}
\frac{1}{\omega-\omega'\pm i\delta}=P\frac{1}{\omega-\omega'}\mp
i\pi\delta(\omega-\omega')
\label{3.36}
\end{equation}
it follows that
\begin{equation}
S_{ij,\eta}(\omega)=\lim_{\delta\rightarrow 0}\frac{i}{2\pi}
\Big(G_{ij,\eta}(\omega+i\delta)-G_{ij,\eta}(\omega-i\delta)\Big).
\label{3.37}
\end{equation}
We can also see that
\begin{equation}
\la B_j(t')A_i(t)\ra=\int_{-\infty}^\infty
\frac{d\omega}{e^{\beta\omega}+\eta}S_{ij,\eta}(\omega)e^{-i\omega(t-t')}
\label{3.38}
\end{equation}
by inserting equation (\ref{3.31}) in this equation and comparing with
(\ref{3.29}).

Together with equation (\ref{3.37}) this yields
\begin{equation}
\la B_j(t')A_i(t)\ra=\lim_{\delta\rightarrow 0}
\frac{i}{2\pi}\int_{-\infty}^\infty
d\omega \frac{G_{ij,\eta}(\omega+i\delta)
-G_{ij,\eta}(\omega-i\delta)}{e^{\beta\omega}+\eta}e^{-i\omega(t-t')}\ .
\label{3.39}
\end{equation}
This is nothing else then equation (\ref{3.18}) with $t=t'$ after a Fourier
transformation
to momentum space. It is  valid for $\eta=\pm 1$.

For $\eta=-1$, this expression diverges in the limit $\omega\rightarrow 0$ and
it is necessary to use eqn (\ref{3.20}). This was first pointed out
in reference \cite{ST65}, see also \cite{RG71}, and can be seen by
decomposing the spectral function (\ref{3.31}) into two terms
referring to $\omega_n\neq \omega_m$ and $\omega_n=\omega_m$ respectively
\begin{equation}
S_{ij,\eta}(\omega)=\tilde{S}_{ij,\eta}|_{\omega_n\neq
\omega_m}+(1+\eta)C^0_{ij}\delta(\omega).
\label{3.40}
\end{equation}
Inserting this in eqn (\ref{3.34}) and taking the limit $\omega\rightarrow 0$
of $\omega G_{ij,\eta}(\omega)$ one finds
\begin{eqnarray}
\lim_{\omega\rightarrow 0} \omega G_{ij,\eta}(\omega)&=&
\lim_{\omega\rightarrow 0}\omega\int_{-\infty}^\infty d\omega'\Big(
\frac{\tilde{S}_{ij,\eta}}{\omega-\omega'}
+\frac{(1+\eta)C^0_{ij}\delta(\omega')}{\omega-\omega'}\Big)\nonumber\\
&=&0+(1+\eta)C^0_{ij}\ .
\label{3.41}
\end{eqnarray}
From this expression, we see that the quantity $C^0_{ij}$ is determined by the
anti-commutator GF ($\eta=+1$)
\begin{equation}
C^0_{ij}=\frac{1}{2}\lim_{\omega\rightarrow
0}\omega G_{ij,\eta=+1}(\omega)\ ,
\label{3.42}
\end{equation}
whose Fourier transform to momentum space is equation (\ref{3.20}).
This completes the proof of the standard spectral theorem.

From equation (\ref{3.41}) an important analytical property follows: the
commutator Green's function ($\eta=-1$) is regular at the origin,
\begin{equation}
\lim_{\omega\rightarrow 0}\omega G_{ij,\eta=-1}=0,
\label{3.43}
\end{equation}
a fact which is necessary to derive the regularity condition (\ref{3.21}).
The
anti-commutator Green's function has a first order pole at $\omega=0$.

\subsubsection*{3.5. The singular value decomposition of $\Gamma$ and its
consequences}
In this section we show that the singular value
decomposition
of the equation-of-motion matrix ${\bf \Gamma}$ obviates the need to use the
anti-commutator
GF when zero eigenvalues occur; the commutator GF suffices.

The standard spectral theorem is of practical use only if the quantity
${\bf R}^0{\bf L}^0$ in eqn (\ref{3.23}) is momentum independent, because
only then can one
arrive at an equation that can be solved by iteration  (see (\ref{3.25}).
If ${\bf R}^0{\bf L}^0$ depends on momentum, the standard
procedure fails because equation (\ref{3.23}) is of the form
\begin{equation}
(1-{\bf R}^0{\bf L}^0){\bf C_{k}}={\bf R}^1{\cal E}^1{\bf L}^1{\bf A_{-1}}.
\label{3.44}
\end{equation}
The term $(1-{\bf R}^0{\bf L}^0)$ is idempotent
and therefore has no inverse; hence, one cannot solve for ${\bf C_k}$.
This arises for instance
for the reorientation of the
magnetization using exchange anisotropies, see Section 4.2.3.

{\small
\noindent$\diamond\diamond\ \ \ \ $

\noindent An idempotent operator $P$ has no
inverse.

\noindent Proof:
 assume the existence of an inverse:
$P^{-1}P=1$ and  idempotence $P=P^2$,

\noindent then $P^{-1}P^2=1$,
implying $P=1$, which is a contradiction.

\noindent$\diamond\diamond$}

The singular value decomposition (SVD) offers a way out of this situation by
providing a transformation that eliminates the null-space; in effect, it
defines a smaller number of Green's functions whose associated equation of
motion matrix, $\gamma$, has no zero eigenvalues, thus dispensing with the
anti-commutator GF as well as reducing the number of equations.

The singular value decomposition states that ... "any $M\times N$ matrix
${\bf A}$ whose number of rows $M$ is greater or equal to its number of
columns, can be written as the product of an $M\times N$ column-orthogonal
matrix ${\bf U}$, an $N\times N$ diagonal matrix ${\bf W}$ with positive or
zero elements and the transpose of an $N\times N$ orthogonal matrix V"..., \cite{PFTV89}

The equation-of-motion matrix can therefore be decomposed as
\begin{equation}
{\bf \Gamma}={\bf UW\tilde{V}}={\bf uw\tilde{v}}.
\label{3.45}
\end{equation}
where ${\bf U}$ and ${\bf V}$ are orthogonal matrices
(${\bf \tilde{U}U}={\bf 1},  {\bf \tilde{V}V=1}$) and
${\bf W}$ is a diagonal matrix with singular values on the diagonal.
${\bf U, V}$ and ${\bf W}$ can be determined very efficiently numerically
\cite{PFTV89}. The matrices ${\bf U}$ and ${\bf V}$
can also be obtained by diagonalising ${\bf \tilde{\Gamma}\Gamma}$ or
${\bf \Gamma\tilde{\Gamma}}$ respectively. The singular values are the
positive square roots of the eigenvalues of these matrices:
\begin{eqnarray}
{\bf \tilde{V}\tilde{\Gamma}\Gamma V}&=&{\bf
\tilde{V}VW\tilde{U}UW\tilde{V}V}={\bf W}^2,\\
{\bf \tilde{U}\Gamma\tilde{\Gamma} U}&=&{\bf
\tilde{U}UW\tilde{V}VW\tilde{U}U}={\bf W}^2.\nonumber
\label{3.46}
\end{eqnarray}
If  ${\bf \Gamma}$ has zero eigenvalues, it has the same number
of zero singular values.
The matrix ${\bf \Gamma}$ is also given by ${\bf uw\tilde{v}}$, where
${\bf u}$ and ${\bf v}$ are obtained from ${\bf U}$ and ${\bf V}$ by omitting
columns corresponding to  singular values zero.
${\bf u}$ and ${\bf v}$ are again orthogonal matrices $({\bf \tilde{u}u}={\bf
1}, {\bf \tilde{v}v=1})$. Note that
${\bf v\tilde{v}}$ is a projector onto the non-null-space and
${\bf v_0\tilde{v_0}}$ a projector onto the null-space
(${\bf v\tilde{v}+v_0\tilde{v_0}=1}$).
The matrix ${\bf w}$ is diagonal having positive  singular values
on the diagonal.

To eliminate the null-space, it suffices to use  the following
transformations:
\begin{eqnarray}
{\bf \gamma}&=&{\bf \tilde{v}\Gamma v},\nonumber\\
{\bf g}&=&\tilde{\bf v}{\bf G}, \nonumber \\
{\bf a}&=&\tilde{\bf v}{\bf A},\nonumber \\
{\bf c}&=&\tilde{\bf v}{\bf C_{\bf k}}.
\label{3.47}
\end{eqnarray}
Multiplying eqn (\ref{3.12}) by  $\tilde{\bf v}{\bf v}=1$ and
${\mathbf\Gamma}={\bf uw}(\tilde{\bf v}{\bf v})\tilde{\bf v}=
({\bf uw}\tilde{\bf v}){\bf v}\tilde{\bf v}=
{\mathbf \Gamma}{\bf v}\tilde{\bf v}$
one obtains
\begin{eqnarray}
   \tilde{\bf v}(\omega{\bf 1}-{\mathbf \Gamma}
   {\bf v}\tilde{\bf v}){\bf G}
     &=& \tilde{\bf v}{\bf A}_{-1}, \nonumber \\
    (\omega {\bf 1} - \tilde{\bf v}{\mathbf \Gamma}{\bf v})\tilde{\bf v}{\bf G}
     &=& \tilde{\bf v}{\bf A}_{-1}, \nonumber \\
(\omega{\bf 1}-{\bf \gamma}){\bf g}&=&{\bf a}.\nonumber
\label{3.48}
\end{eqnarray}
Now we diagonalize ${\bf \gamma}$

\begin{equation}
{\bf l\gamma r}=\omega^1, \ \ \ {\rm where}\ \
\ {\bf l=L^1v}\ {\rm and}\ \ {\bf
r=\tilde{v}R^1}.
\label{3.49}
\end{equation}
${\bf \gamma}$ is a reduced matrix with the same non-zero eigenvalues
$\omega^1$ as the original matrix ${\bf \Gamma}$. Since there are now no
zero eigenvalues, we can apply the spectral theorem with respect to the
non-null-space:
\begin{equation}
{\bf c=r}{\cal E}^1{\bf la},
\label{3.50}
\end{equation}
where ${\cal E}^1$ is the matrix occurring in eqn (\ref{3.23}).
A Fourier transformation to configuration space yields the self-consistency
equations (analogous to eqn (\ref{3.25}):

\begin{equation}
0=\int d{\bf k}({\bf r}{\cal E}^1{\bf l}\tilde{\bf v}{\bf A}_{-1}-\tilde{\bf
v}{\bf C_{\bf k}}).
\label{3.51}
\end{equation}
Again, this can be solved for the correlations in configuration space
${\bf C}$ if one can find a row-vector ${\bf \tilde{v}}_j$ which is ${\bf
k}$-independent, i. e.
\begin{equation}
  \int d{\bf k}\ \tilde{\bf v}_j{\bf C}_{\bf k} =
  \tilde{\bf v}_j \int d{\bf k}\ {\bf C}_{\bf k} =
  \tilde{\bf v}_j {\bf C}.
\label{3.52}
\end{equation}
This equation may be supplemented by the regularity condition
\begin{equation}
\lim_{\omega\rightarrow 0}\tilde{\bf u}_0(\omega{\bf 1}-{\bf uw\tilde{v}}){\bf
G }=\tilde{\bf u}_0{\bf A_{-1}}=0 .
\label{3.52a}
\end{equation}
This is because $\tilde{\bf u}_0{\bf u}=0$ and because the commutator GF is
regular at the origin.

One may be tempted to object that eqn (\ref{3.52}) is no improvement over eqn
(\ref{3.25}) because, in both cases, it is the ${\bf k}$ dependence of a term
containing ${\bf C_k}$ that creates a problem. In practice, however, it is much
better to use SVD because diagonalization of the full matrix ${\bf \Gamma}$
to get ${\bf R}^0$ and ${\bf L}^0$ is fraught with numerical difficulties when
there are non-zero eigenvalues which are very small. Furthermore, the vectors
${\bf R}^0$ and ${\bf L}^0$ are non-orthogonal, whereas the projector
${\bf v\tilde{v}}$ onto the non-null-space is built from orthogonal vectors --
this makes it easier in practice to find a row vector ${\bf \tilde{v}_i}$ that
is independent of the momentum ${\bf k}$. This search is technically
complicated,
and for a more detailed description, we refer the reader to Ref. \cite{FK05}.
Here, we give  a recipe for seeking for appropriate
${\bf \tilde{v}}_j$.

The row vectors in ${\bf \tilde{v}}$ are determined numerically and are
unique up
to a sign change or, for degenerate singular values, up to an orthogonal
transformation of the degenerate vectors. In order to distinguish among the
row vectors of ${\bf \tilde{v}}$, it is very helpful if they are suitably
labelled; e.g. they can often be characterized by a layer index
or a sublattice index. The following procedure is useful:
decompose the ${\bf \Gamma}$-matrix into a reference matrix and the rest,
\begin{equation}
{\bf \Gamma=\Gamma}_{ref}+{\bf \Gamma}_{rest},
\label{3.53}
\end{equation}
where ${\bf \Gamma}_{ref}$  has a block structure determined by the chosen
labels.
With a singular value decomposition,
\begin{equation}
{\bf \Gamma}_{ref}={\bf U}_{ref}{\bf W}_{ref}{\bf \tilde{V}}_{ref},
\label{3.54}
\end{equation}
one can define a block-label operator
\begin{equation}
{\bf P}_{op}:=\sum_{i=1}^{N_B}{\bf V}_{ref}(i)L(i){\bf \tilde{V}}_{ref}(i),
\label{3.55}
\end{equation}
with $L(i)=N_B-i+1$.
In the basis of the singular vectors {\bf v}
(and analogously for {\bf v$_0$}), we define a matrix
\begin{equation}
{\bf P}={\bf \tilde{v}P}_{op}{\bf v}=
\sum_{i=1}^{N_B}{\bf \tilde{v}V}_{ref}(i)\sqrt{L(i)}
\sqrt{L(i)}{\bf \tilde{V}}_{ref}(i){\bf v}={\bf \tilde{S}S}
\label{3.56}
\end{equation}
with
\begin{equation}
{\bf S}= [\sqrt{L(1)}{\bf \tilde{V}}_{ref}(1)\oplus...\oplus
\sqrt{L(N_B)}{\bf \tilde{V}}_{ref}(N_B) ]{\bf v},
\label{3.57}
\end{equation}
where $\oplus$ defines the direct sum.

Now the singular value decomposition of ${\bf S}$,
\begin{equation}
{\bf S=Ly\tilde{Z}},
\label{3.58}
\end{equation}
furnishes a matrix $\bf \tilde{Z}$ that diagonalizes $\bf \tilde{S}S$:
\begin{equation}
{\bf \tilde{S}S}={\bf Zy\tilde{L}Ly\tilde{Z}}={\bf Zy^2\tilde{Z}},
\label{3.59}
\end{equation}
where
${\bf y}^2\approx L(i)$, which labels the blocks.
To each block-label belongs a labelled vector
\begin{equation}
{\bf \tilde{v}}_L={\bf \tilde{Z}\tilde{v}}, \ \ \ \ (\rm L=labelling)
\label{3.60}
\end{equation}
which is the desired result.

A further difficulty is connected with the fact that the computed ${\bf
\tilde{v}}$  will not necessarily be continuous, even if the elements of the
$\bf \Gamma$-matrix are changed continuously (e.g. by varying the momentum
$\bf k$ on which they depend); i.e. vectors at neighbouring values of
${\bf k}$ can have arbitrary phases. This difficulty is overcome by a smoothing
procedure, which consists of the following steps:

(1) Create well-behaved reference vectors ${\bf V}_{ref}(r=1,..,N_r)$
in the momentum range of the first Brillouin zone
for the vectors ${\bf V}^0$ at $k_0$ and ${\bf V}^1$ at $k_1$, etc.
by overlaps as in the labelling procedure.

(2) Interpolate the reference vectors at each k
\begin{equation}
{\bf \tilde{\bar{V}}}_{ref}(k)
=w_l{\bf \tilde{V}}_{ref}(k_l)
+w_h{\bf \tilde{V}}_{ref}(k_h)
\label{3.61}
\end{equation}
with
\begin{equation}
w_l=\cos^2\Big(\frac{\pi}{2}(\frac{k-k_l}{k_h-k_l})\Big)\ \ \ {\rm and}\ \ \
w_h=1-w_l \ . \label{3.62}
\end{equation}

(3) Orthonormalize the reference vectors
\begin{eqnarray}
\cal{Y}&=&\tilde{\bar{\bf V}}_{ref}\bar{\bf V}_{ref},\nonumber\\
\lambda&=&\tilde{\bf T}{\cal Y}{\bf T},\nonumber\\
{\cal Y}^{-1/2}&=&{\bf T}\lambda^{-1/2}\tilde{\bf T},\nonumber\\
\tilde{\bf V}_{ref}&=&{\cal Y}^{-1/2}\tilde{\bar{\bf V}}_{ref}\ .
\label{3.63}
\end{eqnarray}
We now have reference vectors for the non-null and the null-space:
${\bf V}_{ref}=({\bf v}_{ref}, {\bf v}_{0,ref}).$

(4) Match the untreated (or, if necessary, labelled) vectors $\tilde{\bf v}$
to the orthonormalized reference vectors $\tilde{\bf v}_{ref}$.
This is done by seeking a transformation ${\bf Q}$ that rotates the target
(original) vectors among themselves to achieve the best match
\begin{equation}
{\bf \tilde{v}}_S={\bf Q\tilde{v}}.\  \ \ \ \ ({\rm S=smoothed})
\label{3.64}
\end{equation}

${\bf Q}$ is found by a SVD of the overlap matrix of the reference vectors with
the target vectors
\begin{equation}
{\bf {\cal S}}=\tilde{\bf v}_{ref}{\bf v}={\bf \cal L}{\bf
x}\tilde{\bf \cal Z}.
\label{3.65}
\end{equation}
Here ${\bf x}$ is a diagonal matrix of the singular values of the overlap
matrix which are close to 1 by construction (${\bf x\simeq 1}$).
The desired transformation matrix is
\begin{equation}
{\bf Q}={\bf {\cal L}}\tilde{\bf \cal Z},
\label{3.66}
\end{equation}
which is a rotation matrix because
\begin{equation}
{\bf \tilde{Q}Q}={\bf \cal{Z}}\tilde{\bf \cal L}{\bf \cal L}\tilde{\bf \cal
Z}={\bf {\cal{Z}}\tilde{\cal Z}}={\bf 1}.
\label{3.67}
\end{equation}
The overlap matrix of the reference vectors with the smoothed vectors is
close to the unit matrix because the phases of the new vectors have been fixed
by ${\bf \tilde{Q}}$:

\begin{equation}
{\bf \tilde{v}}_{ref}{\bf v}_S={\bf \tilde{v}}_{ref}{\bf v\tilde{Q}}=
{\bf S{\cal Z}}\tilde{\bf {\cal L}}={\bf {\cal L}x\tilde{\cal Z}{\cal
Z}\tilde{\cal
L}}={\bf \cal{L}}{\bf x}\tilde{\bf \cal L}\simeq {\bf 1}.
\label{3.68}
\end{equation}
To summarize, the untreated vectors ${\bf \tilde{v}}$ of the original
problem can be labelled and smoothed by the tranformation
\begin{equation}
{\bf \tilde{v}}_{LS}={\bf Q\tilde{v}}_L={\bf \cal{L}}\tilde{\bf \cal
Z}\tilde{\bf Z}\tilde{\bf v}.
\label{3.69}
\end{equation}
In practice, some of the row vectors of this transformation matrix turn
out to
be momentum-independent and can be used in solving equation (\ref{3.51}).

The procedure described above was successfully applied to Heisenberg
multi-layers with exchange anisotropies, see Section $4.2.3$ and to coupled
ferro- and antiferromagnetic layers, see Section 4.3.2.
We stress once more that the standard spectral theorem fails in these cases.

\subsubsection*{3.6. No advantage to using the anti-commutator
instead of the commutator Green's function}

We begin with the simplest case of a Green's function $G_{\eta}$
which has but a single pole and an inhomogeneity $A_\eta$:
\begin{eqnarray}
 G_{\eta}&=&\la\la A;B\ra\ra, \nonumber\\
 A_{\eta}&=&\la [A,B]_{\eta}\ra,
\label{3.70}
\end{eqnarray}
i.e.
\begin{equation}
 G_{\eta}^\alpha=\frac{ A_{\eta}^\alpha}{(\omega-\omega_{\bf
k})}.
\label{3.71}
\end{equation}
The corresponding correlations  in momentum and configuration space are
\begin{eqnarray}
 C_{\bf k}&=&\la BA\ra, \\
 C&=&\frac{1}{N}\sum_{\bf k} C_{\bf k}.\nonumber
\label{3.72}
\end{eqnarray}
Applying the spectral theorem  gives
\begin{equation}
 C_{\bf k}=\frac{A_{\eta}}
{e^{\beta\omega_{\bf k}}+\eta}\ .
\label{3.73}
\end{equation}
Note that
${\bf A}_{+1}({\bf k})={\bf A}_{-1}+2{\bf C_k}$ and
 ${\bf A}_{+1}({\bf k})$ depends on
${\bf k}$, whereas
${\bf A}_{-1}$ does not.

The commutator ($\eta=-1$) GF yields the correlation in configuration
space
\begin{equation}
 C=\frac{1}{N}\sum_{\bf k}
\frac{ A_{-1}}{e^{\beta\omega_{\bf k}}-1},
\label{3.74}
\end{equation}
whereas  the anti-commutator ($\eta=+1$) GF leads to
\begin{equation}
 C=\frac{1}{N}\sum_{\bf k}\frac{
A_{+1}({\bf k})}{e^{\beta\omega_{\bf k}}+1}=\frac{1}{N}\sum_{\bf
k}\frac{ A_{-1}+2C_{\bf k}}{e^{\beta\omega_{\bf k}}+1}\ ,
\label{3.75}
\end{equation}
which cannot be solved because  $C_{\bf k}$ is unknown.
Putting eqn (\ref{3.73}) with $\eta=-1$ into eqn
(\ref{3.75}) leads again to equation (\ref{3.74}),
\begin{equation}
C=\frac{1}{N}\sum_{\bf k}
\frac{A_{-1}+2\frac{A_{-1}}{e^{\beta\omega_{\bf k}}-1}}
{e^{\beta\omega_{\bf k}}+1}
=\frac{1}{N}\sum_{\bf k}\frac{
A_{-1}}{e^{\beta\omega_{\bf k}}-1}\ ,
\label{3.76}
\end{equation}
which can be solved self-consistently.
This shows that there is no advantage in starting
the calculation with the anti-commutator GF.

One can show this more generally with the eigenvector method of Section 3.3,
see \cite{FK03a}:
starting with the anti-commutator formulation, the spectral
theorem yields
\begin{equation}
{\bf C}_{\bf k}={\bf R{\cal E}LA}_{\eta=+1},
\label{3.77}
\end{equation}
where ${\bf {\cal E}}$ is a diagonal matrix with elements
${\cal E}_{ij}=\delta_{ij}(e^{\beta\omega_i}+1)^{-1}$ and ${\bf A}_{\eta=+1}$
depends on the momentum ${\bf k}$, preventing a direct use of this equation.

Because ${\bf A}_{\eta=+1}$=${\bf A}_{-1}+2{\bf C_{k}}$,
\begin{equation}
{\bf C}_{\bf k}={\bf R{\cal E}L}({\bf A}_{-1}+2{\bf C_k)}
\label{3.78}
\end{equation}
or
\begin{equation}
{\bf C_k}=(1-2{\bf R{\cal E}L)}^{-1}{\bf R{\cal E}LA}_{-1}\ .
\label{3.79}
\end{equation}
Introducing
\begin{equation}
{\bf (R{\cal E}L)}^{-1}={\bf L}^{-1}{\cal E}^{-1}{\bf R}^{-1}={\bf R}{\cal
E}^{-1}{\bf L} \label{3.80}
\end{equation}
in (\ref{3.79}),
\begin{equation}
{\bf C_k=(R(1-2{\cal E})L)}^{-1}{\bf R{\cal E}LA}_{-1}
={\bf R(1-2{\cal E})}^{-1}{\cal E}{\bf LA}_{-1}={\bf R\tilde{\cal E}LA}_{-1},
\label{3.81}
\end{equation}
where $\tilde{\cal E}_{ij}=\delta_{ij}{\cal E}_{ii}/
(1-2{\cal E}_{ii})=\delta_{ij}(e^{\beta\omega_i}-1)^{-1}$.
This is still of no use because of the zero eigenvalues.
But we have shown in Section 3.3 that the term
${\bf R}^0{\bf L}^0$ remedies this:
\begin{equation}
{\bf C_k=R}^1{\cal E}^1{\bf L}^1{\bf A}_{-1} +{\bf R}^0{\bf L}^0{\bf C_k}
\label{3.82}
\end{equation}
which is equation (\ref{3.23}), where ${\cal E}^1$ is the matrix
$\tilde{\cal E}$ leaving out the diverging terms.

\subsubsection*{3.7. The intrinsic energy, the specific
heat and the free energy}

The intrinsic energy is the thermodynamic expectation value of the underlying
Hamiltonian
\begin{equation}
E=\la H\ra=NE_i\ ,
\label{E1}
\end{equation}
where $E_i$ is the intrinsic energy
per lattice site and $N$ is the number of lattice sites.
The specific heat at constant volume is obtained by differentiating the
intrinsic energy with respect to the temperature
\begin{equation}
c_V= \frac{dE}{dT}=-\beta^2\frac{dE}{d\beta}\ .
\label{E2}
\end{equation}
The free energy is obtained by integrating over the intrinsic energy
\begin{equation}
F(T)=E(0)-T\int^T_0dT'\frac{E(T')-E(0)}{T'}\ .
\label{E3}
\end{equation}
{\small\noindent$\diamond\diamond$

\noindent Proof of this formula:\\
From $F=E-TS$ and $S=-\frac{dF}{dT}|_V$ one has
\begin{equation}
E(T)=-T^2\frac{d}{dT}\frac{F(T)}{T}
\label{E4}
\end{equation}
from which one obtains eqn (\ref{E3}) by integration.
Differentiating (\ref{E3}) gives (\ref{E4}).

\noindent$\diamond\diamond$}

In order to see how the intrinsic energy per lattice site can be calculated
explicitly, consider the quantity
\begin{equation}
B_i^{C,A}=\la A_i[C_i,H]_-\ra,
\label{E5}
\end{equation}
where $A_i$ and $C_i$ are the spin operators necessary for
constructing
equations of motion for those Green's functions from which the moments
$\la (S^z)^n\ra $  ($n=1,...,2S$) are calculated. $S$ is the spin
quantum number.
The quantity (\ref{E5}) can on one hand be related to the relevant Green's
functions
and on the other hand be calculated explicitly by evaluating the commutator.
This leads to a set of equations from which, together with equation
(\ref{E1}), the intrinsic energy can be calculated.

The connection to the Green's function results from the
spectral theorem:
\begin{eqnarray}
B_i^{C,A}&=&\la A_i[C_i,H]_-\ra=i\frac{d}{dt}\la A_i(t')C_i(t)\ra
|_{t=t'}\nonumber\\
&=&i\frac{d}{dt}\lim_{\delta\rightarrow 0}\frac{1}{N}\sum_{\bf k}\frac{i}{2\pi}
\int\frac{d\omega}{e^{\beta \omega}-1}\big(G_{\bf k}^{C,A}(\omega+i\delta)-
G_{\bf k}^{C,A}(\omega-i\delta)\big)e^{-i\omega(t-t')}|_{t=t'}\nonumber \\
&=&\lim_{\delta\rightarrow 0}\frac{1}{N}\sum_{\bf k}\frac{i}{2\pi}
\int\frac{\omega d\omega}{e^{\beta \omega}-1}\Big(G_{\bf
k}^{C,A}(\omega+i\delta)-
G_{\bf k}^{C,A}(\omega-i\delta)\Big)\ .
\label{E6}
\end{eqnarray}
In Appendix 7.1, we treat explicitly the cases for spin $S=1/2$ and $S=1$ for
a Heisenberg Hamiltonian with an external field and a single-ion anisotropy.
For S=1/2, one needs $A_i=S_i^-$ and $C_i=S_i^+$;
for S=1,  one needs a)  $A_i=S_i^-$ and
$C_i=S_i^+$ and b) $A_i=S_i^-$ and $C_i=(2S_i^z-1)S_i^+$.

\newpage
\subsection*{4. The GF formalism for Heisenberg films}
This chapter starts in Section 4.1 with the example of a spin $S=1/2$
ferromagnetic Heisenberg monolayer in a magnetic field \cite{EFJK99}. This is
an exercise
in applying the GF formalism in a simple case. The Tyablicov (RPA) and Callen
decouplings are introduced, the limit of mean field theory (MFT) is discussed,
the Mermin-Wagner
theorem is proved for this case, and the effective (temperature-dependent)
single-ion anisotropy is calculated by thermodynamic perturbation theory.

In Section $4.2$, ferromagnetic Heisenberg films with anisotropies and general
spin $S$ are treated. For the single-ion anisotropy, the Anderson-Callen
decoupling is used. The exchange anisotropy is treated by a generalized
Tyablikov decoupling. Susceptibilities are calculated. It is also shown how the
single-ion anisotropy can be treated exactly. As a further application, it is
shown that spin waves are very important for treating a trilayer in which two
ferromagnets are separated by a non-magnetic layer. Finally, the temperature
dependence of the interlayer coupling is discussed.

Section 4.3 deals with a unified treatment of ferromagnetic (FM),
antiferromagnetic (AFM) and coupled ferromagnetic-antiferromagnetic (FM-AFM)
Heisenberg films.

\subsubsection*{4.1. The ferromagnetic Heisenberg monolayer in a
magnetic field}

We choose this example
because it illustrates  the GF formalism in a simple case and allows the
validity
of the different approximations within the formalism to be checked against
`exact' Quantum Monte Carlo (QMC) calculations \cite{TGHS98}.

The Heisenberg Hamiltonian for a ferromagnetic monolayer in a magnetic
field is
\begin{eqnarray}
H&=&-\frac{1}{2}\sum_{<kl>} J_{kl}{\bf S}_k{\bf S}_l-B\sum_lS_l^z\nonumber\\
 &=&-\frac{1}{2}\sum_{<kl>} J_{kl}(S_k^-S_l^++S_k^zS_l^z)-B\sum_lS_l^z.
\label{4.1}
\end{eqnarray}
Here $J_{kl}$ is the exchange interaction strength,  $k$ and $l$ are lattice
site indices, and $<kl>$ means summation over nearest
neighbours only. The magnetic field $B$ is assumed to be in the $z$-direction
perpendicular to the film $xy$-plane . The second
line of eqn (\ref{4.1}) is obtained with the usual definition
$S_k^\pm=S_k^x\pm iS_k^y$ in terms of the components of the spin
operators.

For spin $S=1/2$, the magnetization is obtained from the relation
\begin{equation}
\la S_i^z\ra =1/2-\la S_i^-S_i^+\ra,
\label{4.2}
\end{equation}
and the  correlation $\la S_i^-S_i^+\ra$ is determined via the spectral theorem
from the commutator GF
\begin{equation}
G_{ij,\eta=-1}(\omega)=\la\la S_i^+;S_j^-\ra\ra .
\label{4.3}
\end{equation}
The GF is determined from the equation of motion in energy space
\begin{equation}
\omega\la\la S_i^+;S_j^-\ra\ra=\la
[S_i^+,S_j^-]\ra+\la\la[S_i^+,H]_-;S_j^-\ra\ra\ .
\label{4.4}
\end{equation}
Using spin commutator relations, one obtains
\begin{equation}
[S_i^+,H]_-=BS_i^+-\sum_i J_{il}(S_i^zS_l^+-S_l^zS_i^+).
\label{4.5}
\end{equation}
The equation of motion is then
\begin{equation}
(\omega-B)\la\la S_i^+;S_j^-\ra\ra=2\la
S^z_i\ra\delta_{ij}-\sum_lJ_{il}
\Big(\la\la S_i^zS_l^+;S_j^- \ra\ra -\la\la S_l^zS_i^+;S_j^-\ra\ra\Big)\ ,
\label{4.6}
\end{equation}
which is exact as it stands but, in order to use the equation,
the higher-order Green's
functions on the right hand side must be decoupled.

\subsubsection*{4.1.1. The Tyablikov (RPA)-decoupling}

This decoupling, introduced by Tyablikov \cite{Tya59}, is often called
the random phase approximation (RPA) because it is equivalent to that
approximation in other areas of physics.
It consists in factoring the higher-order Green's
functions:
\begin{eqnarray}
\la\la S_i^zS_l^+;S_j^- \ra\ra&\simeq &\la S_i^z\ra \la\la
S_l^+;S_j^-\ra\ra=\la S_i^z\ra G_{lj} ,\nonumber\\
\la\la S_l^zS_i^+;S_j^-\ra\ra&\simeq &\la S_l^z\ra \la\la
S_i^+;S_j^-\ra\ra=\la S_l^z\ra G_{ij}.
\label{4.7}
\end{eqnarray}
There is no a priori justification for this factorization but it has turned
out to be successful, also in other areas of physics where the
resulting equations can be derived with methods different from Green's
function theory.
In the present context, the quality of this approximation can be checked
against `exact' QMC results \cite{TGHS98}, see Section 4.1.5.

For a ferromagnet, there is translational invariance for the magnetization at
different lattice sites: $\la S_i^z\ra=\la S_l^z\ra=\la S^z\ra$.
After the decoupling, the equation of motion is
\begin{equation}
(\omega -B -\la S^z\ra\sum_lJ_{il})G_{ij}(\omega)+\la S^z\ra\sum_l
J_{il}G_{lj}(\omega)=2\la S^z\ra \delta_{ij}.
\label{4.8}
\end{equation}
A Fourier transform to momentum space (\ref{3.8}) yields
\begin{equation}
\Big(\omega -B -\la S^z\ra (J_0-J_{\bf k})\Big)
G_{\bf k}(\omega)= 2\la S^z\ra,
\label{4.9}
\end{equation}
and the Green's has the pole structure
\begin{equation}
G_{\bf k}(\omega)=\frac{2\la S^z\ra}{\omega-\omega_{\bf k}^{RPA}},
\label{4.10}
\end{equation}
with the dispersion relation
\begin{equation}
\omega_{\bf k}^{RPA}=B+\la S^z\ra (J_0-J_{\bf k}).
\label{4.11}
\end{equation}
For a square lattice with the number of nearest neighbours $z=4$  and a lattice
constant unity, one has
\begin{eqnarray}
J_0&=&\frac{1}{N}\sum_{ij}J_{ij}e^{i({\bf k}=0)({\bf R}_i-{\bf
R}_j)}=zJ=4J,\nonumber\\
J_{\bf k}&=&\frac{1}{N}\sum_{ij}J_{ij}e^{i{\bf k}({\bf R}_i-{\bf
R}_j)}=2J(\cos k_x+\cos k_y).
\label{4.12}
\end{eqnarray}
Applying the spectral theorem (\ref{3.18}) -- there is no zero eigenvalue --
and performing the $\omega$-integration with the relation
\begin{equation}
\frac{1}{\omega-\omega_{\bf k}\pm i\eta}=P\frac{1}{\omega-\omega_{\bf k}}\mp
i\pi\delta(\omega-\omega_{\bf k})
\label{4.13}
\end{equation}
yields for the magnetization $\la S^z\ra$ of the spin $S=1/2$ monolayer
\begin{eqnarray}
\la S_i^z\ra&=&\frac{1}{2}-\la S_i^-S_i^+\ra\nonumber\\
&=&\frac{1}{2}-\lim_{\delta \rightarrow 0}\frac{1}{N}\sum_{{\bf
k}}\frac{i}{2\pi}\int_{-\infty}^{\infty}
\frac{G_{\bf k}(\omega+i\delta)-G_{\bf k}(\omega-i\delta)}
{e^{\beta\omega}-1}\nonumber\\
&=&\frac{1}{2}
-\frac{1}{N}\sum_{{\bf k}}\frac{2\la S^z\ra}
{e^{\beta\omega_{\bf k}^{RPA}}-1}\nonumber\\
&=&\frac{1}{2}
      -\frac{1}{\pi^2}\int_0^{\pi}dk_x\int_0^\pi dk_y
\frac{2\la S^z\ra}
{e^{\beta\omega_{\bf k}^{RPA}}-1},
\label{4.14}
\end{eqnarray}
where the sum over the momenta has been replaced by an integration over the
first Brillouin zone of the square lattice.

With the relation $\frac{2}{e^x-1}=\coth(x/2)-1$, one obtains the following
expression for the magnetization
\begin{equation}
\la S^z\ra=\big[\frac{2}{\pi^2}\int_0^\pi dk_x\int_0^\pi dk_y
\coth(\frac{\beta\omega_{\bf k}^{RPA}}{2})\big]^{-1}.
\label{4.15}
\end{equation}
This equation must be iterated to self-consistency in $\la S^z\ra$, which can
then be compared with QMC (see Section
4.1.5).

\subsubsection*{4.1.2. The Callen decoupling}
In this section, we discuss an attempt of Callen
\cite{Cal63}
to improve the RPA.
We do this because it is the basis of an approximate decoupling of the terms
stemming from the single-ion anisotropy (see Section 4.2.1). This
generalisation of the Tyablikov (RPA) decoupling results from the ansatz
\begin{equation}
\la\la S_i^zS_l^+;S_j^-\ra\ra \simeq \la S_i^z\ra \la\la S_l^+;S_j^-\ra\ra
-\alpha\la S_i^-S_l^+\ra\la\la S_i^+;S_j^-\ra\ra,
\label{4.16}
\end{equation}
with $\alpha=\frac{\la S^z\ra}{2S^2}$; $\alpha\rightarrow 0$ corresponds to the
Tyablikov (RPA) decoupling.
Inserting this expression into the equation of motion and applying the spectral
theorem leads again to a single-pole expression for the Green's function
with a modified dispersion relation. The
spectral theorem yields
\begin{equation}
\la S^z\ra=\Big[\frac{2}{\pi^2}\int_0^\pi dk_x\int_0^\pi dk_y
\coth(\frac{\beta\omega_{\bf k}^{Callen}}{2})\Big]^{-1},
\label{4.17}
\end{equation}
with
\begin{equation}
\omega_{\bf k}^{Callen}=B+\la S^z\ra\Big(J_0-J_{\bf k}\Big)
\Big(1+\frac{\alpha}{\pi^2}\int_0^\pi dk_x\int_0^\pi dk_y
\frac{J_{\bf k}}{J_0}\coth(\frac{\beta \omega_{\bf k}^{Callen}}{2})\Big).
\label{4.18}
\end{equation}
Again, eqn (\ref{4.17}) must be iterated to self-consistency in $\la S^z\ra$.
Although it takes some higher-order correlations are into account,
the Callen approach is worse than RPA for the present case but still much
better than a mean field (MFT) result (see Section 4.1.5).

{\small
\noindent $\diamond\diamond\ \ $

Derivation of the Callen dispersion relation
(\ref{4.18}).

In order to make the Callen decoupling plausible, consider two equivalent
formulas for spin $S=1/2$
\begin{eqnarray}
S^z_i&=&S-S_i^-S_i^+,\nonumber\\
S_i^z&=&\frac{1}{2}(S_i^+S_i^--S_i^-S_i^+).
\label{4.19}
\end{eqnarray}
Multiplying the first equation by $\alpha$ and the second by ($1-\alpha$), one
can
write the Green's function $\la\la S_i^zS_l^+;S_j^-\ra\ra$in the following
form:
\begin{equation}
\la\la S_i^zS_l^+;S_j^-\ra\ra=\alpha S\la\la
S_l^+;S_j^-\ra\ra+\frac{1}{2}(1-\alpha)\la\la S_i^+S_i^-S_l^+;S_j^-\ra\ra
-\frac{1}{2}(1+\alpha)\la\la S_i^-S_i^+S_l^+;S_j^-\ra\ra.
\label{4.20}
\end{equation}
Now factorize the Green's functions on the right hand side as follows:
\begin{eqnarray}
\la\la S_i^+S_i^-S_l^+;S_j^-\ra\ra&\simeq&
 \la S_i^+S_i^-\ra\la\la S_l^+;S_j^-\ra\ra+\la S_i^+S_l^+\ra\la\la
S_i^-;S_j^-\ra\ra+ \la S_i^-S_l^+\ra\la\la S_i^+;S_j^-\ra\ra,\nonumber\\
\la\la S_i^-S_i^+S_l^+;S_j^-\ra\ra&\simeq&
\la S_i^-S_i^+\ra\la\la S_l^+;S_j^-\ra\ra+\la S_i^-S_l^+\ra\la\la
S_i^+;S_j^-\ra\ra+ \la S_i^+S_l^+\ra\la\la S_i^-;S_j^-\ra\ra.\nonumber\\
\label{4.21}
\end{eqnarray}
Approximating the terms non-diagonal in the z-component of the spin by
$\la S_i^+S_l^+\ra\simeq 0$ and using the relation
$\la S_i^+S_i^-\ra=2\la S_i^z\ra+\la S_i^-S_i^+\ra$, we obtain
\begin{equation}
\la\la S_i^zS_l^+;S_j^-\ra\ra \simeq \la S_i^z\ra \la\la S_l^+;S_j^-\ra\ra
-\alpha\la S_i^-S_l^+\ra\la\la S_i^+;S_j^-\ra\ra,
\label{4.22}
\end{equation}
which is expression (\ref{4.16}).
Taking $\alpha=\frac{\la S^z\ra}{S}$
interpolates between the case
$\alpha=1$, where the first of the equations (\ref{4.19}) should be used for
the
decoupling at low temperatures ($\la S^z\ra\simeq S$), and $\alpha=0$,
where the second
formula should be used ($\la S^z\ra\simeq 0$).
For arbitrary spins, arguments in favor of
$\alpha=\frac{\la S^z\ra}{2S^2}$  (which includes the
spin $S=1/2$ case) are given in \cite{Cal63}.

Introducing the decoupling (\ref{4.22}) into the equation of
motion (\ref{4.6}) yields
\begin{eqnarray}
\Big(\omega-B-\la S^z\ra\sum_l
J_{il}-\alpha\sum_lJ_{il}\la S_i^-S_l^+\ra\Big)G_{ij}(\omega)\nonumber\\
+\Big(\la S^z\ra\sum_l J_{il}+\alpha\sum_lJ_{il}\la
S_l^-S_i^+\ra\Big)G_{lj}(\omega)=2\delta_{ij}\la S^z\ra.
\label{4.23}
\end{eqnarray}
A Fourier transform to momentum space leads to
\begin{equation}
\Big[ \omega-B-\la S^z\ra(J_0-J_{\bf k})-\alpha
\frac{1}{N}\sum_{\bf q}(J_{\bf q}
-J_{\bf q+k})\la S^-S^+\ra_{\bf q}
\Big]G_{\bf k}(\omega)=2\la S^z\ra,
\label{4.24}
\end{equation}
where the Green's function is given by
\begin{equation}
G_{\bf k}(\omega)=\frac{2\la S^z\ra}{\omega-\omega_{\bf k}^{Callen}}
\label{4.25}
\end{equation}
with
\begin{equation}
\omega_{\bf k}^{Callen}=B+\la S^z\ra(J_0-J_{\bf
k})+\alpha\frac{1}{N}\sum_{\bf q}(J_{\bf q}-J_{\bf q+k})
\la S^-S^+\ra_{\bf q}.
\label{4.26}
\end{equation}
The spectral theorem then determines
\begin{equation}
\la S^-S^+\ra_{\bf k}=\frac{i}{2\pi}\lim_{\delta\rightarrow
0}\int_{-\infty}^{\infty}\frac{d\omega}{e^{\beta\omega}-1}
(G_{\bf k}(\omega+i\delta)-G_{\bf k}(\omega-i\delta)=2\la S^z\ra\Phi_{\bf k}.
\label{4.27}
\end{equation}
with
\begin{equation}
\Phi_{\bf k}=\frac{1}{e^{\beta\omega_{{\bf
k}}^{Callen}}-1}=\smfrac{1}{2}\Big(\coth(\smfrac{\beta\omega_{\bf
k}^{Callen}}{2})-1\Big). \label{4.28}
\end{equation}
It remains to simplify the term proportional to $\alpha$ in the dispersion
relation:
\begin{eqnarray}
& &2\alpha\la S^z\ra \frac{1}{N}\sum_{\bf q}(J_{\bf q}-J_{\bf q+k})\Phi_{\bf
q}\nonumber\\
&=&2\alpha\la S^z\ra
\frac{1}{N}\sum_{ij}J_{ij}(1-e^{i{\bf k}({\bf R}_i-{\bf
R}_j})\frac{1}{N}\sum_{\bf q}e^{i{\bf q}({\bf R}_i-{\bf R}_j)}
\Phi_{\bf q}\nonumber\\
&=&2\alpha\la S^z\ra(J_0-J_{\bf k})\frac{1}{N}\sum_{\bf q}
\frac{J_{\bf q}}{J_0}\Phi_{\bf q},
\label{4.29}
\end{eqnarray}
where we have made use of
\begin{equation}
\frac{1}{N}\sum_{\bf q}e^{i{\bf q}({\bf R}_i-{\bf R}_j)}\Phi_{\bf q}=
\frac{1}{N}\sum_{\bf q}\frac{J}{zJ}\sum_{<ij>}e^{i{\bf q}({\bf
R}_i-{\bf R}_j)}
\Phi_{\bf q}=\frac{1}{N}\sum_{\bf q}
\frac{J_{\bf q}}{J_0}\Phi_{\bf q} \ ,
\label{4.30}
\end{equation}
and $z$ is the number of nearest neighbours.
This completes the proof for the Callen dispersion relation.

\noindent $\diamond\diamond $}

In Ref. \cite{Cal63}, Callen also derives a closed form expression for the
magnetization for general spin S from the solution of a differential equation.
The result is
\begin{equation}
\la S^z\ra=\frac{(S-\Phi_{\bf k})(1+\Phi_{\bf k})^{(2S+1)}+(S+1+\Phi_{\bf
k})\Phi_{\bf k}^{(2S+1)}}
{(1+\Phi_{\bf k})^{(2S+1)}-\Phi_{\bf k}^{(2S+1)}}\ ,
\end{equation}
a formula which was also found by Pravecki \cite{Pra63}.
For the treatment of general spin S, see also Refs. \cite{TKH62}.

\subsubsection*{4.1.3. Mean field theory (MFT)}
In mean field theory (MFT), which is frequently used as the simplest
approximation, one neglects correlations which lead
to collective excitations (magnons). The essential approximation consists in
writing operator products as
\begin{eqnarray}
S_i^\alpha S_j^\beta& &=(S_i^\alpha-\la S_i^\alpha\ra)(S_j^\beta-\la
S_j^\beta\ra)+S_i^\alpha\la S_j^\beta\ra+\la S_i^\alpha\ra S_j^\beta- \la
S_i^\alpha\ra\la S_j^\beta\ra\nonumber\\
& &\simeq S_i^\alpha\la S_j^\beta\ra+\la S_i^\alpha\ra S_j^\beta- \la
S_i^\alpha\ra\la S_j^\beta\ra,
\label{4.31}
\end{eqnarray}
where the mean field assumptions $S_i^\alpha\simeq \la S_i^\alpha\ra$ and
$S_j^\beta\simeq \la S_j^\beta\ra$ have been made.
Neglecting transverse expectation values ($\la S^\pm_j\ra=0$) as well leads
to the mean field Hamiltonian
\begin{equation}
H^{MFT}=-\sum_{kl}J_{kl}\la S_k^z\ra S_l^z-B\sum_lS_l^z
+\frac{1}{2}\sum_{kl}\la S_k^z\ra\la S_l^z\ra,
\label{4.32}
\end{equation}
where the last term, being a constant, does not influence the equations
of
motion but has to be taken into account when calculating the intrinsic energy.

In Green's function theory, the Hamiltonian $H^{MFT}$ leads without
further approximations to the equations of motion
\begin{equation}
(\omega -B-\la S^z\ra\sum_k J_{ik})G_{ij}=2\la S^z\ra \delta_{ij}\ ,
\label{4.33}
\end{equation}
whose Fourier transform to momentum space is
\begin{equation}
(\omega -B -\la S^z\ra J_0)G_{\bf k}(\omega)=2\la S^z\ra,
\label{4.34}
\end{equation}
where $J_0=zJ$ ($z$ is the number of nearest neighbours; $z=4$ for a square
lattice) and
\begin{equation}
G_{\bf k}(\omega)=\frac{2\la S^z\ra}{\omega-\omega^{MFT}}\ ,
\label{4.35}
\end{equation}
with a momentum-independent dispersion relation
\begin{equation}
\omega^{MFT}=B+\la S^z\ra J_0.
\label{4.36}
\end{equation}
Because there is no momentum dependence in this relation, the $k$-integration
in the spectral theorem is trivial, so that
\begin{equation}
\la S^z\ra=\frac{1}{2}-\la S^-S^+\ra= \frac{1}{2}-\frac{1}{N}\sum_{\bf k}
\frac{2\la S^z\ra}{e^{\beta\omega^{MFT}}-1}
=\frac{1}{2}-\frac{2\la S^z\ra}{e^{\beta\omega^{MFT}}-1}\ ,
\label{4.37}
\end{equation}
from which
\begin{equation}
\la S^z\ra=\frac{1}{2}\tanh(\frac{\beta\omega^{MFT}}{2}).
\label{4.38}
\end{equation}
This result is obtained from the RPA result by setting $J_{\bf k}$ to zero
in eqn (\ref{4.11}), thereby
neglecting the ${\bf k}$-dependence of
the lattice. In MFT, it is only
the number of nearest neighbours $z$
that count. This is also true  for the more complicated cases discussed later.
The neglect of the ${\bf k}$-dependence (this corresponds to the
neglect of magnons) makes MFT  much worse than RPA, as seen in Fig.
\ref{fig1} of Section 4.1.5 .

Because MFT is easily applied, often with qualitatively reasonable
results, we quote a few papers where MFT is extensively used: in
Refs. \cite{JB98} and \cite{HU00} and references therein, the spin
reorientation
transition is treated and effective (temperature-dependent) lattice anisotropy
coeefficients are calculated; in Ref. \cite{JKD05}, coupled ferro-
antiferromagnetic layers are treated.

\subsubsection*{4.1.4. The Mermin-Wagner theorem}
Mermin and Wagner \cite{MW66}  have shown  quite generally that the pure
Heisenberg model (without magnetic field and anisotropies) in less than 3
dimensions does not
exhibit collective order at finite temperatures (for $\la S^z\ra\rightarrow 0$
the Curie temperature
goes to $0$: $ T_{\rm Curie}\rightarrow 0$).

From the expressions derived above, we can see that RPA  obeys the
theorem whereas MFT  violates it.
Expanding the RPA expression for the magnetization (\ref{4.15}) for small
$\la S^z\ra$ and $B=0$, one obtains an expression for the Curie temperature
\begin{equation}
T_{\rm
Curie}^{RPA}=\frac{1}{\sum_{\bf k}\frac{2}{J_0-J_{\bf k}}}
\simeq\frac{1}{\frac{2}{2\pi^2}\int_0^\pi
dk_x\int_0^\pi
dk_y\frac{2}{J_0-J_{\bf k}}}\propto \frac{1}{\int_0^\pi dk_x\int_0^\pi
\frac{1}{k_x^2+k_y^2}}\rightarrow 0 \ .
\label{4.39}
\end{equation}
This is so because the integral diverges at the lower boundary, which can be
seen by expanding
the square lattice dependence of $J_{\bf k}$, eqn (\ref{4.12}), for small
momenta. This means that the Mermin-Wagner theorem is obeyed in RPA.
In three dimensions, the RPA expression gives a finite value for the Curie
temperature and is used in ab initio calculations of the Heisenberg exchange
interaction to determine the Curie temperature, see e.g. \cite{TKDBB03}.

Calculating the Curie temperature from the MFT result (\ref{4.38}) for $B=0$
by expanding for small $\la S^z\ra$ gives a finite Curie temperature
\begin{equation}
T_{\rm Curie}^{MFT}=\frac{1}{4}J_0=\frac{1}{4}zJ=J\ ,
\label{4.40}
\end{equation}
where we have taken $z=4$, the number of nearest
neighbours for a square lattice. This is in clear violation of
the theorem.

\subsubsection*{4.1.5. Comparing with Quantum Monte Carlo calculations}
\begin{figure}[htb]
\begin{center}
\protect
\includegraphics*[bb = 175  65 500 580,
angle=0,clip=true,width=8cm]{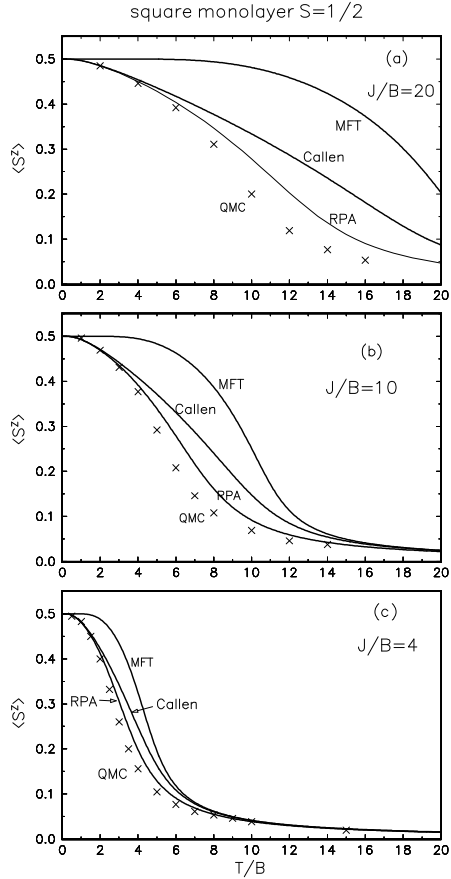}
\protect
\caption{The temperature dependence of the magnetization of a ferromagnetic
Heisenberg monolayer for a square lattice with spin $S=1/2$. Comparison
between the `exact' quantum Monte Carlo (QMC) result \cite{TGHS98} and
the results
obtained with MFT, RPA and Callen decoupling. We have used (a) J/B=20, (b)
J/B=10 and (c) J/B=4 (from reference \cite{EFJK99}). }
\label{fig1}
\end{center}
\end{figure}

In this section, we compare
the temperature dependence of the magnetization of a ferromagnetic spin $S=1/2$
Heisenberg monolayer on a square lattice obtained with the approximations just
described with quantum Monte Carlo (QMC)
calculations \cite{TGHS98}, which are `exact' within their
statistical errors. The results are shown in
Fig.{\ref{fig1}}.
The RPA of  Section 4.1.1 is the best and is a fairly good
approximation. Before QMC calculations were available,
it was not possible to check the quality of RPA.
Although there are additional correlations  taken into account in the
Callen approach of Section 4.1.2, its results for spin $S=1/2$ are not as good
as those from simple RPA.
In Ref. \cite{Cal63}, Callen  argues that his decoupling should give
better results for larger spin values, but there are no QMC calculations
available to support his statements.
The mean field theory (MFT) of Section 4.1.3 yields by far the worst results.
This results from not taking collective excitations
(magnons) into account, which is also the reason for the violation
of the Mermin-Wagner theorem.

We mention that
RPA gives still better results for the magnetization when
higher-order
Green's functions with vertex corrections for their decoupling are included
\cite{JIRK04}. For quantities with transverse correlations, like the intrinsic
energy or the specific heat, one has to go beyond RPA. See Section 5, in
particular Ref. \cite{JIRK04}.

\subsubsection*{4.1.6. The effective (temperature dependent) single-ion
lattice anisotropy}
\begin{figure}[htb]
\begin{center}
\protect
\includegraphics*[bb = 00 00 96 144,
angle=0,clip=true,width=8cm]{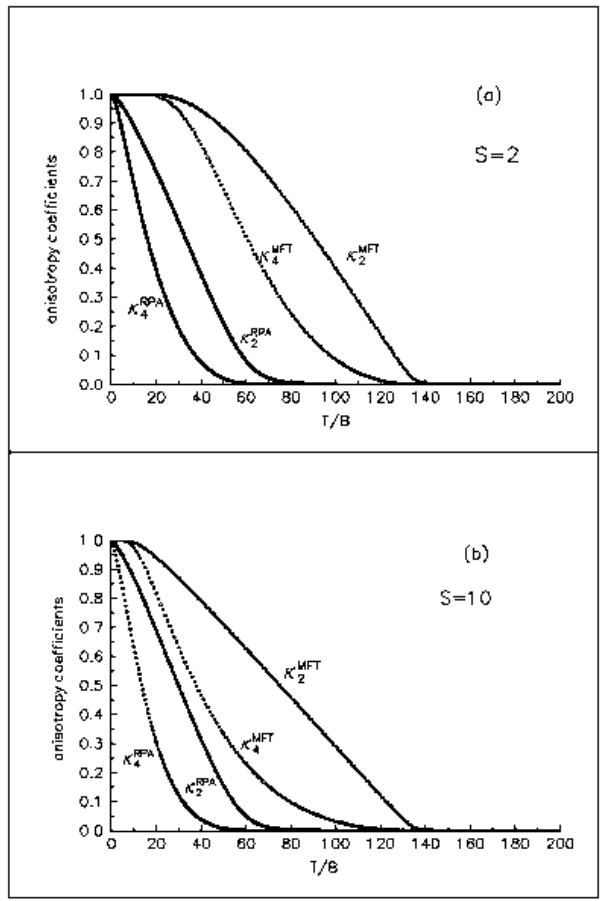}
\protect
\caption{The temperature dependence of the effective lattice
anisotropy coefficients ${\cal K}_2(T)$ and ${\cal K}_4(T)$
of a square Heisenberg monolayer calculated with thermodynamic perturbation
theory for MFT and RPA. We have used $J/B=100$ and (a) $S=2$ and (b) $S=10$.
To allow for comparison between different spin values, we used the scaling
$J\rightarrow J/S(S+1)$ and $B\rightarrow B/S$.
}
\label{fig1a}
\end{center}
\end{figure}

Lattice anisotropy coefficients are defined in an expansion of the free energy
in powers of $\cos\theta$ \cite{Von74}, where $\theta$ is the polar angle
between the magnetization $\la {\bf S}\ra$ and the normal to the film plane
\begin{equation}
F(T,\theta)=F_0(T)-K_2(T)\cos\theta-K_4(T)\cos^4\theta-{\bf B}\cdot{\la {\bf
S}\ra}\ . \label{4.41}
\end{equation}
The anisotropy coefficients can be calculated by thermodynamic perturbation
theory, where the Hamiltonian $H=H_0+V$ is separated into an unperturbed part
$H_0$ consisting of the exchange coupling and the magnetic field and a
perturbation $V_n=-K_n\sum_l(S_l^z)^n$ (n=2,4).
Within first order perturbation theory, effective anisotropy coefficients
can be defined as
\begin{equation}
{\cal K}_n(T)=K_nf_n(T)\ ,
\end{equation}
where the temperature dependence is introduced by the functions $f_n(T)$
which are expressed in terms of expectation values $\la(S^z)^n\ra_0$ for
the unperturbed Hamiltonian
\begin{eqnarray}
f_2(T)&=&\smfrac{1}{2}\Big(3\la(S^z)^2\ra_0-S(S+1)\Big)\ ,\\
f_4(T)&=&\smfrac{1}{8}[35\la(S^z)^4\ra_0-(30S(S+1)-25)\la
(S^z)^2\ra_0+3S(S+1)(S(S+1)-2)]\ .\nonumber \end{eqnarray}
The moments are calculated with RPA and MFT in Ref. \cite{EFJK99} and the
resulting temperature dependent coefficients are shown in Fig. (\ref{fig1a})
for $S=2$ and $S=10$, where $K_n=1$ and a scaling $J\rightarrow J/S(S+1)$ and
$B\rightarrow B/S$ has been used. The resulting behaviour of the ${\cal
K}_n(T)$
calculated by RPA differs markedly from that obtained by MFT particularly at
low temperatures: whereas the ${\cal K}_n(T)$ obtained with MFT show an
exponential decay in this temperature range, those calculated from RPA
decrease
more rapidly and exhibit a nearly linear behaviour.
The ${\cal K}(T)$ calculated with RPA exhibit a much weaker
dependence on the spin $S$ than those calculated with MFT.

In Section 4.2.2 we show that it is better to calculate the
effective
anisotropy coefficients {\em non-perturbatively} by minimizing the free energy
with repect to the reorientation angle.

\newpage

\subsection*{4.2. Ferromagnetic Heisenberg films with
anisotropies for\\ \indent \ general spin S }
An isotropic Heisenberg model in less than three dimensions does not show
spontaneous magnetization at finite
temperature, as explained by the Mermin-Wagner
theorem \cite{MW66}. Such an idealized system does not, however, exist in
nature,
since even the smallest anisotropy leads to a finite magnetization. This can
be caused by
an external magnetic field (as shown in the previous chapter),
single-ion anisotropies, exchange anisotropies, or the magnetic
dipole-dipole interaction.

Many applications of GF-theory deal only with the magnetization in one
direction of space. They treat multi-layers but not all use
the full power of the eigenvector method outlined in Section 3.3.
We mention only a few. Diep-The-Huang et al. \cite{DLN79} treat ferro-
and
antiferromagnetic multilayers but, instead of the eigenvector method, they use
Kramers rule for calculating the GF's. Schiller and Nolting \cite{SN99} treat
sc(100) and fcc(100) ferromagnetic Heisenberg spins with $S=7/2$ using RPA for
the exchange interaction and the Lines decoupling \cite{Lin67} for the
single-ion anisotropy. C. Cucci et al. \cite{CPPR01} consider fcc
(100), (110) and (111) ferromagnetic Heisenberg films using RPA, the Lines
decoupling and the eigenvector method.

In the following, we do not restrict the magnetization to be in one
direction of space because we are  interested in the reorientation of the
magnetization as a function of the temperature and film thickness. Therefore,
we deal from the outset with a multi-dimensional case; the orientation of
the magnetization in one direction and the monolayer then occur naturally
as special cases. An essential complication connected with the reorientation
problem is the occurrence of zero eigenvalues of the equation-of-motion
matrix, which can be handled with the techniques developed in Section
3.

We do not discuss papers dealing with the magnetic reorientation on the basis
of a boson expansion, as e.g. Refs. \cite{EM91a,EM91b} who start with a
Holstein-Primakoff transformation in lowest order, because the validitity of a
linearized spin wave theory is limited to low temperatures only.

 \subsubsection*{4.2.1. The Hamiltonian and
the decoupling procedures}

We consider a spin Hamiltonian consisting of an isotropic Heisenberg exchange
interaction with strength $J_{kl}$
between nearest neighbour lattice sites, an
exchange
anisotropy with strength $D_{kl}$, a second-order single-ion lattice
anisotropy with strength $K_{2,k}$, a
magnetic dipole coupling  with strength $g_{kl}$ and an external magnetic
field ${\bf B}=(B^x,B^y,B^z)$:
\begin{eqnarray}
{\cal
H}=&-&\frac{1}{2}\sum_{<kl>}J_{kl}(S_k^-S_l^++S_k^zS_l^z)
-\frac{1}{2}\sum_{<kl>}D_{kl}S_k^zS_l^z-\sum_kK_{2,k}(S_k^z)^2\nonumber\\
&-&\sum_k\Big(\frac{1}{2}B^-S_k^++\frac{1}{2}B^+S_k^-+B^zS_k^z\Big)\nonumber\\
&+&\frac{1}{2}\sum_{kl}\frac{g_{kl}}{r_{kl}^5}\Big(r_{kl}^2(S_k^-S_l^++S_k^zS_l
^ z )-3({\bf S}_k{\bf r}_{kl})({\bf S}_l{\bf r}_{kl})\Big) .
\label{5.1}
\end{eqnarray}
Here the notation $S_k^{\pm}=S_k^x\pm iS_k^y$ and $B^{\pm}=B^x\pm iB^y$ is
introduced, where $k$ and $l$ are lattice site indices and $<kl>$ indicates
summation over nearest neighbours only.

In order to treat the spin reorientation transition for general spin $S$, we
need the following Green's functions:
\begin{equation}
G_{ij,\eta}^{\alpha,mn}(\omega)=\la\la
S_i^\alpha;(S_j^z)^m(S_j^-)^n\ra\ra_{\omega,\eta}\ ,
\label{5.2}
\end{equation}
where $\alpha=(+,-,z)$ takes care of all directions in space, $\eta=\pm 1$
refers to the anti-commutator or commutator Green's functions respectively,
and $n\geq 1,\ m\geq 0\ (m+n\leq 2S+1)$ are positive integers.
We follow the formalism of Section 3 by evaluating all
formulas for the Hamiltonian (\ref{5.1}).

The exact equations of motion
\begin{equation}
\omega G_{ij,\eta}^{\alpha,mn}(\omega)=A_{ij,\eta}^{\alpha,mn}+\la\la
[S_i^\alpha,{\cal H}]_-;(S_j^z)^m(S_j^-)^n\ra\ra_{\omega,\eta}
\label{5.3}
\end{equation}
with the inhomogeneities
\begin{equation}
A_{ij,\eta}^{\alpha,mn}=\la[S_i^\alpha,(S_j^z)^m(S_j^-)^n]_{\eta}\ra
\label{5.4}
\end{equation}
are given explicitly by
\begin{eqnarray}
\omega G_{ij,\eta}^{\pm,mn}&=&A_{ij,\eta}^{\pm,mn}\nonumber\\
& &\mp\sum_{k}J_{ik}\Big(\la\la S_i^zS_k^\pm;(S_j^z)^m(S_j^-)^n\ra\ra
-\la\la S_k^zS_i^\pm;(S_j^z)^m(S_j^-)^n\ra\ra\Big)\nonumber\\
& &\pm\sum_k D_{ik}\la\la S_k^zS_i^\pm;(S_j^z)^m(S_j^-)^n\ra\ra\nonumber\\
& &\pm K_{2,i}\la\la(S_i^\pm
S_i^z+S_i^zS_i^\pm);(S_j^z)^m(S_j^-)^n\ra\ra\nonumber\\
& &\mp B^\pm G_{ij,\eta}^{z,mn}\pm B^zG_{ij,\eta}^{\pm,mn}\ ,\nonumber\\
\omega G_{ij,\eta}^{z,mn}&=&A_{ij(\eta)}^{z,mn}\nonumber\\
& &+\frac{1}{2}\sum_kJ_{ik}\la\la(S_i^-S_k^+-S_k^-S_i^+);
(S_j^z)^m(S_j^-)^n\ra\ra\nonumber\\
& &-\frac{1}{2}B^- G_{ij,\eta}^{+,mn}+\frac{1}{2}B^+G_{ij,\eta}^{-,mn}.
\label{5.5}
\end{eqnarray}
For the moment, we leave out the terms due to the dipole-dipole
interaction, which we include later.

Once these equations are solved, the
components of the magnetization can be determined from the Green's functions
via the spectral theorem.
A closed system of equations results from
decoupling the higher-order Green's functions on the right-hand sides.
For the exchange interaction and exchange anisotropy terms, we use a
generalized Tyablikov- (or RPA-) decoupling:
\begin{equation}
\la\la S_i^\alpha S_k^\beta;(S_j^z)^m(S_j^-)^n\ra\ra_\eta \simeq\la
S_i^\alpha\ra
G_{kj,\eta}^{\beta,mn}+\la S_k^\beta\ra G_{ij,\eta}^{\alpha,mn} .
\label{5.6}
\end{equation}
The terms stemming from the single-ion anisotropy have to be decoupled
differently,
because  RPA decoupling leads to unphysical results; e.g. for spin $S=1/2$,
the terms due to the single-ion anisotropy do not vanish in RPA as they should
do because, in this case, $\sum_i K_{2,i}\la (S_i^z)^2\ra$ is a constant and
does not influence the equations of motion.
In the appendix of Ref. \cite{FJK00}, we investigate different decoupling
schemes proposed in the literature, e.g. those of Lines \cite{Lin67} or that of
Anderson and Callen \cite{AC64}. These should be reasonable for
single-ion
anisotropies small compared to the exchange interaction. We found the
Anderson-Callen decoupling
to be most adequate in our context.
It treats the diagonal terms as they occur from the single-ion anisotropy in
the same way that Callen \cite{Cal63} used in his attempt to improve the RPA.
Consider eqn (\ref{4.16}) for $i=l$:
add the  term for $\la\la S^z_iS^+_i;...\ra\ra$ and do the
same for the corresponding expressions for $G^{-,mn}$. Using
$S_i^\mp S_i^\pm=S(S+1)\mp S^z_i-S_i^zS_i^z$, one obtains
\begin{eqnarray}
& &\la\la(S_i^\pm S_i^z+S_i^zS_i^\pm);(S_j^z)^m(S_j^-)^n\ra\ra_\eta \nonumber\\
& &\simeq 2\la S_i^z\ra\Big(1-\frac{1}{2S^2}[S(S+1)-\la
S_i^zS_i^z\ra]\Big)G_{ij,\eta}^{\pm,mn}.
\label{5.7}
\end{eqnarray}
This term vanishes for $S=1/2$ as it should.

In Section 4.2.5, we shall demonstrate a procedure for treating the
single-ion anisotropy
exactly by going to higher-order Green's functions. With this,
single-ion anisotropies with arbitrary strength can be treated. This
procedure is, however, tedious to apply for spins $S>1$, whereas there is no
problem when staying at the level of the lowest-order Green's function as
discussed in the present section.

Applying the decouplings (\ref{5.6}) and (\ref{5.7}) and a Fourier transform to
momentum space, one obtains, for a ferromagnetic film with $N$ layers,
$3N$ equations of motion which can be written in compact matrix notation as
\begin{equation}
(\omega{\bf 1}-{\bf \Gamma}){\bf G}^{mn}={\bf A}^{mn}\ .
\label{5.8}
\end{equation}
 ${\bf G}^{mn}$ is a $3N$-dimensional Green's function vector and ${\bf 1}$ is
the $3N\times 3N$ unit matrix. The Green's functions and the inhomogeneity
vectors each
consist of $N$  three-dimensional subvectors which are characterized by the
indices $i$ and $j$, which are now layer indices:
\begin{equation}
{\bf G}_{ij}^{mn}({\bf{k},\omega})\  =
\left( \begin{array}{c}
G_{ij}^{+,mn}({\bf{k}},\omega) \\ G_{ij}^{-,mn}({\bf{k}},\omega)  \\
G_{ij}^{z,mn}({\bf{k}},\omega)
\end{array} \right), \hspace{0.5cm}
{\bf A}_{ij}^{mn} {=}
 \left( \begin{array}{c} A_{ij}^{+,mn} \\ A_{ij}^{-,mn} \\
A_{ij}^{z,mn} \end{array} \right) \;.
\label{5.9} \end{equation}

The equations of motion are then expressed in terms of these layer vectors and
$3\times 3 $ submatrices ${\bf \Gamma}_{ij}$ of the $3N\times 3N$
matrix ${\bf\Gamma}$
\begin{equation}
\left[ \omega {\bf 1}-\left( \begin{array}{cccc}
{\bf\Gamma}_{11} & {\bf\Gamma}_{12} & \ldots & {\bf\Gamma}_{1N} \\
{\bf\Gamma}_{21} & {\bf\Gamma}_{22} & \ldots & {\bf\Gamma}_{2N} \\
\ldots & \ldots & \ldots & \ldots \\
{\bf\Gamma}_{N1} & {\bf\Gamma}_{N2} & \ldots & {\bf\Gamma}_{NN}
\end{array}\right)\right]\left[ \begin{array}{c}
{\bf G}_{1j} \\ {\bf G}_{2j} \\ \ldots \\ {\bf G}_{Nj} \end{array}
\right]=\left[ \begin{array}{c}
{\bf A}_{1j}\delta_{1j} \\ {\bf A}_{2j}\delta_{2j} \\ \ldots \\
{\bf A}_{Nj}\delta_{Nj} \end{array}
\right] \;, \hspace{0.5cm} j=1,...,N\;.
\label{5.10}
\end{equation}
The  ${\bf \Gamma}$ matrix reduces to a band matrix with zeros in the
${\bf \Gamma}_{ij}$ sub-matrices, when $j>i+1$ and $j<i-1$.
The  diagonal sub-matrices ${\bf \Gamma}_{ii}$ are of size $3\times 3$
and have the form
\begin{equation}
 {\bf \Gamma}_{ii}= \left( \begin{array}
{@{\hspace*{3mm}}c@{\hspace*{5mm}}c@{\hspace*{5mm}}c@{\hspace*{3mm}}}
\;\;\;H^z_i & 0 & -H^+_i \\ 0 & -H^z_i & \;\;\;H^-_i \\
-\frac{1}{2}\tilde{H}^-_i & \;\frac{1}{2}\tilde{H}^+_i & 0
\end{array} \right)
\ . \label{5.11}
\end{equation}
where
\begin{eqnarray}
H^z_i&=&B^z_i+\la S_i^z\ra\Big( J_{ii}(q-\gamma_{\bf k})+D_{ii}q\Big)
+(J_{i,i+1}+D_{i,i+1})\la S_{i+1}^{z}\ra+(J_{i,i-1}+D_{i,i-1})\la
S_{i-1}^{z}\ra\nonumber\\
& &+K_{2,i}2\la S_i^z\ra
\Big(1-\frac{1}{2S^2}[S(S+1)-\la S_i^zS_i^z\ra]\Big)\ ,
\nonumber \\
\tilde{H}^\pm_i&=&B^\pm_i+\la S_i^\pm\ra J_{ii}(q-\gamma_{\bf
k})
+J_{i,i+1}\la S_{i+1}^{\pm}\ra+J_{i,i-1}\la
S_{i-1}^{\pm}\ra \ ,
\nonumber\\
H^\pm_i&=&\tilde{H}^\pm_i-\la S_i^\pm\ra D_{ii}\gamma_{\bf k} \ .
\label{5.12}
\end{eqnarray}
For a square lattice and a lattice constant taken to be unity,
$\gamma_{\bf k}=2(\cos k_x+\cos k_y)$ and $q=4$ is the number of nearest
neighbours.

If the
dipole-dipole coupling is small compared to the
exchange interaction, it can be treated in the mean field approximation (see
e.g.
the appendix of \cite{FJKE00} and Section 7.4 of this review). In this case,
the
dipole coupling leads to a renormalization of the magnetic field and one finds
\begin{eqnarray}
B_i^\pm&=&B^\pm +\sum_{j=1}^N g_{ij}\la S_j^\pm\ra T^{|i-j|},\nonumber\\
B_i^z&=&B^z-2\sum_{j=1}^N g_{ij}\la S_j^z\ra T^{|i-j|}\ ;
\label{5.13}
\end{eqnarray}
i.e. there is an enhancement of the transverse fields and a reduction of the
field perpendicular to the film plane.

The lattice sums for a two-dimensional square lattice are given by
($n=|i-j|$)
\begin{equation}
T^n=\sum_{lm}\frac{l^2-n^2}{(l^2+m^2+n^2)^{5/2}}.
\label{5.14}
\end{equation}
The indices $lm$ run over all sites of the $j^{th}$ layer excluding terms with
$l^2+m^2+n^2=0$.

The $3\times 3$
non-diagonal sub-matrices ${\bf \Gamma}_{ij}$ for $j= i\pm 1$ are of the form
\begin{equation}
 {\bf \Gamma}_{ij} = \left( \begin{array}
{@{\hspace*{3mm}}c@{\hspace*{5mm}}c@{\hspace*{5mm}}c@{\hspace*{3mm}}}
-J_{ij}\la S_i^z\ra & 0 & \;\;\;(J_{ij}+D_{ij})\la S_i^+\ra \\
0 & \;\;J_{ij}\la S_i^z\ra & -(J_{ij}+D_{ij)}\la S_i^-\ra \\
\frac{1}{2}J_{ij}\la S_i^-\ra &
-\frac{1}{2}J_{ij}\la S_i^+\ra & 0 \end{array} \right) \;.
\label{5.15}
\end{equation}
Now the system of equations of motion is completely specified.

The case $(K_{ij}\neq 0, D_{ij}=0)$ has been treated in Ref. \cite{FJK00} for a
monolayer and in Ref. \cite{FJKE00} for the multilayer by the eigenvector
method. In this case $\tilde{H}_i^\pm=H_i^\pm$.
In the case ($(K_{ij}= 0, D_{ij}\neq 0)$, treated in \cite{FK03a}, one has
$\tilde{H}_i^\pm\neq H_i^\pm$, which leads to additional dependencies on the
momentum vector ${\bf k}$, requiring a refinement of the treatment.
We discuss these cases separately in the following subsections.

\subsubsection*{4.2.2. Approximate treatment of the single-ion
anisotropy }

For the single-ion anisotropy, one can use eqn
(\ref{3.25}) directly because the
term ${\bf R}^0{\bf L}^0$ turns out to be independent of the momentum {\bf
k}.
The ${+,-,z}$ components of the vector ${\bf C}^{mn}_{\bf k}$ are, however, not
independent, i.e. there are not enough equations to solve for the unknowns.
The remedy is to supplement eqn (\ref{3.25}) with
the regularity conditions (\ref{3.21}).

For illustration, consider the
monolayer. For $D_{ij}=0$ and $K_2\neq 0$ and
$\tilde{H^\pm}=H^\pm$, the eigenvalues of the equation-of-motion matrix
$\bf{\Gamma}$ (\ref{5.11}) and eigenvector matrices ${\bf R}$ and ${\bf L}$
can be determined
analytically. The eigenvalues are $\omega_0=0$, $\omega_\pm=\pm E_{\bf
k}=\pm\sqrt{H^+H^-+H^zH^z}$.
The right eigenvectors are arranged so that the columns 1,2 and 3 correspond
to the eigenvalues 0, $+E_{\bf k}$ and $-E_{\bf k}$ respectively:
\begin{equation}
{\bf R}=\left(\begin{array}{ccc}
\frac{H^+}{H^z} &\frac{-(E_{\bf
k}+H^z)}{H^-}&\frac{(E_{\bf k}-H^z)}{H^-}\\
\frac{H^-}{H^z} &\frac{(E_{\bf
k}-H^z)}{H^+}&\frac{-(E_{\bf k}+H^z)}{H^+}\\
1&1&1
\end{array}\right)\ ,
\label{5.16}
\end{equation}
and the left eigenvectors are arranged in rows 1,2,3 corresponding to the
eigenvalues $0, +E_{\bf k}, -E_{\bf k}$:
\begin{equation}
{\bf L}=\frac{1}{4E_{\bf k}^2}\left(
\begin{array}{ccc}
2H^-H^z&2H^+H^z&4H^zH^z\\
-(E_{\bf k}+H^z)H^-&(E_{\bf
k}-H^z)H^+&2H^-H^+\\
(E_{\bf k}-H^z)H^-&-(E_{\bf
k}+H^z)H^+&2H^-H^+
\end{array}\right)\ .
\label{5.17}
\end{equation}
With the knowledge of ${\bf L^0}$ the regularity conditions (\ref{3.21}) are
\begin{equation}
{\bf L^0A_{-1}}=0=(H^-H^z,H^+H^z,2H^zH^z)\left(\begin{array}{c}
A_{-1}^{+,mn}\\ A_{-1}^{-,mn}\\A_{-1}^{z,mn}
\end{array}\right).
\label{5.18}
\end{equation}
For $m=0,\ n=1$, with $A_{-1}^{+,01}=2\la S^z\ra$,
$A_{-1}^{-,01}=0$ and $A_{-1}^{z,01}=-\la S^-\ra$,
\begin{equation}
\la S^-\ra=\frac{H^-}{H^z}\la S^z\ra=\frac{\Big(\la S^-\ra J(q-\gamma_{\bf
k})+B^-\Big)\la S^z\ra}{B^z+\la S^z\ra J(q-\gamma_{\bf k})+K_22\la S^z\ra
\Big(1-\frac{1}{2S^2}[S(S+1)-\la S^zS^z\ra]\Big)}\ .
\label{5.19}
\end{equation}
Solving for $\la S^-\ra$ and taking the complex conjugate,
\begin{equation}
\la S^\pm\ra=\frac{B^\pm}{B^z+K_22\la S^z\ra
\Big(1-\frac{1}{2S^2}[S(S+1)-\la S^zS^z\ra]\Big)}\la S^z\ra.
\label{5.20}
\end{equation}
Thus, once
$\la S^z\ra $ and $\la S^zS^z\ra$ have been calculated, the
transverse correlations follow from the rgularity condition. Note that the
prefactor $\frac{H^-}{H^z}$ does not depend on the
momentum vector ${\bf k}$.

The lowest spin for which the single-ion anisotropy has an effect is $S=1$
(for $S=1/2$ the anisotropy term is a constant and does not contribute to the
equations of motion). In this case,
only equations of motion with
$(m=0, n=1)$
and $(m=1,n=1)$ are needed to determine the correlations $\la S^z\ra$ and $\la
S^zS^z\ra$.
The regularity conditions with $m+n\leq
2S+1=3$ suffice to express all remaining correlations
as functions
of $\la S^z\ra$ and $\la S^zS^z\ra$ (for more details see Appendix B of Ref.
\cite{FJK00}).

For the monolayer with general spin
$S$, comparison of (\ref{5.19}) with (\ref{5.20}) shows that
\begin{equation}
\frac{H^\pm}{H^z}=\frac{B^\pm}{Z},
\label{A1}
\end{equation}
with
$Z=B^z +K_22\la S^z\ra\Big(1-\frac{1}{2S^2}(S(S+1)-\la S^zS^z\ra)\Big)$.
The regularity conditions (\ref{5.18}) can therefore be written for general
$m,n$ in the form
\begin{equation}
-2ZA_{-1}^{z,mn}=A_{-1}^{+,mn}B^-+A_{-1}^{-,mn}B^+\ .
\label{A2}
\end{equation}
Using the
$z$-component of equation (\ref{3.23}) for the monolayer, one obtains a
relation between the correlations in momentum space

\begin{eqnarray}
\label{A33}
& &
2\frac{B^+B^-}{Z^2}\la(S^z)^m(S^-)^nS^z\ra-\frac{B^-}{Z}\la(S^z)^m(S^-)^n
S^+\ra -\frac{B^+}{Z}\la(S^z)^m(S^-)^nS^-\ra\\
& &=\frac{1}{2}A_{-1}^{+,mn}\frac{E_{\bf k}}{H^z}\frac{B^-}{Z}
\Big[\frac{E_{\bf k}}{H^z}-\coth(\frac{\beta E_{\bf k}}{2})\Big]
+ \frac{1}{2}A_{-1}^{-,mn}\frac{E_{\bf k}}{H^z}\frac{B^+}{Z}
\Big[\frac{E_{\bf k}}{H^z}+\coth(\frac{\beta E_{\bf k}}{2})\Big].\nonumber
\end{eqnarray}
Note that all correlation functions in this equation are written
in a standard form
where powers of $S^z$ are written to the left of the powers of $S^-$:
\begin{equation}
C(m,n)=\langle(S^z)^m(S^-)^n\rangle.
\label{A4}
\end{equation}
The relations $[S^z,(S^-)^n]_-=-n(S^-)^n$ and
$S^-S^+=S(S+1)-S^z-(S^z)^2$ allow us to express all correlations in terms of
the $C(m,n)$:
\begin{eqnarray}
\langle (S^z)^m(S^-)^nS^z\rangle &=&nC(m,n)+C(m+1,n)\ ,\nonumber\\
\langle
(S^z)^m(S^-)^nS^+\rangle&=&\Big(S(S+1)-n(n-1)\Big)C(m,n-1)-(2n-1)C(m+1,n-1)
\nonumber\\ & &-C(m+2,n-1)\ ,\nonumber\\
 \langle (S^z)^m(S^-)^nS^-\rangle&=&C(m,n+1)\ .
\label{A5}
\end{eqnarray}
The inhomogeneities can also be expressed in terms of the
$C(m,n)$ using  binomial series:
\begin{eqnarray}
A_{-1}^{z,mn}&=&-nC(m,n)\ ,\nonumber\\
A_{-1}^{+,mn}&=&\langle\Big[\Big((S^z-1)^m-(S^z)^m\Big)S^-S^++2S^z(S^z-1)^m
+(n-1)(n+2S^z)(S^z)^m
\Big](S^-)^{n-1}\rangle \nonumber \\
& & =S(S+1)\sum_{i=1}^m \left( \begin{array}{c} m\\ i \end{array} \right)
(-1)^iC(m-i,n-1)+(2n+m)C(m+1,n-1)\nonumber\\
& &+\sum_{i=2}^{m+1} \left( \begin{array}{c} m+1\\ i \end{array} \right)
(-1)^{i+1}C(m+2-i,n-1)+n(n-1)C(m,n-1)\ ,
\nonumber \\
A_{-1}^{-,mn}&=&\langle \big[ (S^z+1)^m-(S^z)^m \big](S^-)^{n+1}\rangle
=\sum_{i=1}^m \left( \begin{array}{c} m\\ i \end{array} \right)C(m-i,n+1)\ .
\label{A6}
\end{eqnarray}
The regularity
conditions for all $m$ and $n$ can be written in terms of correlations defined
in the standard form by inserting equation (\ref{A6}) into equation (\ref{A2}):
\begin{eqnarray}
& &2ZnC(m,n)=B^-\Big[
S(S+1)\sum_{i=1}^m \left( \begin{array}{c} m\\ i \end{array} \right)
(-1)^iC(m-i,n-1)\nonumber \\ & &+(2n+m)C(m+1,n-1)
+\sum_{i=2}^{m+1} \left( \begin{array}{c} m+1\\ i \end{array} \right)
(-1)^{i+1}C(m+2-i,n-1)\nonumber \\
& & +n(n-1)C(m,n-1) \Big]
+B^+\sum_{i=1}^m \left( \begin{array}{c} m\\ i \end{array}
\right)C(m-i,n+1)\ \ .
\label{A.7}
\end{eqnarray}

For a given spin $S$ and given values of $C(m,0)$ for $m\leq 2S+1$ this set of
linear equations is solved for $C(m,n>0)$
for all $m+n\leq 2S+1$. The resulting values are then checked for
consistency by insertion into the $2S$ equations (\ref{A33}) using
equations (\ref{A5}) and  (\ref{A6}). The solution consists of
moments $\la(S^z)^p\ra$ ($p$=1,\ldots,2S) for which eqns (\ref{A33}) are
self-consistently fulfilled. Note that
the highest moment $\la (S^z)^{2S+1}\ra$ has
been eliminated in favour of the lower ones through the relation
$\prod_{M_S}(S^z-M_S)=0$.

Multilayers are treated simply by adorning the correlations and
the quantity $Z$ with  a layer index i:
\begin{eqnarray}
Z_i&=&B^z_i
+J_{i,i+1}\la S_{i+1}^{z}\ra+J_{i,i-1}
\la S_{i-1}^{z}\ra\nonumber\\
& &+K_{2,i}2\la S_i^z\ra
\Big(1-\frac{1}{2S^2}[S(S+1)-\la S_i^zS_i^z\ra]\Big)\ .
\label{A.8}
\end{eqnarray}

An alternative method of solution of the present problem is first to eliminate
the null-space by a singular value decomposition (SVD) of  the equation-of-
motion matrix ${\bf\Gamma}$ (\ref{5.11})   and then using
equation (\ref{3.51}) directly. The advantage is to reduce the dimension of
the problem by the number of zero eigenvalues.

The monolayer for $S=1$ is suitable \cite{FK03b} for demonstrating the
procedure because the
SVD of ${\bf \Gamma}$ and the
vector $\tilde{\bf v}$ of eqn (\ref{3.51}) can be obtained analytically.
In order to consider a reorientation of the magnetization in the $xz$-plane, we
put  $B^y=0$, so that $H^\pm=H^x$.
The
${\bf \Gamma}$-matrix can be expressed as a product of three matrices:
\begin{equation}
{\bf \Gamma}={\bf UW\tilde{V}}={\bf uw\tilde{v}}.
\label{5.21}
\end{equation}
Proceeding as in Section 3.5, it is a simple exercise to obtain the three
factors:
\begin{equation}
{\bf W}=\left( \begin{array}{ccc}
               \epsilon_1 &0&0\\ 0&\epsilon_2&0\\ 0&0&0 \end{array}\right);
\ \ \ \ \ \ \ {\bf w}=\left( \begin{array}{cc}
               \epsilon_1 &0\\ 0&\epsilon_2 \end{array}\right),
\label{5.22}
\end{equation}
with $\epsilon_1=\sqrt{H^zH^z+2H^xH^x}$ and
$\epsilon_2=\sqrt{H^zH^z+\frac{1}{2}H^xH^x}$.
We also find
\begin{equation}
{\bf U}=\left( \begin{array}{ccc}
 \frac{-\sqrt{2}}{2}
&\frac{-H^z}{\sqrt{2}\epsilon_2}  &\frac{H^x}{2\epsilon_2}\\
\frac{\sqrt{2}}{2}&\frac{-H^z}{\sqrt{2}\epsilon_2}&\frac{H^x}{2\epsilon_2}\\
0&\frac{H^x}{\sqrt{2}\epsilon_2}&\frac{H^z}{\epsilon_2} \end{array}\right); \ \
\ \ \ \ \
{\bf u}=\left( \begin{array}{cc} \frac{-\sqrt{2}}{2}
&\frac{-H^z}{\sqrt{2}\epsilon_2} \\
\frac{\sqrt{2}}{2}&\frac{-H^z}{\sqrt{2}\epsilon_2}\\
0&\frac{H^x}{\sqrt{2}\epsilon_2}
\end{array}\right),
\label{5.23}
\end{equation}%
and
\begin{equation}
{\tilde{\bf V}}=\left( \begin{array}{ccc}
\frac{-H^z}{\sqrt{2}\epsilon_1}& \frac{-H^z}{\sqrt{2}\epsilon_1}
&\frac{\sqrt{2}H^x}{\epsilon_1}\\
\frac{-\sqrt{2}}{2}& \frac{\sqrt{2}}{2}&0\\
\frac{H^x}{\epsilon_1}&\frac{H^x}{\epsilon_1}&
\frac{H^z}{\epsilon_1} \end{array}\right); \ \ \ \ \
\ \
{\tilde{\bf v}}=\left( \begin{array}{ccc}
\frac{-H^z}{\sqrt{2}\epsilon_1}  &\frac{-H^z}{\sqrt{2}\epsilon_1}
&\frac{\sqrt{2}H^x}{\epsilon_1}\\
\frac{-\sqrt{2}}{2}& \frac{\sqrt{2}}{2}&0
\end{array}\right),
\label{5.24}
\end{equation}
Now everything is specified and one can solve equation (\ref{3.51}).
There is the technical problem that the vectors at neighbouring
$k$-values in general have arbitrary phases but this can be overcome with
the help of the smoothing procedure
described in Section 3.5.

\begin{figure}[htb]
\begin{center}
\protect
\includegraphics*[bb=80 90 540 700,
angle=-90,clip=true,width=10cm]{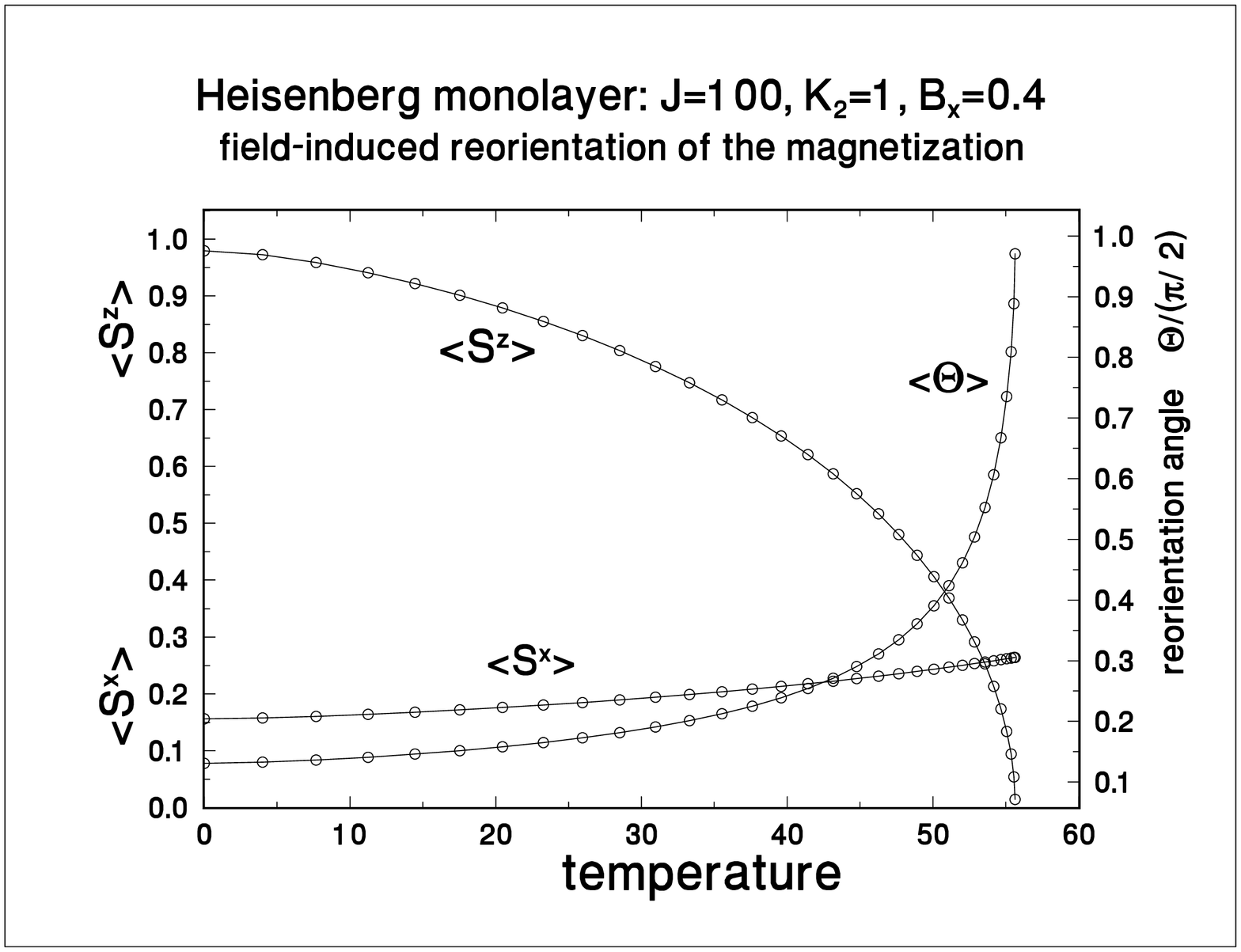}
\protect
\caption{Magnetizations $\la S^z\ra$ and $\la S^x\ra$ and the reorientation
angle $\Theta$ for a spin S=1 Heisenberg monolayer as function of the
temperature.}
\label{fig04}
\end{center}
\end{figure}

In Fig. \ref{fig04} we show as an example the results for the spin $S=1$
Heisenberg
monolayer. It does not matter whether one uses
eqn (\ref{3.25}) or eqn (\ref{3.51}); the results are the same,
as they should be.

This is not the case in later examples (see
Sections 4.2.3 and 4.3.2) where it is necessary to use the
singular value decomposition  to deal with a momentum-dependent
factor ${\bf R}^0{\bf L}^0$.

For spin $S> 1$ and multilayer systems, one can use both methods as long as
${\bf R}^0{\bf L}^0$ is independent of the momentum, but the singular value
decomposition
and the matrix $\tilde{\bf v}$ now have to be determined numerically.

\begin{figure}[htb]
\begin{center}
\protect
\includegraphics*[bb=0 0 97 144,
angle=-90,clip=true,width=12cm]{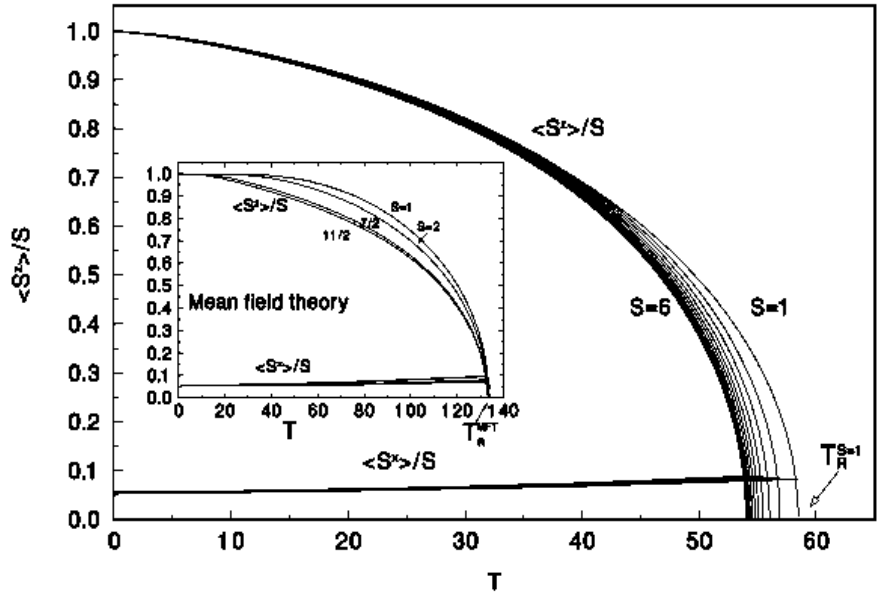}
\protect
\caption{Normalized magnetizations $\la S^z\ra/S$ and
$\la S^x\ra/S $ for a monolayer as functions of the temperature for all
integral and half-integral spin values between $S=1$ and $S=6$ calculated with
Green's function theory. The reorientation temperature $T^S_R$ depends slightly
on S. The inset shows the corresponding results in mean field theory for spins
$S=1, 2, 7/2,$ and $11/2$. The reorientation temperature for MFT does not
depend on $S$.}
\label{fig3}
\end{center}
\end{figure}

\begin{figure}[htb]
\begin{center}
\protect
\includegraphics*[bb=80 90 540 700,
angle=-90,clip=true,width=12cm]{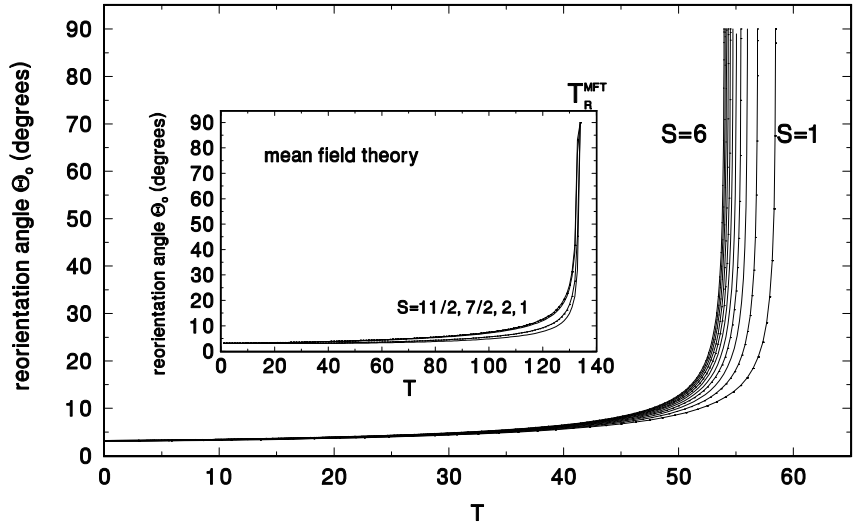}
\protect
\caption{Equilibrium reorientation angle as a function of the
temperature  for the systems of Fig. \ref{fig3} calculated with GFT. The
inset shows the corresponding results for MFT.}
\label{fig4}
\end{center}
\end{figure}

In the next figures, we show further examples from Ref. \cite{FJKE00}.
Fig. \ref{fig3} shows normalized magnetizations $\la S^z\ra/S$ and
$\la S^x\ra/S $ for a monolayer as functions of the temperature for all
integral and half-integral spin values between $S=1$ and $S=6$ calculated with
Green's function theory. The reorientation temperature $T^S_R$ depends slightly
on S. The inset shows the corresponding results in mean field theory for spins
$S=1, 2, 7/2,$ and $11/2$. In this case, the reorientation temperature does not
depend on $S$. Note the very different temperature scale, which is due to the
missing magnon correlations in MFT.
\newpage
\newpage
\newpage
Fig. \ref{fig4} shows the equilibrium reorientation angle as a function of the
temperature  for the systems of Fig. \ref{fig3} calculated with GFT. The
inset shows the corresponding results for MFT.
\newpage

\begin{figure}[htb]
\begin{center}
\protect
\includegraphics*[bb=00 00 82 93,
angle=-90,clip=true,width=12cm]{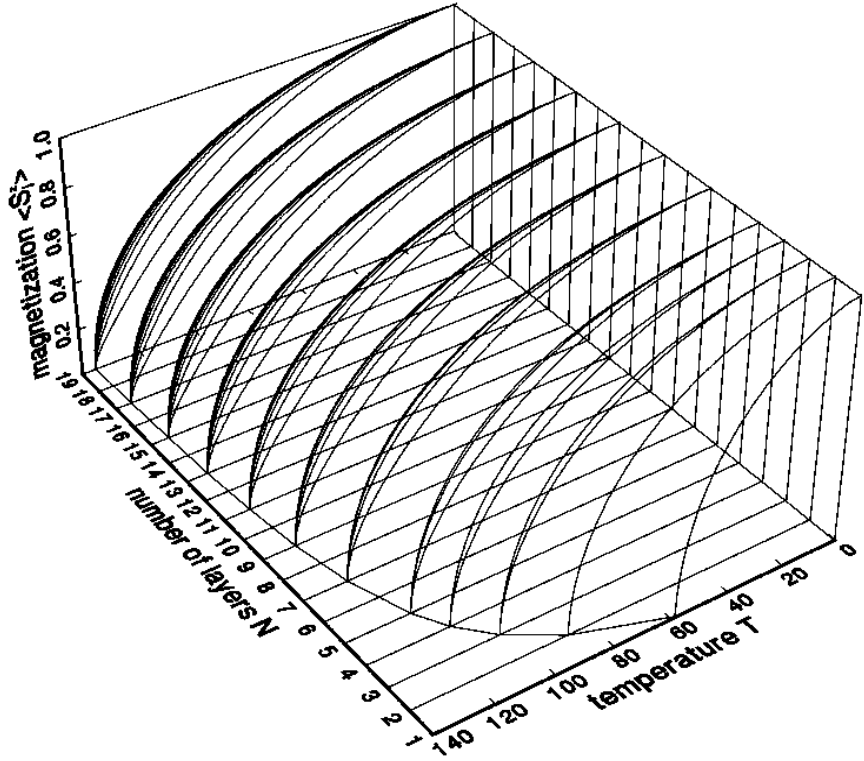}
\protect
\caption{Sublayer magnetizations $\la S_i^z\ra$ as functions of
the temperature for thin ferromagnetic films with $N$ layers and spin $S=1$.
The reorientation temperature $T^N_R$ for the different films can be read off
from the curve in the $N-T$ plane, where $\la S^z_i\ra=0$.}
\label{fig5}
\end{center}
\end{figure}

\newpage
Fig. \ref{fig5} shows the sublayer magnetizations $\la S_i^z\ra$ as functions
of the temperature for thin ferromagnetic films with $N$ layers and spin $S=1$.
The reorientation temperature $T^N_R$ for the different films can be read off
from the curve in the $N-T$ plane, where $\la S^z_i\ra=0$.
\newpage

\begin{figure}[htb]
\begin{center}
\protect
\includegraphics*[bb=80 90 540 700,
angle=-90,clip=true,width=12cm]{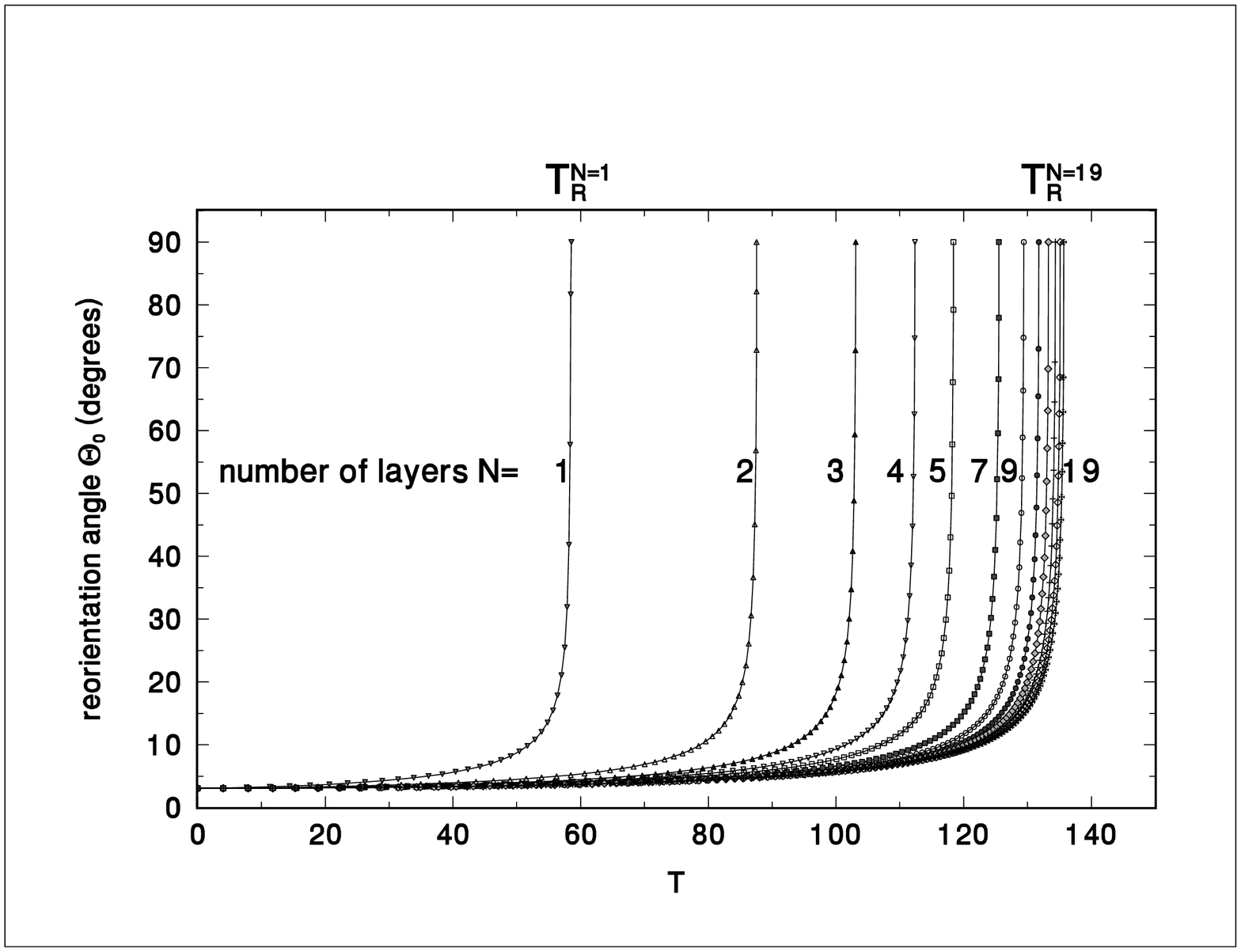}
\protect
\caption{The average equilibrium reorientation angle $\Theta_0$
as a function of the temperature for different film thicknesses. $N$ is
the number of layers in each film and $T^N_R$ are the reorientation
temperatures at $\Theta_0=90^o$.}
\label{fig6}
\end{center}
\end{figure}

In Fig. \ref{fig6}, the average equilibrium reorientation angle $\Theta_0$ is
shown as a function of the temperature for different film thicknesses. $N$ is
the number of layers in each film and $T^N_R$ are the reorientation
temperatures at $\Theta_0=90^o$.

\begin{figure}[htb]
\begin{center}
\protect
\includegraphics*[bb=80 90 540 700,
angle=-90,clip=true,width=12cm]{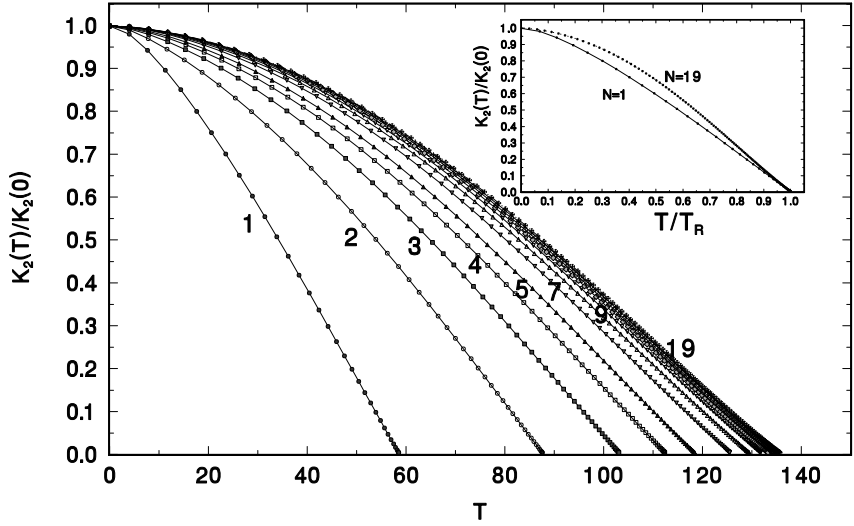}
\protect
\caption{Average effective anisotropy ${\cal K}_2(T)/{\cal K}_2(0)$ as a
function of the temperature and film thickness N. The inset demonstrates the
different functional dependence on T for layers with N=1 and N=19 if the
temperature is scaled with the reorientation temperature.}
\label{fig6a}
\end{center}
\end{figure}

If one is interested in the effective (temperature-dependent) lattice
anisotropy coefficient ${\cal K}_2(T)$, one should not use the thermodynamic
perturbation theory discussed in Section 4.1.6, but rather a
non-perturbative approach in which the free energy is minimized with respect to
the layer-dependent reorientation angles $\theta_i$:
$\partial F_i(T)/\partial \theta_i|_{\theta_{i0}}=0$, where $\theta_{i0}$ are
the equilibrium reorientation angles.
The effective anisotropy of a film consisting of $N$ layers is
\begin{equation}
{\cal K}_2(T)=\sum_{i=1}^N {\cal K}_{2,i}(T)
\end{equation}
with
\begin{eqnarray}
{\cal K}_{2,i}(T)&=&\frac{M_i(T)}{2\sin\theta_{0,i}\cos\theta_{0,i}}\\
& &\Big[\cos\theta_{0,i}(B^x+J_{i,i+1}M_{i+1}(T)\sin\theta_{0,i+1}
+J_{i,i-1}M_{i,i-1}\sin\theta_{0,i-1}+T_i^{\sin})\nonumber\\
& &-\sin\theta_{0,i}(B^z+J_{i,i+1}M_{i+1}(T)\cos\theta_{0,i+1}
+J_{i,i-1}M_{i,i-1}\cos\theta_{0,i-1}-2T_i^{\cos})\Big]\nonumber
\end{eqnarray}
Here $M_i(T)=\sqrt{\la S_i^x\ra^2+\la S_i^z\ra^2}$ and
$\theta_{0,i}=\arctan(\la S_i^x\ra/\la S_i^z\ra)$ are determined from the
magnetization components and
\begin{eqnarray}
T_i^{\sin}=
\sum_{j=1}^N g_{ij}M_j\sin\theta_{0,j}T^{|i-j|}\ ,\nonumber\\
T_i^{\cos}=
\sum_{j=1}^N g_{ij}M_j\cos\theta_{0,j}T^{|i-j|}\ ,
\end{eqnarray}
where $T^{|i-j|}$ are dipole lattice sums, see (\ref{5.14}).

In Fig. \ref{fig6a}, average effective anisotropies
${\cal K}_2(N,T)/{\cal K}_2(N,0)$ of films with different thicknesses N are
shown as functions of the temperature.

Up to now, we have used  $S^+, S^-, S^z$ as the basic operators to
define the Green's functions suitable for treating the reorientation of the
magnetization in the case of uniaxial anisotropies. When there are
anisotropies in all directions of space, it is more natural to start
with the operators $S^x, S^y, S^z$, because this treats the three directions of
space on an equal footing. This is done in Ref. \cite{WDFJK04}, where the
Anderson-Callen decoupling of the single-ion anisotropy terms is invoked for
all
directions of space. A formal advantage is that the equation-of-motion matrix
turns out to be hermitean. Generalising a formula due to Callen \cite{Cal63}
leads to analytical expressions for the first and second moments of the spin
operators. Reorientation transitions and effective
(temperature-dependent) anisotropies are calculated for various 3D and 2D
cases.

The GF theory is used in Ref. \cite{HLT99} to investigate the interplay between
a single-ion easy-plane anisotropy and the dipole-dipole interaction for a
Heisenberg monolayer with the Hamiltonian ($K_2>0$)

\[H=-J\sum_{<ij>}{\bf S}_i{\bf
S}_j+K_2\sum_i(S_i^x)^2-B\sum_iS_i^z+H^{dipole}\].

The Tyablikov decoupling is used for the exchange and the dipole-dipole
interactions and the Anderson-Callen decoupling for the single-ion anisotropy.
An interesting result is that the easy-plane anisotropy alone cannot stabilize
the long-range ferromagnetic order at finite temperatures, one needs the
dipole-dipole interaction in addition in order to do so.

The spin reorientation problem is also investigated in Refs.
\cite{GSL01,GL03}. In these papers, a single-ion anisotropy is used and the
dipole-dipole interaction is approximated by the dipole demagnetization energy.
The exchange interaction and demagnetization energy terms are treated by the
Tyablikov (RPA) decoupling and a decoupling due to Lines \cite{Lin67} is
applied to the single-ion anisotropy terms. However, instead of calculating the
longitudinal and transverse components of the magnetization vector, only the
$z$-component of the magnetization is calculated as a function of the
temperature. For the multi-layer case, the vanishing of the gap in the
corresponding spin-wave
spectrum at a particular temperature is interpreted as the onset of the
reorientation transition. Effective (temperature-dependent) anisotropies are
also calculated within this approximation.

 \newpage

\subsubsection*{4.2.3. Treating the exchange anisotropy}

Although the formalism described in Section 4.2.1 looks very similar for the
single-ion and exchange anisotropy, the direct application of the standard
spectral theorem is not possible because the term ${\bf R}^0{\bf L}^0$ in eqn
(\ref{3.23}) turns
out to be momentum dependent owing to the fact that for the exchange anisotropy
$\tilde{H}_i^\pm\neq H_i^\pm$ in eqn (\ref{5.12}). Thus
the Fourier transform in the second term of eqn (\ref{3.23}) cannot be
performed.

In Ref. \cite{FK03a} we found by intuition a transformation which eliminates
one of the rows of ${\bf R}^0{\bf L}^0$ in the equation
\begin{equation}
{\bf C}={\bf R^1{\cal E}^1L^1A_{-1}}+{\bf R^0L^0C}\ ,
\label{5.25}
\end{equation}
thus allowing the corresponding row to serve as an
integral equation of the eigenvector method.

The transformation is found to be
\begin{equation}
{\bf T}^{-1}=\frac{1}{2}\left(\begin{array}{ccc}
1 & 1 & 0 \\
-1& 1 & 0  \\
0 & 0 & 2
\end{array}\right)\ \ \ \ \ \ \
{\bf T}=\left(\begin{array}{ccc}
1 & -1 & 0\\
1 & 1 & 0 \\
0 & 0& 1
\end{array}\right)
\label{5.26}
\end{equation}
with ${\bf T}^{-1}{\bf T}={\bf 1}$.

Applying this transformation to equation (\ref{5.25})
\begin{equation}
{\bf T}^{-1}{\bf C}={\bf T}^{-1}{\bf R^1{\cal E}^1L^1TT}^{-1}{\bf A_{-1}}+
{\bf T}^{-1}{\bf R}_0{\bf L}_0{\bf TT}^{-1}{\bf C}
\label{5.26a}
\end{equation}
and inserting the analytical eigenvectors ${\bf R}$ and ${\bf L}$
for the monolayer
\begin{equation}
{\bf R}=\left(\begin{array}{ccc}
\frac{H^x}{H^z} &\frac{-(\epsilon_{\bf
k}+H^z)}{\tilde{H}^x}&\frac{(\epsilon_{\bf k}-H^z)}{\tilde{H}^x}\\
\frac{H^x}{H^z} &\frac{(\epsilon_{\bf
k}-H^z)}{\tilde{H}^x}&\frac{-(\epsilon_{\bf k}+H^z)}{\tilde{H}^x}\\
1&1&1
\end{array}\right)\ ,
\label{5.27}
\end{equation}
and
\begin{equation}
{\bf L}=\frac{1}{4\epsilon_{\bf k}^2}\left(
\begin{array}{ccc}
2\tilde{H^x}H^z&2\tilde{H^x}H^z&4H^zH^z\\
-(\epsilon_{\bf k}+H^z)\tilde{H}^x&(\epsilon_{\bf
k}-H^z)\tilde{H}^x&2H^x\tilde{H}^x\\
(\epsilon_{\bf k}-H^z)\tilde{H}^x&-(\epsilon_{\bf
k}+H^z)\tilde{H}^x&2H^x\tilde{H}^x
\end{array}\right)\ ,
\label{5.28}
\end{equation}
the second component of the vector
${\bf T}^{-1}{\bf R}_0{\bf L}_0{\bf TT}^{-1}{\bf C}$ transforms
to zero and one obtains, together with the regularity conditions
(\ref{3.21}), the integral equations for the correlations for each ($m,n$) pair.

The eigenvector method immediately generalizes to the case of $N$ layers if
the transformation $T$ is extended to
$3N$ dimensions  by constructing
$3N\times 3N$-matrices with sub-matrices (\ref{5.26}) on the diagonal.

The intuited transformation (\ref{5.26}) can also be found
systematically enlisting the help of
the singular value decomposition of
the ${\bf \Gamma}$-matrix as described in Sections 3.5 and 4.2.2. This
automatically yields some momentum-independent components of a row vector
$\tilde{{\bf v}}_j$ which enables the Fourier transformation (\ref{3.52}).

This procedure also works for the case of coupled ferro- and
antiferromagnetic layers described in Section 4.3.2.

\begin{figure}[htb]
\begin{center}
\protect
\includegraphics*[bb=80 90 540 700,
angle=-90,clip=true,width=10cm]{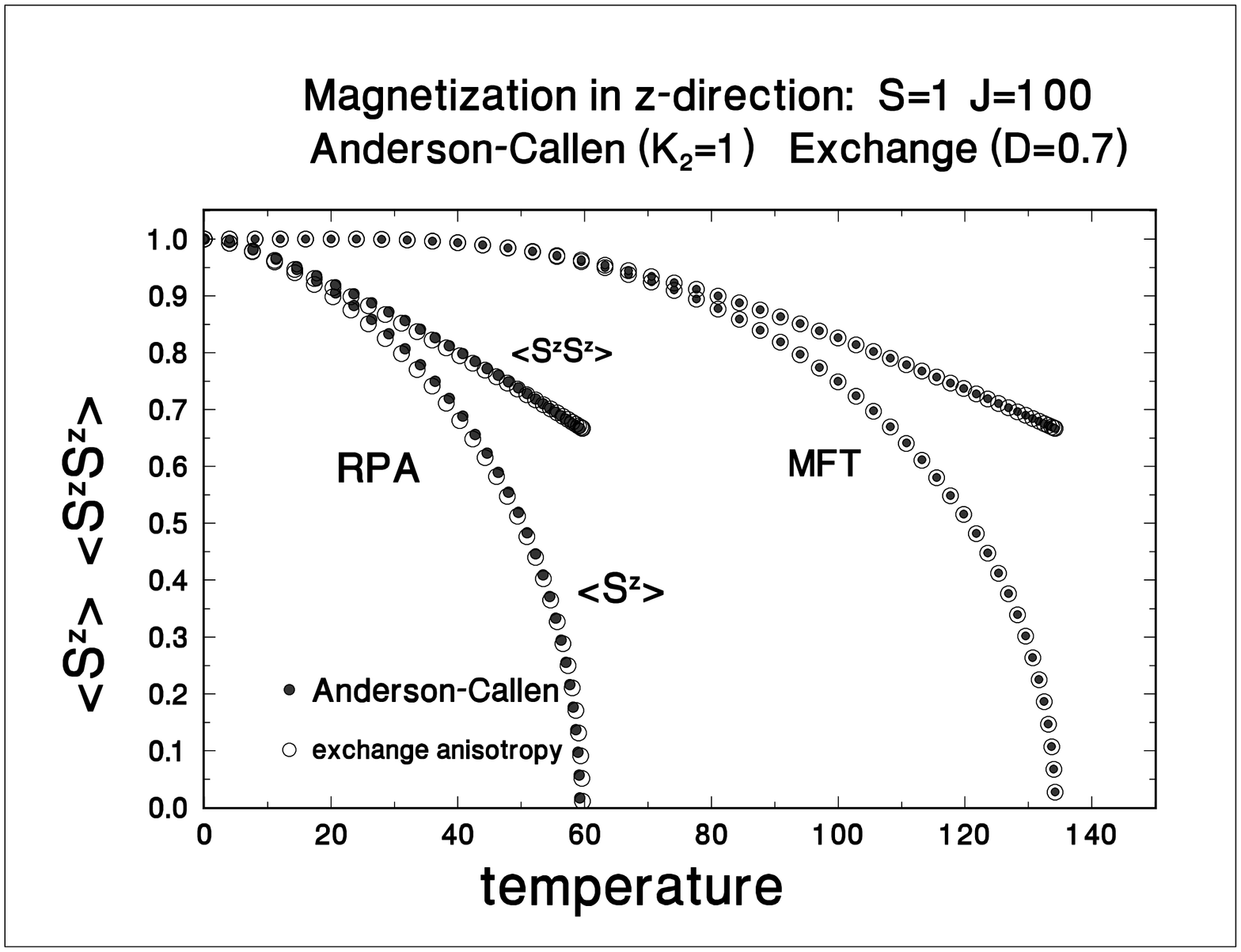}
\protect
\caption{The magnetization $\la S^z\ra$ and its second moment $\la S^zS^z\ra$
of a ferromagnetic spin $S=1$ Heisenberg monolayer for a square lattice as
functions of the temperature, comparing a GFT
calculation using an exchange anisotropy ($D=0$, open circles) with a
single-ion
anisotropy ($K_2=1.0$, solid dots). The corresponding MFT results are also
shown. Note the different Curie temperatures.}
\label{fig7}
\end{center}
\end{figure}

\begin{figure}[htb]
\begin{center}
\protect
\includegraphics*[bb=80 90 540 700,
angle=-90,clip=true,width=10cm]{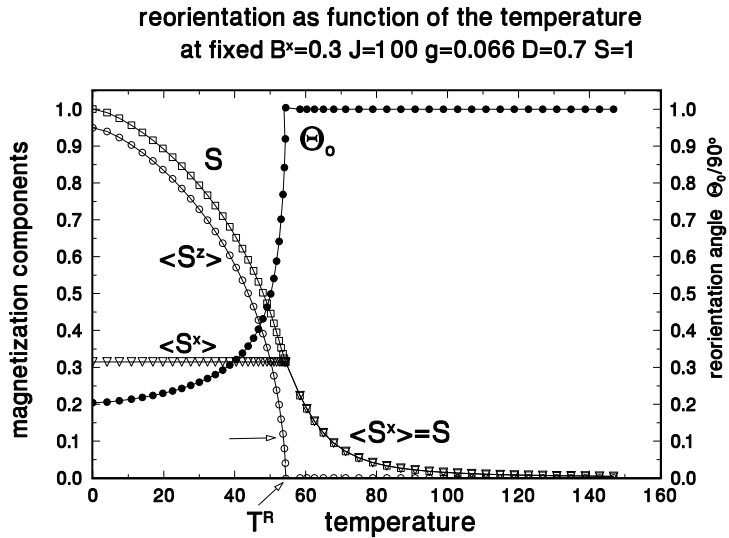}
\protect
\caption{The components of the magnetization $\la S^z\ra$ and
$\la S^x\ra$ and its absolute value $S$ for a fixed magnetic field $B^x=0.3$ as
functions of the temperature for a ferromagnetic spin $S=1$
Heisenberg
monolayer for a square lattice. The exchange interaction strength is $J=100$,
the exchange anisotropy strength is $D=0.7$ and the strength of the magnetic
dipole coupling is $g=0.066$, a value corresponding to ${\rm{Co}}$. Also shown
are the equilibrium reorientation
angle, $\Theta_0$, and the critical reorientation temperature, $T_R$, at which
the in-plane orientation is reached. The small horizontal arrow indicates the
value of $\la S^z\ra$ below which complex eigenvalues occur.}
\label{fig8}
\end{center}
\end{figure}

Some typical results for systems with exchange anisotropy are shown
in Figs. \ref{fig7} and \ref{fig8}. In Fig. \ref{fig7}
the magnetization $\la S^z\ra$ and its second moment $\la S^zS^z\ra$  are
plotted as functions of the temperature for a ferromagnetic spin $S=1$
Heisenberg monolayer for a square lattice with an exchange interaction strength
of $J=100$
and an exchange anisotropy strength of $D=0.7$. It is interesting to note that
there is practically no difference in the magnetization curves when using an
Anderson-Callen decoupled single-ion anisotropy, once its strength is fitted to
an appropriate value, $K_2=1.0$. This makes it difficult to decide which kind
of anisotropy is acting in an actual film.
\newpage
A novel feature occurs with the introduction of the
magnetic dipole
coupling: the eigenvalues and eigenvectors of the
${\bf \Gamma}$-matrix become complex
above a certain temperature, i.e. below a certain value of $\la S^z\ra$.
This behaviour
occurs quite naturally in the theory. It has nothing to do with a damping
mechanism and has to be taken seriously in order to
obtain the results of Fig. \ref{fig8}. Because the ${\bf \Gamma}$-matrix is
real, its
eigenvalues and eigenvectors, if complex, occur pairwise as complex conjugates,
and the integral equations to be solved must be real.

In Fig. \ref{fig8}, the components of the magnetization $\la S^z\ra$
and
$\la S^x\ra$ and the absolute value $S$ for a fixed magnetic field $B^x=0.3$
are shown as
functions of the temperature for a ferromagnetic spin $S=1$ Heisenberg
monolayer for a square lattice. Also shown are the equilibrium reorientation
angle, $\Theta_0$ and the critical reorientation temperature, $T_R$, at which
the in-plane orientation is reached. The small horizontal arrow indicates the
value of $\la S^z\ra$ below which complex eigenvalues occur.
The results for the
exchange anisotropy and the single-ion anisotropy
for spins $S>1$ and multi-layers look very similar,
as seen by comparing references \cite{FJKE00} with \cite{FK03a}.
We therefore do not show the corresponding figures here.

Similar results for general spin S using the
exchange interaction together with the dipole coupling in the mean field
approximation are found in Ref. \cite{WWW04}, but for the monolayer only.
We also mention Ref. \cite{Ilk04}, where the competition of the exchange
anisotropy with the single-ion anisotropy is investigated for a ferromagnetic
S=1 Heisenberg monolayer, and Ref. \cite{TKI06}, where the formalism is applied
to multilayers and higher spin values with parameters that are different at the
surface and the interior of the film. In Ref. \cite{MJB94} the spin
reorientation transition for a ferromagnetic
Heisenberg monolayer with exchange interaction, exchange anisotropy and
dipole dipole interaction is treated with the RPA, where, however, a somewhat
artificial temperature dependence of the exchange interaction had to be used
in order to obtain a favorable comparison with experiment.

\newpage
\subsubsection*{4.2.4. Susceptibilities}
In reference  \cite{JKWO03} Jensen et al.  report
measurements of the parallel and transverse
susceptibilities of a bi-layer Cobalt film having an {\em in-plane} uniaxial
anisotropy. They analyse their results with the help of
a many-body Green's function theory assuming an {\em exchange}
anisotropy and a value for the spin of $S=1/2$.
Here, we generalize their theoretical
model, extending it to multilayers and arbitrary spin. We discuss not
only the exchange exchange anisotropy but also
the {\em single-ion} anisotropy.
 A comparison of the two cases  allows an evaluation
of the robustness of the theoretical conclusions as well as possibly
identifying any qualitative differences which might enable an experiment
to discern which type of anisotropy is acting in a real film.
Accordingly, we investigate the parallel and transverse
susceptibilities for arbitrary spin in multilayer systems. In keeping
with the earlier work \cite{JKWO03}, we
use the Green's function formalism and neglect the dipole-dipole
interaction, since it is nearly isotropic for the in-plane case.
The theory is formulated  in complete analogy to Sections 4.2.2 and 4.2.3
(where
an {\em out-of-plane} magnetization was discussed), the only difference
being that the
applied magnetic fields allow only an {\em in-plane} magnetization. For a
detailed description, refer to references \cite{FK04} and
\cite{FK104}.

The adequate decouplings for the in-plane situation are the same as for the
out-of-plane case:
whereas a RPA decoupling is reasonable for the terms coming from the exchange
interaction and the exchange anisotropy, it leads to incorrect expressions for
the single-ion anisotropy terms. For the latter we therefore use the method
proposed by Anderson and Callen \cite{AC64} at the level of lowest order in the
Green's function hierarchy. This is
certainly an adequate approximation for small anisotropies, as we have shown
in Ref. \cite{HFKTJ02} for the case of an out-of-plane single-ion anisotropy of
a monolayer by comparing with
`exact' Quantum Monte Carlo calculations. In addition to Sections 4.2.2 and
4.2.3, we refer the reader to the literature for a discussion of the
roles of the single-ion- \cite{FJKE00,FJK00,FKS02} and
exchange\cite{FK03a}-anisotropies with
respect to {\em reorientation} of the magnetization of a ferromagnetic film
with an {\em out-of-plane} anisotropy as a function of temperature
and film thickness.

The susceptibilities with respect to the easy ($\chi_{zz}$) and hard
($\chi_{xx}$) axes are calculated as differential quotients
\begin{eqnarray}
\chi_{zz}&=&\Big(\la S^z(B^z)\ra -\la S^z(0)\ra\Big)/B^z\nonumber\\
\chi_{xx}&=&\Big(\la S^x(B^x)\ra -\la S^x(0)\ra\Big)/B^x,
\label{5.29}
\end{eqnarray}
where a value $B^{z(x)}=0.01/S$ turns out to be small enough.
\begin{figure}[htb]
\begin{center}
\protect
\includegraphics*[bb = 80  90 510 410,
angle=0,clip=true,width=8cm]{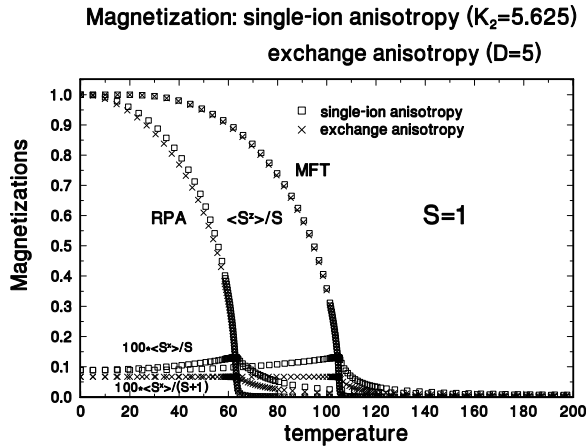}
\protect
\caption{The magnetization $\la S^z\ra/S$  of a
ferromagnetic spin $S=1$ Heisenberg monolayer for a square lattice shown as
a function of the temperature.
Green's function (indicated by RPA) calculations with exchange
anisotropy $D=5$ (crosses) and with single-ion anisotropy ($K_2=5.625$),
(open squares) in the Anderson-Callen approximation are compared.
Also shown  are the
quantities
$100*\la S^x\ra/(S+1)$ for the exchange anisotropy and $100*\la S^x\ra/S$ for
the single-ion anisotropy; the factor 100 is introduced to make the curves
visible.
The corresponding results for
mean field (MFT)  calculations are also displayed.}
\label{sus1}
\end{center}
\end{figure}
\begin{figure}[htb]
\begin{center}
\protect
\includegraphics*[bb = 08  06 178 134,
angle=0,clip=true,width=10cm]{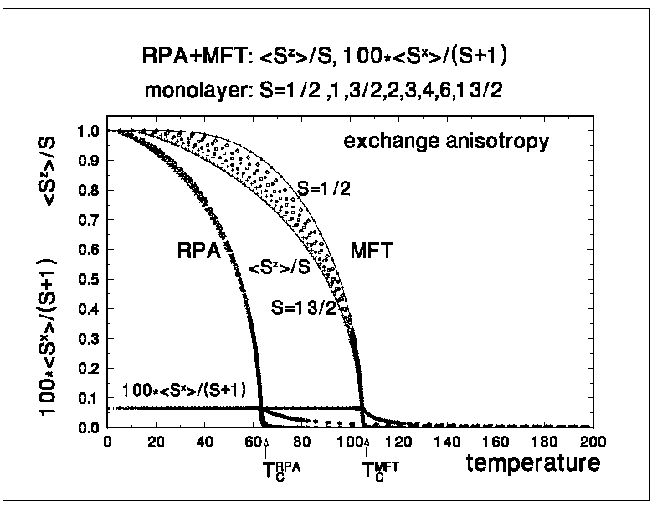}
\protect
\caption{The magnetizations $\la S^z\ra/S$  of
spin $S=1/2,1,3/2,2,3,4,6,13/2$ Heisenberg monolayers for a square lattice
as functions of the temperature, from Ref. \cite{FK04}.
Results from Green's function (RPA) calculations  are compared with
results from mean field theory (MFT) using the exchange
anisotropy strength $D=5$. Also shown is the hard axis magnetization,
which scales to a universal curve $100*\la S^x\ra/(S+1)$,
where the factor 100 is introduced to make the curves visible.}
\label{sus2}
\end{center}
\end{figure}
We compare numerical results obtained with the
single-ion
anisotropy with those from the exchange anisotropy.
As the single-ion anisotropy is not appropriate for $S=1/2$, we show
results for $S\geq 1$. In an attempt to obtain universal curves
(i.e. independent of the spin quantum number S), we scale the
parameters ($B^{x(z)},J,D$) in the Hamiltonian
as $\tilde{B}^{x(z)}/S=B^{x(z)}$, $\tilde{J}/S(S+1)=J$ and
$\tilde{D}/S(S+1)=D$ ($D$ being the strength of the exchange anisotropy).
We also scale the strength of the single-ion anisotropy according to
$\tilde{K_2}/(S-1/2)=K_2$.

In order to compare results obtained with the single-ion anisotropy with
those of
the exchange anisotropy, we set the strength of the single-ion anisotropy to
$K_2=5.625$ for a square lattice monolayer with spin S=1, so that the easy axis
magnetization
$\la S^z\ra/S$ lies as close as possible to the magnetization obtained
with the exchange anisotropy ($D=5$) used in \cite{FK04}. The
exchange interaction parameter is $J=100$ and there is
a small magnetic field in the
$x$-direction, $B^x=0.01/S$, which stabilizes the calculation. The
comparison is shown in Fig. \ref{sus1}.
It is surprising that the results for the easy axis magnetization
$\la S^z\ra$ are very
similar over the whole temperature range although the physical origin for the
anisotropies is very different. An analogous result was observed for the out-of
plane situation discussed in Ref. \cite{FK03a}. For the exchange
anisotropy, the
hard axis magnetization is a constant below the Curie temperature,
whereas for the single-ion anisotropy, it rises
slightly up to the Curie temperature.
In Ref. \cite{FK04}, it is shown analytically that the hard axis magnetization
for the exchange anisotropy
is universal for a scaling $\la S^x\ra/(S+1)$.
For the single-ion anisotropy, a scaling $\la S^x\ra/S$ is found to
be more appropriate.
Comparison
with the corresponding mean field (MFT) calculations, obtained by neglecting
the momentum dependence of the lattice,
shows the well-known shift
to larger Curie temperatures (by a factor of about two for the monolayer with
the present choice of the parameters) owing to the omission of magnon
excitations.

\begin{figure}[htb]
\begin{center}
\protect
\includegraphics*[bb = 80  85 510 410,
angle=0,clip=true,width=8cm]{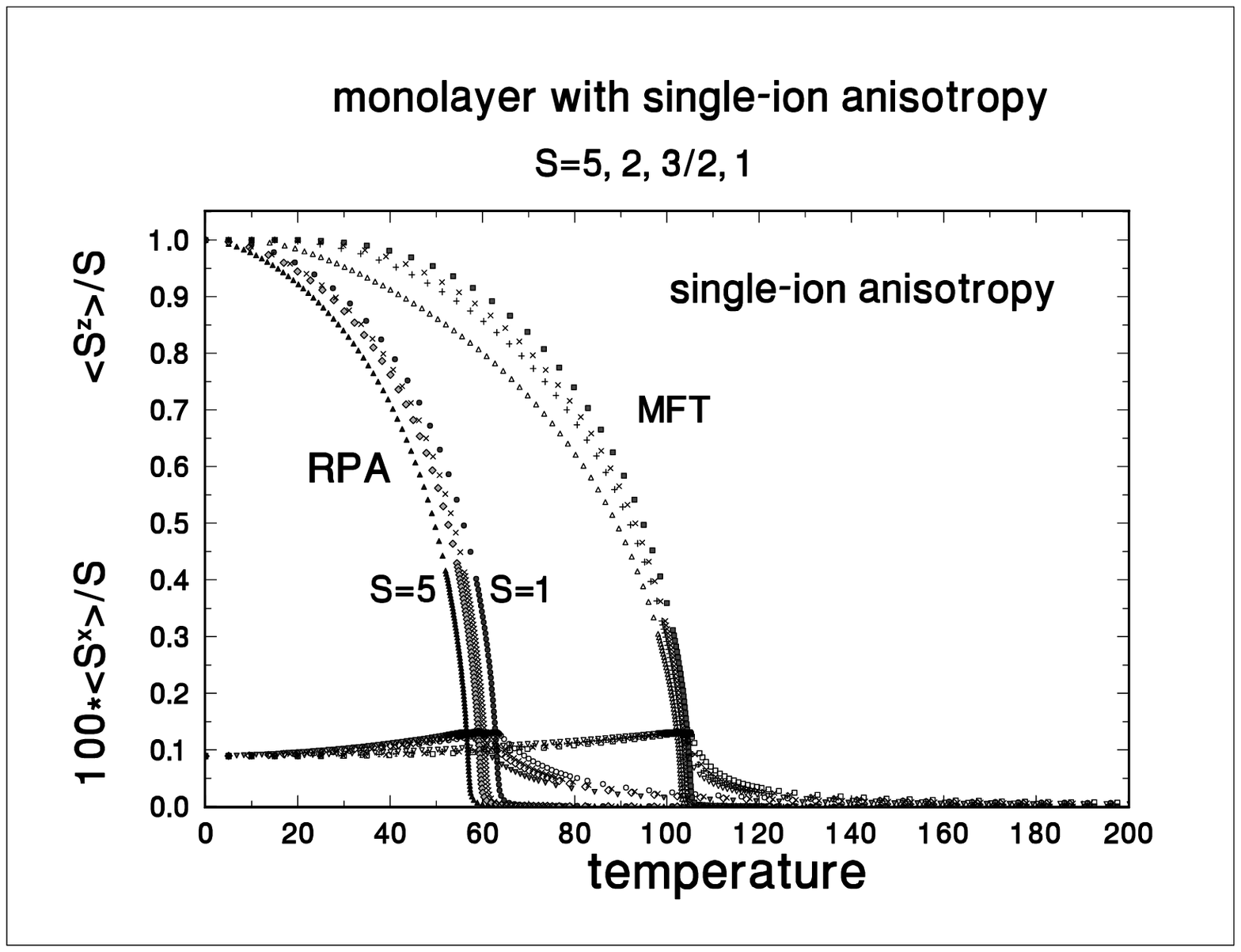}
\protect
\caption{The magnetization $\la S^z\ra/S$  of
ferromagnetic spin $S=1,2,3/2,5$ Heisenberg monolayers for a square lattice
as a function of the temperature
for Green's function (RPA) calculations using  the single-ion
anisotropy strength of $K_2=5.625$
and the corresponding results of
mean field theory (MFT). Also shown are the quantities
 $100*\la S^x\ra/S$;
the factor 100 is introduced to make the curves visible.}
\label{sus3}
\end{center}
\end{figure}
\begin{figure}[htb]
\begin{center}
\protect
\includegraphics*[bb = 80  85 510 410,
angle=0,clip=true,width=8cm]{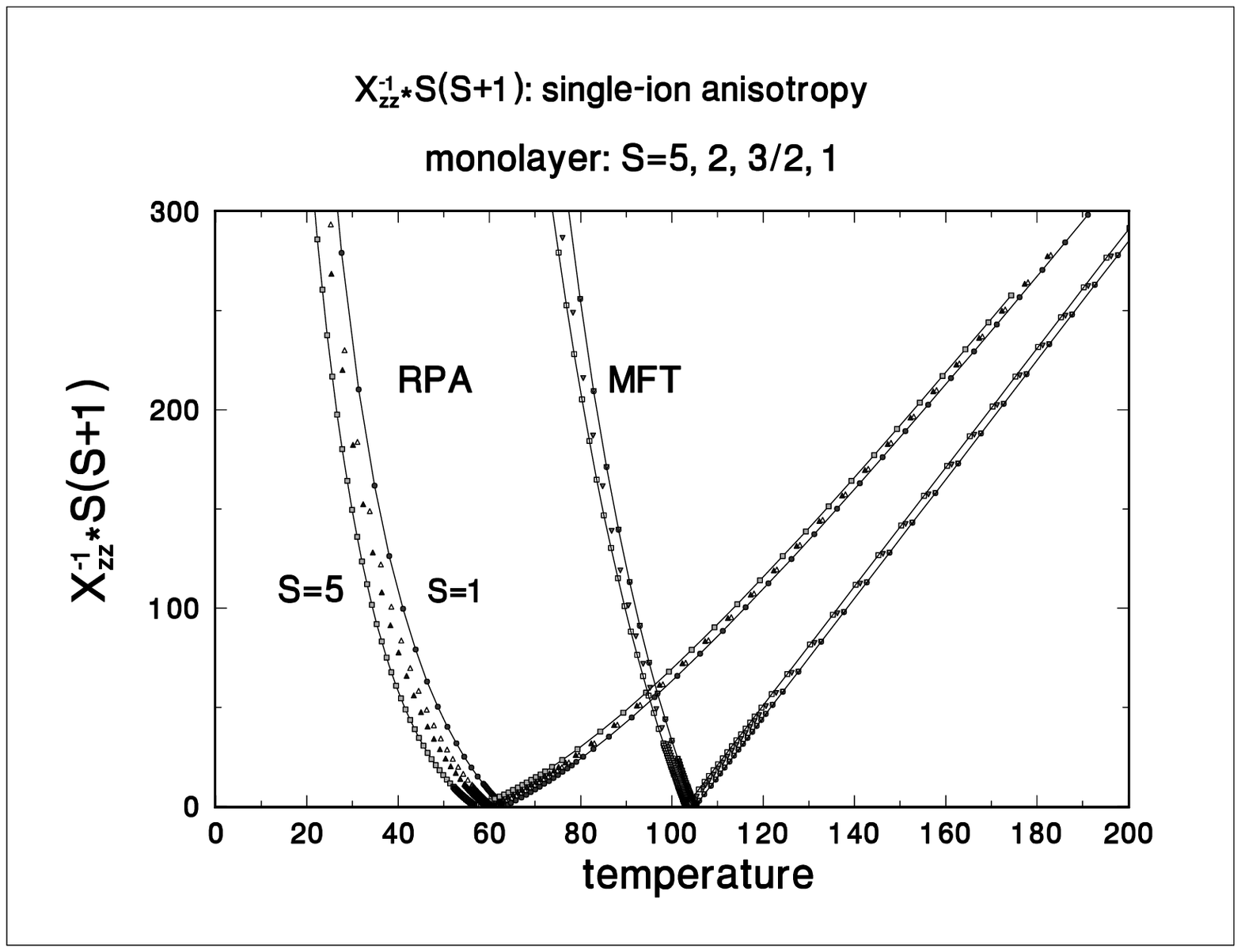}
\protect
\caption{`Universal' inverse easy axis susceptibilities
$\chi_{zz}^{-1}*S(S+1)$ of an in-plane anisotropic
ferromagnetic square lattice Heisenberg monolayer as functions of
the temperature for single-ion anisotropy and
spins $S=5, 2, 3/2, 1$. Compared are
Green's function (RPA)  and
mean field (MFT)  calculations.}
\label{sus4}
\end{center}
\end{figure}
\begin{figure}[htb]
\begin{center}
\protect
\includegraphics*[bb = 80  85 510 410,
angle=0,clip=true,width=8cm]{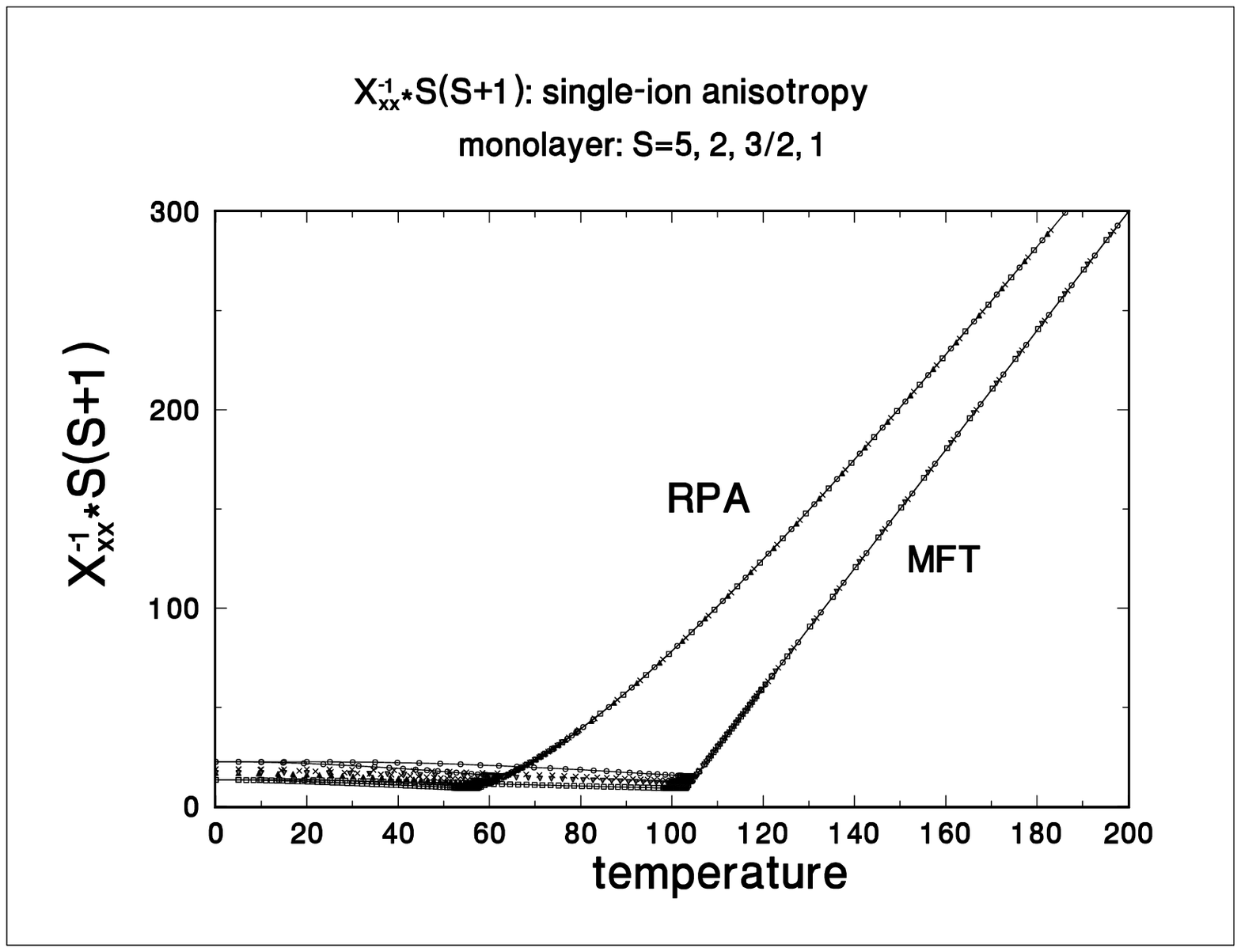}
\protect
\caption{`Universal' inverse hard axis susceptibilities
$\chi_{xx}^{-1}*S(S+1)$ of an in-plane anisotropic
ferromagnetic square lattice Heisenberg monolayer as functions of
the temperature for single-ion anisotropy and
spins $S=5, 2, 3/2, 1$. Compared are
Green's function (RPA)  and
mean field (MFT)  calculations.}
\label{sus5}
\end{center}
\end{figure}
\begin{figure}[htb]
\begin{center}
\protect
\includegraphics*[bb = 80  85 510 410,
angle=0,clip=true,width=8cm]{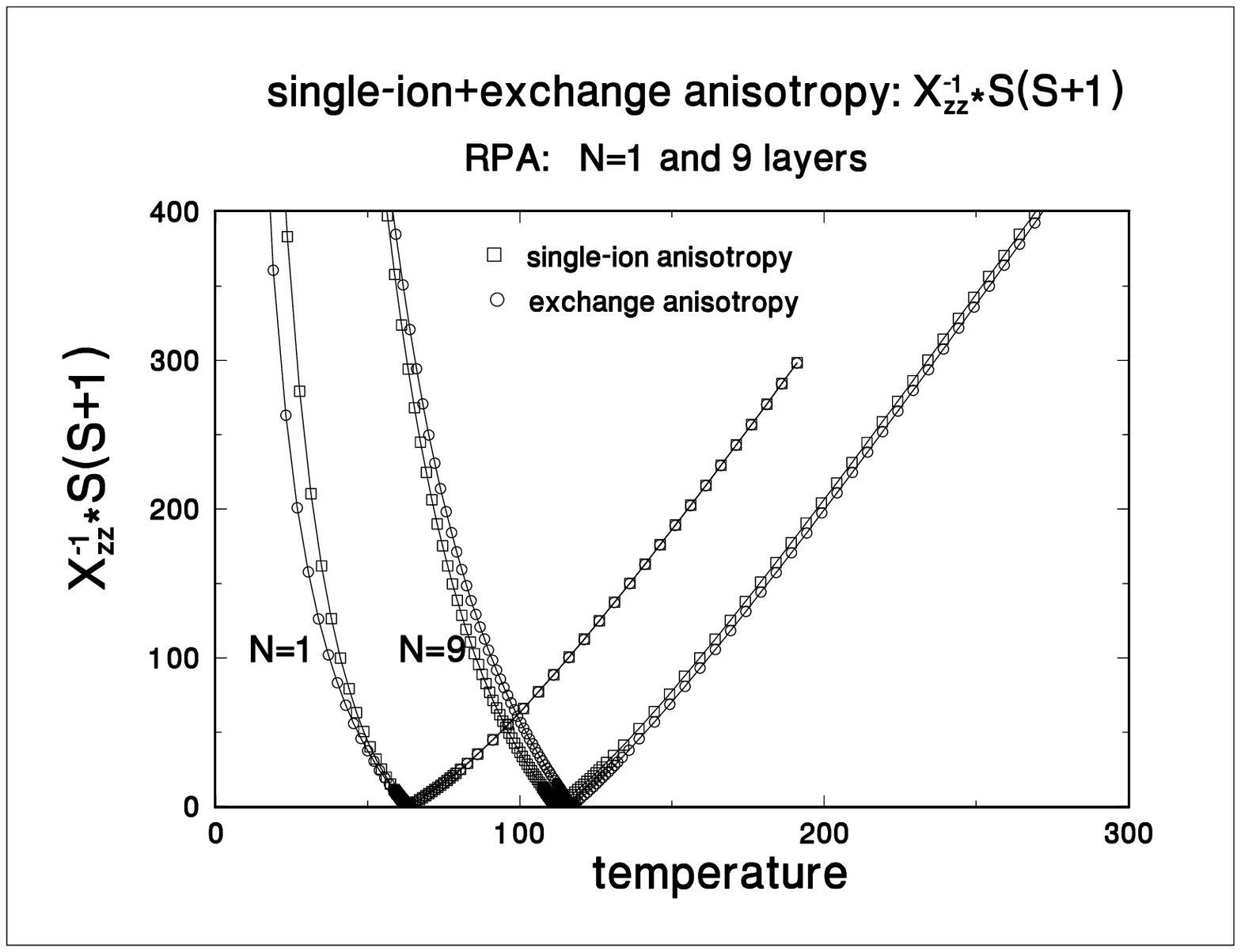}
\protect
\caption{The inverse easy axis susceptibilities $\chi_{zz}^{-1}$ of
ferromagnetic films in RPA
for spin $S=1$ for a monolayer (N=1) and a multilayer (N=9) as functions of
the temperature for single-ion and exchange anisotropies.}
\label{sus7}
\end{center}
\end{figure}

In Figs. \ref{sus2} and \ref{sus3}, we show the easy and hard axes
magnetizations for a monolayer with
different spins $S$. Whereas in Fig. \ref{sus2} one observes a nearly
perfect scaling for
the RPA calculations  with the exchange anisotropy
($S=1/2, 1, 3/2, 2, 3,\ $
$ 4, 6, 13/2$) and a universal Curie temperature
$T_C(S)$  for
RPA and MFT, this is not the case for the corresponding results with
the single-ion anisotropy shown for $S=1, 3/2, 4, 5$ in Fig. \ref{sus3},
although the violation of scaling is not dramatic.

Turning to the inverse easy and hard axes susceptibilities $\chi_{zz}^{-1}$ and
$\chi_{xx}^{-1}$, we find very similar results for the exchange anisotropy and
the single-ion anisotropy. In particular, in the paramagnetic region
($T>T_{\rm Curie}$), the inverse susceptibilities as a function of temperature
are curved
owing to the presence of spin waves, whereas the
corresponding
MFT calculations show a Curie-Weiss (linear in the temperature) behaviour.
There is slightly less universal behaviour
for the single-ion anisotropy (Figs. \ref{sus4} and \ref{sus5}) than for
the exchange anisotropy (Figs.~2 and~3 of Ref. \cite{FK04}).
This is connected with the fact that the exchange anisotropy exhibits
universal values for the Curie temperatures $T_C^{RPA}(S)$ and
$T_C^{MFT}(S)$, which is not strictly the case for the single-ion
anisotropy, (Fig. \ref{sus3}). We were also able to show analytically in
Ref. \cite{FK04} that $\chi_{xx}^{-1}*S(S+1)$ is universal for $T<T_C$ for
the exchange anisotropy; this is not the case for the
single-ion anisotropy.
The only difference is in the curves for the imperfectly scaled
Green's function results for $\chi_{zz}^{-1}$: for the exchange anisotropy, the
curve with the lowest spin value lies to the left of the curves
with the higher spin
values, whereas the converse is true for the exchange anisotropy. This is
not a very pronounced effect and does not lead to a significant
difference between the results for the various anisotropies.
In treating multilayers with the exchange anisotropy in Ref. \cite{FK04},
 we considered only the case  $S=1/2$.
The single-ion anisotropy term in the Hamiltonian is a constant for $S=1/2$;
therefore it is not active when calculating the magnetization,
so we have to use a larger spin here. In the following, we use
spin S=1 as an example but
we also have results for $S>1$ which scale with respect to the
spin in the same way as in the monolayer case.
%
%
The Curie temperatures  for the multilayers $N=2, ..., 19$ (for
N=19 one
is already close to the bulk limit) are only slightly lower for the single-ion
anisotropy than those calculated for the exchange anisotropy.

\begin{figure}[htb]
\begin{center}
\protect
\includegraphics*[bb = 80  85 510 410,
angle=0,clip=true,width=8cm]{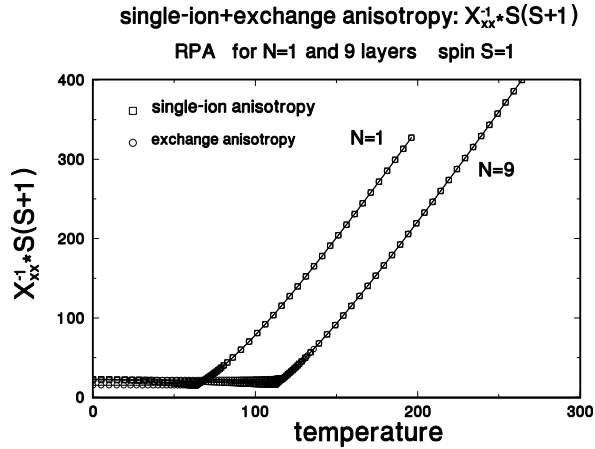}
\protect
\caption{The inverse hard axis susceptibilities $\chi_{xx}^{-1}$ of
ferromagnetic films in RPA
for spin $S=1$ for a monolayer (N=1) and a multilayer (N=9) as functions of
the temperature for single-ion and exchange anisotropies.}
\label{sus8}
\end{center}
\end{figure}

Some results are shown in  Figs. \ref{sus7} and \ref{sus8}.
In order to avoid cluttering the figures, we restrict ourselves to a multilayer
with N=9 layers and spin $S=1$. For $N>9$ the corresponding curves would shift
only
slightly in accordance with the saturation of $T_C$
with
increasing film thickness. We display only the RPA
results for the multilayer
(N=9) and compare with the RPA monolayer (N=1) result. Again, there is no
significant difference in
the results for both anisotropies. We do not plot the corresponding mean field
results, which
are shifted to higher temperatures and, in the paramagnetic region, show only
(a linear in T) Curie-Weiss behaviour, whereas the RPA results have curved
shapes owing to the influence of spin waves, which are
completely absent in MFT.

Although both
kinds of anisotropies are of very different physical origin, it is possible, by
fitting the strengths of the anisotropies properly, to obtain nearly identical
values for the easy axis magnetizations over the complete temperature range for
a spin  $S=1$ monolayer. Using the parameters obtained in this way for
monolayers with
higher spin values and for multilayers, we looked for differences in the
results of calculations
with both kinds of anisotropies.

By using scaled variables we find a fairly
universal behaviour (independent of the spin quantum number S)
of easy and hard axes magnetizations and inverse susceptibilities. Universality
holds better for the exchange anisotropy; e.g. we find a universal
Curie temperature $T_C(S)$ for RPA and MFT. The scaling is not as perfect for
the single-ion anisotropy, but there are {\em no} dramatic deviations which
might enable an experiment to distinguish between the two types of
anisotropies.
It is sufficient to do a
calculation for a particular S and then to apply scaling to obtain the results
for other spin values.
In principle
the measurement of the
hard axis susceptibility together
with the Curie temperature allows one to obtain
information about the parameters of
the model, the exchange interaction and the anisotropy strengths.
One should, however, keep in mind that the quantitative results of the
present calculations correspond to a square lattice. They could change
significantly
for other lattice types. Further changes could result from the
use of layer-dependent exchange
interactions and anisotropies. Such calculations are possible, because the
numerical program is written in such a way that layer-dependent coupling
constants can be used.

A general result is that there are no {\em qualitative} differences
for the calculated observables
(easy and hard axes magnetizations and susceptibilities) between
the single-ion anisotropy on the one hand and the exchange
anisotropy on the other hand.
Therefore, it is not possible for us to propose an
experiment that could decide  which kind of anisotropy is acting in a real
ferromagnetic film.

We mention also a paper by Yablonskyi \cite{Yab91}, who derives analytical
expressions for the static susceptibility and for correlation functions for
ferromagnetic and antiferromagnetic Heisenberg monolayers with general spin (no
anisotropies) on the basis of the Tyablikov (RPA) decoupling.

\newpage .

\subsubsection*{4.2.5. Exact treatment of the single-ion anisotropy}
Up to now we have worked at the level of the lowest-order GF's,
where approximate decoupling schemes lead to closed systems of integral
equations which are solved self-consistently. In this subsection, we
show that a closed system for the terms stemming from the
single-ion anisotropy is attainable without any decoupling
by going to higher-order GF's \cite{FKS02},
generalizing the work of Devlin \cite{Dev71}.
By taking advantage of relations between products of spin
operators \cite{JA2000}, one can show that the
hierarchy of the equations of motion is automatically closed with respect to
the anisotropy terms . In this way, an exact treatment of the
single-ion anisotropy results, i.e. anisotropies of arbitrary strength can
be treated, whereas e.g. the Anderson-Callen deccoupling of the second-order
single-ion anisotropy terms is only reasonable for
anisotropies small compared to the exchange interaction. The terms due to the
exchange interaction must still be decoupled by a generalized RPA.

We develop the general formulation for a spin-Hamiltonian
consisting of an isotropic Heisenberg exchange interaction between nearest
neighbour lattice sites, $J_{kl}$, second-order and
fourth-order
single-ion lattice anisotropies with strengths $K_{2,k}$ and $K_{4,k}$
respectively, a magnetic dipole coupling with
strength $g_{kl}$ and an external magnetic field ${\bf{B}}=(B^x,B^y,B^z)$:
\begin{eqnarray}
{\cal H}&=&-\frac{1}{2}\sum_{<kl>}J_{kl}(S_k^-S_l^++S_k^zS_l^z)\nonumber\\
& &
-\sum_kK_{2,k}(S_k^z)^2-\sum_kK_{4,k}(S_k^z)^4\nonumber\\
& &
-\sum_k\Big(\frac{1}{2}B^-S_k^++\frac{1}{2}B^+S_k^-
+B^zS_k^z\Big) \nonumber\\
& & +\frac{1}{2}\sum_{kl}\frac{g_{kl}}{r_{kl}^5}\Big(r_{kl}^2(S_k^-S_l^+
+S_k^zS_l^z)-3({\bf{S}}_k\cdot{\bf{r}}_{kl})({\bf{S}}_l\cdot{\bf
r}_{kl})\Big)\ ,
\label{FKS1}
\end{eqnarray}

where $S_i^\pm=S_i^x\pm iS_i^y$ and $B^\pm=B^x\pm iB^y$,
$k$ and $l$
being lattice site indices and $\langle kl\rangle$ indicates summation
over nearest neighbours only. We have added  a
fourth-order anisotropy term
for which we  had no decoupling procedure available when working at the
level of the lowest-order GF's.

To allow as general a formulation as possible (with an eye to a future
study of the reorientation of the magnetization), we formulate the
equations of motion for the Green's functions for all spatial
directions:
\begin{eqnarray}
\label{FKS2}
G_{ij}^{+,\mp}(\omega)&=&\la\la S_i^+;S_j^{\mp}\ra\ra_\omega \nonumber \\
G_{ij}^{-,\mp}(\omega)&=&\la\la S_i^-;S_j^{\mp}\ra\ra_\omega  \\
G_{ij}^{z,\mp}(\omega)&=&\la\la S_i^z;S_j^{\mp}\ra\ra_\omega \ . \nonumber
\end{eqnarray}
Instead of decoupling the corresponding equations of motion at this stage,
as we did in our previous work \cite{FJK00,FJKE00}, we add
equations for the next higher-order Green's functions:

\begin{eqnarray}
\label{FKS3}
G_{ij}^{z+,\mp}(\omega)&=&\la\la S_i^zS_i^++S_i^+S_i^z\ra\ra=\la\la
(2S_i^z-1)S_i^+;S_j^{\mp}\ra\ra_\omega \nonumber\\
G_{ij}^{-z,\mp}(\omega)&=&\la\la S_i^zS_i^-+S_i^-S_i^z\ra\ra=\la\la
S_i^-(2S_i^z-1);S_j^{\mp}\ra\ra_\omega \nonumber
\\ G_{ij}^{++,\mp}(\omega)&=&\la\la S_i^+S_i^+;S_j^{\mp}\ra\ra_\omega
\\ G_{ij}^{--,\mp}(\omega)&=&\la\la S_i^-S_i^-;S_j^{\mp}\ra\ra_\omega \nonumber
\\ G_{ij}^{zz,\mp}(\omega)&=&\la\la
(6S_i^zS_i^z-2S(S+1));S_j^{\mp}\ra\ra_\omega \ . \nonumber
\end{eqnarray}
The particular form for the operators used in the definition of the Green's
functions in Eqs.~(\ref{FKS3}) is dictated by expressions coming
from the anisotropy terms. Terminating
the hierarchy of the equations of motion at this level
results in an {\em exact} treatment of the
anisotropy terms for spin $S=1$, since the hierarchy for these terms
breaks off at this stage, as will be shown. The exchange interaction terms,
however, still have to be decoupled, which we do with RPA-like decouplings.

For the treatment of arbitrary spin $S$, it is necessary to use
$4S(S+1)$ Green's functions in order
to obtain an automatic break-off of the equations-of-motion hierarchy
coming from the anisotropy terms. These are functions of the type
$G_{ij}^{\alpha,\mp}$ with $\alpha=(z)^n(+)^m$ and $\alpha=(-)^m(z)^n$, where,
for a particular spin $S$, all combinations of $m$ and $n$ satisfying
$(n+m)=2S$ must be taken into account. There are no Green's functions
having mixed $+$ and $-$ indices because these can be
eliminated by the relation $S^\mp S^\pm=S(S+1)\mp S^z-(S^z)^2$.

Here we treat only the spin $S=1$ monolayer, for which there are
8 exact equations of motion for the Green's functions defined in (\ref{FKS2})
and
(\ref{FKS3}).

The crucial point now is that
the anisotropy terms in these equations can be simplified by using
formulae which reduce  products of spin operators by one order. Such relations
are derived in Ref.~\cite{JA2000}:
\begin{eqnarray}
\label{FKS4}
(S^-)^m(S^z)^{2S+1-m}&=&(S^-)^m\sum_{i=0}^{2S-m}\delta_i^{(S,m)}(S^z)^i,
\nonumber\\
(S^z)^{2S+1-m}(S^+)^m&=&\sum_{i=0}^{2S-m}\delta_i^{(S,m)}(S^z)^i(S^+)^m.
\end{eqnarray}
The coefficients $\delta_i^{(S,m)}$ are tabulated in Ref.~\cite{JA2000}
for general spin.
For spin $S=1$, only the coefficients with $m=0,1,2$ occur:
$\delta_0^{(1,0)}=\delta_2^{(1,0)}=0; \delta_1^{(1,0)}=1, \delta_0^{(1,1)}=0,
\delta_1^{(1,1)}=1, \delta_0^{(1,2)}=1$.

These relations effect the reduction of
the relevant Green's functions coming from the anisotropy terms in
the equations of motion
\begin{eqnarray}
\label{FKS5}
G_{ij}^{(z)^4+,\mp}&=&G_{ij}^{(z)^3+,\mp}=G_{ij}^{(z)^2+,\mp}
=\frac{1}{2}(G_{ij}^{z+,\mp} + G_{ij}^{+,\mp}),\nonumber\\
G_{ij}^{-(z)^4,\mp}&=&G_{ij}^{-(z)^3,\mp}=G_{ij}^{-(z)^2,\mp}
=\frac{1}{2}(G_{ij}^{-z,\mp} + G_{ij}^{-,\mp}),\nonumber\\
G_{ij}^{(z)^2++,\mp}&=&G_{ij}^{z++,\mp}=G_{ij}^{++,\mp},\\
G_{ij}^{--(z)^2,\mp}&=&G_{ij}^{--z,\mp}=G_{ij}^{--,\mp}.\nonumber
\end{eqnarray}
The higher Green's functions coming from the
anisotropy terms are thus expressed in terms of lower-order functions
already present in the hierarchy; i.e. with respect to the anisotropy terms,
a closed system of equations of motion results, so that no decoupling of
these terms is necessary.  In other words, the anisotropy is treated
{\em exactly}. For higher spins, $S>1$, one can proceed
analogously.

No such procedure is available for the exchange interaction terms,
which still have to be decoupled. For spin $S=1$,
an RPA-like approximations  effects the decoupling:
\begin{eqnarray}
\la\la S_i^\alpha S_k^\beta;S_j^\mp\ra\ra&\simeq&
\la S_i^\alpha\ra G_{kj}^{\beta,\mp}+\la S_k^\beta\ra
G_{ij}^{\alpha,\mp}\nonumber \\
\la\la S_k^\alpha S_i^\beta S_i^\gamma;S_j^\mp\ra\ra&\simeq&
\la S_k^\alpha\ra G_{ij}^{\beta\gamma,\mp}+\la S_i^\beta S_i^\gamma \ra
G_{kj}^{\alpha,\mp}.
\label{FKS6}
\end{eqnarray}
Note that we have constructed the decoupling so as not to
break correlations having equal indices, since the corresponding
operators form the algebra characterizing the group for a
spin $S=1$ system.
 For spin $S=1$, this decoupling model leads to 8
diagonal correlations for each layer $i$:
\begin{center}
$\la S_i^+\ra, \la S_i^-\ra, \la S_i^z\ra,$
$ \la S_i^+S_i^+\ra, \la S_i^-S_i^-\ra,
\la S_i^zS_i^+\ra, \la S_i^-S_i^z\ra, \la S_i^zS_i^z\ra.$
\end{center}
These are determined by the $8$ decoupled equations.
Performing in addition a two-dimensional Fourier transformation to momentum
space results in a set of equations of motion which, in compact
matrix notation, are
\begin{equation}
(\omega {\bf 1}-{\bf \Gamma}){\bf G}^{\mp}={\bf A}^{\mp},
\label{FKS7}
\end{equation}
where ${\bf G}^\mp$ and ${\bf A}^\mp$ are 8-dimensional vectors with
components
$G^{\alpha,\mp}$ and $A^{\alpha,\mp}$ and  $\alpha=+,-,z,z+,-z,++,--,zz$;
${\bf 1}$ is the unit matrix. The $8\times 8$ {\it non-symmetric}
matrix ${\bf \Gamma}$ is
\begin{equation}
\scriptsize{
{\bf \Gamma} = \left( \begin{array}
{@{\hspace*{3mm}}c@{\hspace*{5mm}}c@{\hspace*{5mm}}c@{\hspace*{3mm}}
@{\hspace*{3mm}}c@{\hspace*{3mm}}c@{\hspace*{5mm}}c@{\hspace*{3mm}}
@{\hspace*{3mm}}c@{\hspace*{3mm}}c@{\hspace*{3mm}}}
\;\;\;\; H^z_k & 0 & -H^+_k & \tilde{K_2} & 0 & 0 & 0 & 0 \\
 0 & -H^z_k& \;\;\;H^-_k & 0 & -\tilde{K_2} & 0 & 0 & 0 \\
-\frac{1}{2}H^-_k & \;\frac{1}{2}H^+_k & 0 & 0 & 0 & 0 & 0 & 0 \\
\tilde{K_2}-\frac{J_k}{2}\la 6S^zS^z-4\ra & -\la S^+S^+\ra J_k &
\la(2S^z-1)S^+\ra J_k&H^z&0&-H^-&0&-\frac{1}{2}H^+\\
\la S^-S^-\ra J_k & -\tilde{K_2}+\frac{J_k}{2}\la 6S^zS^z-4\ra &-\la
S^-(2S^z-1)\ra J_k & 0 & -H^z & 0 & H^+ & \frac{1}{2}H^- \\
 -\la(2S^z-1)S^+\ra J_k & 0 & 2\la S^+S^+\ra J_k & -H^+ & 0 & 2H^z & 0 & 0 \\
0 & \la S^-(2S^z-1)\ra J_k & -2\la S^-S^-\ra J_k & 0 & H^- & 0 & -2H^z & 0 \\
3\la S^-(2S^z-1)\ra J_k & -3\la(2S^z-1)S^+\ra J_k & 0 & -3H^- & 3H^+ & 0 & 0& 0
 \end{array} \right) \;,}
\label{FKS8}
\end{equation}
with the abbreviations
\begin{eqnarray}
H^\alpha_k&=&B^\alpha+\la S^\alpha\ra J(q-\gamma_{\bf k})\,,
\qquad \alpha=+,-,z \nonumber\\
H^\alpha&=&B^\alpha+\la S^\alpha\ra Jq, \qquad \alpha=+,-,z
\nonumber\\ J_k&=&J\gamma_{\bf k},
\label{FKS9} \\
\tilde{K_2}&=&K_2+K_4.\nonumber
\end{eqnarray}
For a square lattice with a lattice constant taken to be unity,
$\gamma_{\bf k}=2(\cos k_x+\cos k_y)$, and $q=4$ is the number of
nearest neighbours.
For spin $S=1$ and $S=3/2$, the $K_4$ term in the Hamiltonian leads only to a
renormalization of the second-order anisotropy coefficient:
$\tilde{K_2}(S=1)=K_2+K_4$ and $\tilde{K_2}(S=3/2)=K_2+\frac{5}{2}K_4$
respectively. Only in the case of higher spins,
$S\geq 2$, are there non-trivial corrections due to the fourth-
order anisotropy coefficient.

If the theory is formulated only in terms of  ${\bf G^-}$, there is no equation
for determining the $\la S^+S^+\ra$ occuring in the  $\bf \Gamma-$matrix.
It is for this reason that we introduced $G^+$ in Eq.(\ref{FKS2}),
for which the $\bf \Gamma-$matrix is
the same, so that, in general, one can take a linear
combination of ${\bf G^+}$ and ${\bf G^-}$ and their corresponding
inhomogeneities:
\begin{eqnarray}
{\bf G}&=&(1-a){\bf G}^{-}+a{\bf G}^{+},\nonumber\\
{\bf A}&=&(1-a){\bf A}^{-}+a{\bf A}^{+}.\nonumber\\
\label{FKS10}
\end{eqnarray}
Hence, the equations of motion are
\begin{equation}
(\omega {\bf 1}-{\bf \Gamma}){\bf G}={\bf A},
\label{FKS11}
\end{equation}
from which the desired correlations
${\bf C}=(1-a){\bf C}^{-}+a{\bf C}^{+}$
can be determined. The  parameter $a$ is arbitrary ($0<a<1$).

An
examination of the
characteristic equation of the ${\bf \Gamma}$-matrix reveals that 2 of the
eigenvalues are exactly zero, so that the term ${\bf R}^0{\bf L}^0$ is needed
when applying the eigenvector method of Section 3.3.
The eigenvector method then yields for the correlations in
configuration space ($i=1,...,8$):
\begin{equation}
{C}_i=\frac{1}{\pi^2}\int_0^\pi
dk_x\int_0^\pi dk_y \sum_{l=1}^8\Big(\sum_{j=1}^6\sum_{k=1}^6
R_{ij}^1{\cal E}_{jk}^1\delta_{jk}L_{kl}^1{A}_l+\sum_{j=1}^2
R_{ij}^0L_{jl}^0{C}_l\Big).
\label{FKS12}
\end{equation}
Without loss of generality, the field component $B^y$ can be set to zero,
which leads to the symmetry requirements: $\la S^+\ra=\la S^-\ra$,
$\la S^+S^+\ra=\la S^-S^-\ra$ and $\la S^zS^+\ra=\la S^-S^z\ra$; i.e. there
are only 5 independent variables defining  8 correlations ${\bf C}$, i.e.
the system of equations is overdetermined. This problem can be overcome with
a singular value decomposition:
define a vector consisting of the five relevant quantities
\begin{equation}
{\bf v}=
 \left( \begin{array}{c}
\la S^-\ra \\ \la S^z\ra \\ \la S^-S^-\ra\\ \la S^-S^z\ra \\
\la S^zS^z\ra
\end{array} \right).
\label{FKS13}
\end{equation}
Then, the correlations $\bf C$ can be expressed as
\begin{equation}
{\bf C}={\bf u_c^0}+{\bf u_cv}
\label{FKS14}
\end{equation}
with
\begin{equation}
{\bf u_c^0}=
 \left( \begin{array}{c}
2-2a \\ 2a \\ 0 \\ 2-2a \\ -2a \\ 0 \\ 0\\ 0
\end{array} \right)\ ;
{\bf u_c} = \left( \begin{array}
{@{\hspace*{3mm}}c@{\hspace*{5mm}}c@{\hspace*{5mm}}c@{\hspace*{3mm}}
@{\hspace*{3mm}}c@{\hspace*{3mm}}c@{\hspace*{5mm}}c@{\hspace*{3mm}}
@{\hspace*{3mm}}c@{\hspace*{3mm}}c@{\hspace*{3mm}}}
0 & a-1 & a & 0 & a-1 \\
0 & a & 1-a & 0 & -a \\
-a & 0 & 0 & 1 & 0 \\
0 & 1-a & -a & 0 & 3a-3 \\
0 & a & 1-a & 0 & 3a \\
2-2a & 0 & 0 & 2a-2 & 0 \\
0 & 0 & 0 & 2a & 0 \\
6a-4 & 0 & 0 & 6-12a & 0 \\
 \end{array} \right) \;.
\label{FKS15}
\end{equation}
The $8\times5$ matrix ${\bf u_c}$ may be written in terms of its
singular value decomposition:
\begin{equation}
{\bf u_c = U W \tilde{V}},
\label{FKS16}
\end{equation}
where ${\bf W}$ is the $5\times 5$ diagonal matrix of
singular values which here are all $>0$ for $0<a<1$.
${\bf U}$ is an  $8\times 5$ orthogonal matrix
and ${\bf V}$ is a $5\times 5$ orthogonal matrix. From Eqs.~(\ref{FKS12})
and~(\ref{FKS14}) it follows that
\begin{equation}
{\bf u_cv=R^1\;{\bf \cal E\;}^1 L^1\;A\; + R^0L^0(u_cv+u_c^0)-u_c^0}.
\label{FKS17}
\end{equation}
 To get $\bf v$ from this equation, we need only multiply through by
${\bf u_c^{-1}=VW^{-1}\tilde{U}}$, which yields
the system of coupled integral equations
\begin{equation}
{\bf v= u_c^{-1}\Big( R^1\;{\bf \cal E}^1\;L^1\;A +
R^0L^0(u_cv+u_c^0)-u_c^0}\Big), \label{FKS18}
\end{equation}
or more explicitly with $i=1,...,5$
\begin{eqnarray}
v_i&=&\sum_{k=1}^8(u_c^{-1})_{ik}\frac{1}{\pi^2}\int_0^\pi dk_x\int_0^\pi dk_y
\sum_{j=1}^8\Big\{ \sum_{l=1}^6R_{kl}^1{\cal E}_{ll}^1L_{lj}^1A_j\nonumber\\
& &\ \ \  +\sum_{l=1}^2R_{kl}^0L_{lj}^0(\sum_{p=1}^5(u_c)_{jp}v_p+(u_c^0)_j
\Big\} -\sum_{k=1}^8(u_c^{-1})_{ik}(u_c^0)_k.
\label{FKS19}
\end{eqnarray}
This set of equations is not overdetermined (5 equations for 5
unknowns in the
present example ) and is solved by the curve-following method described in
Appendix B.

\begin{figure}[htb]
\begin{center}
\protect
\includegraphics*[bb=16 16 93 120,
angle=-90,clip=true,width=12cm]{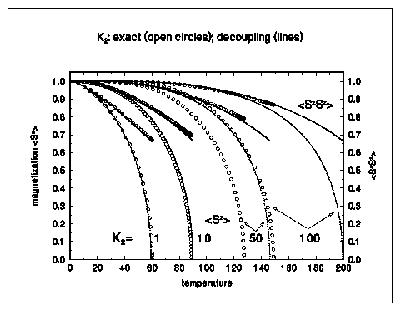} \protect
\caption{The spin $S=1$ monolayer with exchange interaction strength $J=100$.
Comparison of GFT calculations for $\la S^z\ra$ and $\la S^zS^z\ra$ as
functions of the temperature for various anisotropies using the exact treatment
of the anisotropy (open
circles) and the Anderson-Callen decoupling of Section $4.2.1$ (small dots).
}
\label{FKSfig1}
\end{center}
\end{figure}

As an example we investigate the magnetization as a function of the second-order anisotropy
strength and
the temperature for a spin $S=1$ square monolayer,
putting the dipole coupling and the magnetic field equal to
zero. In this case the
magnetization is in the $z$-direction only, $\la S^z\ra$. The results are shown
in
Fig. \ref{FKSfig1} together with those from the Anderson-Callen decoupling.
There is
rather good agreement for small anisotropies, which, however, worsens as
$K_2$ increases.
Another difference concerns the second moments, $\la S^zS^z\ra$, which approach
the value $\la S^zS^z\ra(T\rightarrow T_{Curie})=2/3$ for the Anderson-Callen
decoupling (see Ref.~\cite{FJK00}), whereas
in the exact treatment, the values of
$\la S^zS^z\ra(T\rightarrow T_{Curie})$ are
larger than $2/3$.
Estimates for the Curie temperature, as e.g. in Refs. \cite{EM91a} or
\cite{BM88}, give reasonable values only for small single-ion anisotropies.

\begin{figure}[htb]
\begin{center}
\protect
\includegraphics*[bb=80 90 540 700,
angle=-90,clip=true,width=10cm]{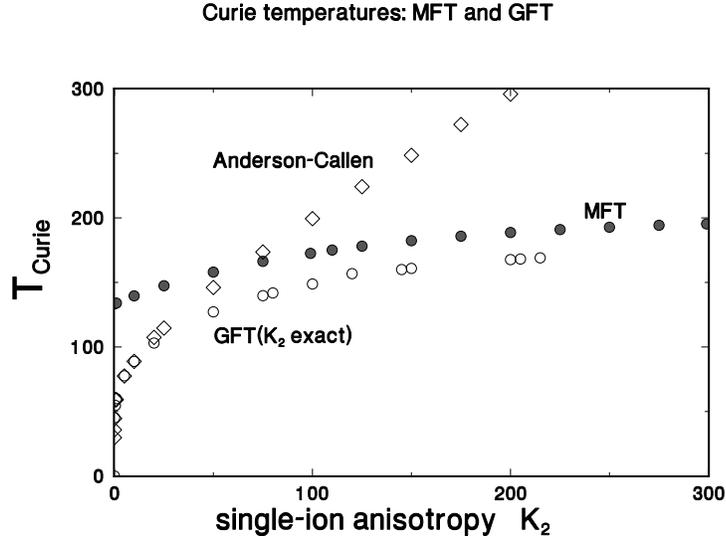}
\protect
\caption{ Comparison of the Curie temperatures calculated with the exact
treatment of the anisotropy, the Anderson-Callen decoupling and MFT. The first
two approaches fulfill the Mermin-Wagner theorem: $T_C\rightarrow 0$ for
$K_2\rightarrow 0$, whereas MFT does not. For large anisotropies, the exact
treatment approaches slowly the MFT result (as also can be shown analytically
\cite{FKS02}), whereas the Anderson-Callen decoupling leads to a diverging
$T_C$}
\label{FKSfig2}
\end{center}
\end{figure}

To show the difference between the new model and the Anderson-Callen
decoupling more clearly, we compare in Fig. \ref{FKSfig2} the Curie
temperatures obtained from MFT,
the Green's function theory with the exact treatment of the anisotropy and the
Green's function theory with the Anderson-Callen decoupling of
Refs.~\cite{FJK00,FJKE00}. For small anisotropy,  there is only
a slight difference between the two GFT results which, in contrast to MFT,
obey the Mermin-Wagner theorem. However, for large
anisotropy, the GFT results deviate from one another significantly:
for $K_2\rightarrow \infty$, the Anderson-Callen
result diverges, whereas the exact treatment approaches the MFT limit. This
is shown analytically in the appendix of Ref. \cite{FKS02}.

Unfortunately, we have not been able to solve the full reorientation problem
with the exact treatment of the single-ion anisotropy with the tools
developed in Section 3.5, because of numerical difficulties.

When using the Anderson Callen decoupling we obtained rather good results when
the external field is in the direction of the anisotropy as long as the
anisotropy is small enough ($K_2\leq 0.1J$). This is seen by
comparing with Quantum Monte Carlo calculations \cite{HFKTJ02}. The
approximation
is much worse when the field is applied perpendicular to the anisotropy.
 A considerable simplification and an
improvement of the results concerning the reorientation
is reported in Ref.
\cite{SKN05}, where the Anderson-Callen decoupling is made in a frame which
is rotated with respect to the original one and in which the magnetization is
in the direction of the new {\em z}-axis.
The reorientation angle is determined from the condition
that the magnetization commutes with the Hamiltonian in
the rotated frame. In this connection see also Ref. \cite{PPS05},
who also apply the approximate Anderson-Callen decoupling in a rotated frame.
In Section 4.4.1 we treat the spin reorientation with an exact treatment of the
single-ion anisotropy by working also in the rotated frame.

\newpage
\subsubsection*{4.2.6. The importance of spin waves in the Co/Cu/Ni trilayer }
\begin{figure}[htb]
\begin{center}
\protect
\includegraphics*[bb=80 90 540
700,angle=-90, clip=false, width=10cm] {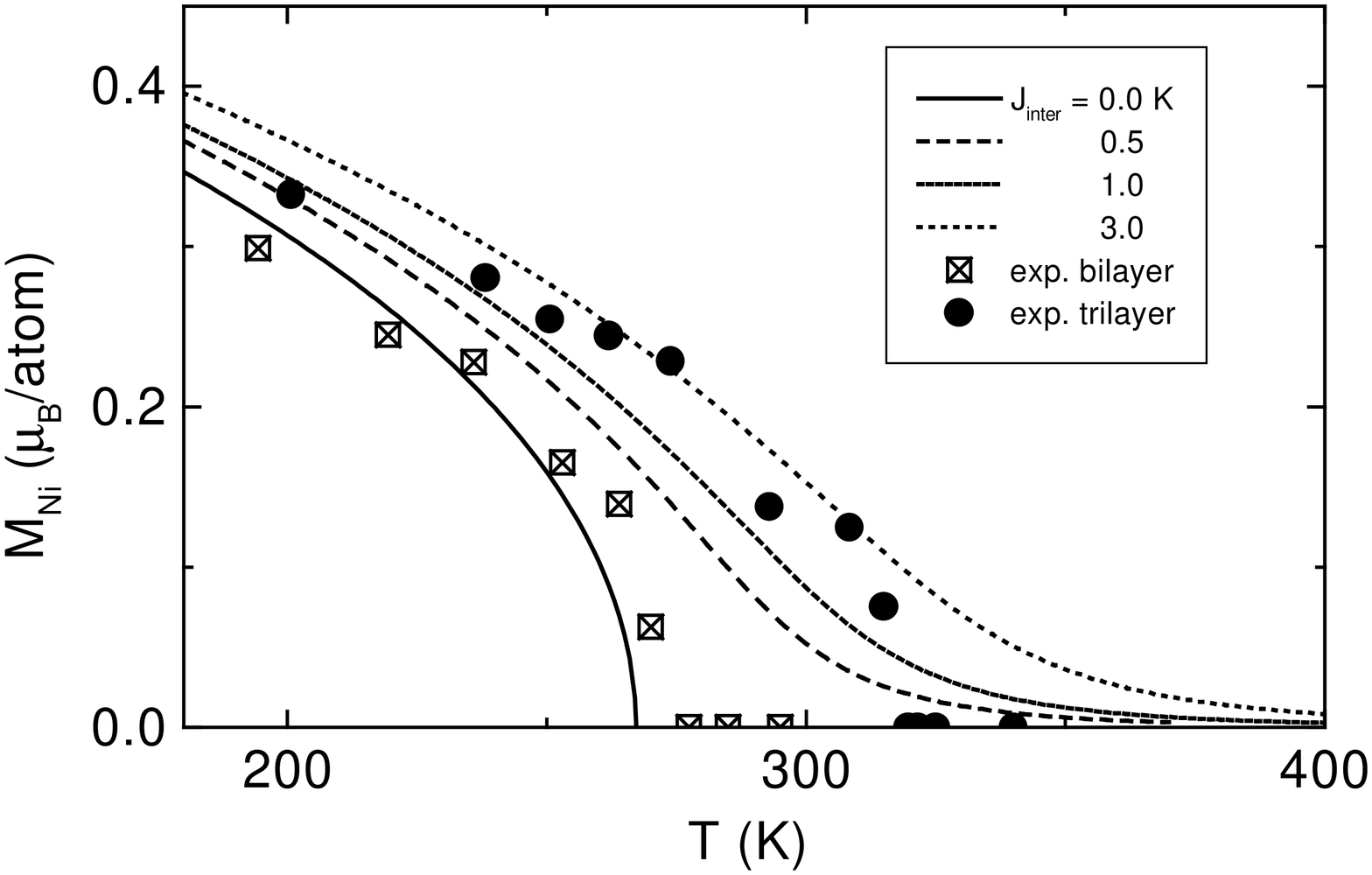}
\protect
\caption{
The measured shift of the Ni magnetization curve for the
trilayer Co/Cu/Ni system (dots) as compared to the Ni magnetization curve for
the bilayer Ni/Cu
system (crosses). Green's function theory (lines) can explain this shift with
realistic
strengths for the interlayer coupling $J_{inter}$ \cite{PJJ99}, whereas MFT
needs unrealistic strong values for $J_{inter}$.}
\label{PRB60f1}
\end{center}
\end{figure}

The importance of spin waves can be demonstrated in Co/Cu/Ni trilayers, where
two magnetic layers are separated by a non-magnetic spacer layer.
In an experiment, the magnetization of Ni in a Ni/Cu bilayer and a Ni/Cu/Co
trilayer is measured as a function of the temperature \cite{PJJ99}.
Figure \ref{PRB60f1} shows
a shift to higher temperatures of the magnetization curve of Ni for the
trilayer system (dots) as compared to the Ni magnetization in the bilayer
system (crosses). This shift is largest at the Curie temperature.
In the figure, results from Green's function theory are also shown.
A Heisenberg exchange
interaction and a dipole-dipole interaction can explain the observed shift with
realistic
strengths \cite{B95} for the interlayer coupling $0.5< J_{inter}<3.0$
\cite{PJJ99}, assuming an in-plane magnetization, whereas MFT (owing
to the neglect of spin waves) needs unrealistic strong values for $J_{inter}$.
For more recent experimental results concerning
Co/Cu/Ni/Cu(100) layers and a comparison with GFT,
see Refs. \cite{JSSBBW05, SSBPBWJ05} and references therein.
\newpage
\subsubsection*{4.2.7. Temperature dependence of the interlayer coupling}
The interlayer coupling between the ferromagnetic layers of Section 4.2.7 is
caused by the spin-dependent reflection of spacer electrons at the
magnetic/non-magnetic interface leading to a spin-dependent interference and to
a renormalisation of the density of states and the free energy within the
non-magnetic spacer. The coupling may then be ferromagnetic or
antiferromagnetic, oscillating with respect to the spacer thickness with a
period  depending on the Fermi surface of the spacer. The
amplitude and the phase of the coupling is determined mainly by the spacer
thickness but may also be influenced
by the interface roughness, disorder etc..

The sources of the temperature dependence of the interlayer coupling are
discussed in Refs. \cite{SN04, SKN05a}:

(i) A part of the temperature dependence is induced by the smearing out of the
Fermi surface of the spacer, as proposed in \cite{BC91}. (`spacer effect')

(ii) The temperature dependence also stems from
altering the properties of the magnetic layers through spin wave
excitations \cite{AMT95}(`magnetic layer effect'), which can affect the
interlayer coupling.

In Ref. \cite{LRKPBBWM02}, ferromagnetic resonance (FMR) experiments lead to an
effective $J_I(T)\sim T^{3/2}$ dependence. Both mechanisms contribute
and the dominant mechanism cannot be deduced directly.
In Ref.\cite{SKN05a}, an alternative analysis of FMR measurements is proposed
that could distinguish between both mechanisms.

\newpage
\subsection*{4.3. Antiferromagnetic and coupled ferromagnetic-
antiferromagnetic Heisenberg films.}
A Green's function theory of antiferromagnetic (AF) and coupled ferro-
and antiferromagnetic
(AF-AFM)-films  relies on periodic structures and therefore requires  the
introduction of sublattices in which periodicity is guaranteed. We start with a
description of an antiferromagnetic
monolayer
in subsection 4.3.1 and follow this in subsection 4.3.2 with a general
formulation in terms
of sublattices, which allows a unified treatment of
FM , AFM and FM-AFM multilayer-systems.

\subsubsection*{4.3.1. The antiferromagnetic spin $S=1/2$ Heisenberg
monolayer.}

According to the Mermin-Wagner theorem \cite{MW66}
the two-dimensional antiferromagnetic or ferromagnetic Heisenberg
monolayers with exchange interaction alone cannot show a finite magnetization.
In order to obtain a finite
magnetization for the antiferromagnet, one can either introduce an artificial
(staggered) field with
opposite directions for the up and down spin sublattices \cite{GHE01} or one
can introduce anisotropies. We use an exchange
anisotropy and demonstrate  how the magnetization of an
antiferromagnet can be calculated with many-body Green's function theory.
The essential step is the introduction of separate sublattices for the up and
down spins.

Consider the  Hamiltonian
\begin{equation}
{\cal{H}}=-\frac{1}{2} \sum_{<kl>}
J_{kl}(S_k^-S_l^++S_k^zS_l^z)-\frac{1}{2}\sum_{<kl>}D_{kl}^zS_k^zS_l^z,
\label{AF1}
\end{equation}
where the exchange interaction and the exchange anisotropy strengths are
negative ($J_{kl}<0$ and $D_{kl}^z<0$).

We only consider the magnetization in $z$-direction. The equation of motion for
the relevant Green's function in energy space
\begin{equation}
G_{ij}^{+-}=\la\la S_i^+;S_j^-\ra\ra
\label{AF2}
\end{equation}
is

\begin{equation}
\omega G_{ij}^{+-}=2\la S^z_i\ra\delta_{ij}+\la\la
[S^+_i,{\cal{H}}];S_j^-\ra\ra. \label{AF3}
\end{equation}
Again, we adopt the Tyablikov (RPA)-decoupling of the higher-order
Green's functions occurring on the right-hand side:
\begin{equation}
\la\la S_i^zS_l^+;S^-_j\ra\ra\approx \la S^z_i\ra\la\la S_l^+;S_j^-\ra\ra\ .
\label{AF4}
\end{equation}
This leads to the equation
\begin{equation}
\big(\omega-\sum_l(J_{il}+D^z_{il})\la S^z_l\ra\big)G_{ij}^{+-}+\la
S^z_i\ra\sum_lJ_{il}G^{+-}_{lj}=2\la S^z_i\ra\delta_{ij}.
\label{AF5}
\end{equation}

We now introduce sublattice
indices ($m,n$) for the up ($u$) and down ($d$) spins. Four
equations
of motion corresponding to the pairs $(i_n,j_m)=(u,u), (d,u), (u,d)$ and
$(d,d)$ result.

Fourier transforms to momentum space for the
sublattices each consisting of $N/2$ lattice sites are

\begin{eqnarray}
& &G_{mn}({\bf k})=\frac{2}{N}\sum_{i_mj_n} G_{i_mj_n}e^{-i{\bf k}({\bf
R}_{i_m}-{\bf R}_{j_n})},\nonumber\\
& &G_{i_mj_n}=\frac{2}{N}\sum_{\bf k} G_{mn}({\bf k})e^{i{\bf k}({\bf
R}_{i_m}-{\bf R}_{j_n})},\nonumber\\
& &\frac{2}{N}\sum_{\bf k}e^{-i{\bf k}({\bf R}_{i_m}
-{\bf R}_{j_n})}=\delta_{i_mj_m},\nonumber\\
& &\frac{2}{N}\sum_{i_m}e^{i({\bf k}-{\bf k'}){\bf R}_{i_m}}=\delta_{{\bf
k}{\bf k'}},
\label{AF6}
\end{eqnarray}
where the subscripts of the Green's functions in momentum space
$G_{mn}({\bf k})$ now denote sublattice indices and not lattice sites.

\begin{figure}[htb]
\begin{center}
\protect
\includegraphics*[bb = 90 100 550 700,
angle=-90,clip=true,width=12cm]{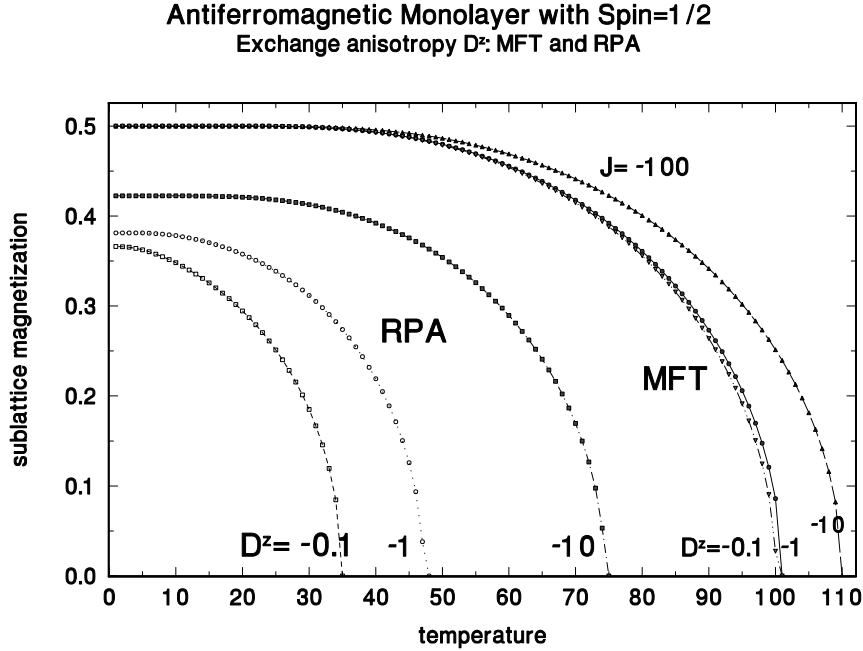}
\protect
\caption{The sublattice magnetization of an antiferromagnetic
Heisenberg monolayer with exchange anisotropy as a function of the
temperature for RPA
and  mean field (MFT) calculations for the parameters $J=-100$ and
$D^z= -0.1, -1.0, -10.0$. }
\label{figAFM1}
\end{center}
\end{figure}

Because  $\la S^z\ra_d=-\la S^z\ra_u$ for an antiferromagnet,
the 4 equations of motion decouple to two identical pairs of equations which
determine $\la S^z\ra_u$ or $\la S^z\ra_d$ respectively.
Before
replacing $\la S^z\ra_d$ by $-\la S^z\ra_u$,
the equations for $G_{uu}^{+-}({\bf k})$ and $G_{du}^{+-}({\bf k})$ are
\begin{eqnarray}
\Big(\omega-\la S^z\ra_u(J_{uu}(0)-J_{uu}({\bf k})+D_{uu}^z(0))&-&\la
S^z\ra_d(J_{ud}(0)+D_{ud}^z(0)\Big)G^{+-}_{uu}({\bf k})\nonumber\\
&+&\la S^z\ra_uJ_{ud}({\bf k})G^{+-}_{du}({\bf k}) =2\la S^z\ra_u\nonumber\\
\Big(\omega-\la S^z\ra_u(J_{du}(0)+D_{du}^z(0))-\la
S^z\ra_d(J_{dd}(0)&-&J_{dd}({\bf k})+D_{dd}^z(0)\Big)G^{+-}_{du}({\bf
k})\nonumber\\
 &+&\la S^z\ra_dJ_{du}({\bf k})G^{+-}_{uu}({\bf k})=0\ .
\label{AF7}
\end{eqnarray}
Restricting the coupling to nearest neighbours only
implies that all interaction terms with equal sublattice indices are zero:
$J_{uu}=D_{uu}^z=J_{dd}=D^z_{dd}=0$. After replacing $\la S^z\ra_d$
by $-\la S^z\ra_u$, the matrix equation is
\begin{equation}
\left( \begin{array}{cc}
\omega+\la S^z\ra_u(J_{ud}(0)+D_{ud}^z(0))\ \ \ \ \la S^z\ra_uJ_{ud}({\bf k})\\
-\la S^z\ra_uJ_{ud}({\bf k})\ \ \ \ \ \ \ \ \ \ \ \ \omega -\la
S^z\ra_u(J_{ud}(0)+D_{ud}^z(0)) \end{array}\right)
\left(\begin{array}{c}
G_{uu}^{+-}({\bf k})\\ G_{du}^{+-}({\bf k})
\end{array}\right)
=\left(\begin{array}{c}
2\la S^z\ra_u\\ 0
\end{array}\right)
\label{AF8}
\end{equation}
For a square lattice with lattice constant $a=1$,
\begin{eqnarray}
J_{ud}({\bf k})=J_{\bf k}&=&\frac{2}{N}\sum_{i_ul_d}J_{i_ul_d}e^{-i{\bf k}({\bf
R}_{i_u}-{\bf R}_{l_d})}=2J(\cos k_x+\cos k_y)\nonumber\\
J_{ud}(0)+D_{ud}^z(0)&=&J_0^z=4(J+D^z)\ .
\label{AF9}
\end{eqnarray}
Eliminating $G_{du}^{+-}({\bf k})$ from the two equations  yields
\begin{equation}
G_{uu}^{+-}({\bf k})=\frac{2\la S^z\ra_u(\omega-\la S^z\ra_uJ_0^z)}{(\omega+\la
S^z\ra_uJ_0^z)(\omega-\la S^z\ra_uJ_0^z)+\la S^z\ra_u^2J_{\bf k}^2}
\label{AF10}
\end{equation}
with the poles
\begin{equation}
\omega_{1,2}=\pm\la S^z\ra_u\sqrt{((J_0^z)^2-J_{\bf k}^2)}
\label{AF11}
\end{equation}
From the spectral theorem, after integrating over the first Brillouin zone
and using the relation $\la S^-S^+\ra_u=1/2-\la S^z\ra_u$ for spin $S=1/2$, the
following equation for the sublattice magnetization $\la S^z\ra_u$ for the
up-spins results:
\begin{equation}
1/2+\frac{1}{\pi^2}\int_0^\pi dk_x\int_0^\pi dk_y \frac{\la
S^z\ra_u^2J_0^z}{\omega_1}\coth(\beta \omega_1/2)=0\ .
\label{AF12}
\end{equation}
This must be iterated to self-consistency in $\la S^z\ra_u$.
Results for $J=-100$ and $D^z=-0.1, 1.0,-10.0$
are shown in figure \ref{figAFM1}.

For the RPA result, the value of
the sublattice magnetization at zero temperature is well below its saturation
value of
$\la S^z\ra_u=1/2$, contrary to the situation for the ferromagnet. This is due
to quantum fluctuations. The mean field limit , obtained by
setting  $J_{\bf k}=0$,
does not show this suppression and also contradicts
the Mermin-Wagner theorem by having a finite N\'{e}el temperature for
$D^z\rightarrow 0$. This theorem ($T_{\rm{N\acute{e}el}}\rightarrow 0$ for
$D^z\rightarrow 0$) is obeyed in the RPA calculation as can be seen by
deriving the
N\'{e}el temperature from equation (\ref{AF12}) by taking first the
limit $\la S^z\ra\rightarrow 0$ and then $D^z\rightarrow 0$.

An analytical approximation to the N\'{e}el
temperature results from
a partial fraction decomposition of the expression obtained
after
expanding the hyperbolic cotangent in eqn (\ref{AF12}) for small sublattice
magnetization. Then the
remaining integrals are expanded around $k_x=k_y=0$ or around
$k_x=k_y=\pi$ repectively with the result
\begin{equation}
T_N\approx \frac{-\pi J}{\ln(1+\frac{\pi^2}{2D^z/J})}.
\end{equation}
The values $T_N(D^z=-0.1)=36.9$, $T_N(D^z=-1.0)=50.6$ and
$T_N(D^z=-10.0)=80.1$  are only slightly higher  (less than 10\%) than the
results of the exact calculations shown in figure \ref{figAFM1}.

The extension to AFM multilayers can be found in Refs. \cite{D91,
WHQW04}.

The two-dimensional spin $S=1/2$ Heisenberg antiferromagnet for a
square lattice with nearest neighbour exchange interaction and dipole-dipole
coupling (no anisotropy) is treated by Pich and Schwabl in Ref. \cite{PS93},
where they use linear spin wave theory by applying the Holstein-Primakoff
transformation \cite{HP40}. They obtain better results
\cite{PS94} for
the N\'{e}el temperature (i.e. closer to experimental data) when applying GFT
along the
lines of Callen \cite{Cal63}. In later papers, they use the same formalism to
treat  two-dimensional
honeycomb antiferromagnets \cite{PS95} and to study the influence of the
dipolar
interaction in quasi-one-dimensional antiferromagnets on a hexagonal lattice
\cite{HPS01}.

\newpage

\subsubsection*{4.3.2. A unified formulation for FM, AFM and FM-AFM
multilayers}

In this section we treat the coupled FM-AFM system in detail \cite{FKJ05},
introducing
sublattices for both the AFM and FM parts. It will then be self-evident that
each part by itself can be described as a special case by choosing the signs of
the parameters appropriately. This shows that FM, AFM and coupled AFM-FM
systems can be handled uniformly within the same formulation.

There is previous work in which Green's
function theory treats the coupling of ferromagnetic layers
to antiferromagnetic layers: in reference \cite{MUH93}, a
bilayer is investigated and
reference \cite{MU93} treats an extension to multilayers. In both cases, only a
collinear magnetization is considered. In reference \cite{WD04}, a
ferromagnetic film is coupled to an antiferromagnetic layer; however, the
orientation
of the magnetization of the antiferromagnet is frozen. Other work considers an
antiferromagnetic coupling between ferromagnetic layers \cite{MU94,L04,NNDN04}.

In our discussion here, we  allow a
non-collinear magnetization, where the
reorientation of the magnetizations of the ferro- and antiferromagnetic layers
is determined by the interlayer coupling as in the MFT approach of
\cite{JKD05}.
We restrict ourselves to Heisenberg systems with spin $S=1/2$ with an exchange
anisotropy. This is not an essential restriction:
references \cite{FK04, FK104} show for ferromagnetic layers
that through an appropriate choice of anisotropy parameters the  exchange- and
single-ion anisotropies yield very similar results and that an appropriate
scaling leads to universal magnetization curves for different spin quantum
numbers. Below, we examine in detail the magnetic
arrangement of the simplest system: a perfectly ordered bilayer
consisting of a FM monolayer that is coupled to an  AFM monolayer.

The starting point is an XXZ-Heisenberg Hamiltonian consisting of an
isotropic Heisenberg exchange
interaction with strength $J_{ij}$ between nearest neighbour lattice sites,
exchange (non-localized)
anisotropies in the $x$- or $z$-directions having strengths $D^x_{ij}$ and
$D^z_{ij}$ respectively and an external magnetic
field ${\bf B}=(B^x,0,B^z)$ confined to the film plane, which is the
$xz$-plane:
\newpage
\begin{eqnarray}
{\cal H}=&-&\frac{1}{2}\sum_{<ij>}J_{ij}(S_i^-S_j^++S_i^zS_j^z)
-\frac{1}{2}\sum_{<ij>}(D^x_{ij}S_i^xS_j^x+D^z_{ij}S_i^zS_j^z)
\nonumber\\
&-&\sum_k\Big(B^xS_i^x+B^zS_k^z\Big).
\label{FA1}
\end{eqnarray}
Again, $S_i^{\pm}=S_i^x\pm iS_i^y$  and $<ij>$ indicates
summation over nearest neighbours only, where $i$ and $j$ are lattice site
indices. Because there is no field perpendicular to the film plane ($B^y=0$),
the reorientation of the magnetization can only occur in the $xz$-plane.
For the FM-AFM bilayer we choose the anisotropy of the ferromagnetic layer in
the $z$-direction, $D_{ij}^z$, and the anisotropy for the antiferromagnetic
layer in the $x$-direction, $D_{ij}^x$.

For $S=1/2$, the required commutator Green's functions are
\begin{equation}
G_{ij}^{\alpha -}(\omega)=\la\la
S_i^\alpha;S_j^-\ra\ra_\omega,
\label{FA2}
\end{equation}
where $\alpha=(+,-,z)$ takes care of all directions in space.
A generalization to spin quantum numbers $S>1/2$ is effected in a
straight-forward way
by introducing $G_{ij}^{\alpha,mn}=\la\la S^\alpha_i;(S_j^z)^m(S_j^-)^n\ra\ra$
with $m+n\leq 2S+1\ (m\geq 0;\ n\geq 1;\  m,n\  {\rm integer})$ as in Section
4.2.1.

The equations of motion for the Green's functions in the energy
representation are
\begin{equation}
\omega G_{ij}^{\alpha -}(\omega)=A_{ij}^{\alpha -}+\la\la
[S_i^\alpha,{\cal H}];S_j^-\ra\ra_{\omega}
\label{FA3}
\end{equation}
with the inhomogeneities
\begin{equation}
A_{ij}^{\alpha -}=\la[S_i^\alpha,S_j^-]\ra=
\left(\begin{array}{c}
2\la S^z_i\ra \delta_{ij}\\
0\\
-\la S^x_i\ra \delta_{ij}
\end{array}
\right),
\label{FA4}
\end{equation}
where $\la ...\ra={\rm Tr}(...e^{-\beta{\cal H}})/{\rm Tr}(e^{-\beta{\cal H}})$
denotes the thermodynamic expectation value.

In order to obtain a closed system of equations, the higher-order Green's
functions on the right hand sides are decoupled as in Section 4.2.1 by a
generalized Tyablikov- (RPA) decoupling
\begin{equation}
\la\la S_i^\alpha S_k^\beta;S_j^-\ra\ra_\eta \simeq\la
S_i^\alpha\ra
G_{kj}^{\beta-}+\la S_k^\beta\ra G_{ij}^{\alpha-} .
\label{FA5}
\end{equation}
After introducing two sublattices per layer,
the resulting equations are Fourier transformed to
momentum space according to eqns. (\ref{AF6}), yielding
\begin{eqnarray}
\omega G_{mn}^{\pm-}&=&\left(\begin{array}{c}2\la S^z_m\ra\delta_{mn}\\ 0
\end{array}\right)\nonumber\\
& &\pm\Big(B^z+\sum_p\la S^z_p\ra(J_{mp}({\bf
0})+D_{mp}^{z}({\bf 0}))\Big)G_{mn}^{\pm-}
\nonumber\\ & &\mp \la S_m^z\ra\sum_p(J_{mp}({\bf k})+\frac{1}{2}D_{mp}^x({\bf
k}))G_{pn}^{\pm-} \nonumber\\
& &\mp\frac{1}{2}\la S^z_m\ra\sum_p D^x_{mp}({\bf k})G^{\mp-}_{pn}\nonumber\\
& &\mp\Big(B^x+\sum_p\la
S^x_p\ra(J_{mp}({\bf 0})+D_{mp}^{x}({\bf 0}))\Big)G_{mn}^{z-}\nonumber\\
& &\pm\la S^x_m\ra\sum_p(J_{mp}({\bf k})+D_{mp}^z({\bf
k}))G_{pn}^{z-},\nonumber\\
\omega G_{mn}^{z-}&=&-\la S^x_m\ra\delta_{mn}
\nonumber\\
& &-\frac{1}{2}\Big(B^x+\sum_p\la
S^x_p\ra(J_{mp}({\bf 0})+D^{x}_{mp}({\bf 0}))\Big)G_{mn}^{+-}\nonumber\\
& &+\frac{1}{2}\la S^x_m\ra\sum_p J_{mp}({\bf k})G^{+-}_{pn}\nonumber\\
& &+\frac{1}{2}\Big(B^x+\sum_p\la
S^x_p\ra(J_{mp}({\bf 0})+D_{mp}^{z}({\bf 0}))\Big)G_{mn}^{--}\nonumber\\
& &-\frac{1}{2}\la S^x_m\ra\sum_pJ_{mp}({\bf k})G_{pn}^{--}.
\label{FA7}
\end{eqnarray}
For a square lattice with lattice constant $a_0=1$, one has
four nearest-neighbour {\em intralayer} couplings with sublattice
indices $n,m$ from the same layer
\begin{equation}
\begin{array}{ll}
J_{mn}(\mathbf{0})=q_0\,J_{mn} \;, &
J_{mn}(\mathbf{k})=\gamma_0(\mathbf{k})\,J_{mn} \;, \\[0.2cm]
D_{mn}^{x,z}(\mathbf{0})=q_0\,D_{mn}^{x,z} \;, &
D_{mn}^{x,z}(\mathbf{k})=\gamma_0(\mathbf{k})\,D_{mn}^{x,z} \;,
\end{array}
\label{FA8}
\end{equation}
with the {\em intralayer} coordination number $q_0=4$ and
the momentum-dependent Fourier factor
\begin{equation} \label{FA9}
\gamma_0({\bf k})= 2(\cos k_x+\cos k_z) \;.
\end{equation}
Correspondingly, for the nearest neighbour {\em interlayer} couplings,
$m$ and $n$  now being sublattice indices from different layers, one has
\begin{equation} \label{FA10}
\begin{array}{ll}
J_{mn}(\mathbf{0})=q_\mathrm{int}\,J_\mathrm{int} \;, &
J_{mn}(\mathbf{k})=\gamma_\mathrm{int}(\mathbf{k})
\,J_\mathrm{int} \;, \\[0.2cm]
D_{m,n}^{x,z}(\mathbf{0})=q_\mathrm{int}\,D_\mathrm{int}^{x,z} \;, &
D_{mn}^{x,z}(\mathbf{k})=\gamma_\mathrm{int}(\mathbf{k})
\,D_\mathrm{int}^{x,z} \;.
\end{array}
\end{equation}
For sc stacking, the {\em interlayer} coordination number and the
corresponding Fourier factor are given by
\begin{equation} \label{FA11}
q_\mathrm{int}=\gamma_\mathrm{int}(\mathbf{k})=1 \;.
\end{equation}

{\noindent}For  fcc or bcc stacking,
\begin{equation}\label{FA11a}
q_\mathrm{int}=4\ \ \  {\rm and}\ \ \
\gamma_\mathrm{int}(\mathbf{k})=4\,\cos(k_x/2)\,\cos(k_z/2).
\end{equation}
The mean field approximation  is obtained by neglecting the Fourier factors,
i.e. $\gamma_0(\mathbf{k})=\gamma_\mathrm{int}(\mathbf{k})=0$.

By choosing the appropriate signs of the exchange interaction and the
exchange anisotropy coupling constants, one can treat ferromagnetic,
antiferromagnetic and mixed systems with coupled FM and AFM layers.

The general formalism is valid for any number of layers and sublattices.
If $Z$ is the total number of sublattices  of the
system, the dimension of the set of equations (\ref{FA7}) is $3Z^2$. We
restrict
ourselves here to the investigation of the bilayer, so that
there are four sublattices and the system of equations
(\ref{FA7}) is of dimension $48$ with a corresponding Green's function vector.
Closer inspection reveals that the system of equations has the following
substructure
\begin{equation}
\left( \omega {\bf 1}-\left( \begin{array}{cccc}
{\bf\Gamma} & 0 & 0 & 0 \\
0 & {\bf\Gamma} & 0 & 0 \\
0 & 0 & {\bf\Gamma} & 0 \\
0 & 0 & 0 & {\bf\Gamma}
\end{array}\right)\right)\left( \begin{array}{c}
{\bf G}_{1} \\ {\bf G}_{2} \\ {\bf G}_3 \\ {\bf G}_{4}
\end{array} \right)=\left( \begin{array}{c}
{\bf A}_{1} \\ {\bf A}_{2} \\ {\bf A}_3 \\
{\bf A}_{4} \end{array}
\right) \ ;
\label{FA12}
\end{equation}
where the diagonal blocks ${\bf \Gamma}$  are identical $12\times 12$
matrices, whose explicit form can be read off from equations (\ref{FA7}). The
sublattice Green's functions ${\bf G}_n\ (n=1,2,3,4)$ are vectors of dimension
$12$ consisting of $4$ subvectors, each of dimension $3$:
\begin{equation}
{\bf G}_n=\left(\begin{array}{c}
{\bf G}_{1n}\\ {\bf G}_{2n}\\ {\bf G}_{3n} \\{\bf G}_{4n}
\end{array}
\right),\ \ {n=1,2,3,4}\ ,
\label{FA13}
\end{equation}
where the 3-component vectors are
\begin{equation}
{\bf G}_{mn}=\left(\begin{array}{c}
{\bf G}_{mn}^{+-}\\ {\bf G}_{mn}^{--}\\ {\bf G}_{mn}^{z-}
\end{array}
\right), \ {m=1,2,3,4}\ .
\label{FA14}
\end{equation}
The inhomogeneity vectors have the same structure:
\begin{equation}
{\bf A}_n=\left(\begin{array}{c}
{\bf A}_{1n}\delta_{1n}\\ {\bf A}_{2n}\delta_{2n}\\ {\bf A}_{3n}\delta_{3n}
\\{\bf A}_{4n}\delta_{4n} \end{array}
\right),\ \ {\bf A}_{nm}=\left(\begin{array}{c}
2\la S^z_m\ra\\ 0 \\ -\la S^x_m\ra
\end{array}
\right), \ \ \ \ \ {m,n=1,2,3,4}\ .
\label{FA15}
\end{equation}
The big equation (\ref{FA12}) of dimension 48 for the bilayer can therefore be
replaced by 4 smaller equations of dimension 12:
\begin{equation}
(\omega{\bf 1}-{\bf \Gamma}){\bf G}_n={\bf A}_n\ \ \ \ {\rm for}\  n=1,2,3,4\
. \label{FA17}
\end{equation}
It turns out that the $12\times 12$ ${\bf \Gamma}$-matrix has 4 zero eigenvalues.
In this case we can use the formalism of Section 3.5, where the
singular value decomposition of ${\bf \Gamma}$ leads to a system of integral
equations for the correlations ${\bf C}_n({\bf k})$ corresponding to the GF's
${\bf G}_n$ (see eqn (\ref{3.51})):
\begin{equation}
0=\int d{\bf k}\Big({\bf r{\cal E}^1l\tilde{v}A}_{-1,n}-{\bf
\tilde{v}C}_n({\bf k})\Big)\ \ \ \ \ n=1,2,3,4.
\label{FA18}
\end{equation}
Section 3.5 explains how to find a ${\bf k}$-independent vector
${\bf \tilde{v}}$ having a layer structure, i.e.
${\bf \tilde{v}}=(0,..,0,{\bf \tilde v}_n,0,.,,0)$.
In this way,
the non-diagonal correlations disappear from those rows in equation
(\ref{FA18})
corresponding to $\tilde{\bf v}_n$ and the ${\bf k}$-integration can be
performed:
$\int d{\bf k}\tilde{\bf v}{\bf C}_n({\bf k})$=
$\tilde{\bf v}\int d{\bf k}{\bf C}_n({\bf k})=\tilde{\bf v}{\bf C}_n$. In the
present case
$\tilde{\bf v}_n$ is given by
\begin{eqnarray}\label{FA19}
\tilde{\bf v}_n
&=&\Big((\frac{1}{\sqrt{2}},-\frac{1}{\sqrt{2}},1)\delta_{1n},
(\frac{1}{\sqrt{2}},-\frac{1}{\sqrt{2}},1)\delta_{2n},\nonumber\\
& &(\frac{1}{\sqrt{2}},-\frac{1}{\sqrt{2}},1)\delta_{3n},
(\frac{1}{\sqrt{2}},-\frac{1}{\sqrt{2}},1)\delta_{4n}\Big)\ ,\ \ \
n=1,2,3,4.
\end{eqnarray}
Putting equation (\ref{FA19}) into equation (\ref{FA18}) yields 4 equations
which contain the
8 magnetization components implicitly. The necessary additional 4 equations are
obtained from the regularity conditions (\ref{3.52a})
\begin{equation}
\int d{\bf k}{\bf L}_0{\bf A}_n=\int d{\bf k}\tilde{\bf u}_0{\bf A}_n=0\ , \ \
\ \  n=1,2,3,4 , \label{FA20}
\end{equation}
which are obtained from the regular behaviour of the commutator
Green's functions at the origin. The $\tilde{\bf u}_0$ are
the
eigenvectors of the singular value decomposition spanning the null-space of the
matrix ${\bf \Gamma}$. The resulting 8 integral
equations are solved self-consistently by the curve-following method described
in detail in the Appendix B.
Note that the $\tilde{\bf u}_0$ are determined numerically only
up to an orthogonal transformation. To ensure proper behaviour as a function of
${\bf k}$, $\tilde{\bf u}_0$ must be calibrated at each ${\bf k}$. A procedure
for doing this is indicated in Section 3.5 and presented in detail in an
appendix of reference \cite{FK05}.

We now present results for the bilayer ferromagnet, the
bilayer antiferromagnet and the coupled ferro- and antiferromagnetic bilayer.
All calculations are for an in-plane orientation of the spins of both layers.
In each case we compare the results of Green's function theory (GFT) with those
of mean field theory (MFT) obtained by putting the
momentum-dependent terms equal to zero.
In order to see the effects of the interlayer coupling most clearly, we use
different exchange interaction strengths for each layer:

(a) FM-FM: $J_{1{\rm FM}}=100, J_{2{\rm FM}}=50$,

(b) AFM-AFM: $J_{1{\rm AFM}}=-100, J_{2{\rm AFM}}=-50$,

(c) FM-AFM: $J_{{\rm FM}}=100, J_{{\rm AFM}}=-50$.

{\noindent}Because of the Mermin-Wagner theorem \cite{MW66},
anisotropies are required in the Green's function description: we take
$D^z=+1.0$ for FM layers and $ D^x=-1.0$ for AFM layers.  These values are
appropriate for 3d transition metal systems.
For a compensated interface, the magnetizations of the FM and AFM layers
are almost orthogonal to each other even at $T=0$ because of the
interface exchange
interaction $J_{\rm int}$. We choose the FM magnetization to be oriented in the
$z$-direction and the AFM magnetization in the $x$-direction.
Our particular choice of the anisotropies supports this arrangement not
only at $T=0$ but also at finite temperatures. For other choices of
anisotropies, the magnetic arrangement could be different.
The
interlayer coupling is assumed to be positive for the ferromagnetic bilayer
and negative for the
antiferromagnetic bilayer. For the coupled FM-AFM system, both signs
are used.
We consider three interlayer coupling constants with strength
$J_{\mathrm int}=30,\ 75,\ 160$, one smaller
than the weakest exchange interaction, one larger than the strongest exchange
interaction and one in between.

\newpage
\subsubsection*{ The ferromagnetic and the antiferromagnetic bilayers}
Results for the FM and AFM bilayers are presented in this subsection
in order to have a basis for discussing the differences from the coupled
FM-AFM bilayer.

\begin{figure}[htb]
\begin{center}
\protect
\includegraphics*[bb = 70 110 430 590,
angle=-90,clip=true,width=13cm]{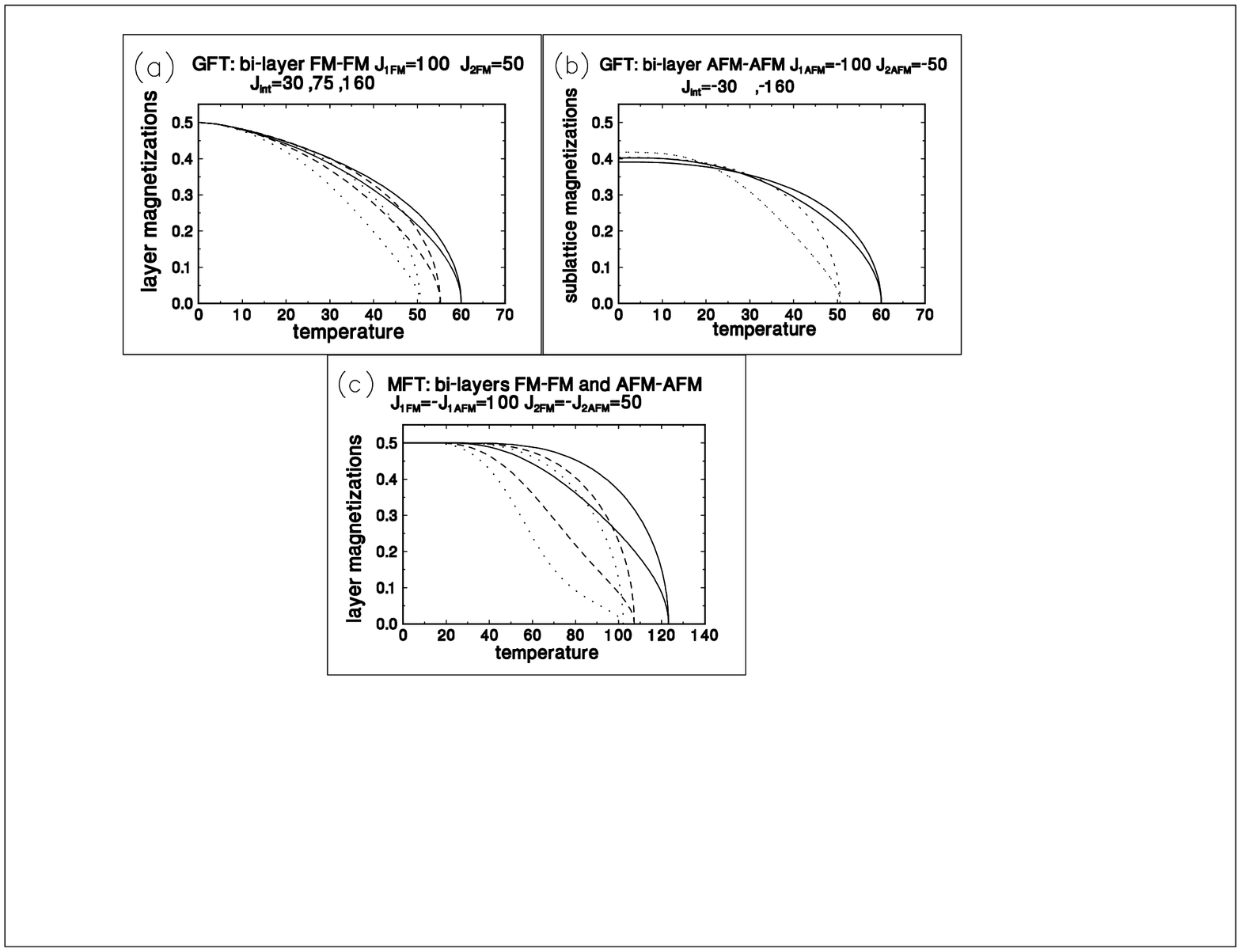}
\protect
\caption{ (a) Green's function theory (GFT) for the ferromagnetic bilayer: The
sublattice magnetizations are displayed as a function of the temperature
for different
interlayer couplings $J_{\rm int}= 30\ ({\rm dotted}),\ 75\ ({\rm
dashed}),\
160 \ ({\rm solid})$. The exchange interaction and anisotropy constants are
$J_{\rm 1FM}=100,\  J_{\rm 2FM}=50,\  D^z_{\rm 1FM}=1.0,\  D^z_{\rm
2FM}=1.0$.
\newline
 (b) GFT for the antiferromagnetic bilayer: The sublattice
magnetizations are displayed as a function of the temperature for
two interlayer couplings $J_{\rm int}= -30\ ({\rm dotted}),\ -160\ ({\rm
solid} )$. The exchange interaction
and anisotropy constants
are $J_{\rm 1AFM}=-100,\ J_{\rm 2AFM}=-50,\ D^x_{\rm 1AFM}=-1.0,
D^x_{\rm 2AFM}=-1.0$.
\newline
(c) Mean field theory (MFT) for the ferromagnetic and antiferromagnetic
bilayers
with identical parameters:
$J_{\rm 1(2)FM}=|J_{\rm 1(2)AFM}|,\ $
$D_{\rm 1(2)FM}=|D_{\rm 1(2)AFM}|,\ $
$J_{\rm intFM}=|J_{\rm intAFM}|$.}
\label{FAfig1}
\end{center}
\end{figure}

In figure \ref{FAfig1}a, we show the sublattice magnetizations of the
ferromagnetic bilayer
as a function of the temperature for three interlayer couplings calculated with
Green's function theory (GFT). The magnetization
profiles are different for the two layers (the magnetization is larger for the
layer with the larger exchange interaction)
but end in a common Curie temperature, which increases with
the strength of the interlayer coupling: $T_{\rm Curie}=50.66,\ 55.24,\
60.04$.

For the antiferromagnetic bilayer, the
parameters are the same as for the ferromagnetic bilayer except for a sign
change. In figure \ref{FAfig1}b we show the sublattice magnetizations of the
antiferromagnetic
bilayer for two interlayer coupling strengths calculated with Green's function
theory. To avoid clutter, we have left out the result for
the intermediate
interlayer coupling strength. The corresponding magnetization curves lie in
between those of the other couplings.
At low temperatures one observes clearly the well-known reduction of the
magnetization due to quantum fluctuations, which are missing
in MFT, see
figure \ref{FAfig1}c. Since $|J_{\rm 1AFM}|>|J_{\rm 2AFM}|$ this reduction is
larger for the first
layer. With increasing temperature the magnetization curves of the
two layers
cross each other  (a fact which was first observed by Diep \cite{D91}) and
finally end in a common N\'eel temperature. A larger
interlayer coupling leads
to a larger suppression  of the magnetization at low temperatures and to a
larger N\'{e}el temperature.
Whereas with the present choice of parameters the magnetization profiles of the
FM and AFM bilayers are rather
different at low temperatures, the critical temperatures turn out to be
identical: $T_{\rm Curie}=T_{\rm N\acute{e}el}$ (cf. figures \ref{FAfig1}(a)
and (b)), a fact already discussed by Lines \cite{Lines64}.

For comparison, we show in figure \ref{FAfig1}(c)  the results of mean field
theory (MFT) with the same
parameters. The magnetization profiles as well as the critical temperatures are
identical for the ferromagnetic and antiferromagnetic bilayers.
As is well known, the Curie (N\'eel) temperatures
$(T_{\rm Curie (N\acute{e}el)}=102.10,\  107.25,\ 123.16\ $) are much larger
(with the present choice of parameters
by about a factor of 2) in MFT
owing to the missing magnon excitations. In MFT the Curie temperature
is not very sensitive to the anisotropies as long as they are much smaller
than the exchange interaction. In GFT, however, the
sensitivity is very much greater because of the
Mermin-Wagner theorem \cite{MW66} ($T_{\rm Curie (N\acute{e}el)}\rightarrow 0$
for $D^{z(x)}\rightarrow 0$). Also, the effect of the
interlayer coupling on the
magnetization profiles is much stronger in MFT than in GFT.

\newpage

\subsubsection*{The coupled ferro-antiferromagnetic bilayer}
This is the most interesting case.
\begin{figure}[htb]
\begin{center}
\protect
\includegraphics*[bb = 100  280 480 550,
angle=-90,clip=true,width=10cm]{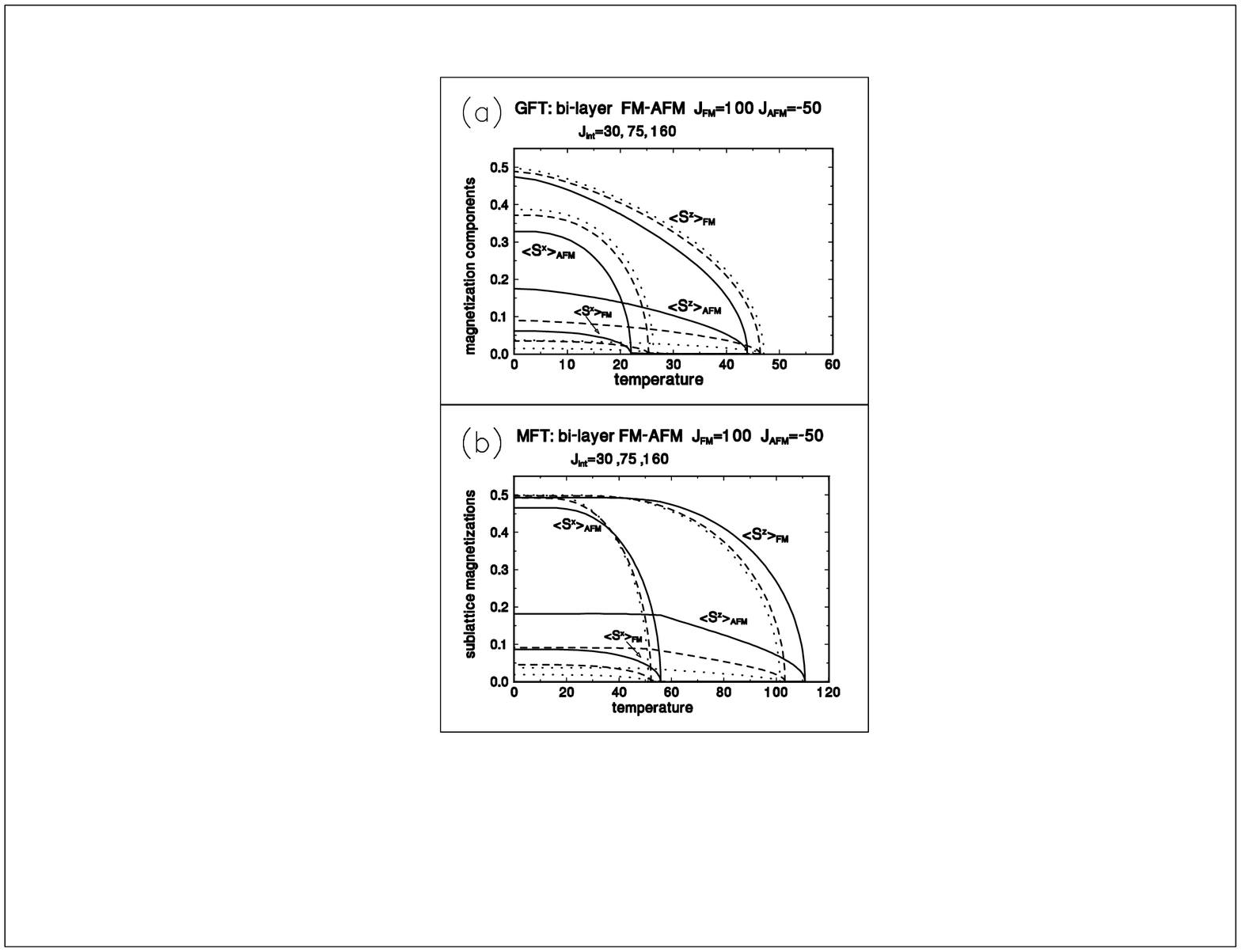}
\protect
\caption{ (a) Green's function theory(GFT): the
sublattice magnetizations of the ferro- and antiferromagnetic
sublattices are displayed as a function of the temperature for different
interlayer couplings $J_{{\rm int}}= 30,\ 75,\ 160$. The exchange interaction
and anisotropy constants are
$J_{{\rm FM}}=100,\ J_{{\rm AFM}}=-50,\ D^x_{{\rm AFM}}=-1.0,\ D^z_{{\rm
FM}}=1.0$.
\newline
(b) Mean field theory (MFT) with the same parameters.
}
\label{FAfig2}
\end{center}
\end{figure}
We consider here two
in-plane magnetization components of each sublattice, thus allowing
noncollinear
magnetizations in both the FM and AFM layers. Our computer code, when
specialized to a single magnetization direction, reproduces
the results of reference \cite{MUH93}.
Without interlayer coupling,
the code also reproduces
 the results for the monolayer ferromagnet and  monolayer antiferromagnet
simultaneously. The choice of
anisotropies supports the orthogonal arrangement of the magnetizations of
the FM and AFM layers favoured by the exchange interaction alone.
The interlayer coupling destroys the perpendicular orientation of the
ferromagnet (in $z$-direction) with respect to the antiferromagnet (in
$x$-direction), even
at temperature $T=0$, as can be seen from figure \ref{FAfig2}. In
this figure,  we show
the sublattice magnetizations calculated with GFT for three
interlayer coupling strengths.
With a positive interlayer coupling, all sublattice magnetizations develop a
positive $z$-component, whereas the $x$-components of the
two
sublattice magnetizations in each layer oppose each other.
With increasing
temperature, all $x$-components decrease until they vanish at a common
temperature  $T^*_{\rm N\acute{e}el}$, slightly above the N\'{e}el
temperature
of the uncoupled AFM.  For $T>T^*_{\rm N\acute{e}el}$ all sublattice
magnetizations point
in the positive $z$-direction. The AFM layer assumes a
ferromagnetic arrangement and remains so until a common critical temperature
$T_C$ is reached, at which the magnetic order vanishes altogether.
\newpage

\subsubsection*{Multilayers}
\begin{figure}[htb]
\begin{center}
\protect
\includegraphics*[bb = 80 95 540 700,
angle=-90,clip=true,width=11cm]{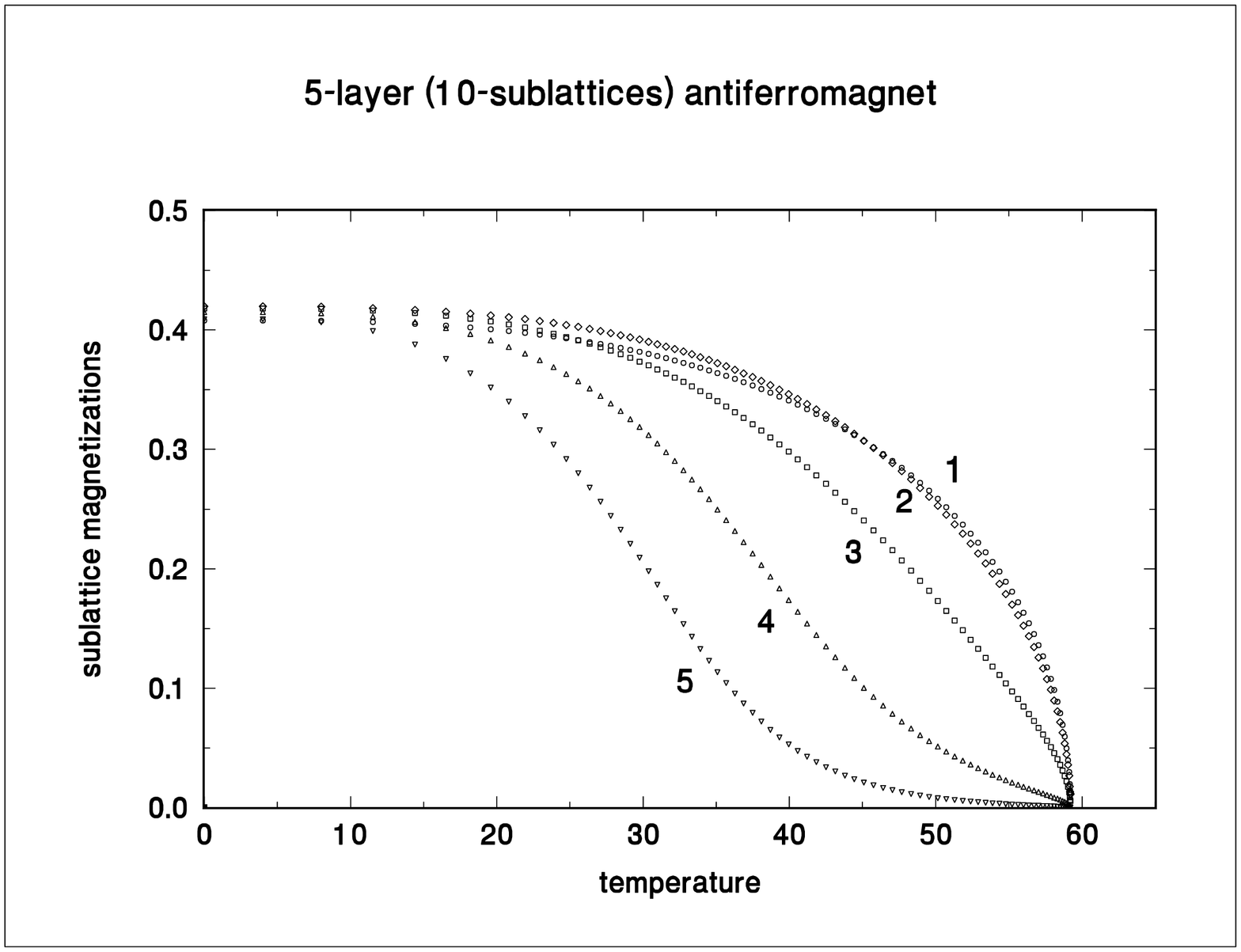}
\protect
\caption{
Green's function theory (GFT): Sublattice magnetizations of a 5-layer (10
sublattices) antiferromagnetic as function of the temperature.
Parameters:
$J_{11}=-100\ J_{12}=-30\
J_{22}=-86,66\ J_{23}=-20\
J_{33}=-73,33\ J_{34}=-10\
J_{44}=-60\ J_{45}=-8.66\
J_{55}=-46,66\  D^x_{ii}=-1.0\ (i=1,...,5). $
}
\label{FAfig24}
\end{center}
\end{figure}

The model is easily extended
to ferromagnetic, antiferromagnetic and coupled ferromagnetic-antiferromagnetic
multilayers with individual parameters for each layer. It is only a question of
computer time. As an example we show in Fig. \ref{FAfig24} the results of a
5-layer
(10 sublattices) antiferromagnet, where each layer has a different exchange
interaction
strength so as not to clutter the diagram.

The theory could possibly serve as a basis for studying the exchange bias
effect, where it seems, however, to be necessary to include interface disorder
\cite{MNLS98, SB98, NCK00} in some way, for instance by introducing more
sublattices per layer with different magnetic arrangements.

 \newpage

\subsection*{4.4. Working in the rotated frame}
In our exposition of GF-theory (e.g. see section 4.2.2), the higher-order
Green's functions are all decoupled in a similar way independently of
whether they are constructed from $S^-$, $S^+$, or
$S^z$ operators or from mixed products of these. This might be
a general weak point in the decoupling procedure---indeed, there is
evidence that this democratic approach ignores essential differences in
the roles of these operators. In particular, GF-estimates of the internal
energy and specific heat are not as reliable as those for the
magnetization and it appears that this might be traceable to an inferior
decoupling of the Green's functions transverse to the $z$-direction, which we
normally choose to be in the direction of the anisotropy.

Some recent publications \cite{SKN05,PPS05} suggest that working in a rotated
coordinate system may provide a way to correct this deficiency, especially
when considering the field-induced
reorientation of the magnetization of a ferromagnetic Heisenberg film. The idea
is that since the decoupling procedure for the single-ion
anisotropy appears to
function better in the direction of the magnetization than in the transverse
direction, it ought to be better to change first to a
rotated coordinate system where the decoupling can be carried out in the
direction of the magnetization only. The angle of the rotation is determined
from the condition that the commutator of $S^z\ '$ with the Hamiltonian
vanish in the rotated frame: $[S^z\ ',H\ ']=0$, where the prime refers to the
rotated frame. This
procedure is remarkably successful \cite{SKN05} in calculating the magnetic
reorientation of
a ferromagnetic film as a function of the external magnetic field in the presence
of a single-ion anisotropy, as can be shown by comparing with the Quantum Monte
Carlo calculations of Ref. \cite{HFKTJ02}. Not only that: the requirement
$[S^z\ ',H\ ']=0$ leads to an equation-of-motion matrix having no
null-space---an enormous simplification of the entire calculation!

Because of the apparent advantages of this new approach, we dedicate an
entire subsection to it. First, we show how to implement the procedure,
applying it to the exact treatment of the single-ion anisotropy; then,
we present some of our own results and those of others \cite{SKN05,JBMD06};
finally, we discuss the method, examining the assumptions and pointing out some
difficulties.

\subsubsection*{4.4.1. The ferromagnetic film with an exact treatment of the
single-ion anisotropy}

In this section, we show how to implement the GF-theory in the rotated
frame for a typical case: the field-induced spin reorientation transition for
spin $S\geq 1$. We go beyond the treatment in \cite{SKN05,PPS05} in that we
treat the single-ion anisotropy exactly \cite{FK06}.

Consider the Hamiltonian (\ref{FKS1}) with a field ${\bf B}=(B_0^x, 0, B_0^z)$
but without the dipole-dipole interaction and the $K_4$-term.
As the external $B_0^x$-field is increased from zero,
the magnetization vector initially in the
$z$-direction rotates by an angle $\theta$ in the $xz$-plane,
so that it points
in the $z'$-direction of a new frame $(x',y',z')$. As in Ref. \cite{SKN05}, we
shall do the calculations in the primed system, in which the magnetization
vector has the components $(0, 0, \la S^z\ '\ra)$. The transformation between
the frames is
\begin{equation}
\left(
        \begin{array}{c}
        \la S^x\ra \\
        \la S^y\ra \\
        \la S^z\ra
\end{array}\right ) =  \left( \begin{array}{ccc}
                             \cos\theta & 0 & \sin\theta \\
                              0         & 1 & 0          \\
                             -\sin\theta& 0 & \cos\theta
\end{array} \right)
\left( \begin{array}{c}
       \la S^x\ '\ra \\
       \la S^y\ '\ra \\
       \la S^z\ '\ra
\end{array} \right).
\label{2}
\end{equation}
Because $\la S^x\ '\ra=\la S^y\ '\ra=0$ in the rotated frame, one need only
calculate $\la S^z\ '\ra$ in order to find the components of the magnetization
in the original frame, once the angle $\theta$ is known.

To get the angle $\theta$, an {\em approximation} is introduced: we
demand that the
commutator of $S^z\ '$ with the Hamiltonian in the rotated system vanish.
 %
%
This implies that  the following Green's function is zero:
\begin{equation}
G_{ij}^{z,-}=\la\la [S_i^z\ ',H']; S_j^-\ ' \ra\ra=0.
\label{11}
\end{equation}
Evaluating the commutator yields a relation between Green's functions
$G_{ij}^{+,-}=\la\la S_i^+\ ';S_j^-\ '\ra\ra$ and
$G_{ij}^{z+,-}=\la\la (2S_i^z-1)S_i^+;S_j^-\ra\ra$,
\begin{equation}
(B_0^x\cos\theta-B_0^z\sin\theta)G_{ij}^{+,-}-K_2\sin\theta\cos\theta
G_{ij}^{z+,-}=0,
\label{12}
\end{equation}
which, after applying the spectral theorem, produces the equation
defining  the reorientation angle in terms of the corresponding diagonal
correlations:
\begin{equation}
(B_0^x\cos\theta-B_0^z\sin\theta)C^{-,+}-K_2\sin\theta\cos\theta
C^{-,z+}=0\ .
\label{13}
\end{equation}

This is a generalization of the angle condition given in
Refs~\cite{SKN05,PPS05}
that can be used for the exact treatment of the single-ion
anisotropy  instead of applying the Anderson-Callen decoupling.
Note that, as used here, the condition on the commutator must
be considered an {\em approximation}. In Refs~\cite{SKN05,PPS05}
the condition is fulfilled automatically because of the use of the
Andersen-Callen decoupling. In general, the condition {\em does not hold}, as
will be shown later.

Following Ref. \cite{SKN05}, we introduce another approximation
that in general also does not hold: we neglect all
GF's not containing an equal number of $S^-\ '$
and $S^+\ '$ operators.

After transforming
the Hamiltonian to the primed system
and making the above approximations, the following
Green's functions are needed:
\begin{eqnarray}
G_{ij}^{+,-}&=&\la\la S_i^+\ ';S_j^-\ '\ra\ra\ , \nonumber\\
G_{ij}^{(z)^n+,-}&=&\la\la (S_i^z\ ')^{n-1}(2S^z_i\ '-1)S_i^+\ ';S_j^-\ '\ra\ra
\ . \label{3}
\end{eqnarray}
The single-ion anisotropy requires that spin $S\geq 1$. Thus, in order to
treat films with $S=1, 3/2, 2, ...$, one
needs the first Green's function and those for $n=1, 2, 3, ...$.
To get the equations of motion, the exchange interaction terms are
treated by a generalized Tyablikov (RPA)-decoupling in which
products of spin operators with equal indices are retained
\begin{equation}
\la\la (S_i^z\ ')^nS_k^+\ ';S_j^-\ '\ra\ra \simeq \la (S_i^z\ ')^n\ra \la
\la S_k^+\ ';S^-_j\ '\ra\ra
+ \la S_k^+\ '\ra \la\la (S_i^z\ ')^n; S_j^-\ '
\ra\ra \ .
\label{4}
\end{equation}
Now in the rotated system, $\la S^+_k\ '\ra=0$; i.e. the second term
vanishes.
After applying the decoupling procedure and performing a Fourier transform to
momentum space, one obtains the following set of equations of motion.
%
\begin{eqnarray}
& &\omega G^{+,-}= 2\la S^z\ '\ra +\la S^z\ '\ra J(q-\gamma_{\bf
k})G^{+,-}\nonumber\\
& &\ \ \ \ \ \ \ \ +(B_0^x\sin\theta+B_0^z\cos\theta)G^{+,-}
+K_2(1-\smfrac{3}{2}\sin^2\theta)G^{z+,-},\nonumber\\
& &\omega G^{z+,-}=(6\la (S^z\ ')^2\ra -2S(S+1))
 -\smfrac{1}{2}J\gamma_{\bf k}\Big(6\la (S^z\ ')^2\ra
-2S(S+1)\Big)G^{+,-}\nonumber\\
& &\ \ \ \ \ \ \ \ +Jq\la S^z\ '\ra G^{z+,-}
+(B_0^x\sin\theta+B_0^z\cos\theta)G^{z+,-}
\nonumber\\
& &\ \ \ \ \
\ \ \ +K_2(1-\smfrac{3}{2}\sin^2\theta)\Big(2G^{(z)^2+,-}-G^{z+,-}\Big)\
, \nonumber\\
& &\omega G^{(z)^2+,-}= 8\la (S^z\ ')^3\ra+3\la(S^z\ ')^2\ra
 -(4S(S+1)-1)\la S^z\ '\ra -S(S+1)\nonumber\\
& &\ \ +J\gamma_{\bf k}\Big(\smfrac{1}{2}S(S+1)+(2S(S+1)-1)\la S^z\
'\ra -\smfrac{3}{2}\la (S^z\ ')^2\ra -4\la(S^z\ ')^3\Big)G^{+,-}
\nonumber\\ & &\ +Jq\la S^z\ '\ra G^{(z)^2+,-}
 +(B_0^x\sin\theta+B_0^z\cos\theta)G^{(z)^2+,-}\nonumber\\
& &\ \ +K_2(1-\smfrac{3}{2}\sin^2\theta)\Big(2G^{(z)^3+,-}-G^{(z)^2+,-}\Big)\ ;
\nonumber\\
& &\omega G^{(z)^3+,-}=10\la (S^z\ ')^4\ra +8\la (S^z\ ')^3\ra
 -(6S(S+1)-5)\la(S^z\ ')^2\ra\nonumber\\
& &\ \ -(4S(S+1)-1)\la S^z\ '\ra -S(S+1)\nonumber\\
& &\ \ +J\gamma_{\bf k}\Big(\smfrac{1}{2}S(S+1)+(2S(S+1)-\smfrac{1}{2})\la
S^z\ '\ra
 +(3S(S+1)-\smfrac{5}{2})\la (S^z\ ')^2\ra\nonumber\\
& &\ \  -4\la (S^z\ ')^3\ra -5\la
(S^z\ ')^4\ra\Big)G^{+,-}\nonumber\\
& &\ \ +Jq\la S^z\ '\ra G^{(z)^3+,-}
+(B_0^x\sin\theta+B_0^z\cos\theta)G^{(z)^3+,-}\nonumber\\
& &\ \ +K_2(1-\smfrac{3}{2}\sin^2\theta)\Big(2G^{(z)^4+,-}-G^{(z)^3+,-}\Big)\ .
\label{5}
\end{eqnarray}
Here $a=1$ is the lattice constant for a square lattice, $q=4$ the number
of nearest neighbours, and $\gamma_{\bf k}=2(\cos{k_x}+\cos{k_y})$.

As they stand, the equations (\ref{5}) do
not form a closed system. This, however, can be achieved by
using formulas derived in Ref. \cite{JA2000} that reduce products of spin
operators by one order (!), allowing the expression of some higher-order
Green's functions in terms of lower order ones:
\begin{eqnarray}
\rm{for}\ \  S=1:\ \ \ &
&G^{(z)^2+,-}=\smfrac{1}{2}(G^{z+,-}+G^{+,-})\ ,\nonumber\\
\rm{for}\ \  S=3/2:\ \ \ &
&G^{(z)^3+,-}=G^{(z)^2+,-}+\smfrac{3}{4}G^{z+,-}\ ,\nonumber\\
\rm{for}\ \  S=2:\ \ \ &
&G^{(z)^4+,-}=\smfrac{3}{2}G^{(z)^3+,-}+\smfrac{7}{4}G^{(z)^2+,-}\nonumber\\
& &-\smfrac{9}{8}G^{z+,-}-\smfrac{9}{8}G^{+,-}\ .
\label{6}
\end{eqnarray}
Inserting these relations into the system of equations (\ref{5}) produces
a closed system of equations.

The equations of motion can be written in compact matrix notation
\begin{equation}
(\omega{\bf 1}-{\bf \Gamma}){\bf G}={\bf A}.
\label{7}
\end{equation}
The quantities ${\bf \Gamma,\ G,\ {\rm and}\  A}$ can be read off from equation
(\ref{5}), where the {\em non-symmetric} matrix ${\bf \Gamma}$ is a
$(2\times 2),\ (3\times 3),\ (4\times 4)$ -matrix for spins $S=1,\ 3/2,\ 2$,
respectively.
The desired correlation vector corresponding to the Green's functions
(\ref{3}),
\begin{equation}
{\bf C}=\left( \begin{array}{c}
\la S^-\ 'S^+\ '\ra \\ \la S^-\ 'S^{(z')^{n-1}}(2S^z\ '-1)S^+\ '\ra
\end{array}
\right),
\label{8}
\end{equation}
is obtained via the spectral theorem. With the eigenvector method of
Section 3.3, the components of the correlation
vector ${\bf C}$ in configuration space are found to be
\begin{eqnarray}
\label{9}
C_i&=&\int d{\bf k}C_i({\bf k})
=\frac{1}{\pi^2}\int_0^\pi dk_x\int_0^\pi
dk_y\sum_{j,k,l=1}^{2S}R_{ij}\epsilon_{jk}L_{kl}A_l\\
& & (i=1,2,..,2S),\nonumber
\end{eqnarray}
where the integration is over the first Brillouin zone and $\bf R(L)$ are
matrices comprising the columns (rows) of the right (left) eigenvectors
of the matrix ${\bf \Gamma}$ and
$\epsilon_{jk}=\delta_{jk}/(e^{\beta \omega_j}-1)$ is a diagonal matrix,
in which $\omega_j$ are the eigenvalues ($j=1,..,2S$) of the
${\bf \Gamma}$-matrix. In sharp contrast to section 4.2.5, there are no
zero eigenvalues of ${\bf \Gamma }$!

Equation (\ref{13}) and the set of integral equations (\ref{9}) have to be
iterated simultaneously to self-consistency
in order to obtain the magnetization $\la S^z\ '\ra$ and
its moments in the rotated system together with the reorientation angle
$\theta$. The curve-following method described in appendix B accomplishes
this with alacrity as before. The components of the magnetizations
in the coordinate system in which
the magnetic reorientation is measured follow from the relations (\ref{2}).
With the formulas from Ref. \cite{JA2000} it would be possible to treat
the fourth-order anisotropy term $-\sum_iK_{4,i}(S_i^z)^4$ exactly.
A generalization to multilayers is also possible.

\subsubsection*{4.4.2. Results of calculations in the rotated frame}
Here we describe results of calculations in the rotated frame, including
results from the method described above.

\begin{figure}[htp]
\begin{center}
\protect
\includegraphics*[bb=80 90 540 700,
angle=-90,clip=true,width=9cm]{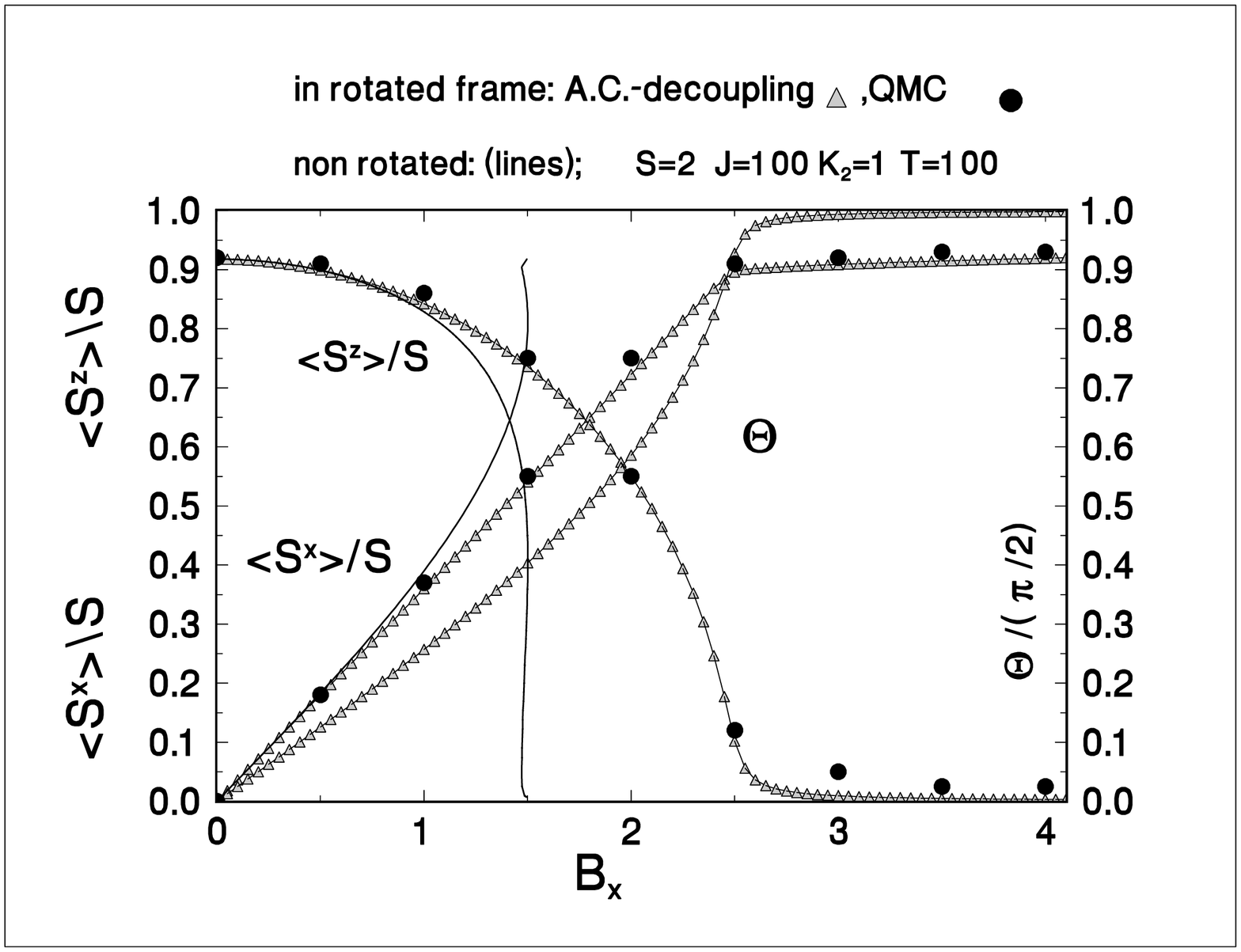}
\protect
\caption{Normalized magnetizations $\la S^z\ra/S$ and $\la S^x\ra/S$ and the
reorientation
angle $\Theta$ for a spin S=2 Heisenberg monolayer for a weak anisotropy as a
function
of the external field: QMC \cite{HFKTJ02}(solid circles), Anderson-Callen
decoupling in the rotated frame \cite{SKN05}(triangles) and in the non-rotated
frame \cite{FJK00}(lines).} \vspace{-0.5cm}
\label{QMCBK1}
\end{center}
\end{figure}

The paper \cite{SKN05} deals with the Heisenberg
ferromagnet with weak single-ion anisotropy in a varying transverse field.
The Anderson-Callen decoupling is used in the rotated frame. The small
anisotropies (e.g. for $S=2$, $K_2=0.01J$) are
appropriate to 3d transition metals. For
the reorientation as a function of the transverse field, there is excellent
agreement with QMC calculations\cite{HFKTJ02}, see Fig. \ref{QMCBK1}. In
particular,
the correlation $\la\,S^x\ra/S$ is a linear function of $B^x$, which is
an improvement over calculations in the original coordinate system
\cite{FJK00},
where the decoupling is performed for GF's corresponding to the components of
the magnetization in the non-rotated frame.

If the anisotropy is treated exactly (see the previous subsection),
the same thing is found, there being  no difference from the
results of Ref.~\cite{SKN05} for weak anisotropies within the line thickness.
This astonishingly good result is perhaps the main point in favour of
working in the rotated system.

\begin{figure}[htp]
\begin{center}
\protect
\includegraphics*[bb=80 90 540 700,
angle=-90,clip=true,width=9cm]{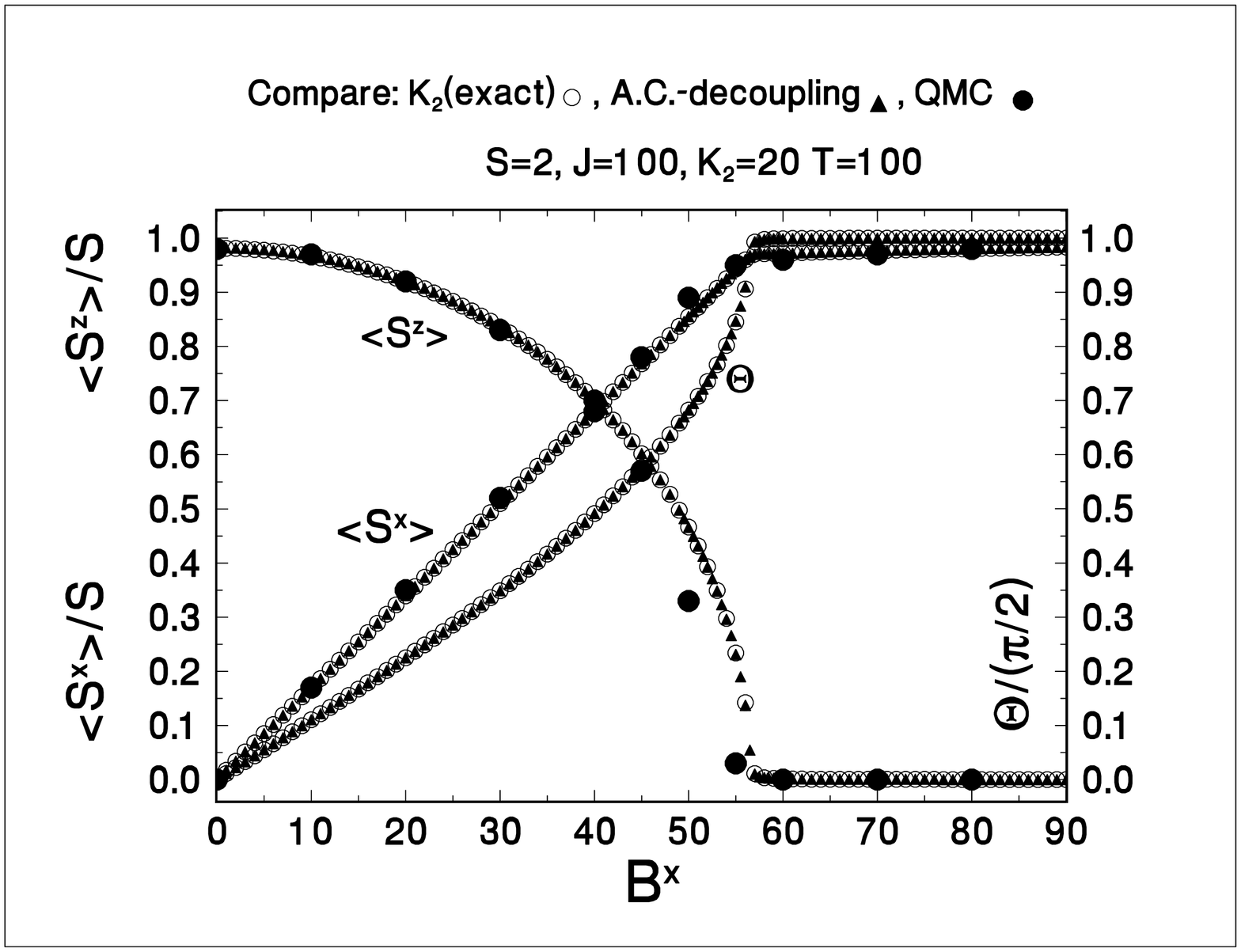}
\protect
\caption{Normalized magnetizations $\la S^z\ra/S$ and $\la S^x\ra/S$ and the
reorientation
angle $\Theta$ for a spin S=2 Heisenberg monolayer as a function of the
external field: QMC \cite{HFKTJ02}(solid circles), Anderson-Callen
decoupling \cite{SKN05}(triangles), present theory \cite{FK06} (open circles).}
\vspace{-0.5cm}
\label{QMCBK2}
\end{center}
\end{figure}

For the lanthanides, where values of the anisotropy
can be of the order of the
exchange interaction, the Andersen-Callen decoupling should break
down and one would expect the exact treatment of the anisotropy
to be superior. Surprisingly, the Anderson-Callen decoupling
in the rotated frame still yields excellent results when compared with the
exact treatment of $K_2$ and with QMC results \cite{HFKTJ02}
for anisotropies up to $K_2\leq 0.2J$. This is seen
in Fig. \ref{QMCBK2} for the magnetic reorientation  induced by the
transverse
$B^x$-field for $K_2=0.2J$ and $T=J=100$. The results of both Green's
function theories (Anderson-Callen decoupled and exact treatment of the
single-ion anisotropy) are nearly identical and deviate only
slightly from the Quantum Monte Carlo results, which can be considered exact
to within the statistical error. The reason for this is that at $T=100$,
the magnetizations from the two theories still lie very close to each
other; at higher temperatures, this is no longer the case and the
results diverge beyond a certain value of $B^x$.

 %
%

\begin{figure}[htp]
\begin{center}
\protect
\includegraphics*[bb=16 18 93 116,
angle=-90,clip=true,width=9cm]{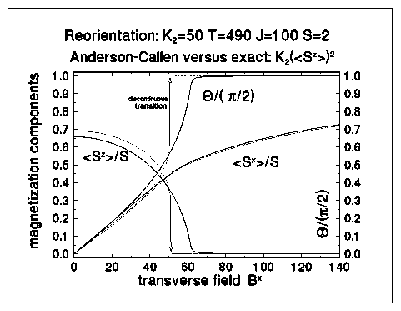}
\protect
\caption{Normalized magnetizations $\la S^z\ra/S$ and $\la S^x\ra/S$ and the
reorientation
angle $\Theta/(\pi/2)$ for a spin S=2 Heisenberg monolayer as function of a
transverse field $B^x$:  Anderson-Callen decoupling
\cite{SKN05}(dotted lines) and the present theory \cite{FK06}(solid lines)
for $ K_2=0.5J\ (T/J=4.9)$.}
\label{B4049}
\end{center}
\end{figure}

Large differences must
also appear as the anisotropy strength is increased,
since it is known that the results from the
Andersen-Callen decoupling do not approach the correct limit. This is
evident from Fig.~\ref{B4049}, where the field-induced reorientation
for a Heisenberg monolayer with $S=2$
from each GF theory is compared  for a temperature
somewhat below the reorientation temperature and for a large anisotropy
$ K_2=0.5J\ (T/J=4.9)$. In this case, implementation of the Anderson-Callen
decoupling along the lines of \cite{SKN05}
leads to a discontinuous transition from an
angle ($\theta/(\pi/2)\approx 0.6$) to full reorientation
($\theta/(\pi/2)=1$), whereas exact treatment of the anisotropy $K_2$
produces a continuous reorientation transition.
 Such discontinuous transitions are also reported in Ref.
\cite{PPS05} for a treatment which is very similar
to that of Ref. \cite{SKN05}. The reason why
discontinuities are not observed in Ref. \cite{SKN05} is that
only very small anisotropies are considered there. We attribute the
discontinuous
transition to the approximate Anderson-Callen decoupling, which is not
justified for large anisotropy.
The difference between the corresponding reorientation fields, $B_{R}$,
increases with
anisotropy. For the present case it is:
$B_R^{\rm K^{\rm exact}_2}-B_R^{\rm A.C.}\simeq 11\  (\rm{for}\ K_2=0.5J)$.
%
%

Unfortunately, we cannot say anything about the accuracy of the model
treating the anisotropy exactly because there are
no QMC calculations available for large anisotropies. The least understood
approximation in this model
is the generalised RPA decoupling of the exchange interaction terms
of the higher-order GF, eqn.(\ref{4}) .
Previous calculations \cite{EFJK99} have shown (by comparing with QMC) that RPA
is a good
approximation for a Heisenberg model (no anisotropy) with a field perpendicular
to the
film plane. To improve the present approach for a field in
the transverse direction, one could resort
to the procedure of \cite{JIRK04} which goes beyond the RPA with respect
to the exchange interaction terms.

We now consider a Heisenberg
{\em antiferromagnet} monolayer with {\em exchange} anisotropy in a transverse
field for $S=1/2$. The Tyablikov decoupling is used. A recent
paper \cite{JBMD06} reports results from an approximate GF treatment where
the sublattices are rotated in such a way as to make the transverse
component of the magnetization in each sublattice vanish. As in the
ferromagnetic case,
it is assumed that $[S^z_i\ ',H']=0$ at this angle, with the consequence that
there
are no zero eigenvalues of the resulting equation-of-motion matrix. The authors
describe their results as unexpected:
the staggered magnetization of the easy axis shows a non-monotonic
behaviour as a function of the transverse field and there is a nonvanishing
easy-axis
magnetization above the N\'eel temperature below a critical
transverse field.

\begin{figure}[htb]
\begin{center}
\protect
\includegraphics*[bb = 80 95 540 700,
angle=-90,clip=true,width=11cm]{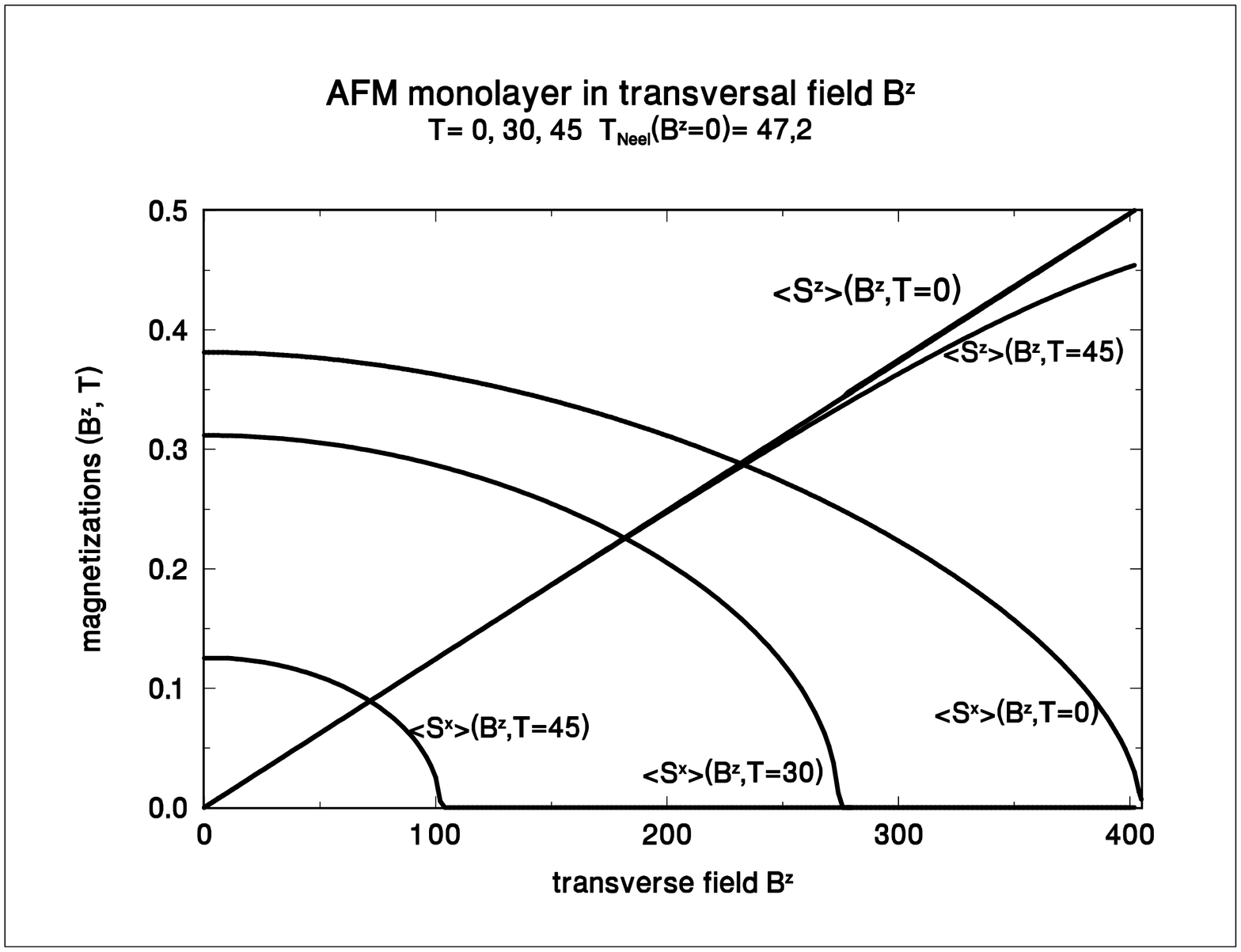}
\protect
\caption{
The magnetization components of a spin $S=1/2$ antiferromagnetic monolayer
(with the easy axis in x-direction) in a transverse field $B^z$  shown as a
function of $B^z$ for different temperatures ($T=0,30,45$). The N\'eel
temperature
$T_{{\rm N\acute{e}el}}(B^z=0)=47,2$ for J=-100 and the exchange-anisotropy
strength $D^x=-0.1$.}
\label{AF1BZT}
\end{center}
\end{figure}

To check the above results, we have computed the
components of magnetization in the {\em non-rotated
frame} directly from equations~(\ref{FA7}) of Section 4.3.2.
Because we have developed  \cite{FK05} a procedure to deal with zero
eigenvalues of the equation-of-motion matrix, we do not need, contrary to Ref.
\cite{JBMD06}, any further approximation apart from the Tyablicov decoupling.
In complete contrast to Ref.~\cite{JBMD06}, our results behave
as one would expect: the easy axis
magnetization decreases monotonically and vanishes as a function of the
transverse field for temperatures above the N\'eel temperature.
Our results are shown in Fig. \ref{AF1BZT}.
We should welcome Quantum Monte Carlo calculations that could
resolve the crass differences between these two sets of results.

\subsubsection*{4.4.3. Discussion}
The most appealing aspect of decoupling in the rotated frame is
not the excellent result for the field reorientation of the
Heisenberg ferromagnet with single-ion anisotropy but rather
the fact that the condition $[S_i^z\ ',H\ ']=0$ leads to an
equation-of-motion matrix devoid of zero eigenvalues.
Ref.~\cite{SKN05} may convey the impression that this condition
is exact; if that were correct, decoupling in the rotated frame
would undoubtedly be the method of choice because of the great
simplification it offers.

But ``if'' stands stiff. We offer
a counter-example as a warning that the approximations used in Refs.
\cite{SKN05, PPS05,JBMD06} should be taken with a grain of salt: an exactly
solvable model demonstrates that $[S_i^z\ ',H\ ']=0$ is not in general valid!

Consider a Hamiltonian having only an external field and a
single-ion anisotropy:
\begin{equation}
H=-\sum_k K_{2,k}(S_k^z)^2-\sum_k(B_0^xS_k^x+B_0^zS_k^z).
\label{D1}
\end{equation}
If it were true that a rotation angle could be found for which
the commutator in the rotated system $[S_i^z\ ',H\ ']$ vanishes,  then
the singular values of the matrix $\la SM'|[S^z\ ',H\ ']|SM\ra $
in the $|SM\ra$ representation would also vanish. A
numerical calculation for $S=1$ shows that this
is {\em not} the case for a finite $K_2$.
Furthermore, the numerical calculation also shows that the
correlations $\la S^-\ 'S^-\ '\ra$ and $\la S^-\ 'S^z\ '\ra$ do not vanish
simultaneously with $\la S^+\ '\ra=\la S^-\ '\ra=0$.
This shows that arguing with the Lehmann representation of the corresponding
Green's functions as in Ref. \cite{SKN05} is not correct because it is
erroneously
assumed that the intermediate energy states $|m\ra$ are eigenstates of the
$z$-component of the angular momentum. They are in fact, however, given by the
superposition
$|m\ra=\sum_Mc_{mM}|SM\ra$, such that e.g. the relevant matrix element
$\sum_{nm}\la n|S^-|m\ra\la m|S^-|n\ra $ does not vanish in general.
In Refs. \cite{PPS05, JBMD06} the Green's function $G_{ij}^{-,-}$ is taken into
account correctly.

Alternatively, consider finding a rotation that diagonalizes
the model Hamiltonian (\ref{D1}). If this were possible, then the commutator
would be zero at the corresponding rotation angle, since two diagonal matrices
commute. For this model, it is possible to show algebraically that
no such angle can be found unless $K_2$ itself vanishes.

At first sight it may seem strange that there is no angle at which
the projection of the spin onto the $z'$ axis is a good quantum
number, for the non-commutativity of $S^z\ '$ with $H'$ implies that
$S^z\ '$ is not a constant of the motion but varies in time.
But there is nothing wrong with this! One cannot argue
that $S^z\ '$ be
time-invariant: the intrinsic anisotropy and the applied external
field favour different directions and they do so according to
completely different mechanisms. It would be wrong to think that there
should be a ``resultant'' direction along which $S^z\ '$ is quantized.
Rather, the time-dependence of the opertor $S^z\ '$ is simply a property of
the Hamiltonian that must be respected.

In conclusion, we regard the procedure of working in the rotated frame
as not yet settled. It may in fact be advantageous if it
succeeds in providing a more uniform way of treating the decoupling.
The practice of employing $[H,S^z]=0$ is very likely much too
severe in general.  The method seems to work for the spin reorientation
problem for the ferromagnet but is questionable for the antiferromagnet in a
transverse field. The embedded
null-space arising from a non-vanishing commutator is more likely an essential
ingredient intimately bound up with the properties of spin. As such,
it could be dangerous to ignore it.

\subsection*{5. Beyond RPA}
Up till now, with the exception of Section 4.2.5,
we did not go beyond the Tyablikov (RPA) decoupling.
In this section, we  develop a formalism for treating the
field induced reorientation of the magnetization for a  spin $1/2$
Heisenberg monolayer with an exchange anisotropy and,
specializing to the magnetization in one direction, we show how
higher-order GF theories discussed in the literature \cite{JIRK04,SFBI00,KY72}
follow quite
naturally as limiting cases of our formalism.

\subsubsection*{5.1. Field-induced reorientation of the magnetization of a
Heisenberg monolayer}

We consider here a spin $S=1/2$  Heisenberg monolayer
with exchange anisotropy in an external field. We go
beyond the Tyablikov (RPA) treatment by decoupling terms due to higher-order
GF's. In the limit of the magnetization
in one direction, we recover the results of Ref.
\cite{JIRK04} for a vanishing
anisotropy and of Ref. \cite{SFBI00} for the one-dimensional chain
in the limit of a vanishing magnetic field.
Without field and anisotropy one recovers the theory of Ref. \cite{KY72}.
The Hamiltonian under investigation is

\begin{equation}
H=-\frac{1}{2}\sum_{lm}J_{lm}(S_m^-S_l^++S_m^zS_l^z)-\frac{1}{2}\sum_{lm}D_{lm}
S_m^zS_l^z-\sum_m(\frac{1}{2}B^-S_m^++\frac{1}{2}B^+S_m^-+B^zS_m^z)\ .
\label{bRPA1}
\end{equation}
The exchange interaction strength is $J_{lm}$, the strength of the exchange
anisotropy is $D_{lm}$ and $B^\pm=B^x\pm iB^y$
with the external magnetic field ${\bf B}=(B^x,B^y,B^z)$.

To get the equations of motion for the spin reorientation
problem, the following first and second-order Green's functions are needed:
\begin{eqnarray}
G_{ij}^{\alpha-(1)}&=&\la\la S_i^\alpha;S_j^-\ra\ra ,\ \ \ \ \
\nonumber\\
G_{ij}^{\alpha-(2)}&=&\la\la [S_i^\alpha,H];S_j^-\ra\ra ,
\ \ \ \ \ \ (\alpha=+,-,z)\nonumber\\
G_{ij}^{zz(1)}&=&\la\la S_i^z;S_j^-\ra\ra , \nonumber \\
G_{ij}^{zz(2)}&=&\la\la [S_i^z,H];S_j\ra\ra.
\label{bRPA2}
\end{eqnarray}
The corresponding 8 equations of motion are
\begin{eqnarray}
\omega G_{ij}^{\alpha-(1)}&=& I_{ij}^{\alpha-(1)}+G_{ij}^{\alpha-(2)} \nonumber\\
\omega G_{ij}^{\alpha-(2)}&=& I_{ij}^{\alpha-(2)}+\la\la
[[S_i^\alpha,H],H];S_j^-\ra\ra,\ \ \ \ \ \
(\alpha=+,-,z)\nonumber\\
\omega G_{ij}^{zz(1)}&=&I_{ij}^{zz(1)}+G_{ij}^{zz(2)},
\nonumber \\
\omega G_{ij}^{zz(2)}&=&I_{ij}^{zz(2)}+\la\la [[S_i^z,H],H];S_j\ra\ra.
\label{bRPA3}
\end{eqnarray}
The double-commutator Green's functions  must be decoupled
in order to obtain  a closed system of equations. After Fourier
transformation to momentum space these are
\begin{equation}
(\omega - {\bf \Gamma}) {\bf G}_{\bf q}={\bf I}_{\bf q},
\label{bRPA4}
\end{equation}
a form that is amenable to the eigenvector
method of Section 3.3.

Generalizing the procedure of Ref. \cite{ST91} to the case where one has
components of the magnetization in all directions of space,
products of three spin operators are decoupled in the following way:

\begin{eqnarray}
& &S_k^zS_l^zS_i^+\approx \alpha^{+-}c_{kl}^{zz}S_i^+
+\alpha^{z-}c_{ki}^{z+}S_l^z+\alpha^{z-}c_{li}^{z+}S_k^z
\ ,\nonumber\\
& &S_l^-S_k^+S_i^+\approx \alpha^{--}c_{lk}^{++}S_l^-
+\alpha^{+-}c_{li}^{-+}S_k^++\alpha^{+-}c_{lk}^{-+}S_i^+
\ ,\nonumber\\
& &S_i^zS_j^+S_l^-\approx \alpha^{zz}c_{jl}^{+-}S_i^z
+\alpha^{+z}c_{jl}^{z-}S_j^++\alpha^{-z}c_{ij}^{z+}S_l^-\ ,
\label{bRPA5}
\end{eqnarray}
where the correlation functions are defined as
$c_{ij}^{\alpha\beta}=\la S_i^\alpha S_j^\beta\ra$. Here we have introduced the
vertex parameters
$\alpha^{+-}$, $\alpha^{--}$, $\alpha^{z-}$
and $\alpha^{zz}, \alpha^{+z}, \alpha^{-z}$, where the indices refer to
the indices of their associated Green's functions
after the decoupling.
In the limiting cases discussed later, we deal only with the magnetization in
one direction, where only the vertex parameters $\alpha^{+-}$ and $\alpha^{zz}$
play a role. We show later how they may be determined by additional
constraints. For the reorientation problem
all 6 vertex parameters could play a role;
however, for simplicity we assume that
$\alpha^{+-}\approx \alpha^{--}\approx \alpha^{z-}$ and
$\alpha^{zz}\approx\alpha^{-z}\approx \alpha^{+z}$, in order not to have
too many additional parameters.

The inhomogeneities in eqn (\ref{bRPA4}) are defined as the Fourier transformed
thermodynamic expectation values of the following commutators
\begin{eqnarray}
& &I_{\bf q}=FT \left(
\begin{array}{c}
\la [S_i^+,S_j^-]\ra\\
\la [S_i^z,S_j^z]\ra\\
\la [S_i^-,S_j^-]\ra\\
\la [S_i^z,S_j^-]\ra\\
\la [[S_i^+,H],S_j^-]\ra\\
\la [[S_i^z,H],S_j^z]\ra\\
\la [[S_i^-,H],S_j^-]\ra\\
\la [[S_i^z,H],S_j^-]\ra
\end{array}
\right)=
\left(
\begin{array}{c}
I_{\bf q}^{+-(1)}\\
I_{\bf q}^{zz(1)}\\
I_{\bf q}^{--(1)}\\
I_{\bf q}^{z-(1)}\\
I_{\bf q}^{+-(2)}\\
I_{\bf q}^{zz(2)}\\
I_{\bf q}^{--(2)}\\
I_{\bf q}^{z-(2)}\\
\end{array}
\right)=
\nonumber \\
& &\left(
\begin{array}{c}
2\la S^z\ra\\
0\\
0\\
-\la S^-\ra\\
zJ(1-\gamma_{\bf q})(c^{+-}_{10}+2c^{zz}_{10})+zD(2c_{10}^{zz}-\gamma_{\bf
q}c_{10}^{+-})+2B^z\la S^z\ra+B^+\la S^-\ra\\
zJc_{10}^{+-}(1-\gamma_{\bf q})+\frac{1}{2}B^+\la S^-\ra+\frac{1}{2}B^-\la
S^+\ra\\
-zJ(1-\gamma_{\bf q})c_{10}^{--}+zD\gamma_{\bf q}c_{10}^{--}-B^-\la S^-\ra\\
-zJ(1-\gamma_{q})c_{10}^{z-}-B^-\la S^z\ra
\end{array}\right)\ .
\label{bRPA6}
\end{eqnarray}
Here,
\begin{equation}
\gamma_{\bf q}=\left\{ \begin{array}{l}
\cos q \ \ \ \ {\rm for\ the\ linear\ chain\ with\ nearest\ neighbours\ } \
 z=2\\
\frac{1}{2}(\cos q_x+\cos q_y) \ \ \ \ {\rm for\ the\ square\
lattice\ with\ \ \ }  z=4.
\end{array}\right.
\label{bRPA7}
\end{equation}
The ${\bf \Gamma}$-matrix has the following form
\begin{equation}
{\bf \Gamma}=\left(\begin{array}{cccccccc}
0&0&0&0&1&0&0&0\\
0&0&0&0&0&1&0&0\\
0&0&0&0&0&0&1&0\\
0&0&0&0&0&0&0&1\\
\Gamma_{51}&0&\Gamma_{53}&\Gamma_{54}&\Gamma_{55}&0&0&\Gamma_{58}\\
0&\Gamma_{62}&0&\Gamma_{64}&0&0&0&0\\
\Gamma_{71}&0&\Gamma_{73}&\Gamma_{74}&0&0&\Gamma_{77}&\Gamma_{78}\\
\Gamma_{81}&0&\Gamma_{83}&\Gamma_{84}&\Gamma_{85}&0&\Gamma_{87}&0\\
\end{array}
\right)\ .
\label{bRPA8}
\end{equation}
Without loss of generality,  the external field may be chosen
such that the reorientation of the magnetization takes
place in the $xz$-plane: ${\bf B}=(B^x,0,B^z)$. Then, $\la S^y\ra=0$ and
$B^+=B^-=B^x$, implying
a number of symmetry relations for the correlation functions,
such as $\la S^+\ra=\la S^-\ra=\la S^x\ra$, $c_{lm}^{++}=c_{lm}^{--}$,
$c_{ij}^{+z}=c_{ji}^{z-}$, etc. The  non-zero matrix
elements are then
\begin{eqnarray}
\Gamma_{15}=& &\Gamma_{26}=\Gamma_{37}=\Gamma_{48}=1\nonumber\\
\Gamma_{51}=& &-\smfrac{1}{2}B^xB^x-B^zB^z-\smfrac{1}{2}DB^xz\gamma_{\bf q}\la
S^x\ra\nonumber\\
& &+zD^2\Big(\smfrac{1}{4}+\alpha^{+-}(c_{20}^{zz}+(z-2)c_{11}^{zz})\Big)
\nonumber \\
& &+zJD\Big(\smfrac{1}{2}(1-\gamma_{\bf q})
+\alpha^{+-}[(2-\gamma_{\bf q})(c_{20}^{zz}+(z-2)c_{11}^{zz})
-\smfrac{1}{2}(c_{20}^{+-}+(z-2)c_{11}^{+-})\gamma_{\bf q}]\nonumber\\
& &-\alpha^{+-}[(z-1)\gamma_qc_{10}^{zz}
-\smfrac{1}{2}(z\gamma_{\bf q}^2-1)c_{10}^{+-}]
\Big)
\nonumber\\
& &+\smfrac{z}{2}(1-\gamma_{\bf q})J^2
\Big(1+\alpha^{+-}[2c_{20}^{zz}+c_{20}^{+-}\nonumber\\
& &+(z-2)(2c_{11}^{zz}
+c_{11}^{+-})-(1+z\gamma_{\bf
q})(2c_{10}^{zz}+c_{10}^{+-})]\Big)\nonumber\\
\Gamma_{53}=& &\smfrac{1}{2}B^xB^x+\smfrac{1}{2}DB^xz\gamma_{\bf q}\la
S^x\ra\nonumber\\
& &+\smfrac{z}{2}\alpha^{+-}DJ\Big((c_{20}^{--}+(z-2)c^{--}_{11})\gamma_{\bf
q}- (z\gamma_{\bf q}^2-1)c_{10}^{--}\Big)\nonumber\\
& &-\smfrac{z}{2}(1-\gamma_{\bf q})
J^2\alpha^{+-}\Big(c_{20}^{--}+(z-2)c_{11}^{--}-c_{10}^{--}
(1+z\gamma_{\bf q})\Big)\nonumber\\
\Gamma_{54}=& &B^zB^x-DB^xz(1+\gamma_{\bf q})\la
S^z\ra+zD^2\alpha^{+-}2(z-1)c_{10}^{z-}\gamma_{\bf q}\nonumber\\
& &+z\alpha^{+-}JD\Big([2-z(1+\gamma_{\bf q}^2)+3\gamma_{\bf
q}(z-1)]c_{10}^{z-}
-c_{20}^{z-}-(z-2)c_{11}^{z-}\Big)\nonumber\\
& &-z(1-\gamma_{\bf q})
J^2\alpha^{+-}\Big(c_{20}^{z-}+(z-2)c_{11}^{z-}-c_{10}^{z-}
(1+z\gamma_{\bf q})\Big)\nonumber\\
\Gamma_{55}=& &2B^z\nonumber\\
\Gamma_{58}=& &-2B^x\nonumber\\
\Gamma_{62}=& &B^xB^x+zB^xD\la S^x\ra\gamma_{\bf q}
-zJD\alpha^{zz}c_{10}^{+-}(1-\gamma_{\bf q})(z\gamma_{\bf q}+1)\nonumber\\
& &+\smfrac{z}{2}J^2(1-\gamma_{\bf q})
\Big(1+2\alpha^{zz}(c_{20}^{+-}+(z-2)c_{11}^{+-}
-(1+z\gamma_{\bf q})c_{10}^{+-})\Big)
\nonumber\\
\Gamma_{64}=& &B^xB^z+zDB^x\la S^z\ra+zJD\alpha^{zz}(1-\gamma_{\bf
q})\Big((z-1)c_{10}^{z-}-(c_{20}^{z-}+(z-2)c_{11}^{z-})\Big)\nonumber\\
& &-zJ^2(1-\gamma_{\bf q})
\alpha^{zz}\Big(c_{20}^{z-}+(z-2)c_{11}^{z-}
-(1+z\gamma_{\bf q})c_{10}^{z-}\Big)
\nonumber\\
\Gamma_{71}=& &\smfrac{1}{2}B^xB^x+\smfrac{z}{2}DB^x\la
S^x\ra\gamma_{\bf q}\nonumber\\
& &+\smfrac{z}{2}DJ\alpha^{+-}\Big((c_{20}^{--}+(z-2)c_{11}^{--})
\gamma_{\bf q}-(z\gamma_{\bf q}^2-1)c_{10}^{--}\Big) \nonumber\\
& &-\smfrac{z}{2}(1-\gamma_{\bf q})J^2\alpha^{+-}\Big(c_{20}^{--}+
(z-2)c_{11}^{--}-(z\gamma_{\bf q}+1)c_{10}^{--}\Big)\nonumber\\
\Gamma_{73}=& &-\smfrac{1}{2}B^xB^x-B^zB^z-\smfrac{z}{2}DB^x\la
s^x\ra\gamma_{\bf q}+zD^2\Big(\smfrac{1}{4}+\alpha^{+-}
(c_{20}^{zz}+(z-2)c_{11}^{zz})\Big)\nonumber\\
& &+zJD\Big(\smfrac{1}{2}(1-\gamma_{\bf q})
+\alpha^{+-}[(2-\gamma_{\bf q})(c_{20}^{zz}
+(z-2)c_{11}^{zz})-(z-1)\gamma_qc_{10}^{zz}\nonumber\\
& &+\smfrac{1}{2}(z\gamma_{\bf q}^2-1)c_{10}^{+-}
-\smfrac{1}{2}(c_{20}^{+-}+(z-2)c_{11}^{+-})\gamma_{\bf q}]\Big)
\nonumber\\
& &+\smfrac{z}{2}(1-\gamma_{\bf q})J^2\Big(1+\alpha^{+-}
[2c_{20}^{zz}+c_{20}^{+-}\nonumber\\
& &+(z-2)(2c_{11}^{zz}+c_{11}^{+-})-(1+z\gamma_{\bf
q})(2c_{20}^{zz}+c_{20}^{+-})]\Big)
\nonumber\\
\Gamma_{74}=& &B^zB^x-zDB^x\la S^z\ra(1+\gamma_{\bf q})
+zD^2\alpha^{+-}2(z-1)\gamma_{\bf q}c_{10}^{z-}\nonumber\\
& &zJD\alpha^{+-}\Big([2-z(1+\gamma_{\bf q}^2)+3\gamma_{\bf q}(z-1)]c_{10}^{z-}
-(c_{20}^{z-}+(z-2)c_{11}^{z-})\Big)\nonumber\\
& &-z(1-\gamma_{\bf
q})J^2\alpha^{+-}\Big(c_{20}^{z-}+(z-2)c_{11}^{z-}
-(z\gamma_{\bf q}+1)c_{10}^{z-}\Big)
\nonumber\\
\Gamma_{77}=& &-2B^z\nonumber\\
\Gamma_{78}=& &2B^x\nonumber\\
\Gamma_{81}=& &\smfrac{1}{2}B^xB^z+\smfrac{z}{2}DB^x\la
S^z\ra\nonumber\\
& & +\smfrac{z}{2}JD\alpha^{+-}(1-\gamma_{\bf q})
\Big((z-1)c_{10}^{z-}-c_{20}^{z-}-(z-2)c_{11}^{z-}\Big)\nonumber\\
& &-\smfrac{z}{2}J^2(1-\gamma_{\bf
q})\alpha^{+-}\Big(c_{20}^{z-}+(z-2)c_{11}^{z-}
-(1+z\gamma_{\bf q})c_{10}^{z-}\Big)
\nonumber\\
\Gamma_{83}=& &\Gamma_{81}\nonumber\\
\Gamma_{84}=& &B^xB^x+zB^xD\la S^x\ra\gamma_{\bf q}
-zJD\alpha^{+-}c_{10}^{+-}(1-\gamma_{\bf q})(z\gamma_{\bf q}+1)\nonumber\\
& &+\smfrac{z}{2}J^2(1-\gamma_{\bf q})
\Big(1+2\alpha^{+-}(c_{20}^{+-}+(z-2)c_{11}^{+-}
-(1+z\gamma_{\bf q})c_{10}^{+-})\Big)
\nonumber\\
\Gamma_{85}=& &-B^x\nonumber\\
\Gamma_{87}=& &B^x\ .
\label{bRPA9}
\end{eqnarray}
We have no explicit calculations with the eigenvector method for the spin
reorientation problem but we show now that, when specialized to one
magnetization direction only, limiting cases of the above equations
lead to results found in
the literature.

{\bf 5.2. Limiting cases}

For a magnetization in only one direction (the $z$-direction), the
equations
of motion reduce to a four-dimensional problem in energy-momentum space:
\begin{equation}
(\omega - {\bf \Gamma}) {\bf G}_{\bf q}={\bf I}_{\bf q}
\label{bRPA10}
\end{equation}
with
\begin{equation}
{\bf G}_{\bf q}=\left(\begin{array}{c}
G_{\bf q}^{+-(1)}\\ G_{\bf q}^{zz(1)}\\ G_{\bf q}^{+-(2)}\\ G_{\bf
q}^{zz(2)} \end{array}
\right)\ \ \ \
\label{bRPA11}
\end{equation}
and
\begin{equation}
{\bf I}_{\bf q}=\left(\begin{array}{c}
I_{\bf q}^{+-(1)}\\ I_{\bf q}^{zz(1)}\\ I_{\bf
q}^{+-(2)}\\ I_{\bf q}^{zz(2)} \end{array}
\right)=\left(\begin{array}{c}2\la S^z\ra \\
0\\ 2B^z\la S^z\ra +zJ(c^{+-}_{10}+2c^{zz}_{10})(1-\gamma_{\bf q})
+zD(2c_{10}^{zz}-\gamma_{\bf q}c_{10}^{+-})\\ zJc_{10}^{+-}(1-\gamma_{\bf q})
\end{array} \right)\ .
\label{bRPA12}
\end{equation}

{\bf 5.2.1. Ferromagnet in a magnetic field, no anisotropy}

In this case, $D=0, B^+=B^-=0, B^z\neq 0$, leading to
the theory of reference \cite{JIRK04}
with the $4\times 4$ $\Gamma$-matrix which is, in the notation corresponding
to eqn. (\ref{bRPA8}),
\begin{equation}
{\bf \Gamma}=\left( \begin{array}{cccc}
0 & 0 & 1& 0 \\ 0 & 0 & 0 & 1 \\
\Gamma_{51}&0& \Gamma_{55}&0\\
0&\Gamma_{62} & 0 & 0
\end{array} \right),
\label{bRPA13}
\end{equation}
where now
\begin{eqnarray}
\Gamma_{62}&=&(\omega_{\bf q}^{zz})^2=\frac{z}{2}(1-\gamma_{\bf
q})J^2\Big(1+2\alpha^{zz}[(z-2)c_{11}^{+-}+c_{20}^{+-}-(1+z\gamma_{\bf
q})c_{10}^{+-}]\Big),\ \nonumber \\
\Gamma_{51}&=&-B^zB^z+(\omega_{\bf q}^{+-})^2,\ {\rm with}\nonumber\\
(\omega_{\bf q}^{+-})^2&=&\frac{z}{2}(1-\gamma_{\bf
q})J^2\Big(1+2\alpha^{+-}[(z-2)(\frac{1}{2}c_{11}^{+-}+c_{11}^{zz})
\nonumber \\ & &+(\frac{1}{2}c_{20}^{+-}+c_{20}^{zz})-
(1+z\gamma_{\bf q})(\frac{1}{2}c_{10}^{+-}+c_{10}^{zz})]\Big),\nonumber\\
\Gamma_{55}&=&2B^z,
\label{bRPA14}
\end{eqnarray}
where $z$ and $\gamma_{\bf q}$ are defined in equation (\ref{bRPA7}) for the
linear chain and the square lattice respectively.
For the linear chain ($z$=2), there are 7 unknowns
($\la S^z\ra, c_{10}^{+-},$
$ c_{20}^{+-}, c_{10}^{zz}, c_{20}^{zz},\alpha^{+-},\alpha^{zz} $)
and, for the square lattice ($z$=4),
two additional unknowns ($c_{11}^{+-},c_{11}^{zz})$.

In both cases, the relations
\begin{eqnarray}
c^{+-}_{00}&=&1/2-\la S^z\ra\nonumber\\
c^{zz}_{00}&=&1/4
\label{bRPA15}
\end{eqnarray}
are valid.

The eigenvector method of Section 3.3 yields 6 equations for the chain
\begin{eqnarray}
c_{j0}^{+-}&=&\smfrac{1}{\pi}\int_0^\pi dq\cos(jq)({\bf R{\cal E}LI})_1\ ,
\nonumber\\
c_{j0}^{zz}&=&\smfrac{1}{\pi}\int_0^\pi dq\cos(jq)({\bf R{\cal E}LI})_2
\label{bRPA16}
\end{eqnarray}
with  $j=0,1,2$.

For the square lattice, 8 equations are obtained ($j=0,1,2$):
\begin{eqnarray}
c_{j0}^{+-}&=&\smfrac{1}{\pi^2}\int_0^\pi dq_x\int_0^\pi dq_y
\cos(q_xj)({\bf R{\cal E}LI})_1\ , \nonumber\\
c_{j0}^{zz}&=&\smfrac{1}{\pi^2}\int_0^\pi dq_x\int_0^\pi dq_y
\cos(q_xj)({\bf R{\cal E}LI})_2\ , \nonumber\\
c_{11}^{+-(zz)}&=&\smfrac{1}{\pi^2}\int_0^\pi dq_x\int_0^\pi dq_y
\cos(q_x+q_y)({\bf R{\cal E}LI})_{1(2)}\ .
\label{bRPA17}
\end{eqnarray}
These equations do not yet suffice to determine the unknowns because
the vertex parameters enter implicitly. In both cases, the
missing condition is supplied by an expression for the intrinsic energy:

\begin{equation}
E_i=\frac{\la H\ra}{N}=-\smfrac{z}{2}J(c_{10}^{+-}+c_{10}^{zz})\ .
\label{bRPA18}
\end{equation}
In order to evaluate $\la H\ra$, eqn (\ref{E6}) for $\la S_i^-[S_i^+,H]\ra$
is compared with the explicit evaluation of
 $\la S_i^-[S_i^+,H]\ra$ to yield
\begin{eqnarray}
E_i&=&-\frac{zJ}{8}-\frac{B^z}{2}+\frac{1}{2}\frac{1}{N}\sum_{\bf
q}\frac{i}{2\pi}\times\\
& &\lim_{\delta\rightarrow 0+}\int_{-\infty}^{\infty}d\omega
\frac{\omega+B^z+\smfrac{1}{2}J(1-\gamma_{\bf q})}{e^{\beta\omega}-1}
\Big(G_{\bf q}^{+-(1)}(\omega+i\delta)-G_{\bf
q}^{+-(1)}(\omega-i\delta)\Big)\ .\nonumber
\label{bRPA19}
\end{eqnarray}
The Green's function $G_{\bf q}^{+-}=G_1$ is the first component of the Greens
function vector, given by
\begin{equation}
G_1=\sum_i R_{1i}\frac{{(\bf LI)}_i}{\omega-\omega_i}\ .
\label{bRPA20}
\end{equation}
Using
\begin{equation}
G_1(\omega+i\delta)-G_1(\omega-i\delta)=-2\pi i\sum_i
R_{1i}\delta(\omega-\omega_i){(\bf LI)}_i
\label{bRPA21}
\end{equation}
in eqn (\ref{bRPA19}), performing the $\omega$-integration and comparing with
eqn
(\ref{bRPA18}) yields the additional equation needed to determine
all unknowns:
\begin{eqnarray}
\label{bRPA22}
& &-B^z\la S^z\ra-\smfrac{z}{2}J(c_{10}^{+-}+c_{10}^{22})\\
&=&-\frac{zJ}{8}-\frac{B^z}{2}+\frac{1}{2}
\left(\begin{array}{c}
\smfrac{1}{\pi}\int_0^\pi dq\\ \smfrac{1}{\pi^2}\int_0^\pi dq_x\int_0^\pi dq_y
\end{array}\right)
\sum_i R_{1i}\frac{\omega_i+B^z+\smfrac{z}{2}J(1-\gamma_{\bf
q})}{e^{\beta\omega_i-1}}{(\bf LI)}_i\ .\nonumber
\end{eqnarray}
The equations (\ref{bRPA15}) together with (\ref{bRPA16},\ref{bRPA17}) and
(\ref{bRPA22}) determine the unknowns, from which one obtains the magnetization
$\la S^z\ra$, the intrinsic energy $E_i$, the susceptibility
$\chi=d\la S^z\ra/ dB^z$ and the specific heat $c_V=dE_i/ dT$.

The numerical results in Ref. \cite{JIRK04} (obtained not with the  present
method but with the standard spectral theorem) demonstrate that RPA is a
rather good
approximation for the magnetization and the susceptibility but that it is
inadequate when transverse correlations play a
role, as is the case for the intrinsic energy and the specific heat.
In this case, it is very important to go beyond RPA.

{\bf 5.2.2. Ferromagnet with no magnetic field and no exchange anisotropy}

In this case, D=0 and
${\bf B=0}$ and, for a linear chain ($z=2, \gamma_{\bf q}=\cos q$), one
obtains  the model discussed by Kondo and Yamaji \cite{KY72}.
Because of the Mermin-Wagner theorem, the magnetization is $\la S^z\ra=0$,
and, for $S=1/2$, isotropy demands that
\begin{equation}
c_{n0}^{zz}=\smfrac{1}{2}c_{n0}^{+-}.
\label{KY1}
\end{equation}
Therefore, one needs either the GF's $\la\la S_i^z;S_j^z\ra\ra$ and
$\la\la [S_i^z,H];S_j^z\ra\ra$
or $\la\la S_i^+;S_j\ra\ra$ and $\la\la [S_i^+,H];S_j^-\ra\ra$.
The first choice
reduces the problem  to  two dimensions:
\begin{equation}
\left(\begin{array}{cc}
\omega &0\\
0&\omega
\end{array}
\right)-
\left(\begin{array}{cc}
0&1\\
(\omega_q^{zz})^2&0
\end{array}
\right)
\left(\begin{array}{c}
G_q^{zz(1)}\\
G_q^{zz(2)}
\end{array}
\right)=\left(\begin{array}{c}
I_q^{zz(1)}\\I_q^{zz(2)}
\end{array}\right)
\label{KY2}
\end{equation}
with
\begin{eqnarray}
I_q^{zz(1)}=0\ ,\ \ \ \ I_q^{zz(2)}=4c_{10}^{+-}(1-\cos q)\ ,\nonumber\\
(\omega_q^{zz})^2=(1-\cos q)J^2\Big(1+2\alpha^{zz}(c_{20}^{+-}-
(1+2\cos q)c_{10}^{+-})\Big).
\label{KY3}
\end{eqnarray}
These equations yield
\begin{equation}
G_q^{zz(1)}=\frac{I_q^{zz(2)}}{\omega^2-(\omega_q^{zz(2)})^2}=
\frac{I_q^{zz(2)}}{2|\omega_q^{zz}|}\Big(\frac{1}{\omega-\omega_q^{zz}}
-\frac{1}{\omega+\omega_q^{zz}}\Big)\ .
\label{KY4}
\end{equation}
The standard spectral theorem produces 3 equations for
determining
the 3 unknowns $\alpha^{zz}, c^{+-}_{10,}, c^{+-}_{20}$. For spin 1/2,
$c_{00}^{zz}=\la S_0^zS_0^z\ra=\smfrac{1}{4}$. This, together with
\begin{equation}
c_{n0}^{zz}=\smfrac{1}{2}c_{n0}^{+-}=\smfrac{1}{\pi}\int_0^\pi dq
\cos(nq)\frac{I_q^{zz(2)}}{|\omega_q^{zz}|}
\coth(\frac{\beta}{2}|\omega_q^{zz}|);\ \ \ \ n=0,1,2\ ,
\label{KY5}
\end{equation}
determines the unknowns.

It is instructive to apply the eigenvector method of Section 3 and
to obtain the same expression from
\begin{equation}
c_{n0}^{zz}=\smfrac{2}{\pi}\int_0^\pi dq \cos(nq) \Big({\bf R{\cal
E}LI}\Big)_1;\ \ \ n=0,1,2.
\label{KY6}
\end{equation}
Here the matrix ${\bf R}$ consists of the right eigenvectors as columns of the
nonsymmetric matrix in eqn ({\ref{KY2}}) and ${\bf L=R^{-1}}$ consists of the
left
eigenvectors as rows and can be calculated as the inverse of ${\bf R}$. Note
that ${\bf R}$ and ${\bf L}$ are not separately orthonormal.
It is only necessary that ${\bf RL=1}$.
One finds
\begin{equation}
{\bf R}=\left(\begin{array}{cc}
1&1\\ \omega_q^{zz}&-\omega_q^{zz}
\end{array}\right),\ \ \
{\bf L}=\frac{1}{2\omega_q^{zz}}\left(\begin{array}{cc}
\omega_q^{zz}&1\\ \omega_q^{zz}&-1
\end{array}\right),\ \ \
{\bf{\cal E}}=\left(\begin{array}{cc}
\smfrac{1}{e^{\beta\omega_q^{zz}}-1}&0\\
0&\smfrac{1}{e^{-\beta\omega_q^{zz}}-1}
\end{array}\right).
\label{KY7}
\end{equation}
Evaluation  of eqn (\ref{KY6}) with these expressions produces
eqn (\ref{KY5}).

Solution of these equations yields the intrinsic energy
per particle
\begin{equation}
E=-\smfrac{1}{2N}\sum_{nm}
J_{nm}(c_{nm}^{+-}+c_{nm}^{zz})=-\smfrac{3}{2}Jc_{10}^{+-},
\label{KY8}
\end{equation}
the specific heat per particle,
\begin{equation}
c_V=\frac{dE}{dT}=\smfrac{3}{2}J\beta^2\frac{d}{d\beta}c_{10}^{+-},
\label{KY9}
\end{equation}
and the susceptibility
\begin{equation}
\chi=\beta\sum_n c_n^{zz}=\smfrac{1}{2}\sum_{n=0}^2c_{n0}^{+-}.
\label{KY10}
\end{equation}

The results of Kondo and Yamaji \cite{KY72}, here reproduced numerically by the
eigenvector method, are largely in agreement with the exact
calculations of Bonner and Fisher \cite{BF64} for a finite number of spins .

The Kondo-Yamaji decoupling is generalized in Refs. \cite{BZSY96,Bao97} in
order to treat the spin S=1 antiferromagnetic Heisenberg chain. It is also
used in Ref. \cite{GP01} for the spin
S=1 low-dimensional quantum XY ferromagnet and in Refs. \cite{YF00,BCL02} for a
kagom\'{e} antiferromagnet.

{\bf 5.2.3. Ferromagnet with exchange anisotropy but no magnetic field}

This case ($D\neq 0, B=0$)
is discussed in Ref. \cite{SFBI00} for the easy-plane XXZ chain,
where 4 vertex parameters are used: $\alpha_1^{+-}, \alpha_2^{+-},
\alpha_1^{zz}, \alpha_2^{zz}$. These are fixed by the exact
expression
for the ground state energy and by assuming that the ratios of corresponding
parameters do not vary with the temperature.

A $2\times 2$-problem results with the equations of motion
\begin{equation}
\left\{\left(\begin{array}{cc}
\omega &0\\
0&\omega
\end{array}
\right)-
\left(\begin{array}{cc}
\Gamma_{51}&0\\
0&\Gamma_{62}
\end{array}
\right)\right\}
\left(\begin{array}{c}
G_q^{+-(1)}\\
G_q^{zz(1)}
\end{array}
\right)=\left(\begin{array}{c}
I_q^{+-(2)}\\I_q^{zz(2)}
\end{array}\right)\ ,
\label{KY11}
\end{equation}
where
\begin{eqnarray}
I_{\bf
q}^{+-(2)}&=&2J(c_{10}^{+-}+2c_{10}^{zz})(1-\gamma_q)+2D(2c_{10}^{zz}-
\gamma_qc_{10}^{+-})\ ,\nonumber\\
I_{\bf q}^{zz(2)}&=&2Jc_{10}^{+-}(1-\gamma_q)\ ,
\label{KY12} \end{eqnarray}
and
\begin{eqnarray}
\Gamma_{51}&=&2D^2(\smfrac{1}{4}+\alpha_2^{+-}c_{20}^{zz})+2JD
\Big(\smfrac{1}{2} (1- \gamma_q)\nonumber\\
& &+\alpha_2^{+-}[(2-\gamma_q)c_{20}^{zz}-\smfrac{1}{2}c_{20}^{+-}\gamma_q]
+\alpha_1^{+-}[-\gamma_qc_{10}^{zz}+\smfrac{1}{2}(2\gamma_q^2-1)c_{10}^{+-}]
\Big) \nonumber\\
& &+(1-\gamma_q)J^2\Big(1+\alpha_2^{+-}(2c_{20}^{zz}+c_{20}^{+-})-
\alpha_1^{+-}(1+2\gamma_q)(2c_{10}^{zz}+c_{10}^{+-})\Big)\ ,\nonumber\\
\Gamma_{62}&=&-2JDc_{10}^{+-}\alpha_1^{zz}(1-\gamma_q)(2\gamma_q+1)\nonumber\\
& &+J^2(1-\gamma_q)\Big(1+2\alpha_2^{zz}c_{20}^{+-}-2\alpha_1^{zz}
(1+2\gamma_q)c_{10}^{+-}\Big)\ .
\label{KY13}
\end{eqnarray}

The thermodynamics of the $S\geq 1$ ferromagnetic Heisenberg chain with
uniaxial single-ion anisotropy using second-order GF's is treated in Ref.
\cite{JIR05}. The antiferromagnetic easy-plane XXZ-model for $S=1/2$ is treated
in Ref. \cite{ISF01}.

\newpage

\newpage
\subsection*{5.3. The Tserkovnikov formulation of the  GF theory}

Until now we have not considered the damping of spin waves. This is because we
have neglected the influence of the self-energy, the imaginary part of which
leads to damping effects. In this Section, we present a formalism which allows
the treatment of damping. In the first subsection we
develop the general
formalism and, in the second subsection, we specialize it to a Heisenberg
monolayer in an external field, evaluating the self-energy approximately.
The formalism follows Tserkovnikov \cite{T70},
who derives
a closed expression for the self-energy without making decoupling assumptions.
For a review of the formalism, see e.g. Ref. \cite{Ku02}; a compact derivation
can be found in the appendix of Ref. \cite{BR94}. The formal derivation
of a Dyson equation for a Heisenberg ferromagnet is given in Ref.
\cite{Pla73}.

\subsubsection*{5.3.1. The general formalism}
The equation of motion for the single-particle Green's function
 $G_1=\la\la A^+;A\ra\ra_\omega$ is
\begin{equation}
\omega G_1=I_1+G_2,\ \ \ {\rm with}\ \ \ I_1=\la[A^+,A]\ra\ \ {\rm and}\ \
G_2=\la\la[A^+,H];A\ra\ra\ .
\label{T1}
\end{equation}
Analogously, the equation for the two-particle Green's
function $G_2$ may be written
\begin{equation}
\omega G_2=I_2+G_3,
\label{T2}
\end{equation}
with $I_2=\la[[A^+,H],A]\ra$. $G_3$ is the three-particle GF, which
Tserkovnikov
obtains by a time derivation with respect to the second operator of $G_2$ as
$G_3=\la\la[A^+,H];[A,H]\ra\ra$.

On the way to deriving an equation for the self-energy, Tserkovnikov
introduces the ansatz
\begin{equation}
\la\la[A^+,H];A\ra\ra=C\la\la A^+;A\ra+\la\la B;A\ra\ra.
\label{T3}
\end{equation}
If one determines $B$ such that $\la [B,A]\ra=0$ the quantity $C$ is determined
by
\begin{equation}
C=I_2I_1^{-1},
\label{T4}
\end{equation}
which can be proved by looking at
$\la[[A^+,H],A]\ra=C\la[A^+,A]\ra+\la[B,A]\ra$.
Introduction of the zero-order Green's function $G_0$ generates a
generalized mean field expression:
\begin{equation}
\omega G_0=I_1+I_2I_1^{-1}G_0\ \ \ \ {\rm or}\ \ \ \
G_0=\frac{I_1}{\omega-I_2I_1^{-1}}\ .
 \label{T5}
\end{equation}
A Dyson equation is now defined for $G_1$:
\begin{equation}
G_1=G_0+G_0MG_1,
\label{T6}
\end{equation}
where the mass operator $M$ is defined by
\begin{equation}
M=I_1^{-1}\la\la B;A\ra\ra G_1^{-1}.
\label{T7}
\end{equation}
The exact single-particle GF may then be written as
\begin{equation}
G_1=\frac{I_1}{\omega-I_2I_1^{-1}-I_1M}
=\frac{I_1}{\omega-I_2I_1^{-1}-\Sigma}\ ,
\label{T8}
\end{equation}
where the self-energy is defined as $\Sigma=I_1M$.

The self-energy can now be expressed by
\begin{equation}
\Sigma=(G_3-G_2G_1^{-1}G_2)I_1^{-1}.
\label{T9}
\end{equation}
The proof of this expression follows from eqns
(\ref{T8},\ref{T1},\ref{T2}):
\begin{eqnarray}
\Sigma&=&\omega-I_1G_1^{-1}-I_2I_1^{-1}\nonumber\\
&=&(G_2G_1^{-1}I_1-I_2)I_1^{-1}\nonumber\\
&=&\Big(G_2G_1^{-1}(G_1\omega-G_2)+G_3-\omega G_2\Big)I_1^{-1}\nonumber\\
&=&(G_3-G_2G_1^{-1}G_2)I_1^{-1}.
\label{T10}
\end{eqnarray}
\subsubsection*{5.3.2. The Heisenberg monolayer in an external
field}

The Green's function $G_0$ of eqn
(\ref{T5}) leading to the generalized mean field expression for the
Heisenberg monolayer
in an external field (see the Hamiltonian of (\ref{4.1})) is obtained from a
Fourier transform to momentum space of
\begin{eqnarray}
I_1&=&\la[S_i^+,S_g^-]\ra=2\la S^z\ra \delta_{ig}\nonumber\\
I_2&=&\la[[S_i^+,H],S_g^-]\ra=\delta_{ig}B2\la S^z\ra+\\
& &+\delta_{ig}\sum_lJ_{il}\Big(2\la S_i^zS_l^z\ra+\la S_l^+S_i^-\ra\Big)
-\sum_l J_{il}\delta_{gl}\Big(2\la S_i^zS_l^z\ra+\la
S_i^+S_g^-\ra\Big),\nonumber
\label{T11}
\end{eqnarray}
and reads
\begin{equation}
G_{{\bf k},0}=\frac{2\la S^z\ra}{\omega -E_{\bf k}^0},
\label{T12}
\end{equation}
with
\begin{equation}
E_{\bf k}^0= B+\frac{1}{2\la S^z\ra}\frac{1}{N}\sum_{\bf q}(J_{\bf q}-
J_{\bf k-q})( 2\psi_{\bf q}^{zz}+\psi_{\bf q}^{+-}),
\label{T13}
\end{equation}
where
\begin{equation}
\psi_{\bf q}^{-+}=\frac{1}{N}\sum_{ij}\la S_i^- S_j^+\ra
e^{i{\bf q}({\bf R}_i-{\bf R}_j)},
\label{T14}
\end{equation}
and
\begin{eqnarray}
\psi_{\bf q}^{zz}&=&\frac{1}{N}\sum_{ij}\la S_i^z S_j^z\ra
e^{i{\bf q}({\bf R}_i-{\bf R}_j)}\nonumber\\
&=&\frac{1}{N}\sum_{ij}\Big(\la S_i^z\ra\la S_j^z\ra+
\la (S_i^z-\la S_i^z\ra)\ra\la (S_j^z-\la S_j\ra)\ra\Big)
e^{i{\bf q}({\bf R}_i-{\bf R}_j)}\nonumber\\
&=&2\la S_i^z\ra\la S^z_j\ra\delta_{\bf
q,0}+\frac{1}{N}\sum_{ij} K_{ij}^{zz}e^{i{\bf q}({\bf R}_i-{\bf R}_j)}.
\label{T15}
\end{eqnarray}
With this expression, the dispersion relation is evaluated as
\begin{equation}
E_{\bf k}^0= B+\la S^z\ra(J_{\bf 0}-J_{\bf k})+
\frac{1}{2\la S^z\ra}\frac{1}{N}\sum_{\bf q}(J_{\bf q}-
J_{\bf k-q})( 2K_{\bf q}^{zz}+\psi_{\bf q}^{+-})\ .
\label{T16}
\end{equation}
The first term corresponds to the Tyablikov (RPA) decoupling (see
eqn(\ref{4.11})), the term proportional to  $ K^{zz}_{\bf q}$ corresponds to
the fluctuations
of the $z$-component of the spin,  and the term proportional to the
transverse component
$\psi_{\bf q}^{-+}$  is similar to the result of the Callen
decoupling but with a different prefactor (see eqn (\ref{4.26})).

In order to describe the damping of magnons, one must go beyond this
generalized mean field approach, approximating  the
self-energy of eqn. (\ref{T10}).
The relevant term, which is the proper part in
a diagram expansion \cite{Ku02} (leading to the name
irreducible GF theory), is
\begin{equation}
\Sigma(t)=G_3I_1^{-1}=\la\la [S_i^+,H];[S_i^-,H]\ra\ra \frac{1}{2\la S^z\ra}.
\label{T17}
\end{equation}
Evaluating the commutators yields
\begin{equation}
\Sigma_{ij}(t)=\frac{1}{2\la S^z\ra}\sum_{lg}J_{il}J_{gj}\la\la
(S_i^+S_l^z-S_l^+S_i^z);(S_g^-S_j^z-S_j^-S_g^z)\ra\ra.
\label{T18}
\end{equation}
A Fourier transformation to momentum space, together with the
formulas of Section 3.4
needed to derive the spectral theorem, allows one to express the self-energy in
terms of the corresponding correlation function
\begin{eqnarray}
\Sigma(\omega)&=&\frac{1}{2\pi}\int_{-\infty}^{\infty}\frac{d\omega'}
{\omega-\omega'}(e^{\beta\omega'}-1)\int_{-\infty}^{\infty}dte^{i\omega't}
\nonumber\\
&\times &\frac{1}{2\la S^z\ra}\frac{1}{N}\sum_{ijgl}J_{il}J_{gj}e^{i{\bf
k}({\bf R}_i-{\bf
R}_j)}\la(S_g^-S_j^z-S_j^-S_g^z)(S_i^+S_l^z-S_l^+S_i^z)\ra.
\label{T19} \end{eqnarray}
In order to proceed, the correlation function
in this expression must be decoupled:
\begin{eqnarray}
& &\la(S_g^-S_j^z-S_j^-S_g^z)(S_i^+S_l^z-S_l^+S_i^z)\ra\nonumber\\
&\simeq&\la S_j^zS_l^z\ra\la S_g^-S_i^+\ra-\la S_j^zS_i^z\ra\la S_g^-S_l^+\ra
-\la S_g^zS_l^z\ra\la S_j^-S_i^+\ra+\la S_g^zS_i^z\ra\la
S_i^-S_l^+\ra\nonumber\\
&=&\psi_{jl}^{zz}\psi_{gi}^{-+}-\psi_{ji}^{zz}\psi_{gl}^{-+}-
\psi_{gl}^{zz}\psi_{ji}^{-+}+ \psi_{gi}^{zz}\psi_{il}^{-+}.
\label{T20}
\end{eqnarray}
The longitudinal
correlation function is approximated by its static value:
\begin{equation}
\la S_j^zS_l^z\ra(t)=\psi_{jl}^{zz}(t)=\psi_{jl}^{zz}(0).
\label{T21}
\end{equation}
The transverse correlation function is given via the spectral theorem by the
single-particle Green's function $G_1$
\begin{eqnarray}
\psi_{gi}^{-+}&=&\frac{i}{2\pi}\int_{-\infty}^{\infty}d\omega
\frac{e^{-i\omega t}}{e^{\beta\omega}-1}
\Big(G_{ig,1}(\omega+i\delta)-G_{ig,1}(\omega-i\delta)\Big)\nonumber\\
&\simeq&\frac{1}{N}\sum_{\bf q}\frac{2\la S^z\ra e^{-iE_{\bf q}^0t}}
{e^{\beta E_{\bf q}^0}-1}e^{-i{\bf q}({\bf R}_i-{\bf R}_j)} .
\label{T22}
\end{eqnarray}
Here, the full Green's functions $G_1$ in the brackets have been approximated
by the zero-order GF $G_0$ (a procedure which can be iterated to
self-consistency) as
\begin{equation}
2i{\rm Im}\frac{1}{N}\sum_{\bf q}G_{\bf q,0}e^{-i{\bf q}({\bf R}_i-{\bf
R}_j)}=\frac{1}{N}\sum_{\bf q}2\la S^z\ra2\pi\delta(\omega-E_{\bf q}^0)
e^{-i{\bf q}({\bf R}_i-{\bf R}_j)}.
\label{T23}
\end{equation}
Now the t- and $\omega'$-integrations in the expression for the self-energy
(\ref{T19}) can be performed and, after a Fourier transform to momentum
space, one obtains
\begin{equation}
\Sigma_{\bf k}(\omega)=\frac{1}{N}\sum_{\bf q}\frac{1}
{\omega-E_{\bf k-q}^0}(J_{\bf q}-J_{\bf q-k})^2 \psi_{\bf q}^{zz}.
\label{T24}
\end{equation}
The single-particle GF is now specified and the magnetization can be
calculated via the spectral theorem. The imaginary part of the
self-energy  describes the damping of magnons. This is the result obtained by
Plakida in Ref. \cite{Pla73}.

We are not aware of a numerical evaluation of the formulas above.
The damping of magnons with the present formalism  is, however, treated by
analytical estimates for a two-dimensional $S=1/2$ Heisenberg antiferromagnet
in Ref. \cite{BM95} and numerically in Ref. \cite{WI99} and is also
treated numerically in Ref. \cite{BR94} for a doped antiferromagnet within the
t-J model.
\newpage
\subsection*{6. Conclusions}
In this review we have given an overview of the formalism of many-body Green's
function theory (GFT) and have applied it mainly to ferromagnetic,
antiferromagnetic and coupled ferromagnetic-antiferromagnetic Heisenberg films.

A prerequisite is that the systems to be examined have periodic structures
in order to be
amenable to the resulting two-dimensional Fourier transform from momentum
to configuration space. Any attempt to deal with local magnetic
impurities would require calculations on a grid in real space, where one is
limited technically by the number of lattice sites which can be taken into
account. In this regard, the situation is the same as for
Quantum
Monte Carlo (QMC) calculations. A Green's function calculation including local
magnetic impurities and a comparison with QMC results is reported in Ref.
\cite{SLS00}.

The crucial approximation in GFT is the decoupling of the higher-order
GF's in the equation-of-motion hierarchy. The Tyablikov (RPA) decoupling yields
reasonable results for the magnetization and susceptibility
in one direction. This is seen by comparison of GFT with
`exact' QMC calculations for simple cases (see Sections 4.1.5 and 4.2.5). If
transverse correlations play a role, e.g. in calculations of the intrinsic
energy
or the specific heat, one has to go beyond RPA (see Section 5). In this case,
third-order GF's have to be decoupled, requiring vertex parameters that
have to be determined by additional constraints. For GF's of even higher
order there is still no systematic procedure for the decoupling.
Therefore, it is very difficult to make progress in this
direction. In
rare cases, e.g. for the single-ion anisotropy terms, it is possible to treat
the corresponding terms exactly by using spin relations that close the
hierarchy of equations automatically with respect to these terms (Section
4.2.5). The
exchange interaction and exchange anisotropy terms, however, have to be
decoupled by generalised RPA procedures at the level of the higher-order GF's.

A particular problem is the occurrence of exact zero eigenvalues of the
equation-of-motion matrix. After application of the the spectral theorem, an
adjunct term taking into account the
corresponding null-space must be retained. If this term is
momentum-independent,
one can apply the standard spectral theorem in which the commutator and
anticommutator GF's have to be used (Section 3.3). If, on the other hand, this
term turns out to be momentum-dependent, the standard spectral theorem fails,
and one must perform a singular value decomposition of the equation-of-motion
matrix in order to eliminate the null-space from the matrix. This not
only reduces the number of integral equations which have to be solved
self-consistently but also makes the use of the anticommutator GF superfluous
(Section 3.3). This procedure is successful in a number of cases (see Sections
4.2.3 and 4.3.2). We were not, however, able to prove that this procedure
works in general. For instance, we could not solve the spin reorientation
problem
of Section 4.2.5 in full because of numerical difficulties
which we think are
related to our inability to eliminate fully the
momentum dependence of the terms connected with the null-space.
We were able, however, to find an approximate solution for the spin
reorientation problem with an exact treatment of the single-ion anisotropy by
working in a rotated frame (see Section 4.4.2). The rotation angle is
determined from the condition that $S^z\ '$ commutes  with the
Hamiltonian in the rotated frame. This treatment simplifies the calculations
because the null-space vanishes. However, we can show by a counter example that
the condition above cannot be fulfilled in general.
For the antiferromagnet in a transverse field we can successfully deal with the
null-space working in the non-rotated frame
(see Section 4.4.2).

The result of this full GF treatment deviates drastically from a GF calculation
\cite{JBMD06} in the rotated frame that employs the additional approximation
that,
as in the ferromagnet, the reorientation angle is determined by the condition
$[S^z\ ',H\ ']\simeq 0$. To resolve the discrepancy between these results, QMC
calculations would be welcome.

In most of the applications, we have considered only
a simple square lattice because  the
double integrals in the Fourier transform from momentum space to real space
can
be transformed into a one-dimensional integral (Appendix C). This reduces the
computer time considerably (by a factor of a few hundreds) because the
Fourier transform must be
calculated many times in the self-consistency procedure.
There is, however, nothing preventing the use of
double integrals directly for other lattice
types if enough  computer time is available.

We are not aware of detailed numerical work which applies the Tserkovnikov
formulation of GFT of Section 6.2 to Heisenberg films. This would allow a
calculation of the damping of spin waves.

We hope that we have succeeded in giving an overview of the present status of
the application
of many-body GFT to Heisenberg films that will stimulate the use of
the reviewed techniques to related problems.

\newpage
\subsection*{7.1. Appendix A: Calculating the intrinsic energy with GFT}

The following Heisenberg Hamiltonian for a monolayer is taken as an example:
\begin{equation}
H=-B\sum_i S_i^z-\frac{1}{2}\sum_{il}
J_{il}(S_i^-S_l^++S_i^zS_l^z)-\sum_iK_{2,i}(S_i^z)^2.
\label{EA1}
\end{equation}
The intrinsic energy per lattice site is given by
\begin{equation}
E_i=-B \la S_i^z\ra -\frac{1}{2}\sum_l
J_{il}(\la S_i^-S_l^+\ra +\la S_i^zS_l^z\ra)-K_{2,i}\la (S_i^z)^2\ra.
\label{EA2}
\end{equation}
In the following, we take $S=1/2$ and  $S=1$ as examples.

\subsubsection*{1. $\rm {\bf S=1/2}$}
In this case, the the single-ion anisotropy term is a constant because
$\la S^zS^z\ra=1/4$.  In order to determine the intrinsic energy within GFT,
one has to calculate the quantities entering eqn (\ref{E6}) of Section 3.7.
Because $(S_i^z)^2=1/4$
and $S_i^-S_i^+=1/2-S_i^z$, the direct commutator yields
\begin{equation}
B_i^{+,-}=\la S_i^-[S_i^+,H]_-\ra=B(\smfrac{1}{2}-\la
S^z_i\ra)+\smfrac{z}{2}J\la S^z_i\ra-
\sum_lJ_{il}(\la S_i^zS_l^z\ra+\la S_i^-S_l^+\ra+\la S_i^zS_i^-S_l^+\ra).
\label{EA5}
\end{equation}
A different expression for $B_i^{+,-}$ can be obtained from
\begin{equation}
B_i^{+,-}=\lim_{\delta\rightarrow 0}\frac{1}{N}\sum_{\bf k}\frac{i}{2\pi}
\int\frac{\omega d\omega}{e^{\beta \omega}-1}\Big(G_{\bf
k}^{+,-}(\omega+i\delta)-
G_{\bf k}^{+,-}(\omega-i\delta)\Big).
\label{EA3}
\end{equation}
where $G_{\bf k}^{+,-}$ is the Fourier transform of the Green's function
\begin{equation}
G^{+,-}_{ij}=\la\la S_i^+;S_j^-\ra\ra.
\label{EA4}
\end{equation}
Equating the expressions (\ref{EA5}) and (\ref{EA3}) yields an expression
for the intrinsic energy if relation (\ref{EA2}),
\begin{equation}
-\sum_lJ_{il}\la S_i^zS_l^z\ra=2E_i+2B\la S_i^z\ra +\sum_l J_{il}\la
S_i^-S_l^+\ra +2K_2\smfrac{1}{4},
\label{EA6}
\end{equation}
is inserted into eqn (\ref{EA5}):
\begin{eqnarray}
E_i=&-&\frac{1}{4}B -\la
S_i^z\ra(\frac{1}{2}B+\frac{1}{4}zJ)+\frac{1}{2}\sum_l J_{il}\la
S_i^zS_i^-S_l^+\ra -K_2\smfrac{1}{4}\nonumber\\
&+&\frac{1}{2}\lim_{\delta\rightarrow 0}\frac{1}{N}\sum_{\bf
k}\frac{i}{2\pi} \int\frac{\omega d\omega}{e^{\beta \omega}-1}\big(G_{\bf
k}^{+,-}(\omega+i\delta)-
G_{\bf k}^{+,-}(\omega-i\delta)\big).
\label{EA7}
 \end{eqnarray}
One now needs an approximation for calculating the Green's function and the
expectation values occuring in equation (\ref{EA7}).
We have done this in Section 4.1.1 in the Tyablikov (RPA) approximation.
%
%
%
%
%
%
%
%
%
%
%
%
%
%
The resulting Green's function is (see eqn (\ref{4.10}))
\begin{equation}
G_{\bf k}(\omega)=\frac{\la [S_i^+,S_i^-]\ra}{\omega-\omega_{\bf k}^{RPA}}=
\frac{2\la S_i^z\ra}{\omega-\omega_{\bf k}^{RPA}}\ ,
\label{EA15}
\end{equation}
with the dispersion relation
\begin{equation}
\omega_{\bf k}^{RPA}=B+\la S_i^z\ra(J_0-J_{\bf k}).
\label{EA16}
\end{equation}
Now from eqn (\ref{EA3}),
\begin{equation}
B_i^{+,-}=\frac{1}{N}\sum_{\bf k}\int \frac{2\la S^z_i\ra\omega
d\omega}{e^{\beta\omega}-1}
\delta(\omega-\omega_{\bf k}^{RPA})=\frac{1}{N}\sum_{\bf k}\frac {2\la
S_i^z\ra\omega_{\bf k}^{RPA}}{e^{\beta\omega_{\bf k}^{RPA}}-1}\ .
\label{EA20}
\end{equation}
The quantity $\sum_lJ_{il}\la S_i^zS_i^-S_l^+\ra$ in equation (\ref{EA7})
is obtained from the Green's function
\begin{equation}
\la\la S_i^+;S^z_jS_j^-\ra\ra,
\label{EA17}
\end{equation}
which has same dispersion relation (\ref{EA16}) but a different inhomogeneity
$\la [S_i^+,S_i^zS_i^-]\ra$; i.e.
\begin{equation}
\la\la
S^+;S^zS^-\ra\ra_{\bf k}
(\omega)=\frac{\la [S_i^+,S_i^zS_i^-]\ra}{\omega-\omega_{\bf k}^{RPA}}.
\label{EA18}
\end{equation}
Applying the spectral theorem and a Fourier
transform one obtains
\begin{equation}
\sum_l J_{il}\la S_i^zS_i^-S_l^z\ra
=\frac{1}{N}\sum_{\bf k}J_{\bf k}
\frac{\la [S^+_i,S^z_iS^-_i]\ra}{e^{\beta\omega_{\bf k}^{RPA}}-1}
=-\frac{1}{N}\sum_{\bf k}J_{\bf k}
\frac{\la S_i^z\ra}{e^{\beta\omega_{\bf k}^{RPA}}-1}.
\label{EA19}
\end{equation}
We can  now evaluate eqn (\ref{EA7}) to
obtain the following expression for the internal energy:
\begin{equation}
E_i=-\smfrac{1}{4}B-\la S_i^z\ra(\smfrac{1}{2}B+\smfrac{1}{4}zJ)+
\frac{1}{N}\sum_{\bf k}(\omega_{\bf k}^{RPA}-\smfrac{1}{2}J_{\bf k})\frac{\la
S_i^z\ra} {e^{\beta\omega_{\bf k}^{RPA}}-1}-\smfrac{1}{4}K_2\ ,
\label{EA21}
\end{equation}
which can be calculated  after the magnetization
has been determined self-consistently from  equation (\ref{4.15})
resulting from the spectral theorem in Section 4.1.1:
\begin{equation}
\la S_i^z\ra=\frac{1}{2}-\la S_i^-S_i^+\ra=\frac{1}{2}-\frac{1}{N}\sum_{\bf
k}\frac{2\la S_i^z\ra}{e^{\beta\omega_{\bf k}^{RPA}}-1}\ .
\label{EA22}
\end{equation}

Knowledge of the intrinsic energy allows a determination of the specific heat
and the free energy via eqns (\ref{E2}) and (\ref{E3}).

\subsubsection*{2. S=1}

For $S=1$ the single-ion anisotropy of eqn (\ref{EA1}) is active.
If the magnetization is in the $z$-direction only, the exact
treatment of the anisotropy of Section 4.2.5 requires the Green's functions:
\begin{eqnarray}
G_{ij}^{+,-}&=&\la\la S^+_i;S^-_j\ra\ra,\nonumber\\
G_{ij}^{z+,-}&=&\la\la(2S_i^z-1)S_i^+;S_j^-\ra\ra\ .
\label{EA23}
\end{eqnarray}
with the exact equations of motion
\begin{eqnarray}
\omega G_{ij}^{+,-}&=&2\delta_{ij}\la
S_i^z\ra-\sum_kJ_{ik}\la\la (S_i^zS_k^+-S_k^zS_i^+);S_j^-\ra\ra
+K_{2,i}G_{ij}^{z+,-}+BG_{ij}^{+,-},\nonumber\\
\omega G_{ij}^{z+,-}&=&\delta_{ij}\la 6S_i^zS_i^z-4\ra+K_{2,i}G_{ij}^{+,-}
-\frac{1}{2}\sum_kJ_{ik}\Big(\la\la(6S_i^zS_i^z-4)S_k^+;S_j^-\ra\ra\nonumber\\
&+&2\la\la S_k^-S_i^+S_i^+;S_j^-\ra\ra-2\la\la
S_k^z(S_i^zS_i^++S_i^+S_i^z);S_j^-\ra\ra \Big)+BG_{ij}^{z+,-}.
\label{EA24}
\end{eqnarray}
We treat the single-ion anisotropy terms exactly, whereas we
introduce RPA-like decouplings for the exchange interaction terms, taking care
not to break terms with equal indices
\begin{eqnarray}
\la\la S_i^zS_k^+-S_k^zS_i^+;S_j^-\ra\ra&\simeq&\la S_i^z\ra G_{kj}^{+,-}-\la
S_k^z\ra G_{ij}^{+,-}\ ,\nonumber\\
\la\la (6S_i^zS_i^z-4)S_k^+;S_j^-\ra\ra&\simeq&\la 6S_i^zS_i^z-4\ra
G_{kj}^{+,-}\ ,\nonumber\\
\la\la S_k^z(S_i^zS_i^++S_i^+S_i^z);S_j^-\ra\ra&\simeq&\la S_k^z\ra
G_{ij}^{z+,-}\ ,\nonumber\\
\la\la S_k^-S_i^+S_i^+;S_j^-\ra\ra&\simeq&0\  (\rm neglect\ of\ transverse
\ correlations)\ . \label{EA25}
\end{eqnarray}
A Fourier transform to momentum space yields
\begin{equation}
\left(\begin{array}{c}
\omega -a\ \ \ \ \ \ \ -b\\
\ \ -c\ \ \ \ \ \ \omega-d
\end{array}\right)
\left(\begin{array}{c}
G_{\bf k}^{+,-}\\ G_{\bf k}^{z+,-}
\end{array}\right)=\left(\begin{array}{c}
A^{+,-}\\ A^{z+,-}
\end{array}\right).
\label{EA26}
\end{equation}
Here
\begin{eqnarray}
A^{+,-}&=&2\la S^z\ra \nonumber\\
A^{z+,-}&=&\la 6S^zS^z-4\ra \nonumber\\
a&=&B+\la S^z\ra (J_0-J_{\bf k})\nonumber\\
b&=&K_2\nonumber\\
c&=&K_2-\frac{1}{2}(\la 6S^zS^z -4\ra)J_{\bf k}\nonumber\\
d&=&B+\la S^z\ra J_0.
\label{EA27}
\end{eqnarray}
For a linear chain, $J_0=2J,\ \  J_{\bf k}=2\cos k$; for a square lattice,
$J_0=4J,\ \  J_{\bf k}=2(\cos k_x+\cos k_y)$.

The eigenvalues of the matrix equations are
\begin{equation}
\omega^\pm =B+\la S^z\ra(J_0-\smfrac{1}{2}J_{\bf k})\pm
\sqrt{K_2^2-\frac{1}{2}(6\la S^zS^z\ra-4)K_2J_{\bf k}+(\frac{1}{2}\la S^z\ra
J_{\bf k})^2}.
\label{EA28}
\end{equation}
The Green's functions are then given by solving eqn (\ref{EA26}).
\begin{eqnarray}
G_{\bf k}^{+,-}
&=&\frac{A^{+,-}(\omega-d)+bA^{z+,-}}{(\omega-\omega^+)(\omega-\omega^-) }\ ,
\nonumber\\
G_{\bf k}^{z+,-}
&=&\frac{A^{z+,-}(\omega-a)+cA^{+,-}}{(\omega-\omega^+)(\omega-\omega^-) }\ .
\label{EA29}
\end{eqnarray}
The spectral theorem then yields two equations determining
$\la S_i^z\ra$ and $\la S_i^zS_i^z\ra$:
\begin{eqnarray}
& & \la S_i^-S_i^+\ra=2-\la S_i^z\ra -\la S_i^zS_i^z\ra=
\frac{1}{N}\sum_{\bf k}\la S^-S^+\ra_{\bf k}= \frac{1}{N}\sum_{\bf
k}\frac{1}{(\omega^+-\omega^-)}\times\nonumber\\
& &\Big\{(A^{+,-}(\omega^+-d)+bA^{z+,-})
\frac{1}{e^{\beta\omega^+}-1}
-(A^{+,-}(\omega^--d)+bA^{z+,-})
\frac{1}{e^{\beta\omega^-}-1} \Big\},\nonumber\\
& &\la S_i^-(2S_i^z-1)S_i^+\ra=\la S_i^z\ra -\frac{1}{2}(6\la S_i^zS_i^z\ra
-4)= \frac{1}{N}\sum_{\bf k}\frac{1}{(\omega^+-\omega^-)}\times\nonumber\\
& &\Big\{(A^{z+,-}(\omega^+-a)+cA^{+,-})
\frac{1}{e^{\beta\omega^+}-1}
-(A^{z+,-}(\omega^--a)+cA^{+,-})
\frac{1}{e^{\beta\omega^-}-1} \Big\}.\nonumber\\
\label{EA30}
\end{eqnarray}
Now, substitution of a) $A_i=S^-_i$,
$C_i=S_i^+$
and  b) $A_i=S_i^-$, $C_i=(2S_i^z-1)S_i^+$ into
 eqn (\ref{E6}) of Section 3.7 and insertion of the GF's
(\ref{EA29}) yields
\begin{eqnarray}
B^{+,-}&=&\la S_i^-[S^+_i,H]\ra
=\frac{1}{N}\sum_{\bf k}\frac{1}{(\omega^+-\omega^-)}\times\nonumber\\
& &\Big\{
(A^{+,-}(\omega^+-d)+bA^{z+,-})\frac{\omega^+}{e^{\beta\omega^+}-1}
-(A^{+,-}(\omega^--d)+bA^{z+,-})\frac{\omega^-}{e^{\beta\omega^-}-1}\Big\},
\nonumber\\
B^{z+,-}&=&\la S_i^-[(2S^z_i-1)S_i^+,H]\ra=\frac{1}{N}\sum_{\bf
k}\frac{1}{(\omega^+-\omega^-)}\times\nonumber\\
& &\Big\{
(A^{z+,-}(\omega^+-a)+cA^{+,-})\frac{\omega^+}{e^{\beta\omega^+}-1}
-(A^{z+,-}(\omega^--a)+cA^{+,-})\frac{\omega^-}{e^{\beta\omega^-}-1}\Big\}.
\nonumber\\
\label{EA31}
\end{eqnarray}
Calculating the commutators directly, inserting eqn (\ref{EA2}) and
eliminating $\sum_{k}J_{ik}\la (S_i^z)^2S_k^z\ra$ by forming the
difference $3B^{+,-}-B^{z+,-}$ and solving for $E_i$ yields
\begin{eqnarray}
E_i=\smfrac{1}{8}(3B^{+,-}-B^{z+,-})-\smfrac{1}{2}(B+K_2) -\smfrac{1}{2}
(B+K_2+ zJ)\la S^z_i\ra)\nonumber\\
+\frac{1}{8}\sum_k J_{ik}\Big(-\la S_k^-S_i^+\ra -2\la S_i^-S_k^+\ra +\la
S_k^-(2S_i^z-1)S_i^+\ra\Big). \label{EA32}
\end{eqnarray}
The first term comes from eqn (\ref{EA31}).
Performing a Fourier transform on the last term gives
\begin{equation}
\smfrac{1}{8}\smfrac{1}{N}\sum_{\bf k}J_{\bf k}(-3\la S^-S^+\ra_{\bf k}+\la
S^-(2S^z-1)S^+\ra_{\bf k}). \label{EA33}
\end{equation}
This together with eqns (\ref{EA30}) and
(\ref{EA27}) yields the final result for the intrinsic energy
 %
%
%
%
%
%
%
\begin{eqnarray}
E_i=& &-\smfrac{1}{2}(B+K_2) -\smfrac{1}{2}
(B+K_2+ zJ)\la S^z_i\ra\nonumber\\
& &+\smfrac{1}{8}\sum_{\bf k}\frac{1}{\omega^+-\omega^-}\Big\{\Big[
\Big(2\la S_i^z\ra(3(\omega^+-B-\la S^z_i\ra J_0)-K_2+\smfrac{1}{2}\la
6S_i^zS_i^z-4\ra J_{\bf k}\Big)\nonumber\\
& &+\la 6S_i^zS_i^z-4\ra\Big(3K_2-(\omega^+-B-\la S_i^z\ra(J_0-J_{\bf k})\Big)
\frac{\omega^+-J_{\bf
k}}{e^{\beta\omega^+}-1}\Big]+\Big[\omega^+\rightarrow\omega^-\Big]\Big\}\ .
\nonumber\\
\end{eqnarray}

For larger values of spin higher-order GF's are needed, but one can
proceed
analogously. The procedure applies of course to other Hamiltonians as well.

\subsection*{7.2. Appendix B: The curve-following procedure}
Consider a set of $n$ coupled equations characterised by $m$
parameters $\{P_i;i=1,2\ldots,m\}$
and $n$ variables $\{V_i;i=1,2,\ldots,n\}$:
\begin{equation}
   S_i({\bf P}[m];{\bf V}[n]) = 0,\  {\rm for}\ \  {i=1,\ldots,n} .
\end{equation}
In our case, the parameters are
the temperature, the magnetic field components, the dipole coupling strengths,
the anisotropy strengths, etc; the variables are the spin-correlations.
The coupled equations $S_i$ are obtained  from the spectral theorem expressions
for the correlations supplemented by the regularity conditions if necessary.

For fixed parameters ${\bf P}$, we look for solutions $S_i=0$
at localised points, ${\bf V}[n]$, in the n-dimensional space. If now one of
the parameters $P_k$ is considered to be an additional
variable $V_\circ$ (e.g. the
temperature),
then the solutions to the coupled equations define
curves in the $(n+1)$-dimensional space ${\bf V}[n+1]$. From here on, we
denote the points in this space by
$\{V_i;i=0,1,2,...\ldots,n\}$. The curve-following
method is a procedure for generating these solution-curves point by
point from a few closely-spaced points already on a curve; i.e. the method
generates a new solution-point from
the {\em approximate} direction of the curve in the vicinity of
a new {\em approximate} point. This is done by an iterative procedure
described below. If no points on the curve are known, then
an approximate solution point and an approximate direction must be
estimated before applying the iterative procedure to obtain the first
point on the curve. A second point can then be obtained in the same
fashion. If at least two solution-points are available, then the new
approximate point can be extrapolated from them and the approximate
direction can be taken as the tangent to the curve at the last point.

The iterative procedure for finding a better point,
${\bf V}$, from an approximate point, ${\bf V}^{\circ}$, is now described.
One searches for the isolated solution-point
in the $n$-dimensional subspace perpendicular to the approximate direction,
which we characterise by a unit
vector, $\widehat{\bf u}$.
The functions $S_i$ are expanded up to first order in the corrections
about the approximate point, ${\bf V^{\circ}}$:
\begin{equation}
   S_i({\bf V}) = S_i({\bf V^{\circ}}) + \sum_{j=0}^n
\frac{\partial S_i^{\circ}}{\partial V_j} \Delta V_j ,
\end{equation}
where $\Delta V_j=V_j-V_j^{\circ}$.
At the solution, the $S_i$ are all zero, whereas at the approximate
point ${\bf V^\circ}$ the functions have non-zero values, $S_i^\circ$; hence,
one must solve for the corrections $\Delta V_j$ for which the left-hand side
in the above equation is zero:
 \begin{equation}
   \sum_{j=0}^n \frac{\partial S_i^{\circ}}{\partial V_j} \Delta V_j =
-S_i^\circ; \{i=1,2,\ldots,n\}.
 \end{equation}
These $n$ equations are supplemented by the constraint requiring the
correction to be perpendicular to the unit direction vector:
\begin{equation}
  \sum_{j=0}^n\widehat{u_j} \Delta V_j = 0.
\end{equation}
This improvement algorithm in the subspace is repeated until each of
the $S_i^\circ$ is sufficiently small. In practice we required that
$\sum_i{(S_i^\circ)^2 \leq \epsilon}$, where we took $\epsilon=10^{-16}$. If there is no convergence,
the extrapolation
step-size used to obtain the original ${\bf V}^{\circ}$ is halved,
a new extrapolated point obtained and the improvement algorithm repeated.

The curve-following method is quite general and can be applied to any
coupled equations characterised by differentiable functions. By utilizing
the information about the solution at neighbouring points, the method is
able to find new solutions very efficiently, routinely converging after
a few iterations once two starting points have been found.
In addition, no single parameter or variable is singled out as "the"
independent variable; instead, the ($n$+1)-dimensional curve can be viewed as
being described parametrically in terms of the distance along the curve. This
vantage point has great practical consequence: solutions in the neighbourhood
of turning points (e.g. hysteresis for $\la S^z\ra$ as a function of field
${\bf B}$) are just as easily determined as in any other region because the
solution is always sought in a subspace nearly orthogonal to the solution
curve.

\newpage
\subsection*{7.3. Appendix C: Reducing a 2-dimensionsl to a 1-dimensional
integral for a square lattice}

In the following we show how the double integral occuring from a
two-dimensional Fourier transform when dealing with a
square lattice (see e.g. eqn (\ref{4.15})) can be transformed into a
1-dimensional integral.
This transformation saves a lot of computer time in many of the applications
discussed in the present review.

\begin{figure}[htb]
\begin{center}
\protect
\includegraphics*[bb = 85  0 575 485,
angle=0,clip=true,width=7cm]{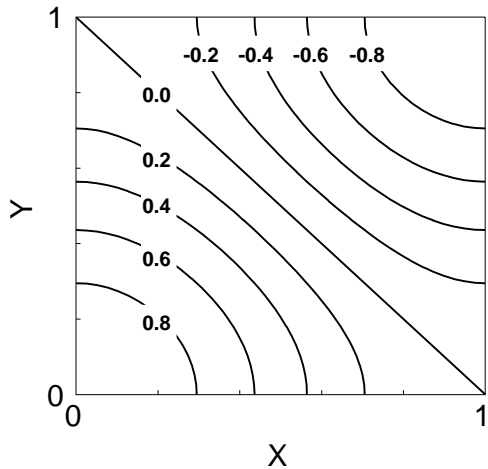}
\protect
\caption{ Contours of constant $\gamma$}
\label{cfig1}
\end{center}
\end{figure}

\begin{figure}[bht]
\begin{center}
\protect
\includegraphics*[bb = 85  0 575 485,
angle=0,clip=true,width=7cm]{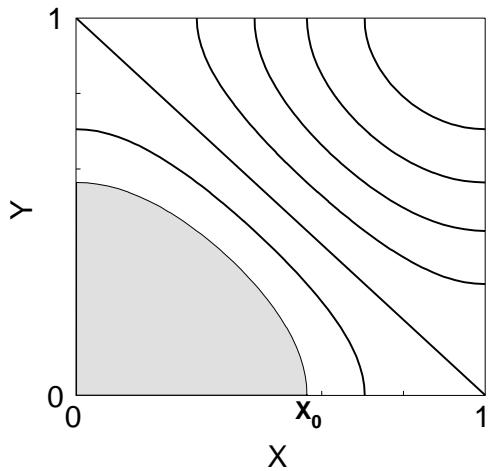}
\protect
\caption{
Area of the unit
   square to the left of a $\gamma$-contour for $\gamma>0$}
\label{cfig2}
\end{center}
\end{figure}

\begin{figure}[htb]
\begin{center}
\protect
\includegraphics*[bb = 85  0 575 485,
angle=0,clip=true,width=7cm]{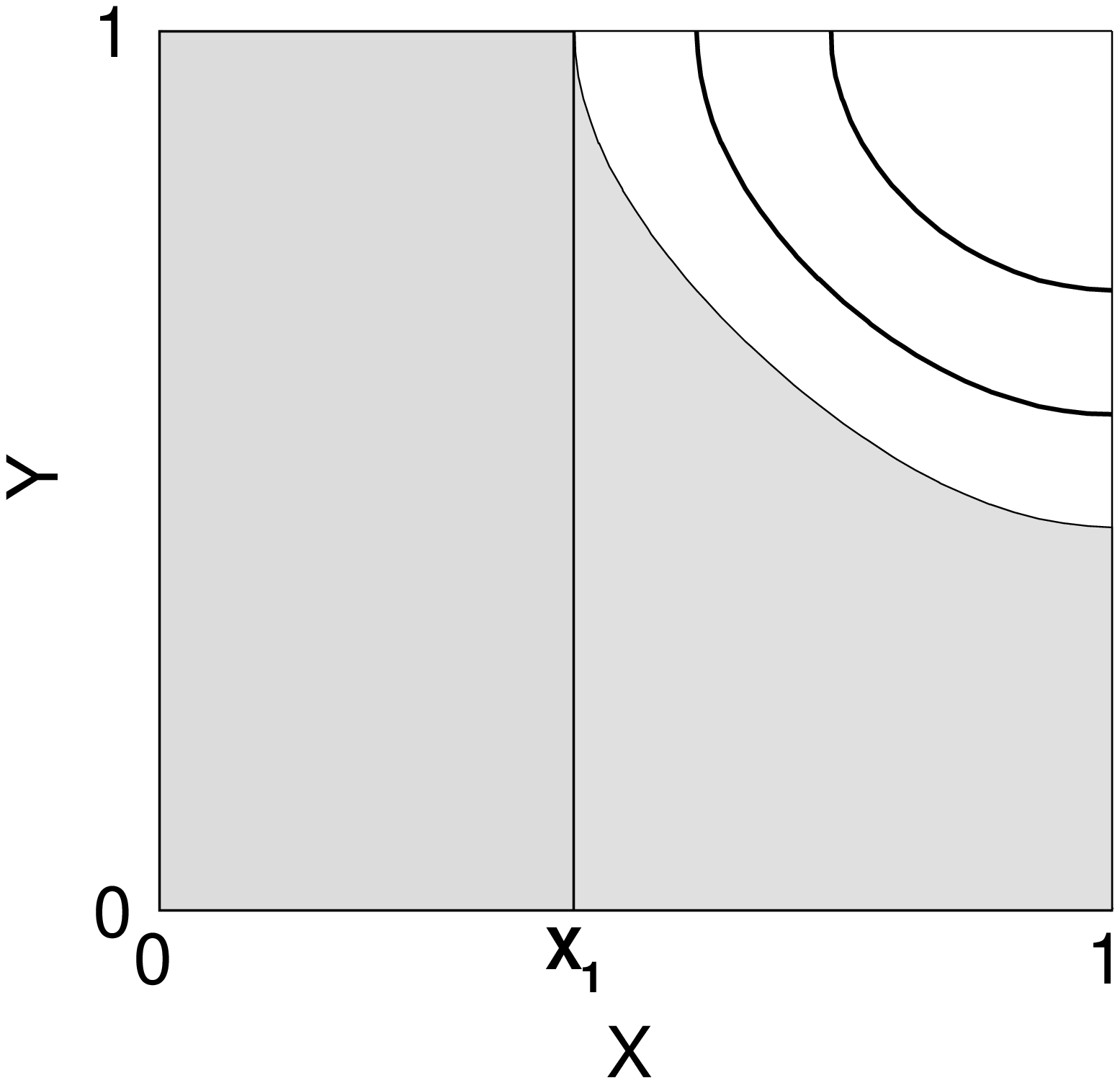}
\protect
\caption{
Area of the unit
   square to the left of a $\gamma$-contour for $\gamma<0$}
\label{cfig3}
\end{center}
\end{figure}

Consider the evaluation of a double integral with the structure
\begin{equation}
I=\frac{1}{\pi^2}\int_{0}^{\pi}\!\int_{0}^{\pi} f(\cos k_x + \cos k_y)\,dk_x\,dk_y.
\end{equation}
By substituting $x=k_x/\pi$ and $y=k_y/\pi$, this can be written as
\begin{equation}
   \int_{0}^{1}\!\int_{0}^{1} f(\cos \pi x + \cos \pi y)\,dx\,dy.
\end{equation}
By making use of the fact that the integrand has the same value for all
values of $x$ and $y$ satisfying the relation
\begin{equation}
   \cos \pi x + \cos \pi y = 2\gamma,
\end{equation}
where $\gamma$ lies in the range $(-1,1)$,
it is possible to reduce the double integral to a single integral over
some suitable variable.  The contours of constant $\gamma$ are shown in
Fig.~\ref{cfig1}.  Each contour is given by an equation
\begin{equation}
   \pi y(x) = \arccos (2\gamma - \cos \pi x).          \label{e:gamcon}
\end{equation}
Define now a function, $A(\gamma)$, which is the area in the unit square in the
$xy-$plane lying to the left of the line $y(x)$ defined by
Eq.~\ref{e:gamcon} for each value of $\gamma$.  From the diagram, it is
evident that the function $A(\gamma)$ is given by
\begin{equation}
   A(\gamma) =\frac{1}{\pi} \int_{0}^{x_0}\! \arccos (2\gamma - \cos\pi x)\,dx
\end{equation}
for $\gamma > 0$ and
\begin{equation}
   A(\gamma) = x_1 + \frac{1}{\pi}\int_{x_1}^{1}\! \arccos (2\gamma - \cos \pi
x)\,dx \end{equation}
for $\gamma \leq 0$, where
$x_0 =\frac{1}{\pi} \arccos (2\gamma - 1)$ and $x_1 =\frac{1}{\pi} \arccos
(2\gamma + 1)$. These areas are shown in Fig.~\ref{cfig2} and
Fig.~\ref{cfig3}

The
double integral may now be written as a single integral over the
variable $A$ over the interval $(0,1)$:
\begin{equation}
   I = \int_{0}^{1}\!f(2\gamma(A))\,dA.
\end{equation}
In order to evaluate the integral numerically, it is only necessary to
have an efficient representation of the function $\gamma(A)$, so that
a quadrature can be used to estimate the integral.  A good strategy is
to compute the function $\gamma(A)$ at a sufficiently large number of points
so that it can be accurately fitted to a cubic spline function.  Thus,
the labour involved in evaluating the integral $I$ is enormously
reduced, since the numerical representation of $\gamma(A)$ need only
be computed once.
%
%
%

%
%
%
%
%
\begin{figure}[htb]
\begin{center}
\protect
\includegraphics*[bb = 0  0 775 485,
angle=0,clip=true,width=11cm]{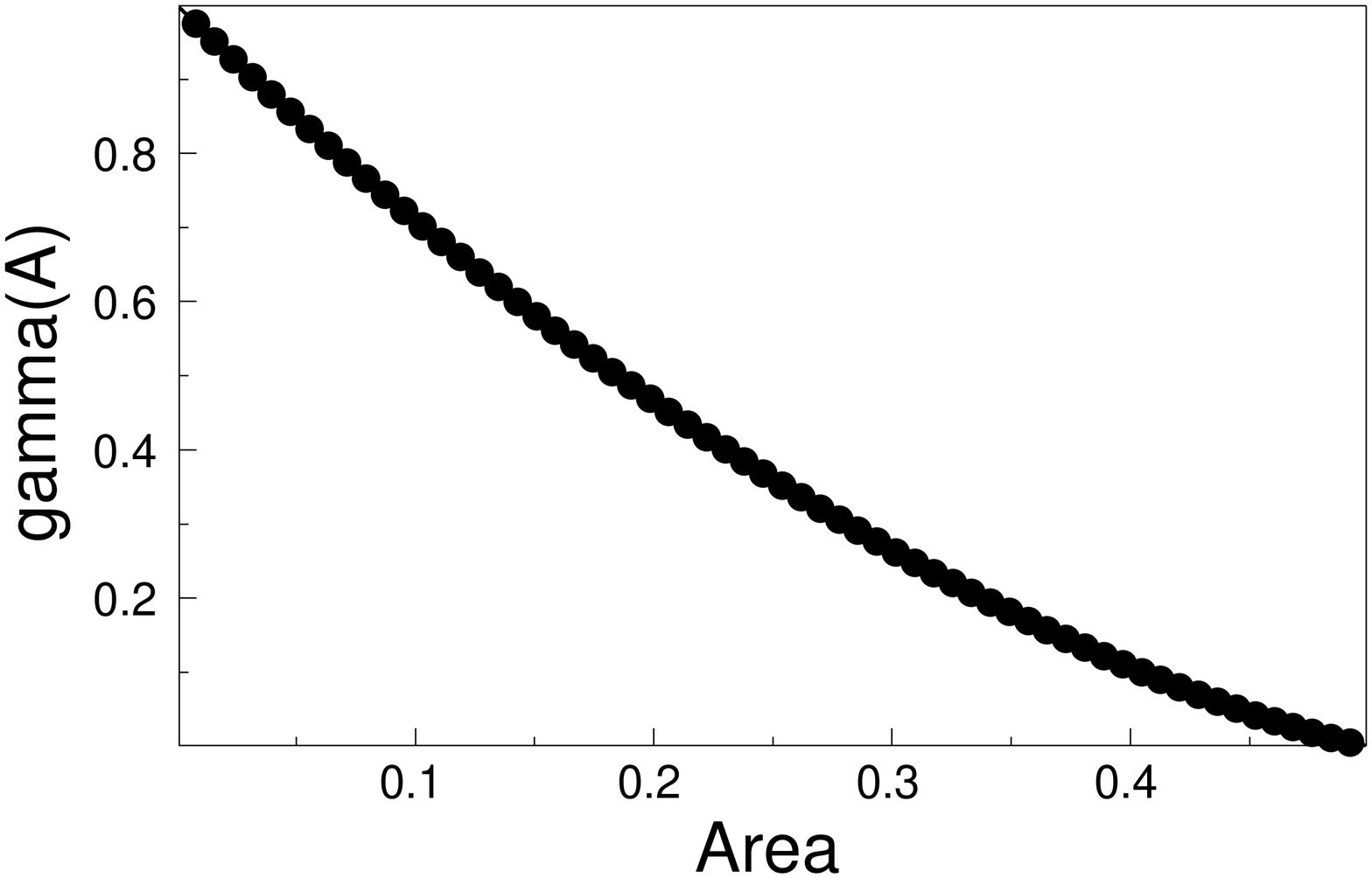}
\protect
\caption{
 $\gamma$ as a
   function of area $A$.}
\label{cfig7}
\end{center}
\end{figure}

The numerical evaluation of $\gamma(A)$ is not without its problems,
since the first derivative of the inverse function $A(\gamma)$ has a
singularity in its first derivative at $\gamma=0$.
Even though we need the inverse function,
$\gamma(A)$, whose derivatives go to zero at $\gamma=0$, there are still
numerical difficulties in representing $\gamma(A)$ by a spline function
in the neighbourhood of $A=0.5$; hence, it is better use a spline
function to represent the function $g(A)$, defined as $(0.5-A)\gamma(A)$,
and to get $\gamma(A)$ from $g(A)/(0.5-A)$ in the neighbourhood of the
singularity, using the value $\gamma(0.5)=0$ at the singularity itself.
The function $g(A)$, fitted to a cubic spline function, yields numerically
stable values of the function and its first two derivatives.
The spline fit to $g(A)$ is obtained from
values of the function tabulated at a set of knots equally spaced
in the range $(0,0.5)$ plus values of the derivative of the function
at $A=0$, (derivative = $-1/(2\pi)$) and $A=0.5$, (derivative=0).
The second derivative of $g(A)$ actually goes smoothly to zero at
$A=0.5$.
Values of $\gamma(A)$ in the range $(0.5,1)$ are obtained from the
fitted values using the symmetry relation
$\gamma(0.5 + u)= -\gamma(0.5 - u)$.  The function $\gamma(A)$ is
shown in Fig.~\ref{cfig7}.

While the above procedure allows one to obtain accurate values of
$\gamma(A)$ over the whole range of area, it does nothing to suppress
the effects of the singularity in the first derivative of $A(\gamma$) at
$\gamma=0$.  These effects are not serious but they demand more effort
from the integrator near $A=0.5$.  They can, however, be minimized
by integrating out the constant part of the function $f(2\gamma)$,
which is just $f(0)$ itself:
\begin{equation}
   I = f(0) + \int_{0}^{1}\!(f(2\gamma(A)-f(0))\,dA.
\end{equation}

We mention that in Ref. \cite{Ilk04} the double integral is transformed
into an elliptic integral of
the first kind
with a transformation found in Ref. \cite{Colpa72}.
\newpage

\newpage

\subsection*{7.4. Appendix D: Treatment of the magnetic dipole-dipole
interaction}

In this appendix, we apply the generalized Tyablikov (\ref{5.6})decoupling to
the magnetic
dipole-dipole interaction. From this result the mean field approximation, as it
is used e.g. in eqn (\ref{5.13}), is obtained by neglecting the momentum
dependence due to the lattice.
After the decoupling procedure, the resulting term
$\la\la[S_i^\alpha,H^{dipole}];(S_j^z)^m(S_j^-)^n\ra\ra$  is added to the
equations of motion
(\ref{5.5}), where
$H^{dipole}$ is the last term in eqn (\ref{5.1}). After a Fourier transform to
momentum space, one has the following additional terms in the equation of
motion:
\begin{equation}
\left(\begin{array}{ccc}
-T_{\bf k}^+ & -T_{\bf k}^- & -T_{\bf k}^z\\
(T_{\bf k}^-)^* & T_{\bf k}^+ & (T_{\bf k}^z)^*\\
T_{\bf k}^{z\pm} & -(T_{\bf k}^{z\pm})^* & 0
\end{array}\right)
\left(\begin{array}{c}
G_{\eta}^{+,mn}\\
G_{\eta}^{-,mn}\\
G_{\eta}^{z,mn}
\end{array}\right)\ ,
\label{DP1}
\end{equation}
where
\begin{eqnarray}
T_{\bf k}^+&=&g\la S^z\ra\Big(T_{20}^0+T_{02}^0
+\smfrac{1}{2}T_{20}^{\bf k}+\smfrac{1}{2}T_{02}^{\bf k}\Big)\ , \nonumber\\
T_{\bf k}^-&=&\smfrac{3}{2}g\la S^z\ra\Big(T_{20}^{\bf k}-T_{02}^{\bf k}
+2iT_{11}^{\bf k}\Big)\ , \nonumber\\
T_{\bf k}^z&=&g\la S^+\ra\Big(T_{20}^{\bf k}+T_{02}^{\bf k}
+\smfrac{1}{2}T_{20}^{0}+\smfrac{1}{2}T_{02}^{0}\Big)\ ,\nonumber\\
T_{\bf k}^{z\pm}&=&\smfrac{1}{4}g\Big(\la S^-\ra(T_{20}^{\bf k}+T_{02}^{\bf k}
-T_{20}^{0}-T_{02}^{0})\nonumber\\
& &-3\la S^+\ra(T_{20}^{\bf k}-T_{02}^{\bf k}-2iT_{11}^{\bf k})\Big)\ ,
\label{DP2}
\end{eqnarray}
and
\begin{equation}
T_{\mu\nu}^{\bf k}=\sum_{lm}\frac{x_l^\mu
y_m^\nu}{(x_l^2+y_m^2)^{5/2}}\exp(ik_xx_l)
\exp (ik_yy_m)
\label{DP3}
\end{equation}
are oscillating lattice sums, which can be evaluated with Ewald summation
techniques as outlined e.g. in Ref. \cite{J97}.

This RPA treatment of the magnetic dipole coupling complicates the calculation
of the magnetization considerably because of the presence of complex and
dispersive (${\bf k}$-dependent) terms;
therefore, we have neglected these terms in the applications and retained the
non-dispersive terms only. This corresponds to a mean field treatment of the
dipole coupling. In this approximation, eqns (\ref{DP2}) reduce to
\begin{eqnarray}
T_{\bf k}^+&=&g\la S^z\ra(T_{20}^0+T_{02}^0)\ ,\nonumber\\
T_{\bf k}^-&=&0\ ,\nonumber\\
T_{\bf k}^z&=&\frac{g}{2}\la S^+\ra(T_{20}^0+T_{02}^0)\ ,\nonumber\\
T_{\bf k}^{z\pm}&=&-\frac{g}{4}\la S^-\ra(T_{20}^0+T_{02}^0).
\label{DP4}
\end{eqnarray}
This simplification takes the dipole coupling into account by an effective
renormalization of the external magnetic field and leads to eqns (\ref{5.13})
of Section 4.2.1.

In order to justify this procedure we have done RPA calculations
for the dipole interaction for two limiting cases: a
perpendicular and an in-plane magnetization. In the appendix of
Ref. \cite{FJKE00}, it is shown that, for these cases, a mean field calculation
is a rather good approximation to the RPA result if the dipole coupling
strength is much smaller than the strength of the exchange interaction, which
is the case for many systems.
We are not aware of a numerical treatment of the dipole coupling for the spin
reorientation problem in GFT taking the dispersive and complex terms of eqn
(\ref{DP2}) into account.

In the present review, we have applied the dipole-dipole interaction only in
cases where the dipole coupling strength is small as compared to the
strength of the exchange interaction, $g/J<<1.$ Ref. \cite{BMW00} reviews
dipolar
effects in quasi-two-dimensional magnetic films,  treating also
cases $g\simeq J$, $g>>J$  and $J=0$  with classical Monte Carlo
simulations.

\newpage


\begin{thebibliography}{99}
\bibitem{Zub60} D.M. Zubarev, Sov. Phys. Usp. {\bf 3}, 320 (1960).
\bibitem{BoTy62} V.L. Bonch-Brevich, S.V. Tyablicov, The Green Function Method
in Statistical Mechanics, North-Holland, Amsterdam, 1962.
\bibitem{GHE01} W. Gasser, E. Heiner, K. Elk, Greensche Funktionen in
der Festk\"orper- und Vielteilchenphysik, Wiley-VHC, Berlin, 2001.
\bibitem{Tya67} S.V. Tyablicov, Methods in the Quantum Theory of Magnetism,
Plenum Press, New York, 1967.
\bibitem{Nol86} W.Nolting, Quantentheorie des Magnetismus, vol.2,
Teubner, Stuttgart, 1986.
\bibitem{Held05} K. Held, Electronic Structure Calculations using Dynamical
Mean Field Theory, cond-mat/0511293, based on the Habilitation thesis/
Stuttgart University, 2004.
\bibitem{JB06} P.J. Jensen, K.H. Bennemann, Surface Science Reports {\bf 61},
129 (2006).
\bibitem{Faz99} P. Fazekas, Lecture Notes on Electron Correlations and
Magnetism, World Scientific, Singapore, 1999.
\bibitem{FJKE00} P. Fr\"obrich, P.J. Jensen, P.J. Kuntz, A. Ecker, Eur. Phys.
J. B {\bf18}, 579 (2000).
\bibitem{ST65} K.W.H. Stevens, G.A. Tombs, Proc. Phys. Soc. {\bf 85}, 1307
(1965).
\bibitem{RG71} J.G. Ramos, A.A. Gomes, Il Nuovo Cimento {\bf 3}, 441 (1971).
\bibitem{PFTV89} W.H. Press, B.P. Flannery, S.A. Teukolsky, W.T. Vetterling,
Numerical Recipes, Cambridge University Press, 1989.
\bibitem{FK05} P. Fr\"obrich, P.J. Kuntz, J. Phys.: Condens. Matter {\bf 17},
1167 (2005).
\bibitem{EFJK99} A. Ecker, P. Fr\"obrich, P.J. Jensen, P.J. Kuntz,
J. Phys.: Condens. Matter {\bf 11}, 1557 (1999).
\bibitem{TGHS98} C. Timm, S.M. Girvin, P. Henelius, A.W. Sandvik, Phys. Rev. B
{\bf 58} (1998) 1464.
\bibitem{Tya59} S.V. Tyablikov, Ukr. Mat. Zh. {\bf 11}, 289 (1959).
\bibitem{Cal63} H.B. Callen, Phys. Rev. {\bf 130}, 890 (1963).
\bibitem{Pra63} E. Pravecki, Phys. Lett. {\bf 6}, 147 (1963).
\bibitem{TKH62} R. A. Tahir-Kheli, D. ter Haar, Phys. Rev. {\bf 127}, 88 and 95
(1962).
\bibitem{JB98} P.J. Jensen, K.H. Bennemann, in Magnetism and Electronic
Correlations in Local-Moment Systems: Rare-Earth Elements and Compounds, ed. M.
Donath, P.A. Dowben, W. Nolting, World Scientific, Singapore, 1998, p.131-141.
\bibitem{HU00} A. Hucht, K.D. Usadel, Phil. Mag. B {\bf 80}, 275 (2000).
\bibitem{MW66} N.M. Mermin, H. Wagner, Phys. Rev. Lett. {\bf 17}, 1133 (1966).
\bibitem{TKDBB03} I. Turek, J. Kudrnovsky, V. Drchal, P. Bruno, S. Bl\"ugel,
phys. stat. sol. (b) {\bf 236}, 318 (2003).
\bibitem{Von74} S.V. Vonsovskii, in Magnetism, vol. 2, Chapter 23 (J. Wiley and
Sons, 1974).
\bibitem{DLN79} Diep-The-Hung, J.C.S. Levy, O.Nagai, phys. stat. sol. (b) {\bf
93}, 351 (1979).
\bibitem{SN99} R. Schiller, W. Nolting, Solid State Commun. {\bf 110}, 121
(1999).
\bibitem{CPPR01} C. Cucci, M.G. Pini, P. Politi, A.Rettori, J. Mag. Mag. Mat.
{\bf 231}, 98 (2001).
\bibitem{FJK00} P. Fr\"obrich, P.J.
Jensen, P.J. Kuntz, Eur. Phys. J. B {\bf13}, 477 (2000).
\bibitem{Lin67} M.E. Lines, Phys. Rev. {\bf156}, 534 (1967).
\bibitem{EM91a} R.P. Erickson, D.L. Mills, Phys. Rev. B {\bf 43}, 11527 (1991).
\bibitem{EM91b} R.P. Erickson, D.L. Mills, Phys. Rev. B {\bf 44}, 11825 (1991).
\bibitem{AC64} F.B. Anderson, H.B. Callen,  Phys. Rev. {\bf
136}, A1068 (1964).
\bibitem{WDFJK04} H.Y. Wang, Z.H. Dai, P. Fr\"obrich, P.J. Jensen, P.J. Kuntz,
Phys. Rev. B {\bf 70}, 134424 (2004).
\bibitem{HLT99} L. Hu, H. Li, R. Tao, Phys. Rev. B {\bf 60}, 10222 (1999).
\bibitem{GSL01} W. Guo, L.P. Shi, D.L. Lin, Phys. Rev. B {\bf 62}, 14259
(2001).
\bibitem{GL03} W. Guo, D.L. Lin, Phys. Rev. B {\bf 67}, 224402 (2003).
\bibitem{FK03a} P. Fr\"obrich, P.J. Kuntz, Eur. Phys. J. B {\bf 32}, 445
(2003).
\bibitem{FK03b} P. Fr\"obrich, P.J. Kuntz, Phys. Rev. B {\bf 68}, 014410
(2003).
\bibitem{WWW04} H.Y. Wang, C.Y. Wang, E.G. Wang, Phys. Rev. B {\bf 69}, 174431
(2004).
\bibitem{Ilk04} V. Ilkovic, phys. stat. sol. (b) {\bf 241}, 420 (2004).
\bibitem{TKI06} S. Tuleja, J. Kecer, V. Ilkovic, phys. stat. sol. (b) {\bf
243}, 1352 (2006).
\bibitem{MJB94} D.K. Morr, P.J. Jensen, K.H. Bennemann, Surface Science {\bf
307-309}, 1109 (1994).
\bibitem{Yab91} D.A. Yablonskyi, Phys. Rev. B {\bf 44}, 4467 (1991).
\bibitem{FKS02} P. Fr\"obrich, P.J. Kuntz, M. Saber, Ann. Phys. (Leipzig)
{\bf 11}, 387 (2002).
\bibitem{Dev71} J.F. Devlin, Phys. Rev. B {\bf 4}, 136 (1971).
\bibitem{BM88} M. Bander, D.L. Mills, Phys. Rev. B {\bf38}, R12015 (1988).
\bibitem{JA2000} P.J. Jensen, F. Aguilera-Granja, Phys. Lett. A
{\bf 269}, 158 (2000).
\bibitem{HFKTJ02} P. Henelius, P. Fr\"obrich, P.J. Kuntz, C. Timm, P.J. Jensen,
Phys. Rev. B {\bf 66}, 094407 (2002).
\bibitem{SKN05} S. Schwieger, J. Kienert, W. Nolting, Phys. Rev.
B {\bf 71}, 024428 (2005).
\bibitem{PPS05} M.G. Pini, P. Politi, R.L. Stamps, Phys. Rev. B {\bf 72},
014454 (2005).
\bibitem{BC91} P. Bruno, C. Chappert, Phys. Rev. Lett. {\bf 67}, 1602 (1991).
\bibitem{AMT95} N.S. Almeida, D.L. Mills, M. Teitelman, Phys. Rev. Lett. {\bf
75}, 733 (1995).
\bibitem{PJJ99} P.J. Jensen, K.H. Bennemann, P. Poulopoulos, M. Farle, F.
Wilhelm, K. Baberschke , Phys. Rev. B {\bf 60}, R14994 (1999).
\bibitem{B95} P. Bruno, Phys. Rev. B {\bf 52}, 411 (1995).
\bibitem{JSSBBW05} P.J. Jensen, C. Sorg, A. Scherz, M. Bernien, K. Baberschke,
H. Wende, Phys. Rev. Lett. {\bf 93}, 039703 (2005).
\bibitem{SSBPBWJ05} A. Scherz, C. Sorg, M. Bernien, N. Ponpandian, K.
Baberschke, H. Wende, J. Jensen, Phys. Rev. B {\bf 72}, 054447 (2005).
\bibitem{SN04} S. Schwieger, W. Nolting, Phys. Rev. B {\bf 69}, 224413 (2004)
\bibitem{SKN05a} S. Schwieger, J. Kienert, W. Nolting, Phys. Rev. B
{\bf 71}, 174441 (2005).
\bibitem{FK06} P. Fr\"obrich, P.J. Kuntz, in Progress in Nonequilibrium Green's
Functions III, J. Phys.: Conf. Series {\bf 35}, 157 (2006).
\bibitem{LRKPBBWM02} J. Lindner, C. R\"udt, E. Kosubek, P. Poulopoulos, K.
Baberschke, P. Blomquist, R. W\"appling, D.L. Mills, Phys. Rev. Lett. {\bf
88}, 167206 (2002).
\bibitem{PS93} C. Pich, F. Schwabl, Phys. Rev. B {\bf 47}, 7957 (1993).
\bibitem{HP40} T. Holstein, H. Primakoff, Phys. Rev. {\bf 58}, 1098 (1940).
\bibitem{PS94} C. Pich, F. Schwabl, Phys. Rev. B {\bf 49}, 413 (1994).
\bibitem{PS95} C. Pich, F. Schwabl, J. Mag. Mag. Mat. {\bf 148}, 30 (1995).
\bibitem{HPS01} M. Hummel, C. Pich, F. Schwabl, Phys. Rev. B {\bf 63}, 094425
(2001).
\bibitem{FKJ05} P. Fr\"obrich, P.J. Kuntz, P.J. Jensen, J.
Phys.: Condens. Matter {\bf 17}, 5059 (2005).
\bibitem{D91} H.T. Diep, Phys. Rev. B {\bf 40}, 4818 (1989)
;  ibid. {\bf 43}, 8509 (1991).
\bibitem{WHQW04} H.Y. Wang, M. Qian, E.G. Wang, J. Appl.
Phys. {\bf 95}, 7551 (2004).
\bibitem{MUH93} A. Moschel, K.D. Usadel, A. Hucht,  Phys. Rev.
B {\bf 47}, 8676 (1993).
\bibitem{MU93} A. Moschel, K.D. Usadel, Phys. Rev. B {\bf 48},
13991 (1993).
\bibitem{WD04} H.Y. Wang, Z.H. Dai, Commun. Theor. Phys.
(Beijing, China) {\bf 42}, 141 (2004).
\bibitem{MU94} A. Moschel, K.D. Usadel, J. Mag. Mag. Mat.
{\bf 136}, 99 (1994).
\bibitem{L04} Li Qing'an, Phys. Rev. B {\bf 70}, 014406 (2004).
\bibitem{NNDN04} V. Than Ngo, H. Viet Nguyen, H.T. Diep, V. Lien Nguyen,
Phys. Rev. B {\bf 69}, 134429 (2004).
\bibitem{JKWO03} P.J. Jensen, S. Knappmann, W. Wulfhekel,  H.P, Oepen, Phys.
Rev. B {\bf 67} 184417 (2003).
\bibitem{FK04} P. Fr\"obrich, P.J. Kuntz, phys. stat. sol.
(b) {\bf 241}, 925 (2004).
\bibitem{FK104} P. Fr\"obrich, P.J. Kuntz, J. Phys. Condens.
Matter {\bf 16}, 3453 (2004).
\bibitem{JKD05} P.J. Jensen, M. Kiwi, H. Dreyss\'e,
 Eur. Phys. J. B {\bf 46}, 541 (2005).
\bibitem{MNLS98} T.J. Moran, J. Nogu\'es, D. Lederman, I.K. Schuller, Appl.
Phys. Lett. {\bf 72}, 617 (1998).
\bibitem{SB98} T.C. Schulthess, W.H. Butler, Phys. Rev. Lett. {\bf 81}, 4516
(1998).
\bibitem{NCK00} U. Nowak, R.W. Chantrell, E.C. Kennedy, Phys. Rev. Lett. {\bf
84}, 163 (2000).
\bibitem{JBMD06} P.J. Jensen, K.H. Bennemann, D.K. Morr, H. Dreyss\'e,
Phys. Rev. B {\bf 73}, 144405 (2006).
\bibitem{Lines64} M. E. Lines, Phys. Rev. {\bf 133}, A841 (1964).
\bibitem{ST91} H. Shimahara, S. Takada, J. Phys. Soc. Jpn. {\bf 60}, 2394
(1991); ibidem {\bf 61}, 989 (1992).
\bibitem{JIRK04} I.Junger, D. Ihle, J. Richter, A. Kl\"umper, Phys. Rev. B
{\bf 70}, 104419 (2004).
\bibitem{JIR05} I.J. Junger, D. Ihle, J. Richter, Phys. Rev. B {\bf 72},
064454 (2005).
\bibitem{SFBI00} C. Schindelin, H. Fehske, H. B\"uttner, D. Ihle, Phys. Rev.
B {\bf 62}, 12141 (2000).
\bibitem{ISF01} D. Ihle, C. Schindelin, H. Fehske, Phys. Rev. B {\bf 64},
054419 (2001).
\bibitem{KY72} J. Kondo, K. Yamaji, Progr. Theor. Phys. {\bf 47}, 807 (1972).
\bibitem{BZSY96} S.Q. Bao, H. Zhao, J.L. Shen, G.Z. Yang, Phys. Rev. B {\bf
53}, 735 (1996).
\bibitem{Bao97} S.Q. Bao, Solid State Communication {\bf 10}, 193 (1997).
\bibitem{BF64} J.C. Bonner, M. E. Fisher, Phys. Rev. {\bf 135}, A640 (1964).
\bibitem{GP01} M.E. Gouvea, A.S.T. Pires, Phys. Rev. B {\bf 63}, 134408 (2001).
\bibitem{YF00} W.Yu, S. Feng, Eur. Phys. J. B {\bf 13}, 265 (2000).
\bibitem{BCL02} B.H. Bernhard, B. Canals, C. Lacroix, Phys. Rev. B {\bf 66},
104424 (2002).
\bibitem{T70} Yu. A. Tserkovnikov, Teor. Mat. Fiz. {\bf 7}, 147 (1970); ibid.
{\bf 12}, 135 (1972).
\bibitem{Ku02} A.L. Kuzemsky, Rivista Nuovo Cimento {\bf 25}, 1 (2002).
\bibitem{BR94} A. Belkasri, J.L. Richard, Phys. Rev. B {\bf 50}, 12896 (1994).
\bibitem{Pla73} N.M. Plakida, Phys. Lett. {\bf 43A}, 481 (1973).
\bibitem{WI99} S. Winterfeldt, D. Ihle, Phys. Rev. B {\bf 59}, 6010 (1999).
\bibitem{BM95} A.F. Barabanov, L.A. Maksimov, Phys. Lett. A {\bf 207}, 390
(1995).
\bibitem{Colpa72} J.H.P. Colpa, Physica {\bf 57}, 347 (1972).
\bibitem{J97} P.J. Jensen, Ann. Physik {\bf 6}, 317 (1997).
\bibitem{BMW00} K. De'Bell, A.B. MacIsaac, J.P. Whitehead, Rev. Mod. Phys.
{\bf 72}, 225 (2000).
\bibitem{SLS00} Y. Song, H.Q. Lin, A.W. Sandvik, J. Phys. Condens. Matter {\bf
12}, 5285 (2000).
%
\end{thebibliography}
\end{document}